\documentclass[11pt,twoside,a4paper,openright]{book}

\usepackage{floatflt} 
\usepackage{pdfsync} 
\usepackage[bookmarks=true, pdfpagelabels=true, colorlinks=true, linkcolor=black, urlcolor=black, citecolor=blue, anchorcolor=blue]{hyperref}

\usepackage[cyr]{aeguill}
\usepackage{booktabs}
\usepackage[utf8]{inputenc}
\usepackage[T1]{fontenc}

\usepackage{amsmath,amssymb,bm}	  
\usepackage{graphics,xcolor}			  
\usepackage{times}
\usepackage{graphicx}
\usepackage{emptypage}
\usepackage{cite}
\usepackage[export]{adjustbox}

\usepackage{empheq}

\usepackage{lscape} 

\usepackage{afterpage} 

\usepackage{etoolbox}

\usepackage[font=small,labelfont=bf]{caption}

\usepackage{multirow}

\usepackage[a4paper,total={12cm,20cm},centering,includehead]{geometry}
\usepackage{fancyhdr}
\usepackage{sectsty}
\sectionfont{\fontsize{16}{18}\usefont{OT1}{phv}{m}{n}\selectfont}
\subsectionfont{\fontsize{13}{15}\usefont{OT1}{phv}{m}{n}\selectfont}
\subsubsectionfont{\fontsize{11}{13}\usefont{OT1}{phv}{m}{n}\selectfont}

\usepackage[Lenny]{fncychap}
\ChNameVar{\fontsize{14}{16}\usefont{OT1}{phv}{m}{n}\selectfont}
\ChNumVar{\fontsize{60}{62}\usefont{OT1}{phv}{m}{n}\selectfont}
\ChTitleVar{\fontsize{24}{26}\usefont{OT1}{phv}{m}{n}\selectfont}

\makeatletter
\renewcommand\ps@plain{\let\@mkboth\@gobbletwo
     \let\@oddhead\@empty
     \def\@oddfoot{\reset@font\hfil}
     \let\@evenhead\@empty\let\@evenfoot\@oddfoot}

\patchcmd{\@makechapterhead}{\vspace*{50\p@}}{}{}{}
\patchcmd{\@makeschapterhead}{\vspace*{50\p@}}{}{}{}

\g@addto@macro\normalsize{%
  \setlength\abovedisplayskip{10pt}
  \setlength\belowdisplayskip{10pt}
  \setlength\abovedisplayshortskip{10pt}
  \setlength\belowdisplayshortskip{10pt}
}

\makeatother

\usepackage{makeidx} 
\makeindex

\newcommand{\nocontentsline}[3]{}
\newcommand{\tocless}[2]{\bgroup\let\addcontentsline=\nocontentsline#1{#2}\egroup}

\newcommand{\boldmathsymbol}[1]{{\ensuremath{\boldsymbol{#1}}}}

\DeclareMathOperator{\arcsinh}{arcsinh}
\DeclareMathOperator{\arctanh}{arctanh}

\newcommand{\rr}{\mathrm}

\newcommand{\munu}{_{\mu\nu}}

\newcommand{\rhofl}{\rho_{\rm fl}}

\newcommand{\zero}{{_0}}

\newcommand{\ud}{\mathrm{d}}

\newcommand{\bmk}{\boldmathsymbol{k}}

\newcommand{\Mpc}{\mathrm{Mpc}}

\newcommand{\Mpl}{M_{\rm pl}}
\newcommand{\gi}[1]{{#1}_\mathrm{g.i.}}

\newcommand{\mpl}{m_{\rm pl}}
\newcommand{\Mp}{M_{\rm pl}}

\newcommand{\beq}{\begin{equation}}
\newcommand{\eeq}{\end{equation}}
\newcommand{\be}{\begin{equation}}
\newcommand{\ee}{\end{equation}}
\newcommand{\bea}{\begin{equation}\begin{aligned}}
\newcommand{\eea}{\end{aligned}\end{equation}}

\newlength{\wsingfig}
\newlength{\wdblefig}
\newlength{\wquadfig}
\newlength{\wtriplefig}
\newcommand{\Eq}[1]{Eq.~(\ref{#1})}
\newcommand{\Eqs}[1]{Eqs.~(\ref{#1})}
\newcommand{\Fig}[1]{Fig.~{\ref{#1}}}

\renewcommand{\Ref}[1]{Ref.~{\cite{#1}}}

\newcommand{\abs}[1]{|{#1}|}

\renewcommand{\ij}{{\ \!}_{ij}}  
\renewcommand{\IJ}{{\ \!}^{ij}}

\newcommand{\gm}{\gamma} 
\newcommand{\Gm}{\Gamma} 
\newcommand{\tgm}{{\tilde\gm}}
\newcommand{\tGm}{{\tilde\Gm}}
\newcommand{\tA}{{\tilde A}}

\newcommand{\rhozero}{\rho_\zero}

\begin{document}  

\setlength{\parindent}{0cm}
\setlength{\parskip}{1.0pt}
%
%
%

\begin{figure}[t!]
\centering
\includegraphics[scale=0.6]{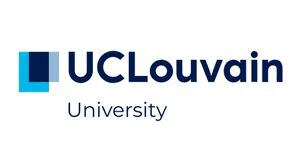}
\end{figure}

\vspace{1.5cm}
\begin{center}

\hspace*{-3.05cm} \parbox{1.5\textwidth}{\fontsize{22}{30}\selectfont\centering{Cosmic inhomogeneities in the early Universe: \\ A numerical relativity approach}}
\end{center}
\vspace{0.6cm}
\begin{center}
Doctoral dissertation presented by \\
\vspace{2mm}
{\Large Cristian Joana}\\
\vspace{2mm}
in fulfilment of the requirements for the degree of Doctor in Sciences.
\end{center}
\vspace{\fill}
\begin{center}
\begin{tabular*}{0.95\textwidth}{l @{\extracolsep{\fill}} r}
{Supervisors: } & \\[3.5pt]
{Prof. Christophe Ringeval} & UCLouvain, Belgium \\
{Prof. S\'ebastien Clesse}    & ULB, Belgium \\
\end{tabular*}

\vspace*{0.5cm}
\textsl{September, 2022}\\[1pt]
\end{center}
\thispagestyle{empty}


\hfill
\newenvironment{acknowledgements}%
    {\cleardoublepage\thispagestyle{empty}\null\vfill\begin{center}%
    \bfseries Acknowledgements\end{center}}%
    {\vfill\null}
\begin{acknowledgements}
I would like to express my gratitude to my superb supervisors, S\'ebastien Clesse and Christophe Ringeval. It has been an absolute pleasure to share the last four years in such a cheerful and exciting academic environment. I sincerely thank you for all your support and trust, counting all the enlightening discussions that have allowed me to progress.  I also warmly thank the members of my defence jury: Katy Clough, Vincent Vennin, Jean-Marc G\'erard and Vincent Lema\^itre for they comments that greatly improved the quality of the manuscript. 
\ \\

I want to thank my colleagues from GRChombo for welcoming me in and for all the outstanding work done by many, which we all share and benefit from. 
I particularly thank Eugene Lim and Katy Clough for fruitful discussions and support, as well as Thomas Helfer, Josu C. Aurrekoetxea, Miren Radia and Tiago Fran\c{c}a for priceless conversations, mutual help with the code and for cherishing a fun atmosphere during the GRChombo meetings.
\ \\

I also thank all the people that made my stay at UCLouvain a lovely experience: my collegues and friends at CURL, Disrael Camargo, Pierre Auclair and Baptiste Blachier, also to the \textit{frag} team for making my lunch breaks fun and chirpy, to the CP3 football group with whom I shared splendid afternoons, and also a special thanks to Carinne Mertens for her excellent job as a CP3/CURL secretary, whose work has immensely simplified and improved the daily lives of the institute's members.
\ \\

I should also appreciate the support of all my friends, with special affection to Jordi Morales, Christian Keup, Karthik Srinivasan and Juan-Antonio Julián. 
Lastly, and most importantly, all my gratitude to my family: to Khawla, with whom I am sharing this fantastic journey, and who has managed to keep me sane all this time; and to my parents and siblings, whose invaluable support and continuous encouragement has always kept me going.
\end{acknowledgements}


\hfill
\newenvironment{Abstract}%
    {\cleardoublepage\thispagestyle{empty}\null\vfill\begin{center}%
    \bfseries Abstract\end{center}}%
    {\vfill\null}
        \begin{Abstract}

%

Cosmic inflation is arguably the most favoured paradigm of the very early Universe. It postulates an early phase of fast, nearly exponential, and accelerated expansion. Inflationary models are capable of explaining the overall flatness and homogeneity of today's Universe at large scales. Despite being widely accepted by the physics community, these models are not absent from criticism. In scalar field inflation, a necessary condition to begin inflation is the requirement of a Universe dominated by the field's potential, which implies a subdominant contribution from the scalar field dynamics. This has originated to large amounts of scientific debate and literature on the naturalness, and possible fine-tuning of the initial conditions for inflation. Another controversial issue concerns the end of inflation, and the fact that a preheating mechanism is necessary to originate the hot big bang plasma after inflation.
\ \\

In this thesis, we present full general relativistic simulations to study these two problems, with a particular focus on the Starobinsky and Higgs models of inflation, being those the most favoured by the latest observations.  First, we consider 
the fine-tuning problem of beginning inflation from a highly dynamical and inhomogeneous "preinflation" epoch in the single-field case. In our second study, we approach the multifield paradigm of inhomogeneous preinflation,  together and consistently,  with the preheating phase.  These investigations further confirm the robustness of these types of models to highly inhomogeneous initial conditions, while putting in evidence 
the non-negligible gravitational effects during preheating.  At the end of the manuscript,  we discuss some of other potential applications of numerical simulations to study the early Universe, including our preliminary investigations on primordial black hole formation in asymmetric three-dimensional configurations.


        \end{Abstract}

\clearpage

\begin{center}

\vspace{121pt}

{\LARGE Associated Publications:}
\vspace{22pt}
\begin{description}
\item[\cite{Joana2020}] 
Joana, C., Clesse, S. (2021) "Inhomogeneous pre-inflation accross Hubble scales in full general relativity", Phys. Rev. D, vol. 103, pp. 083501 (2021). arXiv:2011.12190
\item[\cite{Joana:2022uwc}]
Joana, C. (2022) "Gravitational dynamics of Higgs inflation: Preinflation and preheating with an auxiliary field", Phys. Rev. D, vol. 106, pp. 023504 (2022). arXiv:2202.07604
\end{description}

\vspace{40pt}

{\LARGE Other Publications:}
\vspace{22pt}
\begin{description}
\item[\cite{Andrade2021}]
Andrade, T., Joana C. \textit{et. al.}, (2021) "GRChombo: An adaptable numerical relativity code for fundamental physics", Journal of Open Source Software, 6(68), 3703, https://doi.org/10.21105/joss.03703.
\item[\cite{LISACosmologyWorkingGroup:2022jok}] 
Auclair, P., Joana C. \textit{et. al.}, 
LISA Collaboration (2022), "Cosmology with the Laser Interferometer Space Antenna", arXiv:2204.05434  
\end{description}

\vfill
\begin{tabular}{llr} 
                    &               &                           \\
 \multicolumn{3}{c}{\large Thesis Jury}                                  \\
                    &               &                           \\
\toprule 
Prof. Vincent Lema\^itre       &  President             & Université Catholique de Louvain  \\
Prof. Christophe Ringeval    &  Supervisor            & Université Catholique de Louvain  \\
Prof. S\'ebastien Clesse       &  Supervisor            & Université Libre de Bruxelles   \\
Prof. Jean-Marc G\'erard       &  Secretary             & Université Catholique de Louvain  \\

\multirow{2}{*}{Prof. Vincent Vennin  }
       &   \multirow{2}{*}{Member}                 & Laboratoire de Physique de \ \ \ \\
                                            &      & l’Ecole Normale Supérieure \\
Dr. Katherine Clough       &  Member                & Queen Mary University of London   \\
\bottomrule
\end{tabular}

\vspace{44pt}

\end{center}

\pagestyle{empty}
\tableofcontents

\setlength{\parskip}{8.0pt}  

%
%
%
\chapter{Introduction}

\pagestyle{fancy}
\setcounter{page}{1}

The search for the understanding  of the origin of humanity and our place in the Universe has been  a relevant quest throughout the history of humankind. Starting in the ancient Greek civilization, Anaximander (600 BC) already conceived the Earth as a spherical object where the celestial bodies revolved around it,  a model further developed by his Greek successors giving birth to what today is known as the Ptolemy's geocentric model (200 BC).  In the western civilization, it was not until much later, in the XVII century with the invention of the telescope, that Galileo Galilei and Johannes Kepler gathered strong evidence supporting the heliocentric model revived\footnote{
The first known heliocentric model is by Aristarchus of Samos (300 BC). However, because stellar parallax is only detectable using telescopes, this model was disregarded in favour of the geocentric model by Plato, Ptolemy and their contemporaries throughout the Middle Ages. 
}
by Nicolaus Copernicus. The new point of view situated the Sun at the centre of which the celestial bodies revolved around. Most importantly, the new perspective would motivate the search for natural knowledge by empirical experimentation, which would constitute the basis of the scientific method.  

At the end of the century, in 1687, Isaac Newton published ``the Principia'' \cite{nla.cat-vn2516658} containing the celebrated three Newton's laws of motion of Classical Mechanics and his theory of universal gravitation. Among other achievements, the new mathematical framework was able to satisfactory predict the motions of most of the objects in the Solar system, from the orbital trajectory of the  moons of Jupiter, to the orbit and observed phases of Venus, with remarkable accuracy by the time. Not surprisingly, Newton is commonly considered the ``father of physics", and after him other giants like Faraday, Maxwell or later Kelvin and Boltzmann made outstanding progress in physical sciences by developing the fields of electromagnetism, thermodynamics, optics, etc. The period between the XVII and XIX centuries is known as the ``Age of Enlightenment'' driven by the scientific and technological advances of the time.

Returning to the motion of celestial bodies, the precise calculation of the orbits of all celestial  objects, including the perihelion of Mercury, was only understood in the modern era when Albert Einstein developed his theory of General Relativity in 1916 \cite{Einstein:1915by}. This new paradigm unified the concepts of Space and Time, and most importantly, it removed the concept of an absolute coordinate reference system. 
The XX century was the epoch that gave birth to modern physics with the development of the two most revolutionary breakthroughs of our history: Einstein's theory of General Relativity and Quantum Mechanics \cite{QMPlanck,QMEinstein,PhysRev.28.1049}.
In the early years, Einstein's theory was still conceived in the mental construct of the ``static Universe", which pushed Einstein to incorporate a repulsive cosmological constant to counteract the gravitational pull of distant bodies like galaxies.\footnote{At the time, visible luminous objects in the sky were though to be stars or ``nebulas'' inside our galaxy, the Milky Way. In 1923, Edwin Hubble showed that some of these bodies were, in fact, other galaxies like our own, but significantly further apart. This discovery significantly expanded our notion and conception of the vastness of the Universe.}
A decade later, Georges Lema\^{i}tre (1927) \cite{10.1093/mnras/91.5.483} and Edwin Hubble (1929) \cite{doi:10.1073/pnas.15.3.168} gathered observational evidence that revealed the correlation between the receding velocities $v$ of far-away galaxies  with respect to their measured distance $d$. This empirical observation is known today as the Hubble-Lema\^{i}tre law,  $  v = H_0 d $,  where $H_0$ is the Hubble parameter and is interpreted nowadays as the Universe's expansion rate. This paradigm-shifting discovery revealed the dynamical nature of the Universe, where the kinematics of the cosmos are successfully described  by the Einstein field equations of General Relativity. %
The discovery by Lema\^{i}tre and Hubble gave birth to the Hot Big Bang (HBB) as a cosmological model of the Universe. It inspired the concept of the beginning of the Universe, set at the time of the primordial singularity when, hypothetically, the whole Universe was smaller than a single atom. The Big Bang model soon gained wide acceptance in the scientific community, once it was detected its main prediction: the Cosmic Microwave Background (CMB), a relic bath of photons free-streaming since back the early Universe. The new paradigm disregarded the need for the cosmological constant, and Einstein himself qualified the term as his ``biggest blunder''. Amusingly, a few decades later, the term would resurrect when, in 1998, astrophysical observations of distant supernovas discovered that the expansion of the Universe was being accelerated. Indeed, the cosmological constant is no longer invoked as a mechanism to prevent the gravitational collapse of our Universe, but to explain its current accelerated expansion.  The accelerated expansion of the Universe has been corroborated nowadays by modern CMB experiments. 

\section{The Cosmic Microwave Background}

In 1964, the failed attempt to isolate the source of a surplus excess of microwave radiation affecting the Bell Labs' astronomical equipment, turned into an accidental discovery that revolutionized our understanding of the Universe. The noise puzzled  Penzias and Wilson and, after exhausting every possible explanation, they realized the significance of this observation: the noise was of cosmological origin, sourced by the relic, oldest light in the history of the Universe. Indeed, this light was emitted $400.000$ years after the Big Bang, once the hot dense plasma cooled enough to let the light free-stream from the matter content. 
The two radio astronomers won the 1978 Nobel Prize in physics for this discovery. 

\begin{figure}[htbp!]
\vspace{2.5cm}
\includegraphics[width=0.990\textwidth]{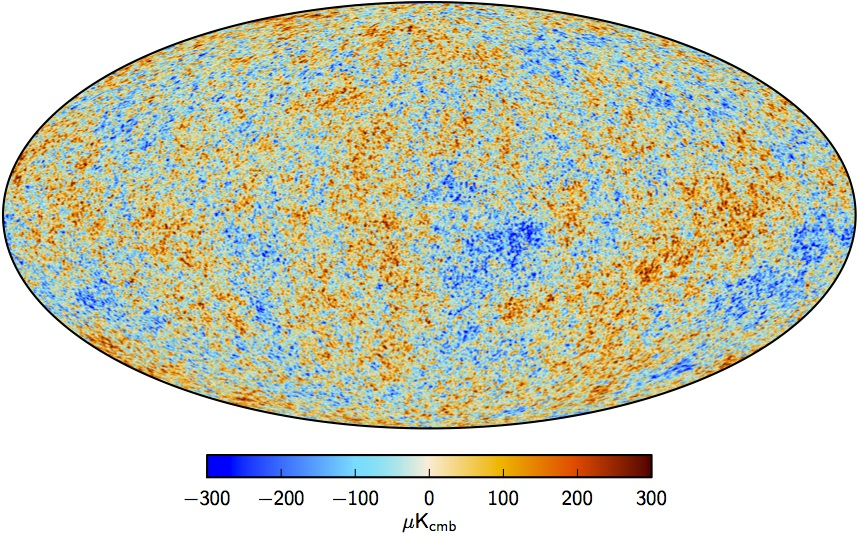}
\caption{Celestial map of the CMB temperature anisotropy measured by the Planck experiment \cite{Akrami:2018odb,Ade:2015lrj}.}
\label{fig:CMB_map}
\vspace{1.5cm}
\end{figure}

Despite the detection being in the midst of the XX century, the precise CMB measurements capable to minutiously capture the temperature anisotropies were not possible until the entry to the new century. Thanks to the technological progress, the launch of satellite experiments such as the WMAP (in 2001-2010) and Planck (in 2009-2013) have provided unprecedented data about the early Universe imprinted in the characteristics of the CMB map (see \Fig{fig:CMB_map}), a picture of the Hot Big Bang plasma from more than 13 billions years ago. 

\begin{figure}[htbp!]
\includegraphics[width=0.50\textwidth]{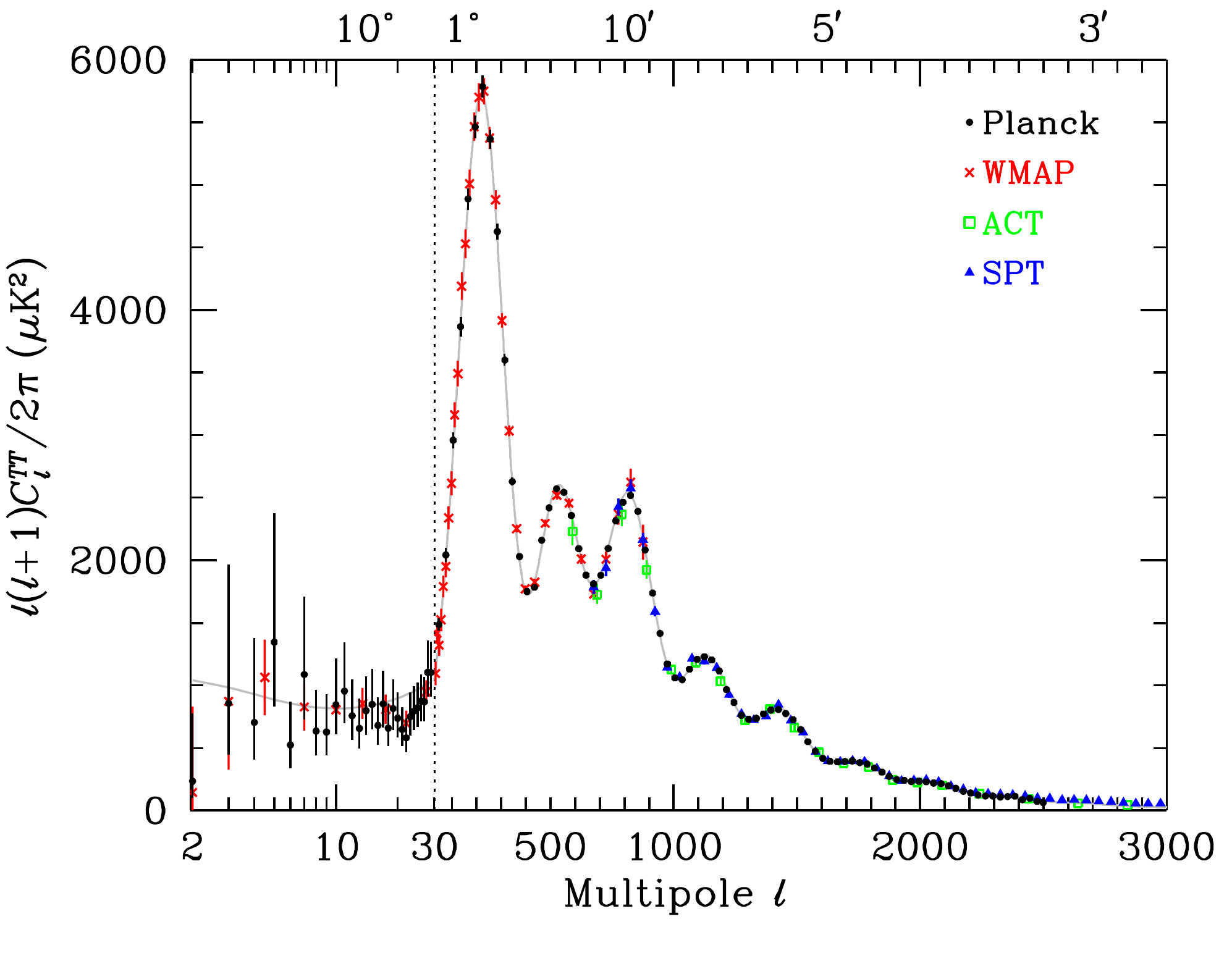}
\includegraphics[width=0.50\textwidth]{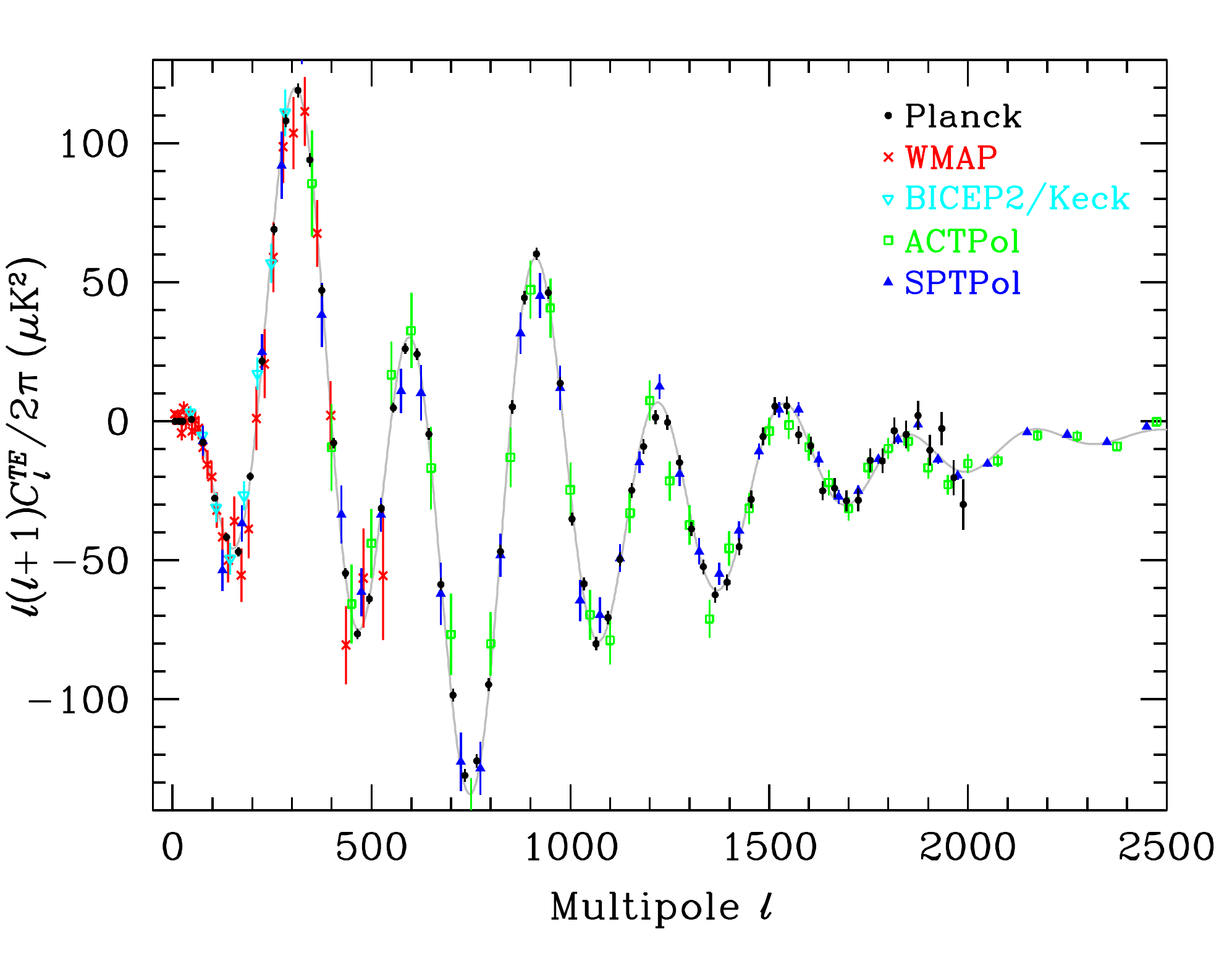}
\includegraphics[width=0.50\textwidth]{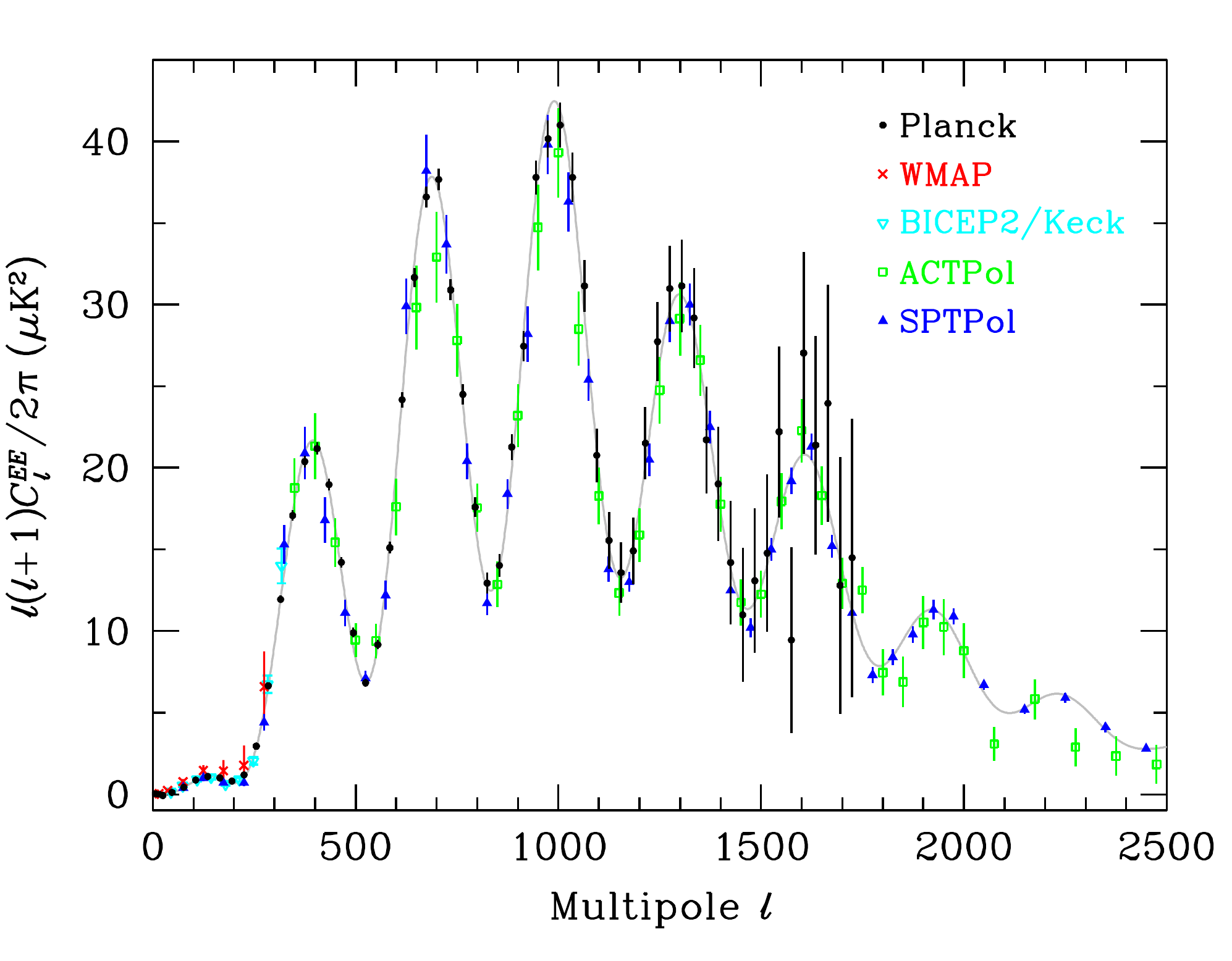}\
\includegraphics[width=0.50\textwidth]{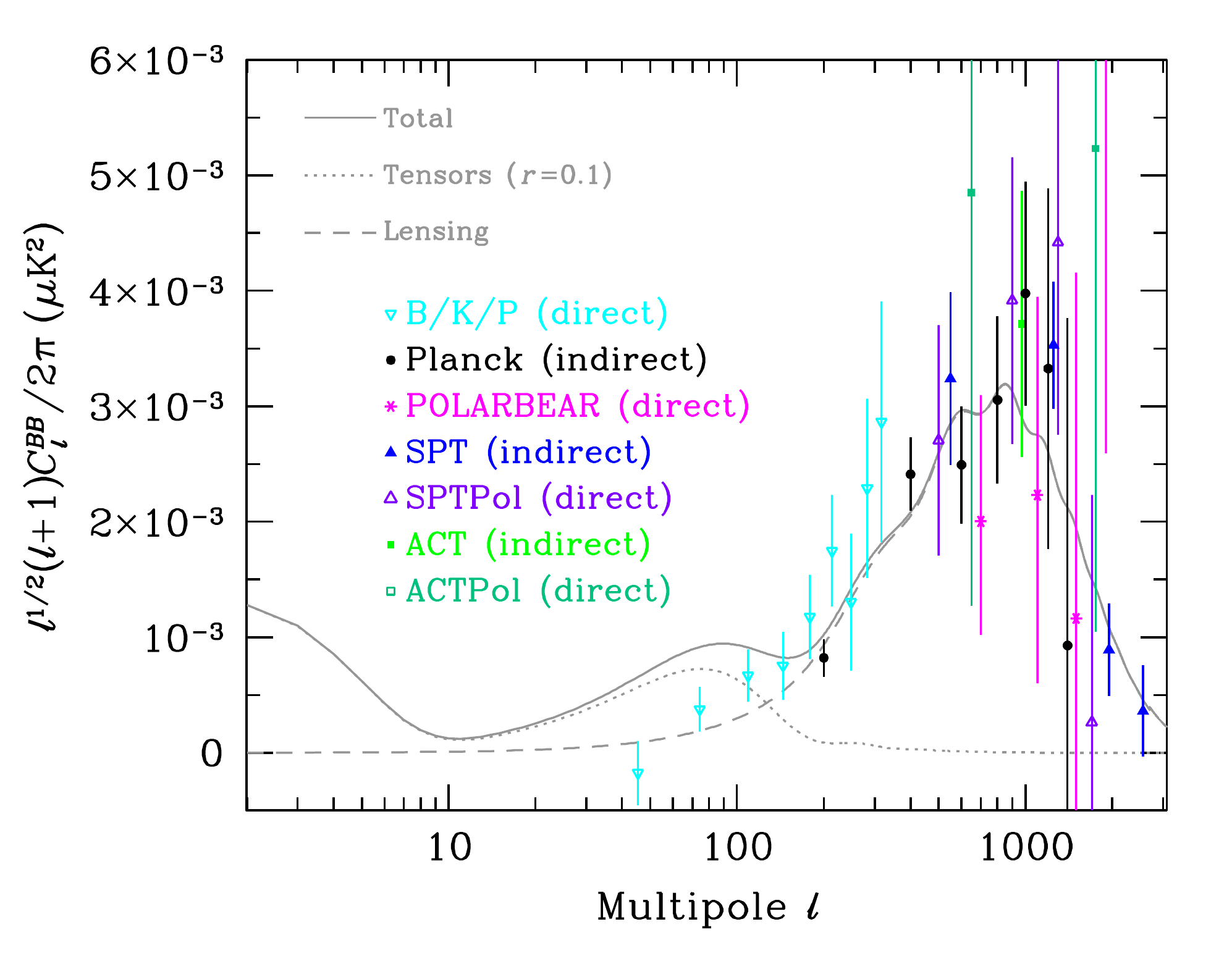}
\caption{CMB temperature anisotropy power-spectrum (top left) and
temperature-polarisation cross-power spectrum (top right), from {\it Planck},
{\it WMAP}, BICEP/Keck, ACT and SPT.  Polarisation $EE$ (bottom left) and $BB$ (bottom right)
power-spectra from several experiments.  The $BB$ spectrum here is scaled by a power of $\ell$ that makes
it possible to see all three of the expected peaks (from reionisation, recombination and lensing). In all figures, the gray line represents the best-fitting model of the $TT$ power-spectrum, which also shows excellent agreement in the predictions of the other measures. This illustrates the high-level precision with which these power spectra have been measured. Figures taken from  ~\cite{PhysRevD.98.030001}. }
\label{fig:EEandBB}
\end{figure}

One important source of information in the CMB comes from the statistics, i.e. the computed power-spectrum, of the temperature anisotropies. The shape of the primordial power-spectrum in terms of the angular size 
is found to be nearly scale-invariant  with a light red-tilt.  The spectra contain a series of peaks, with a predominant one at around $1^\circ$ angle (see \Fig{fig:EEandBB}). These peaks correspond to sound/pressure waves at the time of the CMB. These sound waves produced peaks and valleys in the primordial plasma related to the sound horizon at the time. The predominant peak corresponds to overdensities in the fluid that precisely peaked first at the time when the CMB was released. The following peaks (at higher degree angle) correspond to peaking overdensities that also peaked at even earlier times, in a recurrent harmonic mode. In fact, the time odd peaks (the first, the third, etc) corresponds to hot overdensities in the CMB map, while even peaks (the 2nd, the 4th...) correspond to the underdense regions.  Furthermore, the positions of the peaks, their widths and the ratio between them give us a lot of information related to the evolution of the sound horizon, the ratio of self-interacting (baryonic) matter and  the inert (dark) matter quantities, and even the background geometry of the Universe. Indeed, the best-fitting model that matches the oscillations of the temperature power spectrum tells us that, today, the Universe is only $5\%$ made of ordinary matter and $26.5\%$ of undetected dark matter. The resultant $68\%$ is of a mysterious dark energy, that behaves like a cosmological constant, explaining the current accelerated expansion of the Universe. The nature of both dark matter and dark energy are a subject of intense investigation and debate in the physics community of nowadays. 

A significant remark on the CMB temperature map is it is large degree of homogeneity, where anisotropies are only of $\delta T \sim 10^{-5} $ K. This is paradoxical because when one computes the time of which the CMB was emitted, it turns out that the Universe at that time was constituted by $10^6$ non-causally connected regions, and therefore, it is hard to explain why all of these regions have nearly the same temperature. This paradox is solved by the theory of inflation, which proposes a very early phase of a rapid and accelerated expansion of the Universe, which allows all the regions to be in causal contact by the time of the CMB. While this theory can be undeniably labelled as ``wild'', it has also been very successful in explaining many observed features of the Universe, including the statistical properties of the CMB  anisotropies.

The other relevant set of data coming from the CMB is in the polarization of light. There exist two possible polarization modes known as E (vertical\, /\, horizontal) and B (diagonal) modes. The case of E modes is coupled to scalar perturbations of the CMB, while the B modes contain information on the tensor part (i.e. gravitational waves). Searches of B modes in the CMB polarization map are a hot topic of research, as the existence of primordial tensor modes are one of the ``smoking gun'' predictions of the theory of inflation. Indeed, a clear detection of B modes in the CMB map would be seen, at least by many, as a confirmation test of the theory.

\vspace{0.5cm}

\clearpage

\section{Current and future experiments }

After the great success of the previous CMB experiment, the challenging endeavour of exploring the cosmos has just begun.  Many current and near-future experiments are set to map the large-scale structure (LSS) of the Universe with unprecedented precision. Like the photons in the CMB, the matter forming LSS is formed by the gravitational clustering of the primordial perturbations seed that originated in the early Universe. Thus, experiments such as the Dark Energy Survey~\cite{DES:2016jjg}, Euclid~\cite{EuclidTheoryWorkingGroup:2012gxx}, SKA~\cite{Maartens:2015mra}, among others, are set on providing novel insights into both the early and late cosmology. 

Another remarkable discovery has been the detection of gravitational waves (GW) by the LIGO/Virgo collaboration in 2015~\cite{LIGOScientific:2016aoc}. This discovery not only has proved once more the validity of Einstein's General Relativity (which predicted them), but also confirmed the existence of faraway extremely massive compact objects, orbiting and coalescing each other, thought to be black holes and neutron stars. The ability to measure gravitational radiation has opened a new window to explore the Universe, using a messenger (i.e. the GW) that practically free-streams undisturbed by the medium. This has enormous potential in cosmology, as it opens the door to exploring the pre-CMB early Universe in a transparent manner. In the incoming years, many gravitational wave experiments are scheduled to become operational (e.g. the Laser Interferometer Space Antenna~\cite{LISA:2017pwj}, the Einstein Telescope~\cite{Punturo:2010zz}, TianQin~\cite{TianQin:2015yph}, and the Pulsar-Timing-Array network~\cite{Manchester:2013ndt,McLaughlin:2013ira,Kramer:2013kea,Manchester:2012za}, among others), set to cover the exploration of several frequency ranges in the GW spectra. 

An exciting prospect is the potential detection of sub-Solar mass black holes. The existence of these black holes can not be explained by conventional stellar evolution process, and thus strongly favouring the existence of primordial black holes (PBH). These type of black holes would be formed during the early Universe, much earlier than the formation of stars, and, importantly, they could constitute a large part, if not all, of the dark matter in the Universe \cite{Ivanov:1994pa,Carr:2016drx}.

\clearpage

This thesis is written in the context of paving the path to better understand the physics of the early Universe, as well as providing the bases for new analytical and numerical methods to study cosmological processes beyond the 
traditional assumptions of homogeneity and isotropy.  %
The structure of this thesis is the following:  Chapter~\ref{chap:LambdaCDM} is dedicated to introducing the standard cosmological model, which successfully explains the evolution of the homogeneous Universe as a whole, at very large scales. In Chapter~\ref{chap:inflation} we introduce the inflationary paradigm and the perturbative approach to study the origin of the small cosmological inhomogeneities present in the CMB, and, in Chapter~\ref{chap:NR}  we treat the inhomogeneous universe using the 3+1 formalism of numerical general relativity. Chapters \ref{p1_chap:prepaper1} and \ref{p2_chap:prepaper2} contain the original published work of the thesis, where we tackle theoretical issues such as the initial conditions problem in the context of inflation. Chapter~\ref{ch:addwork} describes additional work which is not-yet-published and whose investigation will be continued after the defense of this thesis. To finalize, conclusions and additional future prospects are summarized in Chapter~\ref{chap:conclusions}.

\clearpage
\section*{Notation and conventions}
\addcontentsline{toc}{section}{Notation and conventions}

In this thesis, we adopt the Planck units in which $G = c = \hbar = 1$ where $G$, $c$ and $\hbar$ are the Newton’s constant, the speed of light and the reduced Planck's constant, respectively. The spacetime metric is chosen with the ``mostly-plus'' signature ${( -+++)}$ and we follow the sign conventions of the textbook of Wald~(1984). 
The Greek indices $\{\alpha, \beta,  \dots, \mu, \nu,\dots \}$ (that run from 0 to 3) are used to denote the spacetime components in tensors, while we use lower-case Latin indices  $\{i, j, k, \dots \}$ (that run from 1 to 3) to denote the spatial components in tensors. Upper-case Latin indices are used to denote an integer number of scalar fields in multifield inflationary scenarios. The Einstein’s convention is adopted throughout the thesis, thereby the summation over repeated indices is assumed. Moreover, $\nabla_\mu$ denotes the 4-dimensional covariant derivative associated with the 4-metric $g_{\mu\nu}$, $D_i$ denotes the 3-dimensional covariant derivative associated with the induced 3-metric $\gamma\ij$ representing the spatial hypersurfaces within the 3+1 decomposition, and $\partial_\mu$ denotes the partial derivative. In addition, square brackets in the subscripts denote that the symmetric relation is implied, e.g.  


\begin{equation*}
 \Sigma_{(ij)} = \frac{1}2 \left( \Sigma\ij + \Sigma_{ji} \right) ~. 
\end{equation*}

A summary of the widely used notation and abbreviations follows:
\begin{align*}
g\munu & : & \text{ spacetime metric} \\
\gamma\munu & : & \text{ spatial metric} \\
\tilde\gamma\munu & : & \text{ conformal spatial metric} 
\end{align*}

\begin{align*}
g & : & \text{ determinant of the spacetime metric } g\munu\\
\gamma & : & \text{ determinant of the spatial metric } \gamma\munu\\
\tilde\gamma & : & \text{ determinant of the conformal spatial metric } \tilde\gamma\munu \\
\nabla_\mu & : & \text{ spacetime covariant derivative associated with }g\munu \\
D_\mu & : & \text{ spatial covariant derivative associated with }\gamma\munu \\
\tilde D_\mu & : & \text{ spatial covariant derivative associated with }\tilde\gamma\munu \\
\eta\munu & : &   \text{flat spacetime or Minkowsky metric} \\
^{(4)} R_{\alpha\beta\gamma\delta}  & : &   \text{four-dimentional spacetime Riemann tensor associated with }g\munu\\
R_{\alpha\beta\gamma\delta}  & : &   \text{spatial Riemann tensor associated with }\gamma\munu \\
^{(4)} R_{\mu\nu}  & : &   \text{four-dimentional spacetime Ricci tensor associated with }g\munu \\
R_{\mu\nu}  & : &   \text{spatial Ricci tensor associated with }\gamma\munu \\
\tilde R_{\mu\nu}  & : &   \text{conformal spatial Ricci tensor associated with }\tilde\gamma\munu \\
K_{\mu\nu}  & : &   \text{extrinsic curvature tensor on spatial hypersurfaces } \\
T\munu & : &   \text{spacetime energy-momentum tensor } \\
H_0  & : & \text{Hubble parameter measured as today} \\
m_{\rm pl}  & : &   \text{Planck mass, }  m_{\rm pl} \equiv \sqrt{\frac{\hbar c}{G}} \\
\Mpl  & : &   \text{Reduced Planck mass, }  \Mpl \equiv \frac{m_{\rm pl}}{\sqrt{8\pi}} \\
\text{ADM} & : & \text{Arnowitt-Deser-Misner} \\
\text{BSSN} & : & \text{Baumgarte-Shapiro-Shibata-Nakamura} \\
\text{CMB} & : & \text{Cosmic Microwave Background} \\
\text{CPT} & : & \text{Cosmological Perturbation Theory} \\
\text{HBB} & : & \text{Hot Big Bang} \\
\text{FLRW} & : & \text{Friedmann-Lema\^{i}tre-Robertson-Walker} \\
\text{PBH} & : & \text{Primordial Black Hole} \\
\end{align*}

%
%
%

\chapter{The  standard model of Cosmology}
\label{chap:LambdaCDM}
\pagestyle{fancy}

In this section, we examine the global time evolution of our Universe. At scales larger than a Megaparsec, the spacetime dynamics are well described by an expanding  Universe which is largely homogeneous and isotropic. Using Einstein's theory of General Relativity, such spacetimes are described by the Friedmann-Lema\^{i}tre solutions of the Einstein equations, and they are the common background used to understand the cosmological evolution of our Universe, from the HBB plasma until today. We also present the original problems of the HBB models, and how inflation provides an elegant paradigm that naturally solves these problems. %

\section{The Einstein equations} \label{sec:EinsteinEquations}

Let us first start by discussing the evolution equations provided by the Einstein's theory of gravitation, also known as General Relativity. The theory can be constructed by invoking the least action principle, in which the gravitational part corresponds to the geometrical curvature of the spacetime metric tensor. Formally, the Einstein-Hilbert action reads 

\beq\label{S_HE}
S = \int \ud^4x \sqrt{-g}  \left( \frac {\Mpl^2}2  R  + \mathcal{L}_\mathrm{matter} \right) ~,
\eeq
where $\mathcal{L}_\mathrm{matter}$ is the Lagrangian of matter content in the universe and  $g$ is the determinant of the metric $g_{\mu\nu}$  and  $\Mpl \equiv {m_{\rm pl}}/{\sqrt{8\pi}}$ is the reduced Planck mass.  
The Ricci scalar curvature is defined as the contraction of the Ricci tensor, i.e.  $R \equiv g_{\mu\nu} \,^{(4)}R^{\mu\nu}$, which is given by 
\beq
\,^{(4)}R_{\mu\nu}\equiv \partial_\rho \Gamma^\rho_{\mu\nu} -  \partial_\nu \Gamma^\rho_{\mu\rho} + \Gamma^\rho_{\mu\nu}
\Gamma^\lambda_{\rho\lambda} -  \Gamma^\rho_{\mu\lambda}\Gamma^\lambda_{\nu\rho}
~,
\eeq
where  the Christoffel symbols are the affine connections with respect to the metric tensor, defined by 
\beq
\Gamma^\rho_{\mu\nu}=\frac12~g^{\rho\lambda}~\left(\partial_\nu g_{\lambda\mu} + \partial_\mu g_{\nu\lambda} - 
\partial_\lambda g_{\mu\nu} \right)
~.
\eeq

The equations of motion are derived by varying the action with respect to the metric, which, after subtracting the surface terms, it gives rise to the Einstein equations
\beq\label{EinsteinEq}
\frac{\delta S}{\delta g_{\mu\nu}} = 0 \quad \rightarrow \quad 
{G}_{\mu\nu}  \equiv \,^{(4)}R_{\mu\nu} - \frac 12 g_{\mu\nu} R + \Lambda g_{\mu\nu} 
= \frac 1 {\Mpl^2} T_{\mu\nu},   
\eeq
where 
\beq
\frac{2}{\Mpl^2\sqrt{-g}} \frac{\partial}{\partial g_{\mu\nu}} { \left(\sqrt{-g} R\right)} = \,^{(4)}R_{\mu\nu} - \frac{1}{2}R\,g_\mathrm{\mathrm{\mu\nu}} + \Lambda g_{\mathrm{\mu\nu}} 
~,
\eeq
and $\Lambda$ is the cosmological constant. In the matter side, the energy-momentum tensor is defined by  
\beq
T_{\mu\nu} \equiv -\frac{2}{\sqrt{-g}}\frac{\partial} {\partial g_{\mathrm{\mu\nu}}}  {(\sqrt{-g} \mathcal{L}_\mathrm{matter})} = g_{\mathrm{\mu\nu}}\mathcal{L}_\mathrm{matter} - 2 \frac{\delta \mathcal{L}_\mathrm{matter}}{\delta g^{\mathrm{\mu\nu}}} 
~.
\eeq

In this theory, the metric tensor is then the fundamental gravitational quantity that defines the geometry of the spacetime. The matter content is encoded in the energy-momentum tensor given described by the Lagrangian of arbitrary matter species. Our cosmological models are typically constructed by taking either scalar fields, or generic types of fluids, as we shall see below.

\section{The homogeneous Universe}

The solution of the Einstein equations for an expanding \textbf{homogeneous} and \textbf{isotropic} Universe is known as the Friedmann-Lema\^{i}tre-\,Robertson-Walker (FLRW) metric. This is constructed by defining the cosmological scale factor $a(t)$ that encodes the variation of measured spatial distances between two instances of the cosmic time $t$ and $t_{\rm ini}$, 
\beq
d(t) = d_{\rm ini} \frac{a(t)}{a_{\rm ini}}~.
\eeq

The same reasoning also applies to photons, where the expansion causes a redshift $z$ in the photon's wavelength $\lambda$ 
\beq
z(t) \equiv  \frac{ \lambda(t) - \lambda_{\rm ini}}{ \lambda_{\rm ini}}  = \frac{a(t)}{a_{\rm ini}} - 1 ~. 
\eeq
As it is illustrated in the above equations, the scale factor at a given time is an arbitrary definition on the scale. However, the variation of this quantity along two different times has a clear physical meaning as it indicates the amount of expansion that the Universe has undergone. 

\begin{figure}[htbp!]
\includegraphics[width=0.990\textwidth]{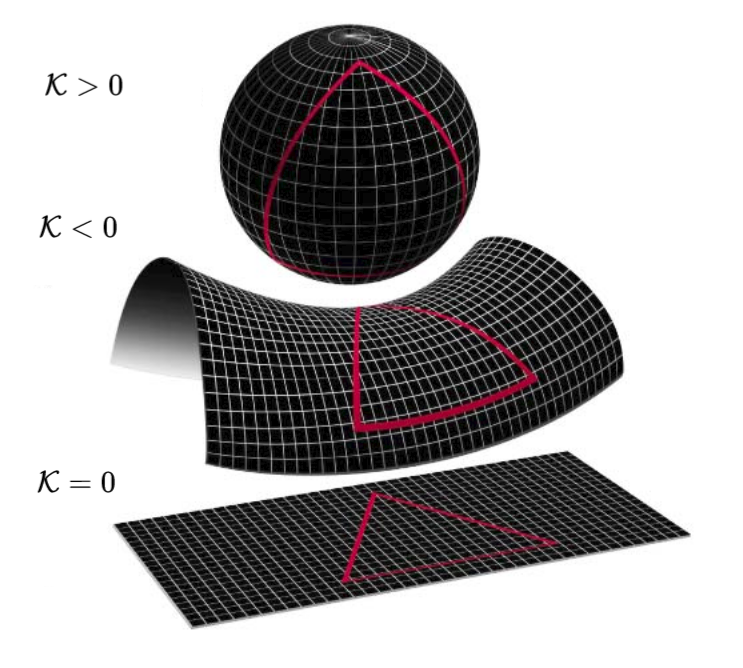}
\caption{Diagrams of three possible geometries of the universe: positively curved, negatively curved and flat from top to bottom, corresponding to a curvature parameter $\Omega_{\mathcal{K}}$ being greater than, less than or equal to 0.
Figure adapted from \cite{wikimedia_commons}. 
}
\label{fig:curvature_universe}
\end{figure}

The FLRW metric in generic coordinates reads 
\beq \label{FRLW metric}
 \ud s^2\equiv g_{\mu\nu}\ud x^\mathrm{\mu}\ud x^\mathrm{\nu} = -\ud t^2+a^{2}(t)
\ud \vec\Sigma^2
~,
 \eeq
where $\vec\Sigma$ represents the spatial metric hypersurface. In spherical coordinates, this is defined by 
\beq
\ud \vec\Sigma^2 =  \frac{\ud r^2}{1- \mathcal{K} r^2}+r^2\left(\ud \theta^2+\sin^2\theta\ud \phi^2\right)
~,
\eeq
with $\mathcal{K}$ being the spatial curvature of the universe normalised at a given time when $a(t_*) = 1$. In other words, its values can contain either $\mathcal{K} = 0$ for the flat configuration, or either $\mathcal{K} = 1$ and $\mathcal{K} = -1$ for the positively or negatively curved spacetime (see Fig.~\ref{fig:curvature_universe}). For the non-flat cases, the scale factor can then be reinterpreted as the hyperbolic radius of the global curvature of the universe. 

On some occasions, it is useful to define the metric in terms of the conformal time $\eta$ instead of the cosmic time $t$  by applying the transformation $ {a(\eta) \ud\eta = \ud t}$, thus the metric reads 
\beq 
 \ud s^2  =  a^{2}(\eta) \left( - \ud \eta^2+  \ud \vec\Sigma^2 \right) ~.
\eeq

We will follow the convention that ``dotted" variables represent derivatives in cosmic time, while the ''prime`` indicates derivatives with respect to the conformal time. For example, the Hubble parameter is given by
\beq  \label{eq:def_H}
\begin{split} 
 H  &=  \frac 1a \frac{\ud a}{\ud t} = \frac{\dot a} a \\
&=  \frac 1{a^2} \frac{\ud a}{\ud \eta} = \frac{ a'} {a^2}  ~,
\end{split} 
\eeq
and the conformal Hubble parameter reads 
\be
\mathcal{H}  =  \frac 1a \frac{\ud a}{\ud\eta} = \frac{a^\prime}{a} = a H
~.
\ee

\section{Spacetime dynamics}

The only energy-momentum tensor compatible with the symmetry (i.e homogeneity and isotropy) is  the one of the perfect fluid, 
\beq
T_{\mu\nu} =  (\rho + p)u_{\mu} u_{\nu}  +  g_{\mu\nu}p
~,
\eeq
where $\rho$  and $p$ are the energy density and pressure density of the fluid, and $u_\mu$ its four-velocity with respect to the comoving frame.

We are now in a position where we can obtain the equations of motion by applying the Einstein equations to the system we just described above. On one hand, the time-time component ($\mathcal{G}_{00}$) results into the 1st Friedmann-Lema\^{i}tre equation 
\beq \label{FriedmannEq}
H^2=\left( \frac{\dot{a}}{a}\right)^2=\frac{\rho}{3\Mp^2}-\frac{\mathcal{K}}{a^2} +\frac{\Lambda}{3} 
~.
\eeq
On the other hand, the spatial components of the Einstein tensor ($\mathcal{G}_{ii}$) yield to the acceleration equation, the 2nd Friedmann-Lema\^{i}tre equation  
\beq \label{AccelerationEq}
\frac{\ddot{a}}{a}=-\frac{1}{6\Mp^2} \left(\rho+3p\right) +\frac{\Lambda}{3}
~.
\eeq

Combining the above equations, by taking a time derivative in eq. (\ref{FriedmannEq}), one can derive the continuity equation,  
\beq \label{ContinuityEq}
\nabla^\mu  T_{\mu\nu}  =  \dot{\rho} + 3H(\rho+p) = 0    ~, 
\eeq
which encodes the conservation of the energy-momentum tensor. 
For a barotropic equation of state, the pressure is simply given by $p = \omega \rho$, where $\omega$ is the fluid's equation of the state. Thus, the equation of motion of the fluid is given by 
\beq \label{Eq::Rho_of_a_inPF}
\rho (a) = \rho_{\rm ini} \left(\frac{ a }{a_{\rm ini}}\right)^{-3(1+\omega)} ~, \qquad \text{with } \ \omega = \text{constant.}
\eeq
In turn, one can also find the expansion history of the universe by solving Eq.~(\ref{FriedmannEq}), which leads to 
\beq \label{expansion_perfectfluid}
a(t) = \left\{
\begin{split}
& ~ a_{\rm ini} \left(t \over t_{\rm ini} \right) ^{2 \over 3(\omega +1)}~, \qquad \,\, \omega \neq -1 \\
& ~ a_{\rm ini}  \exp\left[ {H} \left(t-t_{\rm ini}\right) \right], \quad \omega = -1, \quad  { H} = \text{constant.} 
\end{split} \right.
\eeq

The above equations establish a relationship between the expansion history of the universe and the equation of state of the fluid that dominates the energy budget. Using some physical intuition, we can correctly estimate the equation of state of the relevant types of matter in the cosmological constant: The cases for relativistic and non-relativistic matter particles. Assuming a fluid in a given volume that scales through the expansion like $V\propto a^3$, the case for ultra-relativistic species (e.g. radiation of wavelength $\lambda$), we find that it scales like 
\beq \label{eq:energyradiation}
\rho_{\rm r} \sim \frac 1V \frac {\hbar c}{\lambda} \propto a^{-4}  \quad \Rightarrow \quad \omega = 1/3 ~,
\eeq
alternatively,  the case of non-relativist matter (e.g. ''dust`` particles of mass $m$) is found to scale like 
\beq  \label{eq:energymatter}
\rho_{\rm m} \sim \frac mV  \propto a^{-3}  \quad \Rightarrow \quad  \omega = 0 ~.
\eeq

In a similar manner, one can rewrite Eq. (\ref{FriedmannEq}) by rewriting the curvature and cosmological constant terms in terms of a fluid evolution. Indeed, the curvature term can be replaced by $\rho_\mathcal{K} = -3 \mathcal{K} \Mpl^2/a^2$, which corresponds to a fluid with $\omega = -1/3$, and the cosmological constant term can also be substituted by $\rho_\Lambda =  \Lambda \Mpl^2$, corresponding to a fluid with $\omega = -1$. 
Other parametrisations of the equation of state used in other cosmological contexts are listed in Table~\ref{table:EoS_FLRW}. 
\ \\

\begin{table}[h!]
\centering
\begin{tabular}{c||ccccc}   \hline \hline	
      matter type          &	$\bm \omega$	&	$\rho(a)$	&	$a(t)$	   & $(aH)^{-1}$   & $\ddot{a}/a$			
      \\ \hline \hline
stiff fluid (or kination)      	&	$1$		    &	$a^{-6}$	&	$t^{1/3}$  	& $t^{2/3}$  &  decelerated	
\\
\text{radiation}	        &	$1/3$		&	$a^{-4}$	&	$t^{1/2}$		& $t^{1/2}$  & decelerated
\\
\text{cold matter (dust)}		&	$0$		    &	$a^{-3}$	&	$t^{2/3}$	& $t^{1/3}$  & decelerated	
\\
\text{curvature}		    &	$-1/3$	    &	$a^{-2}$	&	$ t  $   		& constant   & --
\\
$\Lambda$ (de Sitter)		&	$-1$		&	constant	&	$e^{ H t}$	    & $e^{- H t}$  & accelerated
\\ \hline \hline
\end{tabular}
\caption{Relation between the equation of state and the expansion history in a FLRW Universe.
\label{table:EoS_FLRW}
}
\end{table}	

\section{The Hot Big Bang model}

After having introduced the equations that describe the expansion of the FLRW universe, we are in a position where we can model a universe containing $n_{\rm fl}$ number of fluids, with a varying contribution to the total energy budget of the universe, depending on their individual equation of state and the scale factor. For simplicity, let us assume a universe containing three types of fluids: radiation, cold collisionless matter, and one describing a cosmological constant. Furthermore, we further assume that the fluids interact between them only through gravitation, and therefore their temperature is only dependent on the expansion of spacetime. In such a model, the Friedmann-Lema\^{i}tre equation reads
\be
\begin{split}
 H^2(a) & = \frac 1{3\Mpl^2} \left( \frac{\rho_{\rm r, 0}}{a^4} + \frac{\rho_{\rm m, 0}}{a^3} + \rho_{\Lambda} \right)  ~,
 \\
        & = H^2_{0}  \left(  {\Omega_\mathrm{m} \over a^{3} } + { \Omega_\mathrm{r} \over a^{4} } +  \Omega_{\Lambda }\right) ~, \qquad \text{with } ~ \Omega_i = \frac{\rho_{i\>0}}{3\Mp^2H_{0}^{2}} ~,    
\end{split}
 \ee
where $\rho_{\rm r, 0}$ and $\rho_{\rm m, 0}$ are the radiation and matter energy densities as measured in the present times (i.e. $a_0 = 1$), and later expressed in terms of the dimensionless energy density  $\Omega_i$.

\begin{figure}[htbp!]
\includegraphics[width=0.990\textwidth]{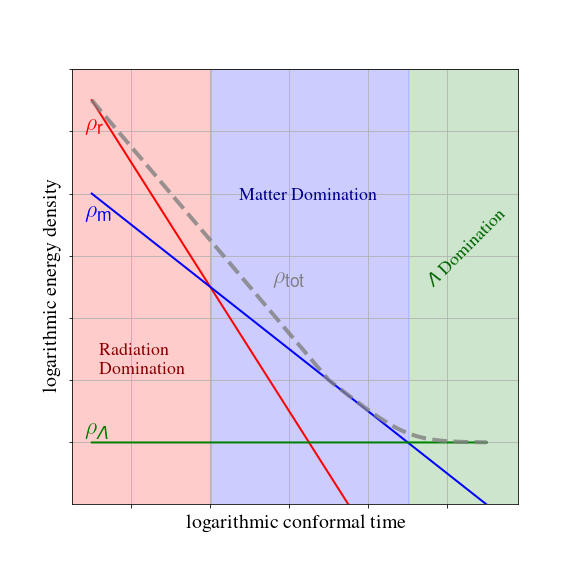}
\caption{Evolution of multi-fluid cosmological model containing radiation, matter and dark energy (or cosmological constant $\Lambda$). 
}
\label{Fig::BBM_simple}
\end{figure}

Today observations indicate that the relation between the energy densities hold 
\beq
\rho_{\Lambda} > \rho_{\rm m, 0} \gg \rho_{\rm r, 0} ~, 
\eeq
however, going backwards in time , as $a(t) \rightarrow 0$,  we can verify from Eqs.~(\ref{eq:energyradiation}) and (\ref{eq:energymatter}) that the Universe undergoes through matter, and radiation domination at sufficient early time.  This is illustrated in Fig. \ref{Fig::BBM_simple}.

\subsection{The ${\Lambda}$CDM model }

Taking into account the evolution of linear perturbations during the cosmological evolution of the Universe assuming linearised Einstein equations, in addition to the thermal description of gases, it is possible to model the particle composition of the Universe from the (unknown) primordial density perturbation through the CMB and structure formation until today. Thus, using observational data from CMB experiments, large scale structure from galaxy surveys, baryon acoustic oscillations (BAO), supernova, and other astrophysical datasets,  one can construct cosmological models where a set of a few free parameters are fit to the aforementioned datasets. As of today, the so-called $\Lambda \mathrm{CDM}$ model represents the best fitting picture of the describing statistics of our Universe. 

The  $\Lambda \mathrm{CDM}$ model consists of six 
 free parameters primarily chosen to avoid degeneracies of the model fit to the data. These parameters are the following: 

\begin{itemize}
 \item \textbf{Baryon density} ($\bm{ h^2\Omega_\mathrm{b} }$):  Composed by the non-relativistic ordinary matter of the Universe, reescaled by the squared dimensionless Hubble parameter $h \equiv H_0 / (100\ \mathrm{km}\ \mathrm{s}^{-1}\ \mathrm{Mpc}^{-1}$).
 \\
 \item \textbf{Cold dark matter density} ($\bm{  h^2\Omega_\mathrm{CDM} }$): Composed by collisionless dark matter, e.g. only interacting via gravitational interactions. 
 \\
 \item \textbf{Sound horizon at CMB} ($\bm{ \theta_{\rm S} }$): Maximal angular distance that sound waves could have travelled prior the last scattering. 
 \\
 \item \textbf{Reionization optical depth} ($\bm{ \tau }$): Its a dimensionless measure of the line-of-sight free-electron opacity to CMB radiation. In other words, a quantification of how much CMB photons are scattered at later times during the reionisation phase (see Sec.~\ref{sec:cosmic_timeline}). It is therefore of astrophysical origin. 
 \\
 \item \textbf{Amplitude of the scalar power spectrum} ($\bm{A_{\rm S} }$): The value  of the scalar power spectrum at the pivot scale $k_* = 0.05\ \mathrm{Mpc}^{-1}$, e.g. $A_S = \mathcal{P}_\zeta (k_*)$ (see Sec.~\ref{sec:perturbed Einstein equations}). 
  \\
 \item \textbf{Scalar spectral index} ($\bm{n_S}$): The slope of the (logarithmic) scalar power spectrum, where $n_s =1$ indicates scale invariance (see Sec.~\ref{sec:perturbed Einstein equations}).  
\end{itemize}

From the free parameters below, under certain assumptions, one can derive other related quantities such as $H_0$, the age of the Universe $t_0$, the densities $\Omega_m$ and $\Omega_\Lambda$, etc. Some of the (a priori) assumptions are a vanishing tensor power spectrum ($A_T = n_T = 0$),  spatial flatness ($\Omega_\mathcal{K} = 0$), dark energy as a cosmological constant ($\omega_\Lambda = -1$),  among others, see \Ref{PhysRevD.98.030001}. However, when relaxing any of these fixed parameters, only one at a time, and promoting them as a free parameter, their favoured value in the extended $(6+1)$-parameter model agrees with the original assumption. The current best-fitting parameters for the model are listed in Table~\ref{table_LCDM}. 

\begin{center}
\centering
\vspace*{10mm}
\begin{table}[htb!]
    \centering
  \begin{tabular}{lc}
    \hline \hline
     Fitting parameters \\  
    \hline
        $\Omega_{\rm b}h^2$       & $0.02227\pm0.00020$\\
        $\Omega_{\rm c}h^2$       & $0.1184\pm0.0012$\\
        $100\>\theta_{\rm s} $            & $1.04106\pm0.00041$\\
        $\tau$                    & $0.067\pm0.013$\\
        $\ln(10^{10}A_{\rm S})$       & $3.064\pm0.024$\\
        $n_{\rm S}$                   & $0.9681\pm0.0044$\\[2mm]
    \hline
    Derived parameters \\  
    \hline
      $h$                       & $0.679\pm0.006$ \\
      $\Omega_{\rm m}$          & $0.306 \pm 0.007$ \\
      $\Omega_{\rm \Lambda}$    & $0.694 \pm 0.007$ \\[2mm]
    \hline 
     Extended fitting parameters (6+1) \\  
    \hline
      $\Omega_{\rm \mathcal{K}}$     & $0.0002 \pm 0.0026$ \\
      $\omega_{\Lambda}$        & $-0.97 \pm 0.05$ \\[1mm]
    \hline
    \hline
    \vspace*{1mm}
  \end{tabular}
  \caption{Best-fitting and derived cosmological parameters, from a combination of {\it Planck\/} data and other datasets (including BAO, supernova and $H_0$ data) 
  \cite{PhysRevD.98.030001}.
  } 
  \label{table_LCDM}
\end{table}
\end{center}

\clearpage
\subsection{The cosmic timeline \label{sec:cosmic_timeline}}

Our current understanding of the particle physics content of the early universe allows us to elaborate a more detailed picture of the timeline history of our universe than just the transitions from radiation to matter domination and now from matter to dark energy. In the following, a summary of the most relevant physics and phase transition is further developed, from the initial times at $t\rightarrow 0$ (at higher energies/temperature) until today.

\begin{itemize}
 \item \textbf{ Prior to radiation domination}   \ ~ (time duration: Unknown).
    \begin{itemize}
    \item  \textbf{ $T > 10^{32}~\mathrm{K}$: The Planck Era}.
    A description of physics beyond these energy scales would require a theory of quantum
    gravity; therefore, the physics of this epoch are entirely unknown.
    \\
    \item \textbf{ $T \approx 10^{26}~\mathrm{K}$: Grand Unified Theory Symmetry
    Breaking - Cosmic  Inflation}. It is speculated that the fundamental energies are unified beyond these energy scales. In theories of inflation, this is usually also the highest energy scale at which inflation could end,
    entering the reheating era when the HBB plasma is formed.
    \end{itemize}
%
\item \textbf{ Radiation domination epoch}   \ ~ (time duration: $\sim 50$ thousand years).
    \begin{itemize}
    \item  \textbf{ $  10^{26}~\mathrm{K} \gtrsim T > 10^{16}~\mathrm{K}$: Reheating, thermalization}.  
    The primordial plasma is thought to be formed by free quarks, leptons, gauge bosons and their antimatter counterparts. 
    Dark matter could also have been formed during this period, whether this is made by exotic particles (quantum fields) and/or of primordial black holes.  
    \\
    \item  \textbf{ $ T \approx 10^{16}~\mathrm{K} \sim ~ ( \mathrm{TeV}) $: Electroweak phase transition}.
    Higgs symmetry breaking occurs, giving mass to the standard model particles.
    \\
    \item  \textbf{ $ T \approx 10^{12}~\mathrm{K} \sim ~ ( \mathrm{MeV}) $: QCD phase transition}.
    Free quarks and gluons from the plasma bound together to form protons, neutrons and other hadrons. 
    Primordial black hole production could also have been enhanced during this phase transition. 
    \\
    \item  \textbf{ $ T \approx 10^{10}~\mathrm{K} \sim ~ (10 ~ \mathrm{KeV}) $: Neutrino decoupling, Big Bang Nucleosynthesis}.
    Freeze-out of the weak interactions, the abundances of protons and neutrons are set and neutrino decouples from the cosmic plasma. Soon after, it begins the matter-antimatter annihilation phase. Protons and neutrons condensed into atomic nuclei forming light elements such as ionised hydrogen and helium. \\
    
    Up to here, all the processes lasted only about 5 minutes since the start of the radiation domination epoch. However, the Universe kept expanding and cooling down in that regime for up to 50 thousand years until it became matter dominated.
    \end{itemize}
\item \textbf{ Matter domination epoch}    \ ~ (time duration: $\sim 7$ billion years).
    \begin{itemize}
    \item  \textbf{ $ T \approx 10^{4}~\mathrm{K} \sim ~ (1 ~ \mathrm{eV}) $: Radiation-matter Equality}.
    Transition to matter domination. Photons continue being tightly coupled to baryonic matter, generating sound waves known as baryonic acoustic oscillations. At that point, the Universe was still opaque to electromagnetic radiation, and was dominated mainly by cold (dark) matter and, in a small proportion, by warm (baryonic) matter.    
    \\
    \item  \textbf{ $ T \approx 10^{3}~\mathrm{K} \sim ~ (0.2~ \mathrm{eV}) $: Recombination - CMB}.
    Free electrons bound to atomic nuclei forming neutral elements and releasing the free streaming photons constituting today's Cosmic Microwave Background. The universe became transparent and dark, a period called the ``dark ages'' starts, before the formation of structure lead to the genesis of the first starts.  
    \\
    \item  \textbf{ $ T \approx 50~\mathrm{K} \sim ~ ( 1 ~ \mathrm{meV}) $: Gravitational bounding, First stars, Reionization phase}.
    The cosmic gas form gravitational bound systems. This first creates dark matter halos that later will make baryonic matter matter would collapse into the first generation of stars, inducing a (re) ionisation phase. These stars were constituted mainly on hydrogen and helium isotrops, which would fuse into (slightly) heavier elements. Next generation of stars (including supernova explosions) would produce the heavier elements we find today in the Universe. 
    \\
    \end{itemize}
\item \textbf{ Accelerated expansion epoch} \ ~ (Started about 8 billions years ago).
    \begin{itemize}
        \item \textbf{$ T \approx 5 ~\mathrm{K}$ : Transition to dark energy domination}
    Around eight billion years ago, matter diluted to the point where it was less dense than the cosmic dark energy. Cosmic deceleration then ceased, and the universe's expansion began to accelerate due to the dark energy. Cosmic structure
    formation on large scales gradually ceased. 
    \\
    \item \textbf{ $ T \approx 2.7 ~\mathrm{K}$ : The present days.}  Galaxy formation, including the Milky Way, and planet Earth. Today. 

    \end{itemize}
\end{itemize}

\begin{figure}[!h]
\vspace*{2.5cm}
\begin{center}
\includegraphics[width=0.990\textwidth]{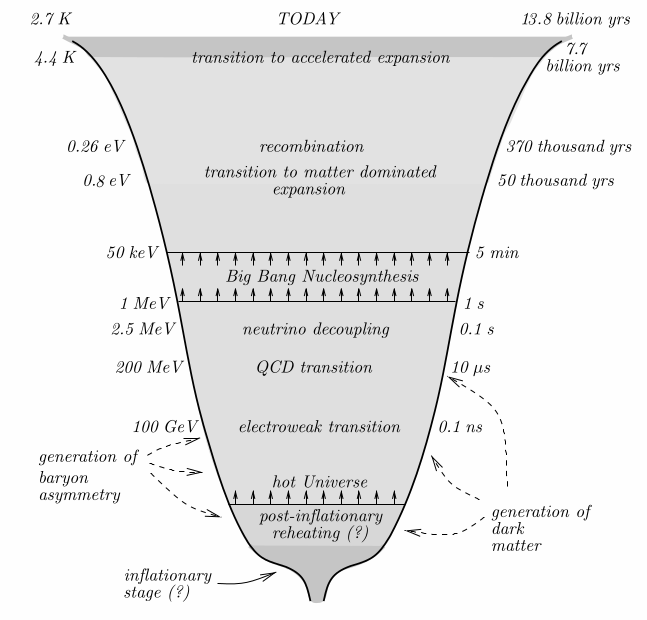}
\end{center}
 \caption{  
  Representation of the different stages during the evolution of the Universe. Figure taken from Ref.~\cite{Rubakov:2017xzr}. 
 \label{fig:cosmo_timeline}
}
\vspace*{2.5cm}
\end{figure}

\section{Physical scales and horizons}

In cosmology, there are two relevant physical scales to consider. On one hand, we have the size of the Hubble radius $R_H \equiv H^{-1}$. This defines the distance at which the (non-local) receding velocity of a comoving particle would be equal to the speed of light as a consequence of the Universe's expansion rate. 
On the other hand,  there is the perturbation wavelength $\lambda = 2\pi / k$ corresponding to a Fourier mode of wavenumber $k$.   Thus,  when considering the evolution  of such perturbations, it will be necessary to distinguish between the two following conditions: 
\begin{equation*}
\begin{split}
 k &\ll H ~,  \quad \text{ super-Hubble modes:  wavelengths larger than the Hubble radius,  }  
\\
k &\gg H ~, \quad \text{ sub-Hubble modes:  wavelengths smaller than the Hubble radius.  }  %
\end{split}
\end{equation*}

Often, in the literature, there is the identification of the Hubble radius with the particle horizon. However, that is not true in general and therefore is misleading. The existence of a particle horizon (or cosmological horizon) is a consequence of two assumptions: First, the universe has a beginning; This is motivated in the HBB model as a consequence of the expansion of the Universe, ultimately leading to the Big Bang singularity at $t \rightarrow 0$. Second, the maximum speed at which information (or events) can propagate is the speed of light (i.e. Einstein's theory of Special Relativity). In other words, only signals emitted within the extent past light-cone, from today to $t\rightarrow 0$, are in causal contact. More formally, the particle horizon can be defined as 
\beq \label{eq:particle_horizon}
 R_P \equiv a(t) \int^t_0 \frac{\ud t'}{a(t')}.
\eeq 

If one assumes that the Universe has undergone through decelerated expansion during the whole past history (i.e. radiation domination followed by matter domination), then the scales factor scales like a power of time, $a(t) \propto t^n$ with $0<n<1$. In that case, using Eqs.~(\ref{eq:def_H})  and (\ref{eq:particle_horizon}), it is straightforward to find that $ R_P \approx \frac n {1-n} R_H$. 
However, as we will discuss below,  in the context of the inflationary paradigm this is no longer true, and therefore the Hubble radius can vastly differ from the particle horizon.

\section{The Hot Big Bang problems}

Despite the success of the HBB model in providing an explanation to cosmological observations, as well as predicting the existence of a relic photon bath (i.e. the CMB), there were a few issues that indicate that this model, as it was originally postulated, is not complete. As we will see, these issues are related to inconsistent or unexplained initial conditions when assuming that the Universe has been all-time under decelerated expansion, where the scale factor scales expands like a power-law, so that  $a\propto t^n$ with $n < 1$.

\subsection{The horizon problem}

The temperature of CMB photons today is measured to be  $T_{\rm CMB, 0} = 2.7260\pm0.0013 ~\mathrm K$. These photons have been traveling since the time of the last scattering surface, which occurred after the recombination phase in the matter domination era. 
Assuming the universe has been matter dominated since the time of the last scattering surface, i.e. neglecting the recent accelerated expansion period. The Hubble radius scales like $R_H \propto a^{3/2}$ and then the size of a Hubble patch  at the recombination epoch is  
\beq
R_{H,\mathrm{rec}}(t_{\rm rec}) \approx R_{H,0}(t_0) \left( \frac {a_{\rm rec}}{a_0} \right) ^{3/2} ~,
\eeq

Since then, these Hubble patches have been expanded, so today, they have a size of  
\beq
R_{H,\mathrm{rec}}(t_{0}) = R_{H,\mathrm{rec}}(t_{\rm rec}) \frac {a_0}{a_{\rm rec}} \approx  R_{H,0}(t_0) \left( \frac {a_{\rm rec}}{a_0} \right)^{1/2} \approx 300~ {\rm Mpc}~.
\eeq
However, the Hubble radius at present is about ${l_{H,0}(t_{0})\approx 4~{\rm Gpc}}$. Under the assumption of an all-time decelerated universe, the Hubble radius is proportional to the particle horizon. Thus, the CMB map that we can observe today contains about $\sim\!10^5$ causally disconnected patches, which is problematic to explain the observed nearly-homogeneous temperature of the CMB photons, as all these Hubble patches should not have been able to thermalise.

\subsection{The flatness problem}

Making use of the dimensionless density parameter for the spatial curvature, $\Omega_\mathcal{K} = -\mathcal{K}/(aH)^2$, using Eq. (\ref{Eq::Rho_of_a_inPF}) we find that this is an increasing quantity,  
\beq
\Omega_\mathcal{K} \propto \left\{
\begin{split}
& ~ a^2 \quad \text{in Radiation domination}, \\
& ~ a\ \ \quad \text{in Matter domination}. \\
\end{split}
\right.
\eeq

Today, the CMB observations set the value of the spatial curvature to 
\beq
 \Omega_\mathcal{K} =  0.0002 \pm 0.0026 ~.
\eeq

Thus, because the spatial curvature at different times is given by 
\beq
{\Omega_\mathcal{K} (t)} = \frac{(a_0H_0)^2}{[a(t) H(t)]^2} ~ {\Omega_\mathcal{K} (t_0)} ~,
\eeq
this implies that at early times the curvature must have been extremely small in order to agree with the observations today. For instance, during the Big Bang Nucleosynthesis its value should be of the order of   $\abs{\Omega_{\mathcal{K},BBN}} \lesssim 10^{-18}$, while back in the Planck epoch this projected value becomes  ${\abs{\Omega_{\mathcal{K}, \rm Planck}} \lesssim 10^{-63}}$, indicating a very flat geometry (fine-tuned?) in the early Universe.

\section{The inflationary paradigm }

Cosmological inflation is a period of rapid, accelerated expansion of the Universe, which is proposed to have happened between the Planck and the radiation dominated era. This idea was proposed independently by Alexei Starobinski \cite{STAROBINSKY198099} and Alan Guth \cite{PhysRevD.23.347}, and it provided a solution for the Horizon and Flatness problems.
\ \\[2mm]
The \textbf{horizon problem} is solved naturally if we allow the universe to expand sufficiently to allow the whole sky to be in causal contact during recombination, thus allowing for homogeneity in the CMB temperature. Assuming exponential (quasi) de Sitter expansion, the scale factor scales like $a(t) \simeq a_{\rm ini} e^{{\cal H} \Delta t} $, while the Hubble radius becomes nearly constant. The condition to solve the horizon problem is none other than to request that the size of the observable universe today ($d_{H_0}$) must be smaller than the size of the causal region at the beginning of inflation ($d_{H_{\rm ini}}$),
\be
d_{H_0} (t_0) \frac{a_{\rm end}}{a_0} < d_{H_{\rm ini}} (t_{\rm ini}) \frac{a_{\rm end}}{a_{\rm ini}} = d_{H_{\rm ini}} (t_{\rm ini}) e^{{ N} t} ~,
\ee
where $a_{\rm end}$ and $a_{\rm ini}$ are the scale factor at the end and the beginning of inflation. Plugging in number, we find that the necessary amount of expansion given by the number of efolds is 
\be
{ N}(t) = \ln \left[ {a(t) \over a_{\rm ini}}   \right] \gtrsim 60 ~.
\ee
The \textbf{flatness problem} is also solved straightforwardly, given that the spatial curvature during the de-Sitter phase scales like $|\Omega_\mathcal{K} | \propto a^{-2}$, and therefore, it decreases exponentially during inflation. This can explains the current observations if inflation lasted a period larger than ${N} \gtrsim 30 - 40$ \text{efolds}.
%
%

%
%
%

\chapter{Scalar-field inflation}
\label{chap:inflation}
\pagestyle{fancy}

This chapter introduces the dynamics of scalar field cosmologies commonly used in inflationary theory. We review the case of single field slow-roll inflation at the background level, following up with the treatment of perturbations. After that, we analyze the connection between inflationary perturbations with the characteristics of the primordial density perturbations and we explain how these predictions are used to constrain inflation models with observations. For simplicity, throughout this chapter we will assume a universe with a flat geometry (i.e. ${\mathcal{K} = 0}$).  

\section{Scalar-field dynamics \label{sec_SFI} }

The following Einstein-Hilbert action describes cosmological models dominated by a single scalar field  
\be
 S = \int d^4 \sqrt{-g}\left[ \frac{\Mp^2}{2} R - \frac12 g^{\mu\nu} \partial_\mu\varphi \partial_\nu\varphi - V(\varphi) \right] ~,
\ee
where $\varphi$ corresponds to the scalar field, or \textit{inflaton}, and $V(\varphi)$ is the scalar field potential. The energy-momentum tensor is then given by 
\beq
T_{\mu\nu} = \partial_\mu \varphi\, \partial_\nu \varphi - \frac{1}{2} g_{\mu\nu}\, \partial_\lambda \varphi \, \partial^\lambda \varphi - g_{\mu\nu} V(\varphi) ~.
\eeq
Assuming a FLRW metric background, the scalar field equation of motion are given by the Klein-Gordon equation
\beq \label{SF_KK}
\ddot \varphi + 3 H\dot\varphi - a^{-2} (D_i\varphi)^2/2 + \frac{\ud}{\ud\varphi} V(\varphi) = 0  ~.
\eeq
and the energy density and pressure are defined as 
\begin{align} \label{sf_rho_p}
 \rho &\equiv T^0_0  = \frac12\dot\varphi^2 + \frac {a^{-2}}2 (D_k\varphi)^2 + V(\varphi) ~,
 \\
  p &\equiv  \frac 13 T^i_i= \frac12\dot\varphi^2 - \frac {a^{-2}}6 (D_k\varphi)^2 - V(\varphi)
 ~.
\end{align}

In analogy to the perfect fluid case, we can define an effective equation of state for the system where
\beq
\omega = \frac{p}\rho = \frac{\frac12\dot\varphi^2 - \frac {a^{-2}}{6} (D_i\varphi)^2 - V(\varphi)}{ \frac12\dot\varphi^2 + \frac {a^{-2}}{2}(D_i\varphi)^2 + V(\varphi)} ~, 
\eeq
and it is straightforward to see that to obtain  a period of accelerated expansion, involving a negative pressure dominated period, it suffices that 
\beq
\omega \simeq -1 \Longleftrightarrow  
\left\{
\begin{split}
V(\varphi) &> \dot{\varphi}^2 ~, \\ 
V(\varphi) &> \left(\partial_k \varphi\right)^2 ~.
\end{split} 
\right.
\eeq

For the time being, we will consider the dynamics deep within the inflationary period, which justifies the quasi-homogeneous treatment of the field, 
i.e. ~${\partial_k\varphi \partial^k\varphi \ll V(\varphi)}$, because, during inflation, thermal and quantum field perturbations quickly redshift as a consequence of the rapid expansion of the Universe.

\section{Slow-roll inflation: background dynamics}

Let us then consider the case where the background dynamics are dominated by the potential, implying that the kinetic term is small, satisfying 
\be
\dot\varphi^2 \ll V(\varphi)  ~.
\ee
Under these conditions, 
the Universe's expansion rate is governed by the following equations 
\begin{align}
H^2 &= \frac 1 {3\Mpl^2} \left[ \frac12\dot\varphi^2 + V \right] \simeq \frac{V }{3\Mpl^2} 
~, \\
\frac{\ddot a} a &= \frac 1 {3\Mpl^2} \left[ - \dot\varphi^2 + V  \right] \simeq \frac{V }{3\Mpl^2}
~.
\end{align}

Assuming a small kinetic acceleration for the field,  $\ddot\varphi \ll 3H\dot\varphi$, 
the Klein-Gordon equation for the evolution of the field simplifies to 
\be
    3H\dot\varphi = \frac{\ud }{\ud \varphi} V ~,
\ee
which is expressed in terms of the number of efolds reads 
\be
  \frac{\ud \varphi}{\ud N} = - \Mpl^2 \frac 1V \frac{\ud V} {\ud \varphi} ~.
\ee
We observe that the dynamics of the fields are thus governed by the shape of the potential, which leads to a large number of efolds when the logarithm of the potential is sufficiently flat. When that is the case, the slow-roll regime becomes a dynamical attractor \cite{Peter:2013avv}. This can be verified by computing the so-called ``slow-roll parameters''. Even though there exist multiple definitions for such parameters, the most commonly used are those defined in terms of the Hubble-flow functions, which can be approximated in terms of the potential and its derivatives, 
\begin{align}
\epsilon_1 & \equiv  -\frac{\dot{H}}{H^2} \  = \  \frac{\Mpl^2}2 \left(  \frac 1V \frac{\ud  V}{\ud \varphi} \right)^2 + \mathcal O(\epsilon^2)~ , 
\\
\epsilon_2 & \equiv  \frac{\dot{H}}{H\dot{H}} - 2\frac{\dot{H}}{H^2} \ = \ 
2\Mpl^2 \left[  \left(\frac 1V \frac{\ud  V}{\ud \varphi}\right)^2 -   \frac 1V  \frac{\ud^2  V}{\ud \varphi^2}   \right] + \mathcal O(\epsilon^2) ~ , 
\\
\epsilon_{n+1} &\equiv \frac{\ud \ln |\epsilon_n|}{\ud N} ~.
\end{align}
A value of $\epsilon_1 < 1$ implies that the Universe is in accelerated expansion, i.e. $\ddot a/a > 0$. A small value in higher order terms is commonly used to ensure slow-roll. As shown below, these parameters are also useful to compute observable predictions for a given model.

\section{Formalism of cosmological perturbations}

Let us now consider the dynamics of perturbations during inflation. %
Deep within slow-roll inflation, linear field perturbations become quickly redshifted due to the expansion rate, described by a nearly constant Hubble parameter. In this phase, the Universe is dominated by a large vacuum-energy where the vacuum-expectation-value is given by the scalar-field potential $\rho \simeq V(\varphi)$, 
which in turn,  defines the energy scale of inflation. In these circumstances, quantum effects cannot be neglected as they become the primary source of field excitations. 
The evolution of these modes can be described by the formalism of Cosmological Perturbation Theory (CPT). This approach is based on the linearized Einstein equations, which provide an accurate description as long as the energy perturbations are small, i.e. ${\delta\rho/\rho\ll1}$.

At the linear level, perturbations consisting of \textit{scalar}, \textit{vector} and \textit{tensor} modes decouple from each other. Thus, before commencing  the  calculations, it is convenient to decompose the system into these three sectors. In the literature, this procedure is known as the Scalar-Vector-Tensor (SVT) decomposition of the metric \cite{Peter:2013avv}.

\subsection{Scalar-Vector-Tensor decomposition}

Considering perturbation on a Universe where background dynamics follow the FLRW solution, 
\be
g_{\mu\nu} = g_{\mu\nu}^{\rm FLRW} + \delta g_{\mu\nu} ~, 
\ee
the most general metric that includes perturbations read  
\beq\label{general metric decomposition}
\mathrm{d}s^2 = a^2(\eta)\left[ -(1+2A)\mathrm{d}\eta^2 + 2 \bar B_i \mathrm{d}x^i\mathrm{d}\eta + \left(\gamma_{ij}+ \bar E_{ij}\right)\mathrm{d}x^i\mathrm{d}x^j \right],
\eeq

The vector component can be further decomposed in 
\beq
\bar B_i=D_i B + B_i, \quad \mathrm{with} \quad D^i {B}_i=0.
\eeq
In the same way, the tensor field $E_{ij}$ is decomposed into
\beq
\begin{split}
\bar E_{ij} = -2C\gamma_{ij} + 2D_iD_jE + 2D_{(i}{E}_{j)} + 2{h}_{ij},\\[3mm]
 \mathrm{with} \qquad  D^i {E}_i=0, \quad  \mathrm{and} \quad  D^i{h}_{ij}=0 ~, \  \gamma\IJ{h}\ij=0 ~,
\end{split}
\eeq
and, here, $h_{ij}$ is a pure transverse-traceless tensor. Thus, this new metric contain 10 new degrees of freedom corresponding to 4 scalars ($A$, $B$, $C$, $E$), plus  2 vectors with two polarizations each ($B_i$, $E_i$), and  1 tensor with also two polarizations ($h_{ij}$).

\subsection{Gauge freedom} \label{par:gauge issue}

Let us consider a generic infinitesimal gauge transformation given by 
\beq\label{generic gauge transformation}
x^\mu \rightarrow \tilde{x}^\mu = x^\mu - \xi^\mu,
\eeq
where $\xi^\mathrm{\mu}=(\xi^0,\bar\xi^i)$ is an arbitrary four-vector. Once more, the vector component $\bar\xi^i$ can then be decomposed in
\be
 \bar \xi^i = D^{i} \xi + \xi^i \quad \mathrm{with } ~ D^i\xi_i = 0~.
\ee
Hence, this transformation represents a total of 4 degrees of freedom (two from  $\{ \xi^0, \xi\}$  and two in $\{ \xi^i \}$).  
Under a gauge transformation,  the perturbation quantities transform like ~\cite{Mukhanov:1990me}
\begin{gather}
\label{scalar perturbation transforms}
{A} \rightarrow A + \mathcal{H}\xi^0 + \xi^{0\prime}, \quad  {B} \rightarrow B - \xi^0 + \xi^\prime, \quad {C} \rightarrow C + \mathcal{H}\xi^0, 
\quad  {E} \rightarrow E + \xi,
\\
\label{vector perturbation transforms}
 B^i  \rightarrow \tilde B^i  +  \xi^{i\prime}, \quad  E^i \rightarrow E^i + \xi^{i}~,
\\ \label{tensor perturbation transforms}
 h_{ij}  \rightarrow h\ij  ~,
\end{gather} 
where we remind that the prime denotes the derivative with respect to the conformal time $\eta$. Given an arbitrary perturbed quantity $\delta Q$, this transforms like 
\be  \label{transformation_perturbationQ}
\delta Q \rightarrow \delta Q + \mathcal{L}_\xi Q
\ee
where $\mathcal{L}_\xi$ is the Lie derivative with respect to the vector $\xi^\mu$.

The scalar quantities $A$ and $B$ are interpreted as the lapse and shift functions between two time-hypersurface. %
The quantity $C$ relates to the scalar gravitational curvature of the same hypersurface, so that 
\be
R \equiv \gamma\IJ R\ij = \frac 4 {a^2} D_i D^i C ~.
\ee
All these quantities depend on the geometry of the chosen time-hypersurface, and therefore they are subject to variations under gauge transformations. 
On the other hand, $h\ij$ is a gauge invariant quantity which corresponds to gravitational waves. By taking into account Eqs.~(\ref{scalar perturbation transforms}-\ref{vector perturbation transforms}), other gauge invariant quantities can be constructed; Common examples are the Bardeen potentials %
defined as 
\beq \label{Bardeen potentials}
\Phi \equiv A + \mathcal{H}\left(B-E^\prime\right)  + \left(B-E^\prime\right)^\prime, 
\quad \Psi \equiv C - \mathcal{H}(B-E^\prime)
~ .
\eeq

In addition, because of \Eq{transformation_perturbationQ},  any perturbation from a scalar quantity $\varphi$ can be redefined in a gauge invariant form as
\be
\delta\gi\varphi= \delta\varphi + \varphi^\prime(B-E^\prime)~.
\ee

On occasions, it is useful for performing calculations to fix the gauge. A common choice, is the Newtonian gauge (N.G.) or \textit{shear-less} gauge, where one fixes $B = E =0$. In such a gauge, the above quantities reduce to 
\be
 \Phi =   A \Big|_{\rm N.G.}~,  \qquad    \Psi =   C \Big|_{\rm N.G.}~, \quad \mathrm{and} \quad \delta\gi\varphi= \delta\varphi \Big|_{\rm N.G.}~. 
\ee

Other useful gauge invariant combinations are 
\be
\zeta \equiv C - \frac H {\dot\varphi} \delta\varphi  = C \Big|_{\rm (U.F.G)}~,
\ee
which corresponds to the curvature perturbations in the uniform-field gauge (U.F.G), i.e. where one fixes $\delta\varphi = 0$. And finally, the curvature perturbation in the uniform density gauge (U.D.G), which fixes $\delta\rho = 0$,  
\be
\tilde\zeta \equiv C - \frac H {\dot\rho} \delta\rho  = C \Big|_{\rm (U.D.G)} ~.
\ee

These two quantities are convenient for studying inflationary perturbations. As we will see below, the amplitude of curvature perturbations freezes at super-Hubble scales. Thus, these gauge invariant quantities can be used to match (at super-Hubble scales) the  $\zeta$ perturbations (from inflation) to the  $\tilde \zeta$ perturbations at later times in the radiation domination. Remarkably, both quantities coincide at super-Hubble scales, 
\be
 \zeta \cong \tilde \zeta \qquad {\rm at } \quad \frac k{\cal H} \ll 1.
\ee

\section{The perturbed Einstein equations} \label{sec:perturbed Einstein equations}

We are considering now linear perturbations at first order in the Einstein equations, such as {${\delta G^\mu_{\nu} = \Mpl^{-2}\delta T^\mu_{\nu}}$}, for a system concerning a single scalar field. As mentioned before, the equations for scalars, vectors and tensors decouple from each other, allowing us to consider them separately. For simplicity, we fix the gauge freedom into the Newtonian gauge by imposing $E = B = 0$. Nonetheless, we prescribe the dynamics using gauge-invariant quantities as much as possible, so that it facilitates the transcription into other gauges.

\subsection{Scalar perturbations}

Taking only the  scalar part of the metric, the line element reads 
\beq
\begin{split}
\label{scalar_metric decomposition}
\mathrm{d}s^2_{\rm scalar} = & a^2(\eta)   \Big\lbrace  -(1+2A)\mathrm{d}\eta^2 + 2\partial_i B\mathrm{d}x^i\mathrm{d}\eta  \\ 
 & + \left[(1-2C)\delta_{ij} + 2 \partial_i\partial_j E \right]\mathrm{d}x^i\mathrm{d}x^j   \Big\rbrace.
\end{split}
\eeq
Fixing the gauge to the Newtonian gauge, allow us to rewrite the metric in gauge invariant quantities as $ {\Phi = A,}\ \Psi = C,\ B = 0,\ E = 0$. Furthermore, we notice that  the non-diagonal ($i\neq j$) part of the $ij$-Einstein equations reads 
\be
 \text{ for $i\neq j$:  }\qquad  D_iD_j \left(\Psi - \Phi\right) = \delta T_{ij} = 0 ~,
\ee
using the fact that  $T_{\mu\nu}$ for a single scalar-field has vanishing non-diagonal terms. This results in a further simplification as it implies $\Phi = \Psi$.  This allows us to write the remaining Einstein equations in the following form, 
\begin{eqnarray}  
D_kD^k \Phi - 3\mathcal{H}\Phi^\prime - 3\mathcal{H}^{2}\Phi &=& \frac{1}{2\Mp^2} \Big(-\varphi^{\prime 2}\Phi +\varphi^\prime \delta\varphi_\mathrm{g.i.}^{\prime} 
 \qquad \nonumber \\ & & \hspace*{20mm}+ \frac{\ud V} {\ud \varphi}a^2  \delta\varphi_\mathrm{g.i.}   \Big), \label{eq:00 component-Einstein equation} 
 \\
(\Phi^\prime + \mathcal{H}\Phi) &=& \frac{1}{2\Mp^2}\varphi^\prime\delta\varphi_\mathrm{g.i.} , \label{eq:0i component-Einstein equation} 
\\
\Phi^{\prime\prime}+3\mathcal{H}\Phi^{\prime} + (2\mathcal{H}^\prime + \mathcal{H}^{2})\Phi &=& - \frac{1}{2\Mp^2}\Big(\varphi^{\prime 2}\Phi -\varphi^\prime \delta\varphi_\mathrm{g.i.}^{\prime} 
 \nonumber \\ & & \hspace*{20mm}+  \frac{\ud V} {\ud \varphi}a^2  \delta\varphi_\mathrm{g.i.}  \Big). \label{eq:ii component-Einstein equation}
\end{eqnarray}
In the next step, we combine the above equations with the perturbed Klein-Gordon equation,  
\begin{equation}\label{kgp}
\delta \varphi '' + 2 \mathcal H \delta \varphi' - \nabla^2 \delta \varphi + a^2 \delta \varphi \frac {d^2V}{\ud\varphi^2}
= 2 (\varphi'' + 2 \mathcal H \varphi' ) \Phi + \varphi' ( \Phi' +3 \Psi')~,
\end{equation}
and, after some algebra, one can find a single evolution equation for $\Phi$ which fully represents the only remaining scalar degree of the freedom
\beq\label{eq:Phi equation of motion}
\Phi^{\prime\prime} + 2\left(\mathcal{H}-\frac{\varphi^{\prime\prime}}{\varphi^\prime}\right)\Phi^\prime - D_k D^k \Phi + 2\left(\mathcal{H}^\prime - \mathcal{H}\frac{\varphi^{\prime\prime}}{\varphi^\prime}\right)\Phi = 0. 
\eeq

As we will see below in section~\ref{sec:quant_scalar}, the above equation can be rewritten in a simpler form by making use of the so-called Mukhanov-Sasaki variable,  
\be
 v = - \frac{a\sqrt 2} \Mpl  \left(\delta \varphi   + \varphi^\prime \frac\Phi{\mathcal{H}} \right) \qquad  \text{and} \qquad   z  \equiv  a \frac {\varphi'} { \mathcal H }  ~,
\ee
which turns \Eq{eq:Phi equation of motion} into a parametric oscillator
\be \label{eq:MSasaki_1}
  v^{\prime\prime }  + \left( D_kD^k - \frac{z^{\prime\prime}}{z} \right) v = 0 ~. 
\ee

\subsection{Vector perturbations}

The metric with only vector perturbations in the Newtonian gauge reads
\begin{equation}
\ud s^2_\mathrm{vector} = a^2 (\eta) \left\{  - \ud \eta^2  + 2 B_i \ud x^i \ud\eta  +  \left[\delta_{ij} +  2 \partial_{(i} E_{j)} \right] \ud x^i \ud x^j  \right\}~. 
\end{equation}
Because the perturbed energy-momentum tensor $\delta T\munu$ for a scalar field does not contain any source of vector perturbations, the first-order Einstein equations are 
\begin{align}
 \delta G^0_i |_{\rm vector}  &= - \frac {1} {2a^2}  D_kD^k \left(E^\prime_i - B_i \right)  = 0 ~, \\
 \delta G^i_j |_{\rm vector}  &= \frac{\delta^{ik}}{a^2} D_{(k} \left\{ \left[ E_{j)}^\prime - B_{j)} \right]^\prime  + 2 \mathcal{H} \left[ E_{j)}^\prime - B_{j)} \right]\right\}  = 0 ~. 
\end{align}
Using the gauge invariant quantity $  \Phi_i \equiv E_i^\prime - B_i$ in the previous equations, we obtain  
\begin{align}
  D_kD^k \Phi_i  = 0 ~, \\
 \Phi_i' + 2 \mathcal H \Phi_i = 0~.
\end{align}
which  the later equation corresponds to a quickly decaying modes with no sources. Since ${\Phi_i \propto a^{-2} }$, and because $a$ grows nearly exponentially during inflation, it is justified to neglect vector perturbations during inflation. 
\subsection{Tensor perturbations}

The metric for the tensor perturbations reads
\begin{equation}
 \ud s^2_\mathrm{tensor} = a^2(\eta) \left[ - \ud \eta^2 + (\delta_{ij} + h_{ij})  \ud x^i \ud x^j \right]~,
\end{equation}
where we note that the metric perturbations $h_{ij}$ are already gauge invariant. It is convenient to express the two degrees of freedom  in  $h_{ij}$ as
\begin{equation}
 h_{ij} = a^2 \left(
\begin{array}{lll}
h_+ & h_\times & 0 \\
h_\times & h_+  & 0 \\
0 & 0 & 0\\  \end{array} \right)
~,
\end{equation}
where $ \alpha = +,\times $ label the two polarizations. 
In a similar manner as for the scalar perturbations, one finds that the first order perturbed Einstein equations are
\begin{equation}
h_{\alpha}'' +  \left ( 2 \mathcal H  + D_kD^k  \right)  h_{\alpha} = 0,
\end{equation}
which  can also be rewritten in the following form
\be \label{eq:MSasaki_2}
 {\rm u}^{\prime\prime}_\alpha  + \left( D_kD^k   - \frac{a''}{a}\right) {\rm u}_{\alpha} = 0  ~, \qquad  \text{with} \quad   {\rm u}_\alpha = \frac{a\Mpl}2  h_\alpha ~. 
\ee

\section{Quantum fluctuations during inflation}

\begin{figure}[htbp!]
\includegraphics[width=0.990\textwidth]{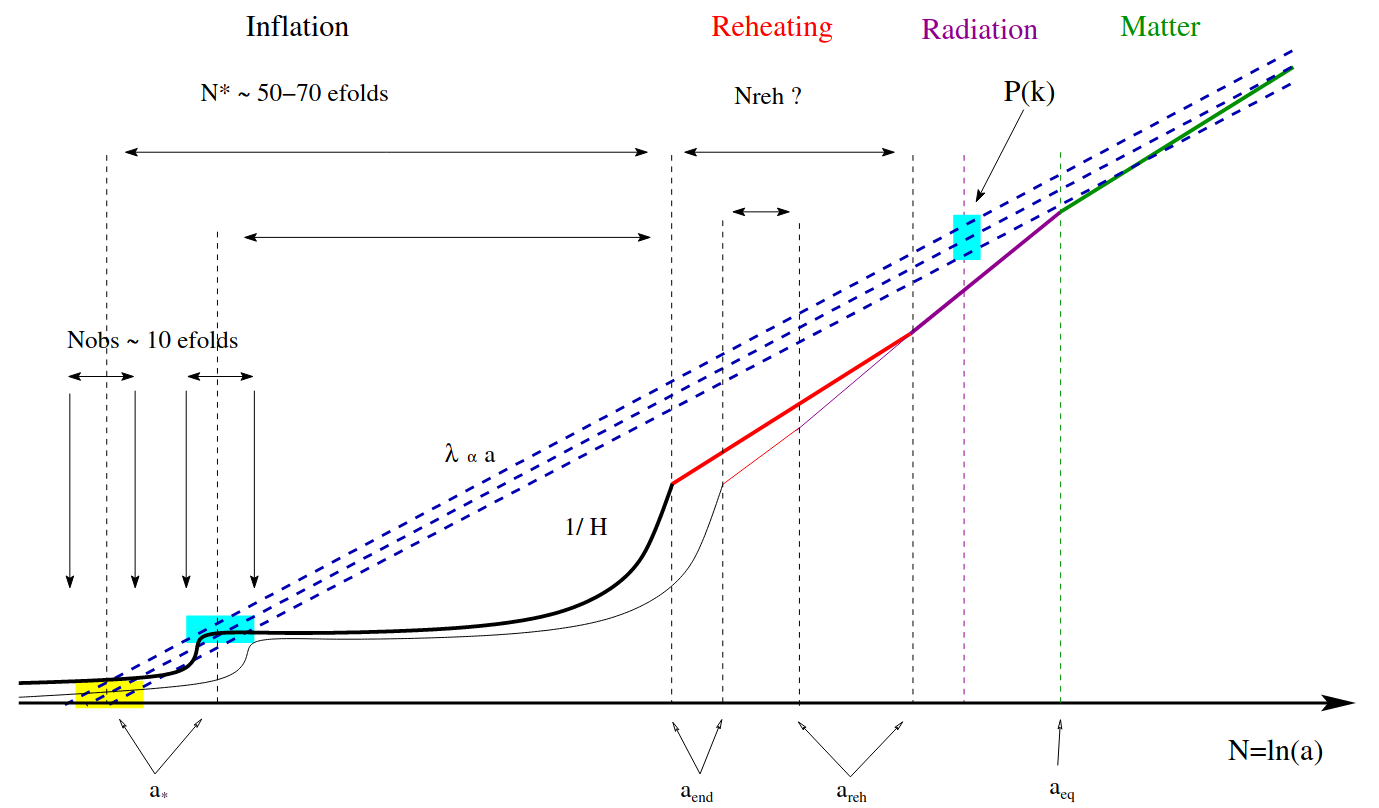}
\caption{This diagram illustrates the evolution of inflationary perturbations (blue dashed lines) and the Hubble radius (solid lines) in the cosmic history of the Universe. The horizontal axis represents the number of efold since the onset of the perturbations, and the vertical axis shows the logarithmic measure of lengths. Once perturbations exit the Hubble volume during inflation, their amplitude remains constant until they re-enter again the causal volume at a later stage. This figure is taken from Ref.~\cite{Ringeval:2007am}. 
}
\label{fig:modes_reentry}
\end{figure}

One of the most important successes of inflation is that when combined with quantum mechanics, it provides a natural explanation of the CMB anisotropies and the large-scale structure of the Universe. The previous section has prepared the ground for studying these primordial  quantum fluctuations generated during inflation. We have found that the evolution of scalar and tensors modes, \Eqs{eq:MSasaki_1} and (\ref{eq:MSasaki_2}), correspond to a parametric oscillator whose amplitude ``freezes'' on super-Hubble scales (see below). On the other hand, in the absence of vector sources, we have seen that vector modes can be safely neglected after a few efolds of inflation. Hence, the ``big picture'' of inflation is that modes are generated quantum mechanically at small sub-Hubble scales, and then they are quickly stretched by the expansion beyond the Hubble radius and freeze in amplitude. Once inflation finishes, the Universe begins the phase of decelerated expansion, allowing super-Hubble modes to ``re-enter'' the causal domain,  perturbing the post-inflationary Universe. See Fig.~\ref{fig:modes_reentry} for an illustrative diagram.

In this section, we consider the quantum generation and evolution of scalar and tensor perturbations that will allow us to formulate predictions on the primordial power spectrum.

\subsection{Scalar perturbations: \label{sec:quant_scalar} }  %
Knowing that the scalar perturbations consist of only one degree of freedom, it is convenient to describe the system in the comoving curvature gauge  which sets $
\delta\varphi = 0$. In this gauge, the scalar curvature is defined in a gauge invariant quantity as 
\be
\zeta \equiv C + \mathcal{H}\frac{\delta\varphi}{\varphi^{\prime}}~. 
\ee
After a rather lengthy analytical expansion \cite{Maldacena:2002vr}, 
 the action at second order in perturbations reads
\be
\begin{split}
\delta S^{(2)}_{\rm scalar}  & =  \frac 12 \int \ud^4 x ~ a^3 \frac{\dot\varphi^2}{H^2} \left[ \dot\zeta^2 + a^{-2}(\partial_i \zeta)^2 \right]
\\  
& =    \frac 12 \int \ud \eta \int  \ud^3 x  \left[  (v^{\prime})^2 + (\partial_i v)^2  + \frac {z^{\prime\prime}}z v^2 \right] ~, 
\end{split}
\ee
where the {Mukhanov-Sasaki variable} has been introduced, which in this gauge is simply given by   
\be
\begin{split}
v & \equiv   z \zeta    ~, 
\end{split}
\qquad  \text{with} \quad   z  \equiv  a \frac {\dot\varphi} { H } = a \sqrt {2 \epsilon_1} \Mpl 
~.
\ee
In Fourier space, quantum fluctuations of this scalar degree of freedom can be treated by promoting the Mukhanov-Sasaki variable to a quantum operator using the usual annihilation and creation operators, 
\be
  v \quad \rightarrow \quad  \hat v = \int \frac {\ud {\bm k}^3}{(2\pi)^3} \left[ v_{\bm k}(\tau) \hat a_{\bm k} e^{ikx} +  v_{\bm k}^* (\tau) \hat a_{\bm k}^\dagger e^{-ikx}\right] ~,
\ee
where $\hat a_{\bm k}$ and $\hat a_{\bm k}^\dagger$ satisfy the conventional commutation relation 
$[\hat a _{\bm k},\hat a _{\bm k}^+ ] = i \delta^{(3)} (\bm k - \bm k' )$. 
The vacuum state is then defined by $\hat a _{\bm k} | 0 \rangle = 0$.  The equation of motion for each Fourier mode is then given by 
\beq
\label{eq:MS_equation}
v^{\prime\prime}_\bmk +\left(k^2- \frac{z^{\prime\prime}}{z}\right)v_\bmk = 0~,
\eeq
recovering \Eq{eq:MSasaki_1}, where the effective mass term at first order in slow-roll parameters is given by 
\be \label{eq:zpp}
\frac{z^{\prime\prime}}{z} = 
\frac{\nu_\mathrm{S}^2  -1/4}{\eta^2}
~, \qquad \text{with } \quad \nu_\mathrm{S} = \frac 32 + \epsilon_1 - \frac 12 \epsilon_2  + \mathcal{O}(\epsilon^2)~. 
\ee

Taking into account that during slow-roll inflation, in the $k/\mathcal{H} \rightarrow \infty$ limit, the solution of Eq.~(\ref{eq:MS_equation}) reduces to the Bunch-Davies vacuum i.e. ${v_\bmk = e^{ik\eta}/\sqrt{2k}}$, this fixes the general solution to 
\be
v_\bmk (\eta) = \frac{\sqrt{-\pi\eta}}{2}    e^{i\frac\pi2 (\nu_\mathrm{S} + \frac 12)} H_{\nu_\mathrm{S}}^{(1)}(-k\eta) ~, 
\ee
where $H_{\nu_\mathrm{S}}^{(1)}$ is the Hankel function of the first kind. At super-Hubble scales, that solution becomes 
\be
v_\bmk (\eta) =  e^{i\frac\pi2 (\nu_\mathrm{S} - \frac 12)}   \frac  {\Gamma(\nu_\mathrm{S})}{\Gamma(3/2)}   \frac{   2^{\nu_\mathrm{S} - \frac 32}  }{\sqrt{2k}}  \left(-k\eta\right)^{-\nu_\mathrm{S} + \frac 12}  \quad \text{for } \  \frac k {\mathcal H} \ll 1 ~.
\ee

To recover an expression for the curvature perturbation, we first need to find the explicit time-dependent function of $z(\eta)$. This can be done by integrating \Eq{eq:zpp} and choosing a normalization such that at the time of Hubble crossing it satisfies  $\zeta(\eta_*) = a_* \sqrt{2\epsilon_{1*}} \Mpl$. This yields to the following expression
\be
z(\eta) = a_* \sqrt{2\epsilon_{1*}} \Mpl \left(  \frac {\eta}{\eta_*} \right)^{1/2 - \nu_\mathrm{S}} ~,
\ee
where the asterisk symbol ($*$) denotes the quantities at Hubble crossing. We are now in a position where we can compute the scalar curvature perturbation 
\be
\zeta_\bmk  = \frac {v_\bmk (\eta)} {z(\eta)}  %
~,
\ee
which in the  $k/\mathcal{H} \rightarrow 0$ limit $\zeta_\bmk$ tends to a constant.

We can now compute the two-point correlation function of scalar perturbations, which is given by 
\be
\begin{split}
 \langle \zeta_\bmk \zeta_{\bmk'} \rangle  & =  \left| \zeta_\bmk \right|^2 \delta \left( \bmk + \bmk' \right) 
 \\
                                           & =   (2\pi)^3 \delta \left( \bmk + \bmk' \right)  P_\zeta (k) ~,
\end{split}
\ee
where $ P_\zeta (k)$ is the scalar power-spectrum, which can be re-expressed in a dimensionless quantity, so that  
\be
    \mathcal P_\zeta (k) =  \frac{k^3}{2\pi^2}  P_\zeta (k)  
    ~.
\ee

In the limit when  $\epsilon_1, \epsilon_2, ... \rightarrow 0$ (and thus $\nu_\mathrm{S} = 3/2$), the power spectrum reads
\be
 \mathcal{ P}_{\zeta 0} =  \frac 1 {8\pi^2} \frac{H_*^2}{ \Mpl^2 \epsilon_{1*}} ~, 
\ee
where we have used the approximation $\eta_* \simeq - (1 + \epsilon_{1*}) / (aH_*)$.  

Instead, expanding it in slow-roll parameters around the pivot scale $k_*$, one gets~\cite{Peter:2013avv}
\be
\mathcal{ P}_{\zeta} (k) =  \mathcal{ P}_{\zeta 0} \left[ a_0^{(\rm S)} + a_1^{(\rm S)} \ln \left( \frac k {k_*} \right) + \  ... \ \right]  ~,
\ee
where the zeroth and first expansion coefficients are
\begin{align}
 a_0^{(\rm S)} & = 1 - (2C+1)\epsilon_{1*} - C\epsilon_{2*} + \mathcal{O} (\epsilon^2) ~,
 \\
 a_1^{(\rm S)} & =  -2\epsilon_{1*} - \epsilon_{2*} + \mathcal{O} (\epsilon^2) ~,
\end{align}
with $C\equiv \gamma_\mathrm{Euler} + \ln 2 - 2$ and $\gamma_{\mathrm{Euler}} \simeq 0.5772$ is the Euler's constant. 

At first order in slow-roll parameters, the deviation of the scalar power-spectrum from scale invariance is given by the scalar spectral index, 
\be
 n_\mathrm{S} - 1 \equiv \frac{\ud \ln \mathcal P_{\zeta} }{\ud \ln k} = -2\epsilon_{1*} - \epsilon_{2*} + \mathcal O (\epsilon^2)~.
\ee

\subsection{Tensor perturbations: } 

The case for tensor is analogous to scalars only that now the equation of motion is given by 
\be
\label{eq:MS_T_equation}
{\rm u}^{\prime\prime}_{\bmk\alpha} +\left[k^2- \frac{a^{\prime\prime}}{a} \right] {\rm u}_{\bmk\alpha} = 0 ~, 
\ee
where the effective mass at first order in slow-roll parameters  is given by 
\be
\frac{a^{\prime\prime}}{a} =
\frac{\nu_\mathrm{T}^2  -1/4}{\eta^2}
~, \qquad \text{with } \quad \nu_\mathrm{T} \simeq \frac 32 + \epsilon_1 ~. 
\ee

In that case, taking into account the two possible polarizations, the dimensionless tensor power-spectrum reads 
\be
    \mathcal P_h (k) =  \frac{k^3}{2\pi^2}  \sum_{\alpha = +,\times } P_{h\,\alpha} (k)  ~,
\ee
and after the expansion in slow-roll parameters, one gets
\be
\mathcal{P}_{h} (k) =  \mathcal{P}_{h 0} \left[ a_0^{(\rm T)} + a_1^{(\rm T)} \ln \left( \frac k {k_*} \right) + \ ... \  \right] ~,
\ee
where now 
\be
 \mathcal{P}_{h 0} =  \frac{2 H_*^2}{ \pi^2 \Mpl^2} ~,
\ee
and the coefficients are 
\begin{align}
 a_0^{(\rm T)} & = 1 - (2C+1)\epsilon_{1*}  + \mathcal{O} (\epsilon^2) ~,
 \\
 a_1^{(\rm T)} & =  -2\epsilon_{1*}  + \mathcal{O} (\epsilon^2) ~.
\end{align}

Finally, the tilt of the tensor power-spectrum is then given by  
\be 
n_\mathrm{T} \equiv \frac{\ud \ln \mathcal P_{h} }{\ud \ln k} =  -2\epsilon_{1*} + \mathcal{O} (\epsilon^2) ~.
\ee

\section{Contact with observations}

Assuming the slow-roll regime, the amplitude and the spectral tilt of the scalar and tensor power spectra can thus be derived at first order in slow-roll parameters. These can be expressed in terms of the potential and its derivatives, providing a prediction that can be tested in the CMB anisotropy observations. The latest measurements  by the Planck experiment \cite{Akrami:2018odb} set the scalar power-spectrum amplitude and spectral index to 
\be
\begin{split}
A_\mathrm{S} &= \mathcal{P}_\zeta (k_*) = (2.098 \pm 0.101) \times 10^{-9}~, \\
n_\mathrm{S} &= 0.968 \pm 0.004 
~,
\end{split}
\ee
corresponding to a deviation from scale-invariance exceeding the 5$\sigma$ level.

So far, no significant evidence for primordial tensor perturbations has been found yet, which has already ruled out some models. This is done by the combined prediction using the spectral index $n_{\mathrm{S}}$, and the tensor-to-scalar ratio $r$, which is defined as 
\begin{equation}
r \equiv \frac{\mathcal P_h}{\mathcal P_\zeta} \simeq -8 n_\mathrm{T} \simeq 16 \epsilon_{1 *}~.
\end{equation}
Planck data sets the upper bound limits on the tensor-to-scalar ratio to 
\be
r < 0.07~. 
\ee

The most favoured models by to-date observations are those with a plateau-shape potential, as it is the case of Starobinsky inflation (or $R^2$ inflation), and Higgs inflation with non-minimal coupling to gravity.  
These models predict a $n_\mathrm{S} \simeq 0.965$ and $r \simeq 0.0035$ assuming a minimal number of efolds such that ${N_* = 55}$. These values are in very much in agreement with the latest observations\cite{Akrami:2018odb}.  More details about Starobinsky and Higgs inflation can be found in the Appendix~A.

\begin{figure}[h!]
\centering
\includegraphics[width=0.990\textwidth]{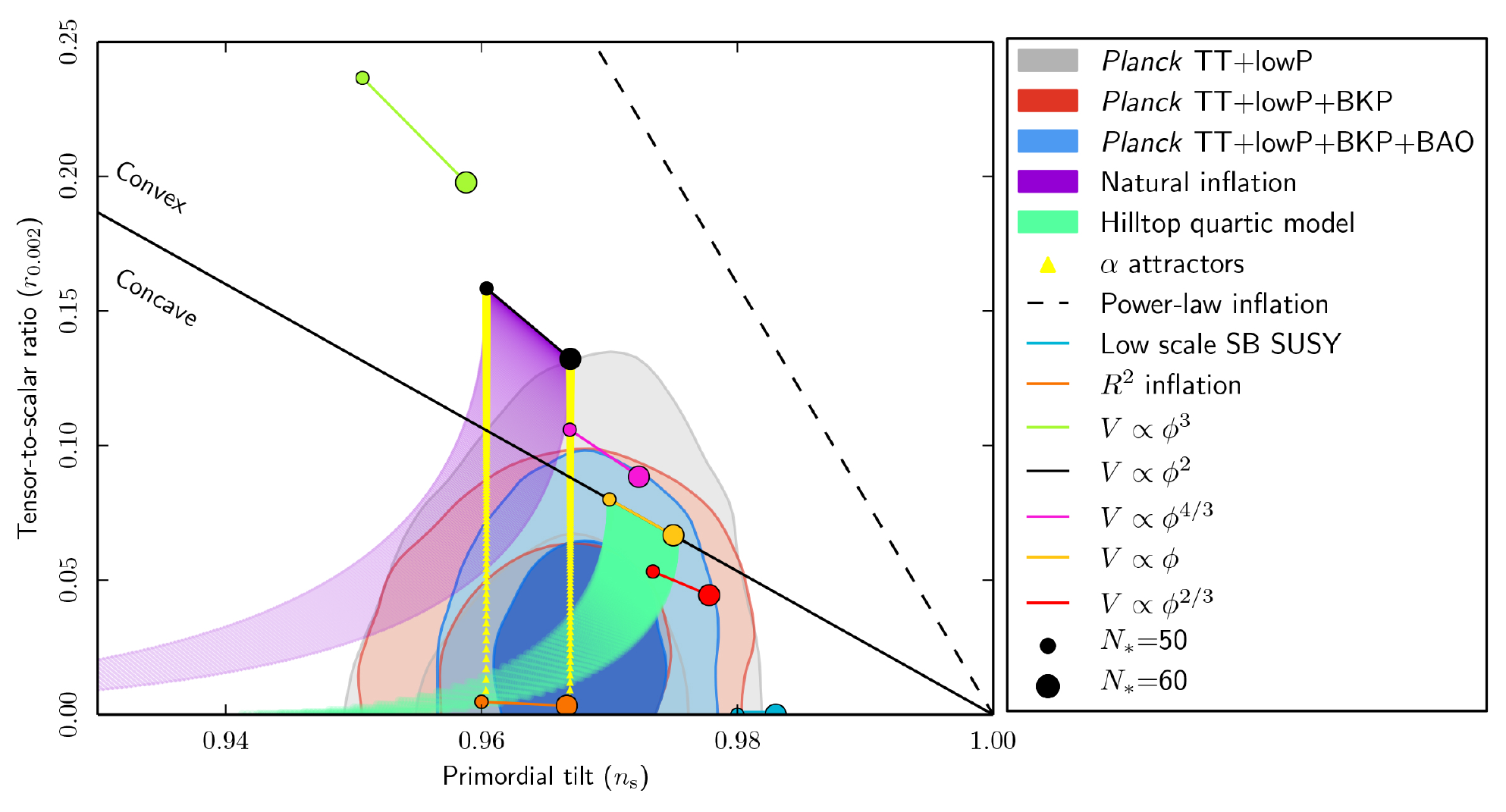} 
\caption{
\label{fig:inflation_constraints}
Constraints for the spectral index and the scalar-to-tensor-ratio with reference $k_* = 0.002\ \Mpc^{-1}$ constructed in a combination of data sets from Planck, BICEP2/Keck Array and  BAO. The legend shows the prediction of a selection of inflationary models. Figure taken from Ref.~\cite{Akrami:2018odb}.
}
\end{figure}

\clearpage

\section{Scalar fields with non-minimal coupling} \label{sec:nonminimalSFs}

In this section, we introduce the covariant formalism for gravitating scalar fields, which is later used in Chapter \ref{p2_chap:prepaper2} in the context of Higgs inflation. In here, though, we do not assume any particular inflationary potential, and we allow the scalar fields to be non-minimally coupled to gravity, with a term proportional to $\sim \xi \bar R \bar\phi^2$. These non-minimal coupling terms are the ones naturally generated by the one-loop order of quantum corrections to the Einstein-Hilbert action of a \textit{minimally} coupled scalar field~\cite{Steinwachs2014}, where $\xi$ is a free parameter corresponding to the coupling strength.

Let us begin by considering a universe containing an arbitrary number of scalar fields $\bar \phi^I$, labelled by Latin capital letters $I, J, K = 1, 2, ... , N$, which can be non-minimally coupled to gravity with coupling strength   $\xi_I$. The action now reads
\be
\begin{split}
S = \int d^4 x & \sqrt{-\bar{g} } \Big[ f (\bar\phi^I ) \bar{R} 
 - \frac{\Mpl^2}{2} \delta_{IJ} \bar{g}^{\mu\nu} \partial_\mu \bar\phi^I \partial_\nu \bar\phi^J - U(\bar\phi^I ) \Big] ~ ,
\end{split}
\label{eq:action_J}
\ee
where $U(\bar\phi^I)$ is the scalar field potential, and $f(\bar\phi^I)$ is a function of the fields such 
\beq \label{eq:f_func}
 f (\bar\phi^I ) = \frac {\Mpl^2} 2 \left( 1 + \sum_K^N \xi_K \bar\phi_K^2 \right) ~.
\eeq

This action is written in what is known as the Jordan frame, because the non-minimal couplings between the fields and the Ricci scalar are written explicitly. Note that the evolution equations of such a system do not correspond to the Einstein equations, as derived in section \ref{sec:EinsteinEquations}. However, it is possible to make a transformation into a new frame, the so-called Einstein frame, where both the metric and scalar fields have been redefined in such a way that non-minimal couplings terms are hidden in the new transformed variables. We are using the notation where the variables with an upper-bar or ``hat'' are described in the Jordan frame, while those without upper-bar are those redefined in the Einstein frame. 

The transformation into the Einstein frame consists of rescaling of the metric tensor, such as 
\bea
\bar{g}_{\mu\nu} (x) \rightarrow g_{\mu\nu} (x) = \frac{2}{M_{\rm pl}^2} f \big(\bar\phi^I\big) \> \bar{g}_{\mu \nu} (x) ~.
\eea

Rewriting the action using the new metric, but keeping for the moment the scalar fields in the original Jordan-frame form, reads 
\beq 
\begin{split}
S = \int d^4 x \sqrt{-g} \Big[ R & - \frac{\Mpl^2}{2} {\cal G}_{IJ} (\bar\phi^K ) g^{\mu\nu} \partial_\mu \bar\phi^I \partial_\nu \bar\phi^J  - V (\bar\phi^I ) \Big] \, ,
\end{split}
\label{eq:action_Einstein}
\eeq 
where ${\cal G}_{IJ} (\phi^K)$ is a field-space metric containing the mixing with the non-minimal coupling, 
\beq
{\cal G}_{IJ} (\bar\phi^K ) = \frac{ M_{\rm pl}^2 }{2 f (\bar\phi^K) }\left[ \delta_{IJ} + \frac{3}{ f (\bar\phi^K) } \frac{\partial f  (\bar\phi^K) } {\partial \bar\phi^I} \frac{\partial f (\bar\phi^K) } {\partial \bar\phi^J} \right] ~,
\label{eq:G_IJ}
\eeq
and the field potential has been redefined as 
\beq
V (\bar\phi^I) = \frac{\Mpl^4}{ 4 f^2 (\bar\phi^I) } U(\bar\phi^I ) .
\label{VE}
\eeq

After this transformation, we recover the metric equations of motion to be the Einstein equations, where now the energy-momentum tensor reads 
\beq
T_{\mu\nu} = {\cal G}_{IJ} \partial_\mu \bar\phi^I \partial_\nu \bar\phi^J - g_{\mu\nu} \left[ \frac{1}{2} {\cal G}_{IJ} \partial_\alpha \bar\phi^I \partial^\alpha \bar\phi^J + V (\bar\phi^I ) \right] ~,
\label{eq:Tmn_J}
\eeq
and the evolution of the fields is described by 
\beq
g^{\mu\nu} \nabla_\mu\nabla_\nu \bar\phi^I + g^{\mu\nu} \Gamma^I_{JK} \partial_\mu \bar\phi^J \partial_\nu \bar\phi^K - {\cal G}^{IJ} \frac{\partial }{\partial {\bar\phi}^J} V (\bar\phi^K) = 0 ~.
\label{eq:eomsf_J}
\eeq
We note the additional terms containing the Christoffel symbols  $\Gamma^I_{JK} (\phi^L)$, which are constructed from the field-space metric ${\cal G}_{IJ}$.
\ \\

In addition, one can also choose to transform the scalar-fields $\bar\phi^I$ into a canonically normalized form in the Einstein frame, say
$\Phi^I$. One has to solve the following system of equations 
\beq \label{eq:convert_frame}
\frac{\Mpl^2}{2} {\cal G}_{IJ} g^{\mu\nu} \partial_\mu \bar\phi^I \partial_\nu \bar\phi^J = {\delta}_{IJ} g^{\mu\nu} \partial_\mu \Phi^I \partial_\nu \Phi^J ~.
\eeq

This transformation further simplifies the action in Eq.~(\ref{eq:action_Einstein}), recovering the classical  Klein-Gordon equations for the evolution of the fields,  
\beq
\Box \Phi^I -\frac{\partial }{\partial \Phi^I} V (\Phi^K) = 0 ~.
\eeq
However, such transformation cannot be found analytically for the whole-field space when the manifold represented by  ${\cal G}_{IJ}$ is curved. For the single scalar-field case, though, this is never a problem.

\section{Before and after inflation }  %

\subsection*{The initial conditions issue}

As we have seen, in the theories of inflation, the Universe undergoes an early phase of nearly exponential, accelerated expansion. Inflation naturally solves the horizon and the flatness problems inherent in the original Big Bang model. In the simplest case, the inflaton corresponds to of scalar field that slowly rolls down a logarithmically flat potential. Quantum fluctuations during inflation predict adiabatic and nearly scale-invariant curvature power spectrum, which matches with observations from the CMB. In addition, these fluctuations provide the seeding mechanism for structure formation in our Universe. The latest results from Planck~\cite{Akrami:2018odb,Ade:2015lrj} favour the plateau-like potentials ~\cite{Martin:2013nzq}, and particularly, the Starobinsky and Higgs inflation model.

Still, inflation takes place when the scalar field potential dominates the energy density of the Universe, and thus, the scalar field's kinetic and gradient energies must be subdominant. At first glance, it seems that to explain the homogeneity of the CMB, one introduces a mechanism (inflation) that already requires a significant amount of homogeneity, to begin with. This is probably why the initial conditions for inflation have been a subject of debate and controversy for the last thirty years. In chapters~5 and 6, we will aim to answer the following question: Can generic (inhomogeneous) preinflationary scenarios successfully lead to the beginning of inflation?

\subsection*{The reheating phase}

The original Hot big bang model rapidly acquired broad community acceptance once the CMB was discovered. Indeed, the CMB was a consequence of the existence of a hot primordial plasma at an earlier stage of the Universe's history. Therefore, the attempts to solve the issues of homogeneity and flatness, as well as explaining the origin of primordial perturbations by including the inflationary phase, must be linked to a mechanism to create such a primordial plasma. %
Indeed, the rapid and accelerated expansion in scalar field inflation is driven by the scalar-field condensate in its ground state, or ``vacuum", and, thus, by the potential. This scenario is very much the opposite of the hot dense plasma which it is supposed to be linked with. The (p)reheating phase is thus a mechanism where the energy stored in the inflaton condensate is efficiently transferred to large amounts of particle production, creating the plasma. A common preheating mechanism is based on the parametric resonances occurring when the inflaton field rapidly oscillates around the potential's minimum after the end of inflation.

\begin{figure}[b!]
\centering
\ \\[10mm]
\includegraphics[width=0.80\textwidth]{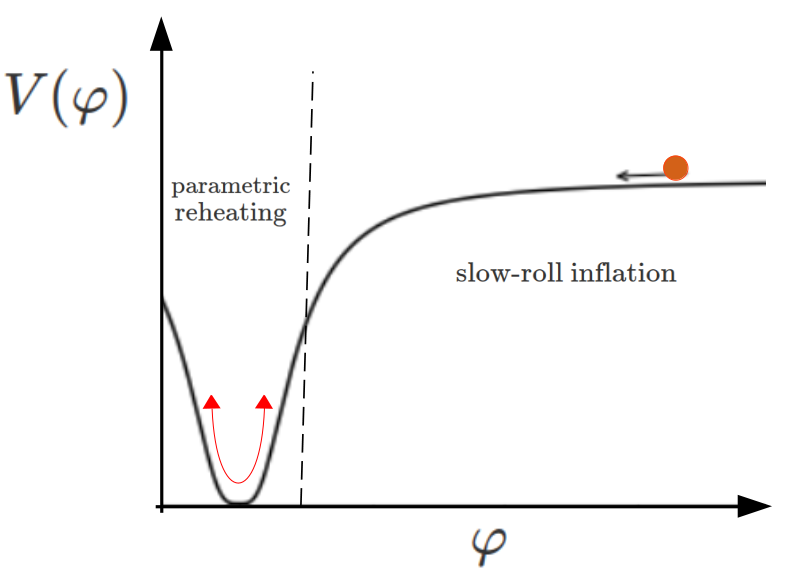} 
\caption{
\label{fig:reheating_diagram}
Illustration of the inflaton trajectory during slow-roll inflation at the potential's plateau and the posterior preheating phase once the inflaton reaches the potential's minimum. 
} 
\end{figure}

In the last decade, the phase of preheating has been studied exhaustively using numerical lattice simulations under the assumption of linearized Einstein equations, which ignores the backreaction effects of metric fluctuations. In this thesis, in Chapter~\ref{p2_chap:prepaper2},  we present our investigations of the preheating mechanism in full gravitational glory assuming the Higgs inflation model. 
\ \\

The questions of the initial conditions and the preheating are very hard to tackle purely by analytical or perturbative methods, as these method do not account for relevant non-perturbative phenomena. Thus, numerical methods are needed to consider them, specially with a full gravitational treatment. In the next chapter, we present the formalism that we will use for numerical General Relativity simulations.

%
%
%

\chapter{Inhomogeneous Cosmology}
\label{chap:NR}
\pagestyle{fancy}

We now proceed to study cosmological processes in full General Relativity. This is necessary for scenarios where the non-linear dynamical equations of Einstein equations become relevant, for example when concerning inhomogeneous systems with large energy fluctuations and/or systems with strong gravity regions. We discuss the 3+1 formalism, which yields the well-established Arnowitt-Deser-Misner (ADM) equations \cite{Arnowitt:1962hi}, and its later adaptations to make them suitable for numerical simulations, solvable by Cauchy integration within numerical accuracy. Finally, we discuss its implementation to study fundamental physics and cosmology, focusing on scalar field systems in the context of inflation.

\section{Foliations of spacetime \label{sec_foliation}}

\begin{figure}[tb!] 
\vspace*{5mm}
\centering
\includegraphics[width=0.990\textwidth]{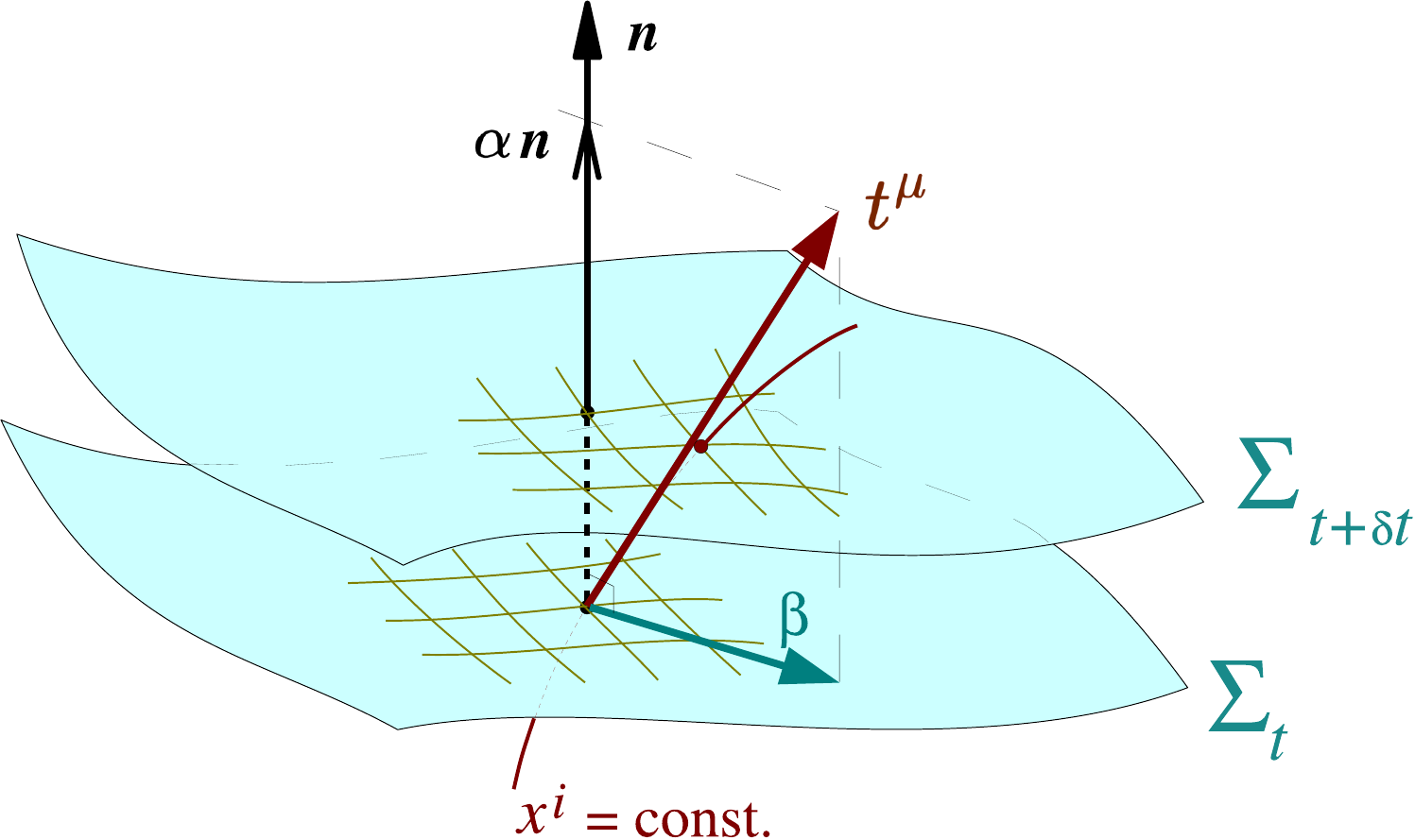} 
\caption{ Representation of the foliation of spacetime in the (3+1) formalism. The two hypersurfaces $\Sigma_t$ and $\Sigma_{t+\delta t}$ represent the system at two different times. Figure adapted from Ref.~\cite{Gourgoulhon:2007ue}. 
\label{Fig:Foliation}
} 
\vspace*{11mm}
\end{figure}

Assuming that spacetime is a 4-dimensional manifold $\mathcal M$ represented by the metric $g_{\mu\nu}$, we want to reformulate the dynamical system in a 3+1 form such that the spacetime manifold is split into a collection of 3-dimensional, spatial and non-intersecting hypersurfaces $\Sigma_t$. That way, we can parametrize a time curve as
\be
t^\mu = \alpha n^\mu + \beta^i e^\mu_i \qquad e^\mu_i = \left(\frac{d\bar x^\mu}{dx^i} \right)_t  
~, 
\ee
where $n^\mu$  and $e^\mu_i$ are the unitary normal and tangential vectors with respect $\Sigma_t$, respectively. The lapse  $\alpha$ and shift functions $\beta^i$  define the spacetime foliation and correspond to the gauge choice (see illustration in \Fig{Fig:Foliation}).
In that sense, the 4-dimensional metric  $g^{\mu\nu}$ is split in  the following manner 
\be \label{splitmetric}
g^{\mu\nu} = -  n^\mu n^\nu  + \gamma^{ij} e_i^\mu e_j^\nu ~, 
\ee
where $\gamma^{\mu\nu} \equiv \gamma^{ij} e_i^\mu e_j^\nu  $ is the 3-dimensional metric (or \textit{3-metric})  of  $\Sigma_t$.  By construction, $n^\mu$ is defined timelike i.e. $n^\mu n_\mu = -1$, and reads 
\be
   n_\mu = (-\alpha, \vec 0)   ~,\qquad  n^\mu = \left( \frac 1 \alpha, -\frac {\beta^i} \alpha  \right)
   ~,
\ee

and, without loss of generality, the line element is now given by  
\be
ds^2 = - \alpha^2 dt^2 + \gamma_{ij}(dx^i + \beta^i dt)(dx^j + \beta^j dt)  ~.
\ee

The 4-dimensional Riemann curvature $ ^{(4)}R_{\mu\nu\delta\gamma} $ can be expressed in terms of spatial quantities: the so-called intrinsic and extrinsic curvature. The intrinsic curvature is simply given by the 3-dimensional Riemann tensor $ ^{(3)}R_{\mu\nu\delta\lambda}$ associated with the metric $\gamma_{\mu\nu}$, while the extrinsic curvature is defined by measuring the variation of $n^\mu$ after parallel transport throughout the metric,   
\be
K_{\mu\nu}  \equiv - \gamma_\mu^\sigma \nabla_\sigma n_\nu   = - ( \nabla_\mu n_\nu  + n_\mu n^\sigma \nabla_\sigma n_\nu )
~.
\ee
By definition, the extrinsic curvature is a purely spatial and tangent tensor so that $n^\mu K_{\mu\nu} =0$. Remarkably,  it can also be defined as the Lie derivative of the 3-metric with respect to the vector $n^\mu$~\cite{Gourgoulhon:2007ue}, 
\be \label{eq:K_LieDerivGamma}
K_{\mu\nu} = -\frac12{\cal L}_n \gamma_{\mu\nu} ~. 
\ee

\section{Dynamics of spacetime \label{ADM_form} }

The next step is to redefine the Einstein equations within the framework of the 3+1 formalism and to define a set of constraint equations that ensures that General Relativity is satisfied, as well as a set of evolution equations that allow us to evolve the variables ($\gamma_{\mu\nu},\ K_{\mu\nu)}$ forward in time. To do so,  we project the 4-dimensional Riemann tensor using the splitting of the metric as it is defined in \Eq{splitmetric}. 

After some elegant algebra \cite{Alcubierre:1138167,Gourgoulhon:2007ue}, one finds the so-called \textit{Gauss} equations,
\be \label{GaussCodazzi}
\gamma^\alpha_\mu  \gamma^\beta_\nu  \gamma^\gamma_\sigma \gamma^\delta_{\lambda}\ ^{(4)}R_{\alpha\beta\gamma\delta}  = \ ^{(3)}R_{\mu\nu \sigma\lambda} + K_{\mu \sigma}K_{\nu \lambda} + K_{\mu \lambda}K_{\nu \sigma}
~, 
\ee
the \textit{Codazzi-Mainardi} equations,
\be \label{GaussMainardi}
\gamma^\alpha_\mu  \gamma^\beta_\nu \gamma^\gamma_\sigma n^\delta\ ^{(4)}R_{\alpha\beta\gamma\delta} =  D_\mu  K_{\nu \sigma} -  D_\nu K_{\mu \sigma}
~, 
\ee
and the \textit{Ricci} equations, 
\be \label{RicciEq}
\gamma^\alpha_\mu  \gamma^\beta_\nu n^\gamma n^\delta\ ^{(4)}R_{\alpha\beta\gamma\delta} = {\cal L}_nK_{\mu\nu } + K_{\mu \sigma}K^\sigma\,_\nu + \frac1\alpha D_\mu D_\nu\,\alpha ~.
\ee

The projection of the matter sector  described by an arbitrary energy-momentum tensor, yields 
\begin{align} 
   \rho &\equiv n^\alpha n^\beta T_{\alpha\beta}   ~,  \label{Ham_RHS} \\
   S_\mu &\equiv \gamma_{\mu}^\alpha n^\beta  T_{\alpha\beta}  ~,     \label{Mom_RHS} \\
   S_{\mu\nu} &\equiv \gamma_{\mu}^\alpha \gamma_{\nu}^{\beta}  T_{\alpha\beta}  ~,     \label {eq:Smunu}\\
   S &\equiv \gamma^{\mu\nu} S_{\mu\nu} ~.   \label{eq:S_trace}
\end{align}

Noticing that the projections of the Einstein tensor can be expressed in terms of the twice-contracted Gauss and Codazzi-Mainardi equations, 
\be  \label{Ham_LHS}
\begin{split}
  2 n^\alpha n^\beta G_{\alpha\beta}  & =   \gamma^{\mu\sigma} \gamma^{\nu\lambda}\ ^{(4)}R_{\mu\nu\sigma\lambda}    \\
                           & = R + K^2 - K_{\alpha\beta}K^{\alpha\beta}    ~,    
\end{split}
\ee
and  
\be   \label{Mom_LHS}
\begin{split}  
  2 \gamma^\alpha_\mu n^\beta G_{\alpha\beta}  & =  - \gamma^{\alpha}_\mu \gamma^{\beta\gamma}  n^\lambda\  ^{(4)}R_{\alpha\beta\gamma\lambda}  \\
                                        & =  D_\alpha K_\mu^\alpha - D_\mu K      ~,
\end{split}
\ee
where $K \equiv K^i_i = \text{Tr} ( K_{ij})$. Thus, we can construct the Hamiltonian and Momentum constraint  equations by substituting in the Einstein equations the combination of \Eq{Ham_LHS} with (\ref{Ham_RHS})  and (\ref{Mom_LHS}) with (\ref{Mom_RHS}). This gives us

\begin{empheq}[box=\fbox]{align}
  {\cal H}  & \equiv R + K^2 - K_{ij}K^{ij} - \frac 2{\Mp^2} \pi\rho  = 0   ~,    \label{Hconts} \\
  {\cal M}_i & \equiv  D_j K^j_{\ i} -  D_i K  - \frac 1{\Mp^2} \pi S_i  = 0   ~.    \label{Mconts}  
\end{empheq}

These equations, (\ref{Hconts}) and (\ref{Mconts}), give us the necessary conditions so that a 3-dimensional hypersurface $\Sigma_t$ is a valid foliation that is embedded in the {4-dimensional} manifold $\mathcal{M}$. In other words, these conditions need to be satisfied for a gravitational system at any given time hypersurface. In addition, these equations are used to compute the valid initial data for numerical simulations.

The evolution of the 3-metric is given by \Eq{eq:K_LieDerivGamma}, by noticing that 
\begin{align}
 \begin{split}
  K_{ij} & = -\frac12{\cal L}_n \gamma_{ij}  \\
	 & = -\frac12{\cal L}_{(\frac t\alpha -\frac \beta\alpha) } \gamma_{ij}\\
	 & = -\frac1{2\alpha}\left(\partial_{t} \gamma_{ij} - {\cal L}_{\beta} \gamma_{ij} \right) ~,
 \end{split} 
\end{align}
which yields 
\be
\boxed{  \label{eq:Evo_3metric}
 \partial_{t} \gamma_{ij} - {\cal L}_{\beta} \gamma_{ij} = - 2 \alpha K_{ij} 
 ~.
 }
\ee

The evolution of the extrinsic curvature is derived using the Ricci equation (\ref{RicciEq}). The 4-Riemann tensor can be rewritten as 
\be
\gamma^\alpha_\mu  \gamma^\beta_\nu n^\lambda n^\delta\ ^{(4)}R_{\alpha\beta\lambda\delta}  = 
  \gamma^{\lambda\alpha}  \gamma^\delta_\mu \gamma^\beta_\nu    \ ^{(4)}R_{\alpha\beta\lambda\delta}  - \gamma^\beta_\mu \gamma^\alpha_\nu\ ^{(4)}R_{\alpha\beta}
\ee
where the first term can be substituted by the Gauss equation (\ref{GaussCodazzi}). Now, the second term can also be eliminated using the alternative form of the Einstein equations, 
\be
  \gamma^\beta_\mu \gamma^\alpha_\nu \ ^{(4)}R_{\alpha\beta} = \frac 1 {\Mpl^2}  \gamma^\beta_\mu \gamma^\alpha_\nu \left(T_{\alpha\beta} - \frac 12 g_{\alpha\beta} T \right) ~, 
\ee
with $  T \equiv  g^{\mu\nu} T_{\mu\nu}   $. 

Putting everything together, using the definition of $\rho$, $S_{\mu\nu}$ and $S$ in Eqs.~(\ref{Ham_RHS}-\ref{eq:S_trace}), the  evolution equation for $K_{\mu\nu}$ reads

\be  \boxed{
\begin{split}   \label{eq:Evo_Extrinsic}
\partial_{t} K_{\mu\nu} - {\cal L}_{\beta} K_{\mu\nu} = - D_\mu D_\nu\,\alpha  &+ \alpha \left[\ ^{(3)}R_{\mu\nu} + K K_{\mu\nu} - 2 K_{\mu\lambda}K^\lambda\,_\nu \right]   
\\      & 
- \frac\alpha{\Mpl^2} \left[ S_{\mu\nu} + \frac12 \gamma_{\mu\nu}(\rho - S) \right]  ~.
\end{split}   }
\ee

The framework developed in this section yielding to the Hamiltonian and Momentum constraints \Eqs{Hconts} and (\ref{Mconts}), and the evolution equations of $\gamma_{\mu\nu}$ and $K_{\mu\nu}$ in \Eqs{eq:Evo_3metric} and (\ref{eq:Evo_Extrinsic}) are usually known as the ADM equations after the work of Arnowitt, Deser and Misner in \Ref{Arnowitt:1962hi} (see also \cite{Anderson:1998we}). In their final form, all these equations are defined by purely spatial quantities, and therefore, they can be more conveniently rewritten using only the (Latin) spatial indices.

\section{Numerics and stability}      

The ADM system of equations is, in principle, already in a form apt for numerical implementation: it contains a set of constraint equations which ensures the validity of initial and evolved data with respect to General Relativity, and also contains a complete set of evolution equations that  can be used  to evolve the gravitating system forward in time.

However, the current formulation is \textit{not} well-possed, and numerical integration of non-trivial situations typically fail due to the unbounded growth of constraint-violating modes arising from numerical errors. The desired property of \textit{well-possedness} requires a system of evolution equations that are \textit{strongly  hyperbolic}\footnote{
Assuming a system of equations like $\partial_t  \vec u + M^i\partial_i \vec u = \vec s(\vec u) $ where $\vec u$ is a vector of the evolved variables, $M^i$ the characteristic matrix and $\vec s$ a source term. When $M^i$ has a complete set of real eigenvalues, then the system is said to be weekly-hyperbolic. If, in addition,  $M^i$ also has a complete said of eigenvectors, then the system is known to be strongly hyperbolic~\cite{Alcubierre:1138167}. 
}. Instead, the ADM formalism in 3 or higher spatial dimensions is found to be \textit{weekly-hyperbolic} due to the propagation of an unphysical scalar mode present in the (crossed) second-order partial derivatives of the metric tensor \cite{2016bookShibata}.  For a more detailed discussion on this problem,  we refer the reader to    %
the references listed at the end of this section. 

Indeed, taking a closer inspection into \Eq{eq:Evo_Extrinsic}, it reveals that the evolution of the extrinsic curvature is dependent on the Ricci tensor, which, expressed in terms of second-derivatives of the metric, reads

\begin{equation} \label{eq::3R_problem}
\begin{split} 
^{(3)} R\ij = &- \frac 12 \partial_k\partial^k \gamma\ij + \frac12 \gamma^{kl}\left( \partial_i\partial_l \gamma_{kj} + \partial_k\partial_j \gamma_{il} - \partial_i\partial_j \gamma_{kl}  \right)  \\ 
&+ \gamma^{kl} \left[ \Gamma^m_{il}\Gamma^{\ }_{mkj}  - \Gamma^m_{ij}\Gamma^{\ }_{mkl} \right]~.
\\[0.5cm]
\end{split}
\end{equation}
%

The first term corresponds to a strongly hyperbolic wave equation and therefore is not an issue for numerical integration. The same is true for the last term in square brackets, as this consists of the product of several first-derivatives of the metric encoded in the Christoffel symbols. The problematic term is the one denoted within the round brackets, because this one contains the second cross-derivatives of the metric and, as mentioned before, they contain hidden the unphysical scalar mode. Solving the problem, thus, will consist on finding smart reformulations for efficiently removing or replacing these terms.  

In the literature, there exist several reformulations of the Einstein equations that intend to alleviate this problem. In the next section, we describe the commonly used Baumgarte-Shapiro\,-\,Shibata-Nakamura (BSSN) formalism,   which   is   later used in this thesis. We refer the interested reader to the standard numerical relativity books (Alcubierre 2008, Baumgarte \& Shapiro 2010, Shibata 2015) for an additional and more detailed discussion on these numerical aspects.


\section{The BSSN scheme \label{sec_BSSN}}

The Baumgarte, Shapiro, Shibata and Nakamura (BSSN) formalism was fully developed in 1999~\cite{Baumgarte:2010ndz,2016bookShibata}, and consists of a modification of the ADM equations presented in section \ref{ADM_form}. This adaptation rewrites the system of equations into a strong hyperbolic set of equations well suited for numerical simulations. In short, these modifications are based on a conformal decomposition of the 3-metric and extrinsic curvature, which facilitates the isolation of a scalar mode in \Eq{eq::3R_problem}, and then substitute the problematic cross derivatives of the metric with  terms containing a new evolved variable,  $\tGm^i$,  that encodes the metric's first derivatives. Then, the evolution equations of $\tGm^i$ are derived, in which, importantly, the unphysical scalar mode is abstent.

The conformal decomposition of the 3-metric is as follows:
\be
\tilde\gamma\ij =  \chi \gm\ij  \qquad \text{with } \quad \text{det}(\tilde\gamma\ij) = 1  ~,
\ee
where $\tilde\gamma\ij$ is the conformal 3-metric and $\chi$ is the conformal factor. Additionally, the extrinsic curvature is decomposed into its trace $K$ and the conformal traceless part $\tilde{A}\ij$, 
\be
K_{ij} = \chi \left( \tilde{A}\ij +\frac13\tgm\ij K \right) ~.
\ee

In terms of these quantities, the Hamiltonian and momentum constraints read
\begin{empheq}{align} %
\mathcal{H} & \equiv  \ ^{(3)}R - \tilde A\ij \tA\IJ + \frac 23 K^2 - \frac{2}{\Mpl^2} \rho  = 0
~, \label{eq:HamiltonianConst_BSSN} \\
\mathcal{M}^i & \equiv  \partial_j \tA\IJ +  \tilde\Gamma^i_{nm} \tA^{nm}  -\frac32 \tA\IJ  \frac {\partial_i \chi}{\chi} - \frac 23 \tilde\gamma\IJ\partial_j K - \frac 1{\Mpl^2} S^i      = 0 
~, \label{eq:MomentumConst_BSSN}
\end{empheq} 
where $\tilde\Gamma^l\ij$ are the Christoffel symbols with respect to the conformal metric.

The evolution equations of the evolved variables, after the conformal decomposition, are 
\begin{empheq}[box=\fbox]{align}
\qquad 
\partial_t\chi  & = \frac{2}{3}\,\alpha\,\chi\, K - \frac{2}{3}\,\chi \,\partial_k \beta^k + \beta^k\,\partial_k \chi ~ , 
\label{eqn:dtchi2} \\[2mm]
\partial_t\tilde\gamma_{ij} & = -2\,\alpha\, \tA_{ij}+\tilde \gamma_{ik}\,\partial_j\beta^k+\tilde \gamma_{jk}\,\partial_i\beta^k 
\nonumber \\ &\hspace{1.3cm} 
-\frac{2}{3}\,\tilde \gamma_{ij}\,\partial_k\beta^k +\beta^k\,\partial_k \tilde \gamma_{ij} ~ , \label{eqn:dttgamma2} \\[2mm]
\partial_t K & = -\gamma^{ij}D_i D_j \alpha + \alpha\left(\tilde{A}_{ij} \tilde{A}^{ij} + \frac{1}{3} K^2 \right) 
\nonumber \\ &\hspace{1.3cm} 
+ \frac 1{2\Mp^2} \,\alpha(\rho + S)  + \beta^i\partial_iK   \label{eqn:dtK2} ~ , 
\\[2mm]
\partial_t\tA_{ij} & = \left[- \chi D_iD_j \alpha + \chi \alpha\left( \,^{(3)} R_{ij} -\frac 1 {\Mp^2}\, \,S_{ij}\right)\right]^\textrm{TF}   \quad
\nonumber \\  &\hspace{1.5cm}
+ \alpha (K \tA_{ij} - 2 \tA_{il}\,\tA^l{}_j) 
\nonumber \\  &\hspace{1.5cm}
+ \tA_{ik}\, \partial_j\beta^k + \tA_{jk}\,\partial_i\beta^k 
\nonumber \\  &\hspace{1.5cm}
-\frac{2}{3}\,\tA_{ij}\,\partial_k\beta^k+\beta^k\,\partial_k \tA_{ij}\, \label{eqn:dtAij2} ~,
\end{empheq} 
where the superscript $^{\rm TF}$ denotes the trace-free parts of the corresponding tensor within the brackets. 
\ \\

These modifications allow us to split the Ricci tensor present in \Eq{eqn:dtAij2} in the form {${ ^{(3)}R\ij = R\ij^\chi + \tilde R\ij}$}, so that 
\be
\begin{split}
  R\ij^\chi \equiv & \frac 1{2\chi} \left( \tilde D_i \tilde D_j \chi + \tilde\gamma\ij \tilde D_m\tilde D^m\chi \right)
            \\ &- \frac 1{4\chi^2} \left( \tilde D_i \chi \tilde D_j \chi + 3 \tilde\gamma\ij \tilde D_m \chi \tilde D^m\chi     \right)  ~,
\end{split} \label{Ricciconf}
\ee
and 
\be
\begin{split}
 \tilde R\ij \equiv & - \frac12\tilde\gamma^{mn}\partial_n\partial_m  \tilde\gamma\ij + \tilde\gamma_{m(i}\partial_{j)}\tilde\Gamma^m + \tGm^m \tGm_{(ij)m} 
 \\ &  + \tilde\gm^{mn}  \left( 2\, \tilde\Gamma^q_{m(i}\, \tGm_{j)qn}^{\ }   + \tilde\Gamma^q_{in}\tGm_{qm j}^{\ } \right)  ~,  
\end{split} \label{Riccitilde}
\ee
where we have introduced three new evolved auxiliary variables, which correspond to the contracted conformal connection 
\be
\tGm^i \equiv \tilde\gamma^{mn}\tGm^i_{mn} = -\partial_j \tilde\gamma\IJ~. 
\ee

We note that \Eq{Riccitilde} is now written mostly in terms of first derivatives of evolved variables, except the first term which is conveniently in the form of a strongly-hyperbolic wave equation.

The evolution equations of $\tGm^i$ are constructed from \Eq{eqn:dttgamma2}, which yields  
\be
\partial_t\tilde\Gamma^i = - \partial_j \left( 2\,\alpha\, \tA^{ij}+\tilde \gamma_{im}\,\partial_j\beta^m+\tilde \gamma_{jm}\,\partial_i\beta^m \right) ~. 
\ee
This equation contains the divergence of the traceless extrinsic curvature, i.e. $\partial_j\tA\IJ\propto\partial_j \partial_t \tilde\gamma\IJ$, and thus it would reintroduce the problem this formulation intends to solve.  However, we can make use of the momentum constraint (\ref{eq:MomentumConst_BSSN}) to substitute this term, so that 
\be
\partial_j \tA\IJ  = -  \tilde\Gamma^i_{nm} \tA^{nm}  + \frac32 \tA\IJ  \frac {\partial_i \chi}{\chi} + \frac 23 \tilde\gamma\IJ\partial_j K + 
\frac 1{\Mpl^2} S^i  ~.    
\ee

In the final form, the evolution equations of $\tGm^i$ become 
\be \boxed{
\begin{split}
\partial_t \tilde \Gamma^i  = \ & 2\,\alpha\left(\tilde\Gamma^i_{jk}\,\tA^{jk}-\frac{2}{3}\,\tilde\gamma^{ij}\partial_j K - \frac{3}{2}\,\tA^{ij}\frac{\partial_j \chi}{\chi}\right)   -2\,\tA^{ij}\,\partial_j \alpha 
-  \frac 2{\Mp^2} \,\alpha\,\tilde\gamma^{ij}\,S_j
\\
&  +\beta^k\partial_k \tilde\Gamma^{i}  +\tilde\gamma^{jk}\partial_j\partial_k \beta^i +\frac{1}{3}\,\tilde\gamma^{ij}\partial_j \partial_k\beta^k  
+ \frac{2}{3}\,\tilde\Gamma^i\,\partial_k \beta^k -\tilde\Gamma^k\partial_k \beta^i  ~ . \label{eqn:dtgamma2}
\end{split}
}
\ee

Note that in this reformulation, the evolution equations do not contain terms with derivatives of $\tilde\gamma\ij$ or $\tA\ij$, except for advection terms and the well-behaving wave equation $\partial_k\partial^k \tilde\gamma\ij$. This is the key of the success of the formalism that greatly improves the numerical stability of the system.

The set of evolution equations (\ref{eqn:dtchi2})-(\ref{eqn:dtAij2}) and (\ref{eqn:dtgamma2}) are known as the BSSN equations, and they are well-suited for the numerical integration. This is the underlying framework used in the numerical investigations presented in the following sections of this thesis.


\section{Numerical Cosmology}    

We now want to use the ADM/BSSN formalism to describe cosmological scenarios, thus we proceed with identifying the numerical quantities commonly used in cosmology, putting particular emphasis on the terms beyond the symmetries of the FLRW Universe. 

nI this exercise, we want to map the BSSN variables into the cosmological variables such as the scale factor $a$, the Hubble rate $H$ and the equation of state $\omega$. In FLRW setups,  all these quantities are defined globally so they are all functions of time only. This allows taking these quantities as background quantities where perturbations evolve on top. However, in our studies such a background cannot be uniquely defined so for the moment we will use local quantities, and therefore they are now functions of the spatial grid coordinates and the temporal slicing.  With this in mind, and without loss of generality, let us fix  $\beta^i = 0$ but keeping $\alpha$ as an arbitrary gauge choice. Then, the cosmological scale factor can be identified with the conformal factor so that $a^2 (\vec x, t)= \chi^{-1} (\vec x, t)$ and, using Eq.~(\ref{eqn:dtchi2}), the Universe's expansion rate reads
\be
H \equiv \frac{\dot a}{a} =  -\frac12 \frac{\partial_t \chi}\chi = -\frac{1}3 \alpha K ~. \label{eq:HubbleParam}
\ee
In the gravitational sector, the energy associated with the gravitational vector and tensor modes is given by 
\bea
  \rho_{\rm shear} = \frac {\Mpl^2}{2} \tilde A\ij \tilde A\IJ \propto \partial_t\tgm\ij \partial_t\tgm\IJ ~,
\eea
and the curvature's contribution to the energy budget is written in terms of the Ricci scalar (of the 3-metric) 
\bea  
\rho_R = \frac{\Mpl^2}{2} \ ^{(3)}R ~.
\eea

After these definitions, one can rewrite the Hamiltonian constraint of Eq.~(\ref{eq:HamiltonianConst_BSSN}) into the inhomogeneous analogue version of the Friedmann-Lema\^{i}tre equation, 
\be
\Rightarrow  \qquad H^2  = \frac 1 {3\Mpl^2} \left( \rho + \rho_{\rm shear} -  \rho_R  \right) ~.
\ee

In addition, the acceleration equation can be thus derived by taking the time derivative of \Eq{eq:HubbleParam} in combination with Eq.~(\ref{eqn:dtK2}), the result reads %
\begin{align}
\frac{\ddot a}{a} &= -\frac\alpha3\left[ \frac{\dot\alpha}\alpha K - D^iD_i\alpha +  {\alpha} \frac 2{\Mpl^2} \left( \rho_{\rm shear} +  \frac14 T \right) \right] ~,\label{eq:adotdot} \\[2mm]
& \text{with } \qquad T \equiv \rho+S =  3\rho \left(\frac13 + \omega \right) ~. \nonumber
\end{align}

By interpreting Eq.~(\ref{eq:adotdot}) in the Eulerian synchronous gauge (i.e.  $\alpha = 1$),  we can infer that the conditions needed for a positive accelerated expansion of the Universe are 
\begin{gather}
\omega_\varphi  < -\frac13 ~, \label{eq:dec.cond1}
\\
\rho_{\rm shear}  <  \left| \frac 34 \rho \left( \frac13 + \omega\right) \right|  ~.
\label{eq:dec.cond2}
\end{gather}

As a result, one finds that the Universe's expansion rate is governed by the interplay between the energy density $\rho$, the equation of state $\omega$ and the energy associated with the gravitational tensor-vector modes  $\rho_{\rm shear}$.

In occasions, we will want to make a connection to global quantities, \textit{ergo} to larger scales. In chapters \ref{p1_chap:prepaper1} and \ref{p2_chap:prepaper2}, we will use the physical volume average overall the simulation grid and treat these average measures as the ``background" quantities at a given time hypersurface. This is motivated in our studies because of the use of periodic boundary conditions and it will allow us to compare the evolution of these average quantities in comparison to FLRW ones. However, we should stress that this averaging procedure is not unique~\cite{Buchert:1999pq,Buchert:1999mc} and that it is strongly dependent on the spacetime foliation, and therefore on the gauge choice.  In our studies, this choice is justified as the system either converges (inflation onset)  or begins (preheating onset) in, approximately, a FLRW Universe.

\section{Dynamics of scalar fields}

Let us consider the generic case of an arbitrary number of scalar fields in the Jordan frame $\phi^I$, with an arbitrary non-minimal coupling $\xi_I$. However, we compute their evolution in the Einstein frame (thus, obeying the Einstein equations, see Sec.~\ref{sec:nonminimalSFs}).  
The energy-momentum tensor is given by 
\beq
T_{\mu\nu} = {\cal G}_{IJ} \partial_\mu \phi^I \partial_\nu \phi^J - g_{\mu\nu} \left[ \frac{1}{2} {\cal G}_{IJ} \partial_\alpha \phi^I \partial^\alpha \phi^J + V (\phi^I ) \right] ~,
\eeq

and the Klein-Gordon equation reads 
\be \label{eq:JordanKG}
g^{\mu\nu} \nabla_\mu\nabla_\nu \phi^I  = \mathcal{G}^{IJ} \frac{\partial V(\phi^K)}{\partial \phi^J} + \Gamma^I_{JK} g^{\mu\nu} \nabla_\mu\phi^J \nabla_\nu \phi^K   ~.
\ee

Now we want to reformulate the equations of motion consistently within the 3+1 formalism. Hence, we manipulate first the right-hand-side in Eq.~(\ref{eq:JordanKG}), 
\be
g^{\mu\nu} \nabla_\mu\nabla_\nu \phi^I  = \left( n^\mu n^\nu + \gamma^{\mu\nu} \right) \nabla_\mu\nabla_\nu \phi^I ~,
\ee
where
\be  \label{Eta1}
 \gamma^{\mu\nu} \nabla_\mu\nabla_\nu \phi^I  = D_\mu D^\mu \phi  + K  \left( n^\mu \partial_\mu \phi^I\right) ~,
\ee
and
\be
\begin{split} \label{Eta2}
 n^\mu n^\nu \nabla_\mu\nabla_\nu \phi^I  & = n^\nu \partial_\nu \left(n^\mu \partial_\mu \phi^I\right) - \frac 1\alpha \gamma^{\mu\nu} \partial_\mu \alpha \partial_\nu \phi^I ~.
\end{split}
\ee

We can now define the scalar field momentum as a free (evolved) variable, 
\be
  \Pi^I  \equiv  n^\mu \partial_\mu \phi^I ~,
\ee
which can be substituted in Eqs.~(\ref{Eta1}) and (\ref{Eta2}), as well as in the last term of the left-hand-side  of  Eq.~(\ref{eq:JordanKG}), i.e. 
\be
\Gamma^I_{JK} g^{\mu\nu} \nabla_\mu\phi^J \nabla_\nu \phi^K  =   \Gamma^I_{JK} \left( \Pi^I\Pi^J + \gamma^{\mu\nu} \partial_\mu\phi^I \partial_\nu\phi^J \right)~.
\ee
The complete system of evolution equations now reads
\begin{align}
\partial_t {\Pi}_{\rm M}^I &= \beta^i\partial_i {\Pi}_{\rm M}^I + \alpha\partial^i\partial_i \phi^I + \partial_i \phi^I \, \partial^i \alpha 
+\alpha \Big( K {\Pi}_{\rm M}^I- \gamma^{ij}\Gamma^k_{ij}\partial_k \phi^I  \Big) \\ \nonumber
& + \alpha \Big[ \Gamma^I_{JK} \left( - \Pi_{\rm M}^J\Pi_{\rm M}^K  + \gamma\IJ\partial_i \phi^J \partial_j \phi^K \right)
- {\cal G}^{IJ} \frac{d }{d {\phi}^J} V (\phi^K) \Big]  ~ ,  \\
\partial_t \phi^I &= \alpha {\Pi}_{\rm M}^I +\beta^i\partial_i \phi^I  ~ .
\end{align} 

These equations simplify in the case of canonically normalized  scalar-fields (or minimally-coupled scalar-fields, e.g. $\xi_I = 0$) where $\Gamma^I_{JK} = 0$ and ${\cal G}_{IJ} = \delta_{IJ}$.

\chapter{ Initial conditions in single-field Starobinsky/Higgs inflation}
\label{p1_chap:prepaper1}
\pagestyle{fancy}

\section{Preamble}

In Chapter~2 we have introduced the basics of the inflationary paradigm, focusing on the description of the slow-roll regime and the mechanism for the generation of scalar and tensor perturbations with a nearly scale-invariant power spectrum. A different question, though,  is how feasible it is for the inflationary regime to begin in the first place, and if it requires any kind of fine-tuning for inflation to start. We investigate this question in Ref.~\cite{Joana2020}, using full General Relativity simulations based on the formalism described in Chapter~4. We focus on the Higgs/Starobinsky models, which are favoured by the Planck observations~\cite{Akrami:2018odb}. These models fall in the categories of plateau-shape potentials at super-Planckian field values,  $\varphi \gg \Mpl$. See the Appendix~A for more in-depth details of these models. 

The paper includes a detailed introduction and summary of previous works on the topic. However, I would like to highlight two relevant publications that have served as a starting point for our work. The first is the work by K.~Clough and E.~Lim \textit{et. al.} in Ref.~\cite{Clough:2016ymm} where they also use numerical General Relativity simulations to investigate the robustness of inflation to initial inhomogeneous field configurations for some canonical small-field and large-field inflation models. They focus on initial conditions consisting of gradients (i.e. with vanishing kinetic energy) at length scales slightly larger than the Hubble radius. Regarding the large-field models,  the three key findings relevant to our work are:

\begin{itemize}
	\item
	 \textit{There is no ``overshooting" problem:} the scalar field hardly leaves the inflation-allowed region of the potential during the preinflationary epoch. This is not necessarily true for small-field models where the preinflationary evolution of the field can easily lead the field closer to the non-inflationary region, reducing significantly the number of efolds of inflation, or even preventing it. 
\\
	\item
	\textit{Local contracting regions do not prevent inflation:}  They tested scenarios containing locally expanding and contracting regions. As long as the simulated region was expanding in average, i.e. $\langle K \rangle < 0$, inflation ends up taking place.  
\\
	\item
	\textit{ Black hole formation does not prevent inflation:} large super-Hubble fluctuations re-entering at Hubble scales might form black holes. However, the event-horizon radius of such black holes is always smaller than the Hubble radius. Thus, the region outside those black holes does still begin inflation. 
\end{itemize}

The second relevant publication is Ref.~\cite{Chowdhury:2019otk} by D.~Chowdhury, J.~Martin, C.~Ringeval and V.~Vennin. In this paper, the authors perform analytical and computational studies (in linearized gravity) of several categories of inflationary models, including Starobinsky-like potentials. They particularly investigate initial conditions corresponding to strong (homogeneous) kination. These conditions assume homogeneous field values but with large field velocities that strongly violate the slow-roll conditions. They show that, for many of the considered models (Starobinski and Higgs models included), the slow-roll regime is a dynamical attractor that leads to enough efolds of inflation, even when starting from a strong kination phase.

In our work, we combine and expand the previous studies by considering initial conditions containing either large gradients (i.e. with an equation of state $\langle \omega \rangle \approx - 1/3$) and/or strong and inhomogeneous kination regions (i.e. $\langle \omega  \rangle \approx 1 $). Furthermore, we study those initial scenarios at both super-Hubble and sub-Hubble scales, and perform a detailed analysis of the dynamics, taking metric and scalar-field fluctuations into account.

The article is attached below and has been published in \textit{Physical~Review~D}. It is reproduced as published. Some of its content might be slightly redundant with the (more detailed) introductory chapter of this thesis, we have chosen to keep it that way to facilitate the readability.

\clearpage

\section*{Research article}
\addcontentsline{toc}{section}{Research article}

 \vspace*{10mm}
 
 \begin{center}
 \textbf{ \Large 
 Inhomogeneous preinflation across  Hubble scales in full general relativity
 }
\\[5mm]
 Cristian Joana,\ \  S\'ebastien Clesse
\end{center}

\vspace{5mm}

\hspace*{0.025\textwidth} 
 \parbox[t]{0.95\textwidth}
 {
We use of the 3+1 formalism of numerical relativity to investigate the robustness of Starobinsky and Higgs inflation to inhomogeneous initial conditions, in the form of either field gradient or kinetic energy density.  Sub-Hubble and Hubble-sized fluctuations generically lead to inflation after an oscillatory phase between gradient and kinetic energies.  Hubble-sized inhomogeneities also produce contracting regions that end up in the formation of primordial black holes, subsequently diluted by inflation.  We analyse the dynamics of the preinflationary and the generation of vector and tensor fluctuations. Our analysis further supports the robustness of inflation to any size of inhomogeneity, in the field, velocity or equation-of-state.  
At large field values, the preinflation dynamics only marginally depends on the field potential and it is expected that such behaviour is universal and applies to any inflation potential of plateau-type, favoured by CMB observations after Planck.
}

\vspace{10mm}

\tocless\section{Introduction \label{p1_sec:intro}}

In the inflationary paradigm, the Universe undergoes an early phase of nearly exponential, accelerated expansion.  Inflation naturally solves a series of problems of the standard cosmological model, among others the flatness and the horizon problems.   It is usually driven by one or several scalar fields that slowly roll along an almost flat direction of their potential.   Quantum fluctuations during inflation provide adiabatic and nearly scale-invariant curvature fluctuations, whose primordial power spectrum is today well constrained by cosmic microwave background (CMB) observations.   The latest results from Planck~\cite{Akrami:2018odb,Ade:2015lrj} favour single-field inflation with a plateau-like potential~\cite{Martin:2013nzq}, such as the Higgs/Starobinsky inflation model.  
Despite those great successes, the naturalness of the inflationary scenario has been questioned for about thirty years.  Indeed, inflation explains why the Universe is homogeneous over about $10^5$ Hubble volumes at the time of the last scattering.  But this would only push backwards the fine-tuning issue if the triggering of inflation requires homogeneous initial conditions over several Hubble volumes.  In such a case, the appealing and naturalness of inflation would be strongly reduced.

The question of how homogeneous must be the Universe prior to inflation has been addressed by several authors, with apparently contradictory results, so that the initial fine-tuning issue has been unclear until recently.   Linear density fluctuations certainly do not prevent the onset of inflation~\cite{Brandenberger:1990wu,Brandenberger:1990xu,Alho_2014, ALHO2011537}.  But dealing with the fully relativistic non-linear dynamics of large inhomogeneities, including the backreactions on the Universe expansion, is a much more complex problem.  This requires to go beyond the linear theory of cosmological perturbations, for instance by using the gradient expansion formalism~\cite{Deruelle:1994pa, Azhar_2018}  or non-perturbative approximations to capture some of the nonlinear backreactions \cite{Bloomfield_2019}. 
In this context, methods of numerical relativity are well-suited~\cite{Goldwirth:1989vz,Laguna:1991zs,KurkiSuonio:1993fg} but their use has been for a long time limited by computational resources.   
Recently, numerical relativity in 3+1 dimensions has been used to study the early universe cosmology in the context of inflation \cite{East:2015ggf,Clough:2016ymm,Aurrekoetxea_2020,PhysRevD.100.063543}, and possible alternatives \cite{PhysRevD.78.083537,Ijjas_2020}. Particularly, the problem of initial conditions for inflation has been considered for several inflaton models and scalar field initial configurations~\cite{East:2015ggf,Clough:2016ymm,Aurrekoetxea_2020}.
Despite these progresses, the required degree of homogeneity and the question of whether inflationary patches can emerge from a landscape of non-linear scalar field fluctuations have been only solved in some specific cases, and in general it is still controversial. First works obtained that inflation cannot start from sub-Hubble non-linear fluctuations, for instance in~\cite{Goldwirth:1989vz,Goldwirth:1990pm,Goldwirth:1989pr} by using numerical relativity in spherical symmetry, and in~\cite{Deruelle:1994pa} by using the gradient expansion formalism.  An opposite result was obtained in~\cite{KurkiSuonio:1993fg} by using the first numerical relativity simulations in 3+1 dimensions (see also~\cite{Brandenberger:2016uzh}), which has been confirmed more recently in~\cite{East:2015ggf}. 

In summary, despite recent progress, the mechanisms leading (or not) initial non-linear inhomogeneities to inflate are not yet fully understood, as well as their model dependence.
This work aims to contribute to paving the way to a better understanding of the fully relativistic non-linear preinflation dynamics, and thereby  a better view of viable and theoretically motivated inflationary models.

In this paper, we investigate the inhomogeneous scalar field dynamics and the possible onset of inflation, with the use of numerical relativity in 3+1 dimensions.   For this purpose, we rely on the \texttt{GRCombo} code~\cite{Clough_2015}, based on the Baumgarte-Shapiro-Shibata-Nakamura (BSSN) formalism \cite{PhysRevD.52.5428,Baumgarte_1998,10.1143/PTPS.90.1}.  It is used to solve the full Einstein field equations together with the Klein-Gordon equation for a scalar field.
The BSSN formalism has been proved to be stable and efficient for a variety of problems going from the dynamics of black hole binaries to cosmological problems such as the gravitational collapse of cosmic strings
\cite{Helfer_2019}, the non-linear collapse of matter inhomogenities~\cite{Rekier:2014rqa,Rekier:2015isa} and their backreactions on the Universe expansion.

Our analysis focuses on the Higgs/Starobinsky inflation model.  This choice is well motivated for several reasons.  First, the potential has a single parameter, fixed through the CMB power spectrum normalization.   This restricts the parameter space to explore to the initial conditions of the field.  Second, it is the best favoured (and the simplest) inflation model after Planck~\cite{Martin:2013nzq}.   Third, the model has been considered in~\cite{Aurrekoetxea_2020}, which allows us to compare some of our results to the literature.
In particular, we reproduce the case of a Gaussian field fluctuation on top of a background field value lying in the slow-roll region.   But compared to previous work, our analysis has been extended to study more exhaustively the dynamics of the preinflation era and determine where inflation can take place and for which fluctuation sizes.  Not only we consider the case of an inhomogeneous initial field, but also cases with inhomogeneous field velocity and equation of state that had not been considered so far.  Some universal behaviours in the field and space-time dynamics are identified, depending on the characteristic fluctuation sizes.  Our results further support the robustness of inflation against various configurations of initial conditions.  And for sub-Hubble and Hubble-sized fluctuations, they better emphasize a universal behaviour in the form of an oscillating equation of state, a signature of the respectively quick or slow wobbling between field gradient and kinetic terms that alternatively dominate the total energy density.

For each considered case, we have monitored the evolution of all geometrical {quantities, the scalar field} and its velocity.  We identify the conditions under which inflation can be triggered in some parts of the lattice, whereas other parts can undergo a gravitational collapse leading to black hole formation.   Finally, we clarify and explain why previous works have led ostensibly to different conclusions, which is related to the time at which the initial Hubble scale is defined.  We emphasize that the level of initial inhomogeneity is naturally restricted if the energy density is initially dominated by field gradients.   Inflation is generally the natural outcome, except in regions where the field Laplacian is maximal that can start contracting and collapse into preinflation black holes.

The paper is organized as follows:  In Section~\ref{p1_sec:prevwork} we review previous results in the topic of the inhomogeneous initial conditions of inflation.  The Higgs and Starobinsky models are introduced in Section~\ref{p1_sec:potential}. The BSSN formalism of numerical relativity is detailed in Section~\ref{p1_sec:BSNN} and
its link to coarsed-grained cosmology and metric perturbations is explained in Section~\ref{p1_sec:cors_universe}. 
Section~\ref{p1_sec:inital_conditions} describes the considered initial conditions.  Our results are presented in Section~\ref{p1_sec:results} and their implications are discussed in Section~\ref{p1_sec:discussion}.  Our conclusion and the perspectives of this work are presented in {Section}~\ref{p1_sec:ccl}.  
In appendix, we provide more technical details on the convergence tests to check the stability and validity of our simulations.
\ \\

\tocless\section{Summary of previous work  \label{p1_sec:prevwork}}

There exist only a few references having investigated the initial inhomogeneity problem of inflation.   In this section, we give a brief and general overview of previous works on this topic.
The question of how generic or fine-tuned are \textit{homogeneous} initial conditions leading to inflation is another related issue, also controversial, and we will make some connections to this problem, in particular to the slow-roll attractor solution and the dynamics of the preinflation phase in the presence of a large kinetic term, for plateau inflation.  We let the interested reader to refer to the recent literature, see e.g.~\cite{Linde:2017pwt,Finn:2018krt,Chowdhury:2019otk,Tenkanen:2020cvw}.

The first attempts to study the problem of \textit{inhomogeneous} initial conditions using numerical relativity are due to Goldwirth and Piran in 1989 and 1990~\cite{Goldwirth:1989vz,Goldwirth:1989pr,Goldwirth:1991rj}.    The numerical method was introduced in~\cite{Goldwirth:1989vz} and their results were presented in~\cite{Goldwirth:1989pr,Goldwirth:1991rj} for five scalar field potentials:  large-field inflation with quadratic and quartic potentials, and small-field inflation with a quartic or a Coleman Weinberg (CW) potential~\footnote{Large-field and small-field inflation are respectively called with their old denomination \textit{chaotic} and \textit{new} inflation in~\cite{Goldwirth:1989pr,Goldwirth:1991rj}.}.   They focused on spherically symmetric cosmologies and found that only sufficiently large inhomogeneities, at least of the size of the Hubble radius, can lead to large-field inflation.  They gave an estimate of the fluctuation size in units of $H^{-1}$ preventing the onset of inflation.  For all these models, the onset of inflation required homogeneity over several horizon sizes, at the noticeable exception of the CW potential.  However, for small-field inflation, the onset of inflation requires tiny values of the mean scalar field and its fluctuations.  Thus, inflation seems to be less natural in small-field inflation than in large-field inflation.  Sub-Hubble fluctuations typically did not lead to inflation, because the slow-roll regime cannot be reached before the mean-field value reaches the bottom of the potential.   They nevertheless point out that for sub-Hubble fluctuations, inflation could nevertheless be triggered if scalar field oscillations are damped down to a sufficiently large homogeneous field value.  
{But} due to obvious computational limitations, there were not able to cover full ranges of fluctuation sizes and amplitude.  They did not either cover the interesting parameter range for the considered models since CMB observations were not available at that time.  Furthermore, spherical symmetry does not capture all the possible general relativistic effects, and so the application of their results to more general fluctuations can be questioned.  Finally, they did not consider cases where the initial density is strongly dominated by kinetic or gradient terms.  As a consequence, the analytical approximation proposed by Goldwirth in~\cite{Goldwirth:1990pm}, stating that the comoving size of inhomogeneities $\Delta$ needs to be such that
\be   
a \Delta > \sqrt{ \frac{ 8 \pi}{3 } } \frac{\delta \varphi}{H_{\rr{inf}} \mpl} 
\ee
where $a$ is the initial scale factor, assumes that the energy density is dominated by the potential and the field gradients are suppressed.  If this is true at the onset of inflation, this assumption can be violated in the preinflation era, with $H_{\rr{inf}} \ll H_{\rm ini} \lesssim m_{\rm Pl} $.  

An impressive study of the problem, using 3+1 numerical relativity, has been achieved in 1993 by Kurki-Suonio, Laguna and Matzner.   As in previous works, they used the York's procedure \cite{York:1971hw,York:1972sj} to solve the initial condition problem, for a large-field quartic potential.   For inhomogeneous runs, they correctly relate the Hubble horizon size to the total density, including the field gradients and velocities.  For their initial conditions, the density is actually dominated by kinetic terms, and $H\sim m_{\rm pl}$ initially.  They run a few 3D simulations, both for super-Hubble and slightly sub-Hubble fluctuations.   In the latter case, they first observe field oscillations and when the fluctuations become super-Hubble, expansion occurs in a homogeneous way and inflation can take place before the field reaches the bottom of the potential.  They conclude that inflation can arise from non-linear field fluctuations of the order of the Hubble horizon and beyond.  However, they restricted their analysis to a potential shape -- and set of parameters -- that is now strongly disfavoured by CMB observations and only considered initial conditions with sub-dominant field gradients compared to the kinetic and potential terms.  Nevertheless,  their analysis certainly remains very impressive given that it was performed in 3D, with limited computational resources compared to the ones at disposal nowadays.

As an alternative to full numerical relativity methods, Deruelle and Goldwirth have used in 1995 a long wavelength iteration scheme ~\cite{Deruelle:1994pa}, also referred as gradient expansion formalism, in order to determine how large the initial homogeneities of a massive scalar field can be without preventing inflation to set in.  They found that homogeneity over patches of the order of, or larger than the local Hubble radius, is a general condition needed for inflation, but such a method does not allow to study the evolution of strong initial inhomogeneities.  It nevertheless provides an understanding of the factors controlling the system behavior, without assuming any spatial symmetry.  

After a gap of about twenty years, the problem of inhomogeneous initial conditions for inflation has seen a renewed interest in the recent literature
\cite{Ijjas_2013,Guth_2014,Ijjas_2016}\cite{Chowdhury:2019otk}. In parallel, inhomogeneous initial field values in multi-field inflation models were considered in~\cite{Easther:2014zga}, but without including gravitational backreactions.  Their analysis particularly focused on hybrid models of inflation, extending the work of~\cite{Clesse:2009ur,Clesse:2008pf}.  

Fall 2015, East, Kleban, Linde and Senatore have released a new analysis of the initial homogeneity problem~\cite{East:2015ggf} based on numerical relativity.  They considered three scalar field potentials:  a constant, a smooth step and a notch potential, the latter describing a family of cosmological attractors.  They studied initial conditions dominated by the field gradient energy and used 3+1 lattice simulations in full general relativity, thus extending the initial work of Kurki-Suonio et al.  They found that field fluctuations initially smaller than the Hubble radius but contained in a flat region of the potential can lead to inflation, after the gradient and kinetic field energy is diluted by expansion.  They also found that, at the same time underdense regions lead to inflation, overdense regions can collapse and form preinflation black holes (PIBHs).  

In~\cite{Clough:2016ymm}, Clough, Lim and DiNunno also used numerical relativity in 3+1 dimensions to study the robustness of initial conditions leading to inflation, for different inflationary models.  In particular, they compared large field to small field scenarios.  Their results suggest that it is much less natural to get inflation in small-field models even when the gradient energy is subdominant initially.   This result is nevertheless mitigated by the fact that initial field values outside the slow-roll regions of the potential can lead to inflation.  For large-field models with relatively flat initial hypersurfaces, they also confirmed that PIBH formation does not prevent the onset of inflation in other regions.  In a following paper~\cite{Clough_2018}, the authors considered inhomogeneities in the metric sector with tensor modes, while keeping the scalar field rather homogeneous. They noticed a reduction on the duration of inflation for small-field models, however suppressed for large tensor modes due to a large Hubble fiction. Large-field models are not strongly affected by the tensor perturbations. Gravitational collapse due to tensor modes was also reported.

Most recently, the effect of the potential shape on initial scalar-field gradients has been further explored in~\cite{Aurrekoetxea_2020}.  Convex potentials are found to be more robust than concave ones for sub-Planckian characteristic scales, for which the field can be dragged-down the bottom of the potential with a significant loss of efolds.  Super-Planckian scales can more generically lead to long enough inflation, also for concave potentials. They suggest that the onset (or not) of sufficient inflation can be inferred from an analytical criterion, consisting in finding the critical scalar field amplitude for which the drag-down of the potential overcomes the pull-back effect of the gradient pressure.  

Some of these considerations have been summarized early 2016 by Brandenberger in a short and general review on the issue of the initial conditions for inflation~\cite{Brandenberger:2016uzh}, reporting on the possible solutions to an initial fine-tuning problem, and emphasizing that slow-roll appears to be a local attractor for large-field models, on the contrary to small-field models.

In summary, the current status of the paradigm is that in large field and plateau potentials, non-linear initial field Hubble-sized fluctuations do not prevent the onset of inflation.  The preinflation dynamics of sub-Hubble and Hubble-sized fluctuations is nevertheless not yet fully understood, in particular in the case of PIBH formation.  It is also uncertain if these conclusions still apply for initially inhomogeneous field velocity or any other type of initial conditions.  The dynamics of the Higgs/Starobinsky model that is the simplest and one of the best favoured model nowadays, were not explored in detail until now. 
\ \\

\tocless\section{Higgs/Starobinsky inflation  \label{p1_sec:potential}}

The Higgs inflation model~\cite{Bezrukov:2007ep} identifies the inflaton to the Standard Model Brout-Englert-Higgs field, but it is non-minimally coupled to gravity in order to provide a sufficiently flat potential to realise inflation.   It is the simplest inflationary model which relates to the standard model of particle physics. The Lagrangian includes an extra term $\xi H^\dagger H R$, where $H$ is the Higgs field, $R$ the Ricci scalar, and $\xi$ is the only parameter of the model.   This term is generated automatically by quantum corrections in curved space-time.  In the Einstein frame, the action is the one of a minimally coupled scalar field with the following potential,
\be
V(\varphi) = \Lambda^4 \left( 1 - {\rm e}^{-\sqrt{2/3}  \varphi  / \Mpl} \right)^2~,
\ee
where $\Mpl$ is the reduced Planck mass, $\Lambda^4 \equiv \Mpl^4 \lambda / (4 \xi^2)$ becomes the unique potential parameter, $\lambda$ being a constant characterizing the model.  The CMB observations allow to fix $\Lambda \simeq 3.1 \times 10^{-3} \Mpl $~\cite{Martin:2013nzq}.  
At large field values, the potential has a plateau allowing for slow-roll inflation.   Slow-roll conditions are violated, and inflation ends at $\varphi_{\rm end} = 1.83 \Mpl$.  About $\Delta N_* \simeq 62 $ e-folds before the end of inflation, observable scales leave the Hubble radius, at a field value $\varphi_* \approx 5.48\ \Mpl$.

The Starobinsky model of inflation is a scalar-tensor theory with $f(R) = R + \epsilon R^2 / \Mpl^2 $, which in the Einstein frame has the same linear dynamics and effective field potential as the one of Higgs inflation.  Therefore, even if the non-linear dynamics of scalar field and BSSN variables during the preinflation era differ between the Jordan frame and the Einstein frame for the Starobinsky model, showing that inflation is reached in the Einstein frame is a sufficient condition to guarantee that inflation is also reached in the Jordan frame.
Although these models have different reheating mechanisms, and eventually distinguishable observable predictions, we are not interested in this issue in this paper.  Finally, let us point out that the Higgs/Starobinsky model is the
best favored slow-roll inflation model after Planck~\cite{Martin:2013nzq}.
\ \\

\tocless\section{BSSN formalism of numerical relativity  \label{p1_sec:BSNN} }

In this work, we solve the BSSN formulation of the Einstein equation using \texttt{GRChombo}~\cite{Clough_2015}, a multipurpose numerical relativity code. In the context of the 3+1 decomposition of General Relativity, the line element can be written as
\be\label{p1_timeline}
\rr d s^2  = - \alpha^2 \rr d t^2 + \gamma_{ij}(\rr d x^i + \beta^i \rr d t)(\rr d x^j + \beta^j \rr d t)
\ee
where $\gamma\ij$ is the metric of the 3-dimensional hypersurface, and the   lapse and shift gauge parameters are given  by $\alpha(t)$ and $\beta^i(t)$ {respectively}.  A further conformal decomposition of the 3-metric follows,
\be
\gm\ij = \frac1\chi \tgm\ij = \psi^4\tgm\ij \quad \text{with } \text{ det}(\tgm\ij) =  1 ~,  
\ee
and here, $\chi$ and $\psi$ are two different parametrisations of the metric conformal factor. While the former is used during the temporal integration, the latter is preferred when constructing the initial conditions. The extrinsic curvature is thus split in $\tA\ij$ and $K$, respectively, the conformal traceless part and its trace,  
\be
K\ij = \frac1\chi \left( \tA\ij +\frac13\tgm\ij K\right)~.
\ee
In addition, the first spatial derivatives of the metric are considered as dynamical variables
\be
\tilde\Gamma^i \equiv \tgm^{jk} \tilde\Gamma^i_{jk} = - \partial_j\tgm\ij ~,
\ee
where $ \tilde\Gamma^i_{jk} $ are the Chritoffel symbols associated to the conformal metric $ \tilde\gamma_{ij} $.
\ \\

\tocless\subsection{Evolution equations}

The evolution equations for the BSSN variables are then given by 
\begin{align} 
\partial_t\chi  & = \frac{2}{3}\,\alpha\,\chi\, K - \frac{2}{3}\,\chi \,\partial_k \beta^k + \beta^k\,\partial_k \chi ~ , \label{p1_eqn:dtchi2} \\[2mm]
\partial_t\tilde\gamma_{ij} & = -2\,\alpha\, \tA_{ij}+\tilde \gamma_{ik}\,\partial_j\beta^k+\tilde \gamma_{jk}\,\partial_i\beta^k 
-\frac{2}{3}\,\tilde \gamma_{ij}\,\partial_k\beta^k +\beta^k\,\partial_k \tilde \gamma_{ij} ~ , \label{p1_eqn:dttgamma2} \\[2mm]
\partial_t K & = -\gamma^{ij}D_i D_j \alpha + \alpha\left(\tilde{A}_{ij} \tilde{A}^{ij} + \frac{1}{3} K^2 \right) 
+ \frac 1{2\Mp^2} \,\alpha(\rho + S)  + \beta^i\partial_iK   \label{p1_eqn:dtK2} ~ , 
\\[2mm]
\partial_t\tA_{ij} & = \left[- \chi D_iD_j \alpha + \chi \alpha\left( R_{ij} -\frac 1 {\Mp^2}\, \,S_{ij}\right)\right]^\textrm{TF}  + \alpha (K \tA_{ij} - 2 \tA_{il}\,\tA^l{}_j) 
\nonumber \\  &\hspace{1.3cm}
+ \tA_{ik}\, \partial_j\beta^k + \tA_{jk}\,\partial_i\beta^k 
-\frac{2}{3}\,\tA_{ij}\,\partial_k\beta^k+\beta^k\,\partial_k \tA_{ij}\, \label{p1_eqn:dtAij2} ~, \\
\partial_t \tilde \Gamma^i  = \ & 2\,\alpha\left(\tilde\Gamma^i_{jk}\,\tA^{jk}-\frac{2}{3}\,\tilde\gamma^{ij}\partial_j K - \frac{3}{2}\,\tA^{ij}\frac{\partial_j \chi}{\chi}\right)   -2\,\tA^{ij}\,\partial_j \alpha 
-  \frac 2{\Mp^2} \,\alpha\,\tilde\gamma^{ij}\,S_j
\\
&  +\beta^k\partial_k \tilde\Gamma^{i}  +\tilde\gamma^{jk}\partial_j\partial_k \beta^i +\frac{1}{3}\,\tilde\gamma^{ij}\partial_j \partial_k\beta^k  
+ \frac{2}{3}\,\tilde\Gamma^i\,\partial_k \beta^k -\tilde\Gamma^k\partial_k \beta^i  ~ . \label{p1_eqn:dtgamma2}
\end{align} 
where the superscript $\rm{TF}$ denotes the trace-free parts of tensors.  The 3+1 decomposition of the energy-momentum tensor $T^{\mu\nu}$ gives
\bea \label{p1_3+1sources}
    \rho  &=& n_\mu n_\nu T^{\mu\nu} ~,\\
    S_i  &=& -\gamma_{i\mu} n_\nu T^{\mu\nu} ~,\\ 
    S_{ij}  &=&  \gamma_{i\mu} \gamma_{j\nu} T^{\mu\nu} ~,\\ 
    S &=& \gamma\IJ S\ij ~,
\eea
where $n_\mu=(-\alpha,\vec 0)$ is the unit normal vector to the three-dimensional slices.

The Hamiltonian and Momentum constraints, 
\begin{align}
\mathcal{H} & = R + K^2-K_{ij}K^{ij}-16\pi \rho = 0\, , \label{p1_eqn:Ham}  \\
\mathcal{M}_i & = D^j (K_{ij} - \gamma_{ij} K) - 8\pi S_i =0\, .  \label{p1_eqn:Mom}
\end{align}
are only solved explicitly when constructing initial data.  They are also monitored during the time evolution in order to ensure that there is no  significant deviations from General Relativity. 
\ \\

\tocless\subsection{Scalar field equations}
For a single scalar field $\varphi$,  the energy-momentum tensor is given by
\begin{equation}
T_{\mu\nu} = \partial_\mu \varphi\, \partial_\nu \varphi - \frac{1}{2} g_{\mu\nu}\, \partial_\lambda \varphi \, \partial^\lambda \varphi - g_{\mu\nu} V(\varphi) \,
\end{equation}
where $V(\varphi)$ is the scalar field potential. The scalar field dynamics is governed by the  the Klein-Gordon equation, split into two first order equations for the field and its momentum $\Pi_{\rm M}$
\begin{align}
\partial_t \varphi &= \alpha \Pi_{\rm M} +\beta^i\partial_i \varphi \label{p1_eq:dtvarphi} ~ , \\
\partial_t \Pi_{\rm M} &= \beta^i\partial_i \Pi_{\rm M} + \alpha\partial^i\partial_i \varphi + \partial_i \varphi \, \partial^i \alpha \\
& \ +\alpha \left( K\Pi_{\rm M}-\gamma^{ij}\Gamma^k_{ij}\partial_k \varphi - V'(\varphi) \right) \label{p1_eq:dtPi} ~ ,
\end{align} 
where the superscript ($'$) denotes the derivative with respect to the field. 
\ \\

\tocless\subsection{Gauge choice and singularity avoidance }%

The gauge parameters are initially set to $\alpha=1$ and $\beta^i=0$ and then evolved in accordance with the \textit{moving puncture gauge} \cite{Baker_2006, Campanelli_2006}, for which evolution equations are
\begin{eqnarray}
\partial_t \alpha &=& -\eta_\alpha \alpha K +  \,\beta^i\partial_i \alpha \ , \label{p1_eqn:alphadriver}\\
\partial_t \beta^i &=& B^i\, \label{p1_eqn:betadriver},\\
\partial_t B^i &=& \frac34\, \partial_t \tilde\Gamma^i - \eta_B\, B^i\ \,, \label{p1_eqn:gammadriver}
\end{eqnarray}
where the constants $\eta_\alpha$ and $\eta_B$ are conveniently chosen to improve the numerical stability.  This way, $\alpha$ and $\beta^i$ are boosted in the problematic regions with strongly growing extrinsic curvature and spatial derivatives of the three-metric $\tilde \gamma_{ij}$.  
The goal of this gauge is to prevent the code from resolving the central singularity of any black hole that may eventually form.
\ \\

\tocless\section{Link to coarsed-grained Cosmology  \label{p1_sec:cors_universe} }

One can map the BSSN variables into more usual cosmological variables (scale factor $a$, Hubble rate $H$ and equation of state $w$) in the separate Universe assumption, corresponding to homogeneity or the super-Hubble limit of field and metric fluctuations.  If one assumes $\beta^i = 0$ but keep $\alpha$ as an arbitrary gauge choice,  the scale factor can be defined as  $a^2 = \chi^{-1}$ and using Eq. (\ref{p1_eqn:dtchi2}), the inhomogeneous analogue of the first Friedmann equation reads
\be
H \equiv \frac{\dot a}{a} = -\frac{1}3 \alpha K ~, \label{p1_eq:HubbleParam}
\ee
where a dot denotes the derivative with respect to cosmic time.  By taking its time derivative and using Eq.~(\ref{p1_eqn:dtK2}), one gets the equation for the acceleration of the expansion of Universe,
\begin{align}
\frac{\ddot a}{a} &= -\frac\alpha3\left( \frac{\dot\alpha}\alpha K - D^iD_i\alpha +  \alpha \left(\mathbf{\aleph} + 4\pi T \right) \right) ~,\label{p1_eq:adotdot} \\[2mm]
\mathbf{\aleph} &\equiv  \tA\ij \tA\IJ ~,\\
T &\equiv \rho+S =  3\rho \left(\frac13 + \omega_\varphi\right) ~.
\end{align}
The first two terms are gauge related terms {that are vanishing in the \textit{synchronous} gauge where $\alpha = 1$}, $\mathbf{\aleph}$ is a new geometrical term that vanishes in the homogeneous and isotropic case, and $T$ is the trace of the energy-momentum tensor.
The later can be written in terms of the scalar field equation of state,  $\omega_\varphi (t)$, which value depends on the dominant term in the scalar field's energy density
\begin{align}
    \rho &\equiv  \frac12 \Pi_{\rm M}^2 + \frac12 \partial_i \varphi\, \partial^i\varphi  + V(\varphi) \\
    &=\rho_{\rm kin}  + \rho_{\rm grad}  + \rho_{\rm V} ~,
\end{align}
corresponding respectively to kinetic, gradient and potential energy density. Thus, one can identify three limiting cases,
\be
 \omega_\varphi (t) \simeq \left\{
 \begin{array}{rl}
      1   &  \quad\rightarrow \quad \rho \simeq \rho_{\rm kin} %
      \\
      -1/3  & \quad\rightarrow \quad \rho \simeq \rho_{\rm grad} %
      \\
      -1  & \quad\rightarrow \quad \rho \simeq \rho_{\rm V}
 \end{array} \right. ~.
\
\ee

By interpreting Eq. (\ref{p1_eq:adotdot}), and neglecting the gauge effects, (i.e.  $\dot \alpha =0$, and assuming $D^iD_i\alpha \approx 0$),  we can infer that the conditions for allowing a positive accelerated expansion of the Universe, ($\ddot a > 0$),  are
\begin{gather}
\omega_\varphi  < -\frac13 ~, \label{p1_eq:dec.cond1}
\\
\mathbb{\aleph} <  \left| 12\pi\rho \left( \frac13 + \omega_\varphi\right) \right|  ~.
\label{p1_eq:dec.cond2}
\end{gather}

As a result, one finds that the expansion rate of the Universe is governed by the interplay of the energy density $\rho$, the equation of state $\omega_\varphi$ and a geometrical parameter $\mathbb{\aleph}$. We remark that, even in scenarios where $\rho > \mathbb{\aleph} > 0$, the  $\mathbb{\aleph}$-term will become dominant when $\omega_\varphi \approx -\frac13$.  
\\

We can interpret $\mathbb{\aleph}$ as an energy density associated to perturbations in the spatial metric.  One sees that Eq.~(\ref{p1_eqn:dttgamma2}) reduces to 
\be
\partial_t \tilde \gamma_{ij} = -2 \alpha \tilde A_{ij}
\ee
and therefore,
\be
 \mathbb{\aleph} \propto \dot\tgm\ij \dot\tgm\IJ
\ee
which is constituted by vector and tensor modes. The $\dot\tgm\ij$ is not related to scalar perturbations because $\tilde A_{ij}$ is traceless by definition.  Within a flat background, one would interpret them as a cosmic shear and gravitational waves.
In the absence of source terms in the evolution equation of $\tilde A_{ij}$ (i.e. $R\ij^{TF}, S\ij^{TF} \approx 0$, in Eq.~\ref{p1_eqn:dtAij2}) they are quickly decaying as the universe expands, 
where cosmic shear goes like $\mathbb{\aleph}_{\rm shear} \propto a^{-6}$, and gravitational waves like $\mathbb{\aleph}_{\rm GW} \propto a^{-4}$. 
However, such definitions are ill-defined in highly non-linear systems, the source terms of $\mathbb{\aleph}$ are no longer negligible and ultimately $\mathbb{\aleph}$ cannot be defined as a gauge-invariant quantity.  So we already point out that the importance and evolution of $\mathbb{\aleph}$ can only be revealed in fully relativistic $3+1$ simulations. 
\\

Because $\mathbb{\aleph}$ is strictly defined in by the (traceless) extrinsic curvature, in this paper we also refer to its sub-Hubble modes as extrinsic curvature modes (ECMs). 
\ \\

\tocless\section{Initial conditions  \label{p1_sec:inital_conditions}}

Our simulations of the preinflationary era have been performed for various sets of initial conditions (ICs).  
In all cases, we follow most of previous works and assume initial conformal flatness, i.e. $ \tgm\ij(t_0) = \delta\ij$ and $\tilde\Gamma^i(t_0) = 0$. Under this assumption, and by setting $\tA\ij(t_0) = 0$, the constraint equations (\ref{p1_eqn:Ham})-(\ref{p1_eqn:Mom}) greatly simplifies and can be rewritten as 
\begin{align}
 {\cal H}^* &= - \nabla^2 \psi + \psi^5 \left( \frac23 K^2 - 16\pi\rho  \right) ~, \label{p1_eq:HamBBSN} \\
\tilde {\cal M}_i^* &= \frac23 \partial_i K + 8\pi S_i   \label{p1_eq:MomBBSN} \\
&\text{with } ~ S_i = -\Pi_{\rm M}\partial_i \varphi 
~. \nonumber
\end{align}

Defining the perturbed energy density as $\delta \rho \equiv \rho - \rho_V$,  where initially $\delta \rho \gg \rho_V$, we  
then consider the following types of ICs:  
\begin{enumerate}
    \item When the energy density is dominated by field gradients, i.e. $\delta \rho (t_0) = \rho_{\rm grad}$, 
    \item When the energy density is dominated by inhomogeneous field velocities,  i.e. $\delta\rho (t_0) = \rho_{\rm kin}$ 
    \begin{enumerate}
    \item with sub-dominant mean value,  $  \langle \Pi_{\rm M} \rangle  \sim 0  $ 
    \item with dominant mean value, $ | \langle \Pi_{\rm M} \rangle | \simeq   \Pi_{\rm M}  $
    \end{enumerate}
\end{enumerate}
With these choices, the momentum density initially vanishes, $S_i =0$,  and therefore the momentum constrain is trivially satisfied if one considers an  homogeneous $K$.  Here, we have use the $\langle ... \rangle$ brackets to represent the mean value of a given function (i.e, $\theta$), averaged across the physical volume ${\mathcal V}$ represented {by} the whole lattice.  {For instance,}
\be \label{p1_eqn:Kini}
\langle \theta \rangle \equiv \frac{1}{{\cal V}} \int \theta \, \rr d {\cal V}~,
\ee
{which takes into account the inhomogeneous physical volume of lattice cells that depends on the local value of the conformal factor at a given time.}  

We make use of periodic boundary conditions. The lattice can then either represent an initially flat, topologically close and compact Universe, or a small region of a much larger classical patch.  

In any case, we consider an initial configuration of the Universe constituted of inhomogeneous scalar energy density $\langle \rho \rangle \gg \langle \rho_V\rangle$, compensated in Einstein's Equations by the scalar curvature of the metric. The initial energy density of vector and tensor modes of the metric is chosen to be zero (i.e. $\mathbf{\aleph} (\vec x) = 0$).

Below, we detail the methods used to solve the Hamiltonian constraint initially and the chosen sets of initial conditions.


{
\begin{figure}[bh!]
\begin{center}

\vspace*{15mm}

\includegraphics[width=0.80\textwidth]{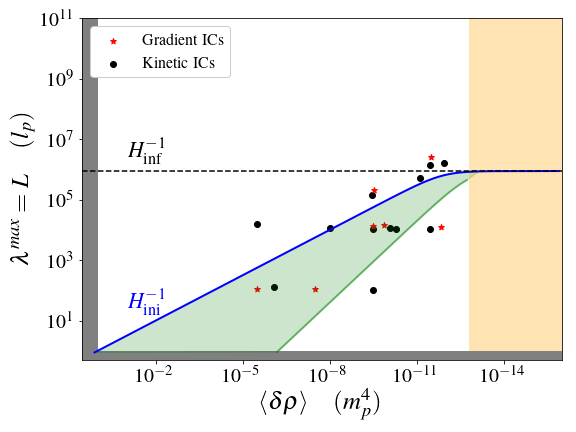}
\caption{ Representation of the considered initial conditions in terms of the maximal physical mode (delimited by the lattice size $L$) and the characteristic mean value of the inhomogeneous energy density. Black and red dots represent one simulation with respectively gradient or kinetic dominated initial conditions. The size of the Hubble scale at origin is represented in the continuous blue line, while the horizontal dotted line indicates the Hubble length at inflation onset at field values in the potential plateau. 
The green shadowed area indicates the sub-Hubble region where dominant extrinsic curvature modes are produced at later simulated time.
The orange shadowed region indicates the inflationary domain. 
Black shadowed areas delimit the super-Planckian energy scales, excluded in our simulations.\label{p1_fig:scale_diagram}}
\end{center}
\end{figure}
}

\clearpage
\tocless\subsection{Gradient-dominated initial conditions}

We consider different configurations for the initial inhomogeneities of the scalar field $\varphi$.
The Hamiltonian constraint, Eq.~(\ref{p1_eq:HamBBSN}), is solved with an iterative method to obtain the corresponding initial distribution of the conformal factor $\psi$ on the lattice.
Like in previous works \cite{Clough:2016ymm, Aurrekoetxea_2020,  East:2015ggf}, the homogeneous value of the extrinsic curvature is chosen such that $K = -  \sqrt{ 24\pi\langle \rho \rangle}$.  Note that when calculating the average, the physical volume is dependent of the conformal factor $\psi$. The negative sign in $K$ reflects an initially expanding Universe.  The field velocity initially vanishes everywhere so that the momentum constraint, Eq.~(\ref{p1_eq:MomBBSN}), is trivially satisfied.

The initial scalar field configuration is chosen as follows,
\be
\begin{split}
\varphi(t_0,\, & \vec x)  =  \bar\varphi_0  
                        + \Delta_e^\varphi \exp\left[ \frac{ -(\vec x - \vec\mu)^2}{ \sigma^2  }\right]  \\
                        & + \frac{\Delta_{cos}^ \varphi}3  \left( \cos \frac{2\pi x}L + \cos \frac{2\pi y}L +\cos\frac{2\pi z}L \right)~.
\end{split}
\ee
Such a pattern can represent a Gaussian field fluctuation of amplitude $\Delta_{\rm exp}^\varphi$ at the centre of the lattice, on top of an inhomogeneous field value $\Delta_{\rm cos}^\varphi$. Here, $\vec\mu$ and $\sigma$ denote the mean and variance of the Gaussian mode. Unless otherwise specified, we choose $\vec\mu = (L/2, L/2, L/2),~ \sigma = L/6$, with $L$ being the physical size of the lattice. 
\ \\

\tocless\subsection{Kinetic-dominated initial conditions \label{p1_sec::KDics}}

For this set of initial conditions, we fix a homogeneous scalar field by imposing
$\varphi (\vec x)  = \varphi_0$. The other ICs can be solved in three different ways:
\begin{enumerate}
\item
Analogously to what is done for gradient initial conditions, one can choose a initial inhomogeneous configuration for $\Pi_{\rm M}$ and solve equation (\ref{p1_eq:HamBBSN}) to obtain $\psi$ by allowing an homogeneous $K$.  This method is prone to fail to converge due to the existence of non-unique solutions.
\item
By solving $\Pi_{\rm M}$ from an arbitrary configuration in the energy density, at given homogeneous field value $\varphi = \varphi_0$,
\begin{align}
    \Pi_{\rm M} &= \pm \sqrt{2(\rho - \rho_{V}) } \\
    \rho_V &= V(\varphi_0) = \rm {const.}
\end{align}
Often this method will lead to initial conditions with only either positive or negative $\Pi_{\rm M}$ regions. 
\item
By setting an inhomogeneous conformal factor $\psi$ and then solving $\Pi_{\rm M}$ with
\be
    \Pi_{\rm M}^2 = - \frac{\psi^{-5}}\pi  \nabla^2\psi(x) +  \frac1{24} K^2(x) - \rho_V %
\ee
where $K$ here is given by
\begin{align}
    K &= \sqrt{ 24(\Psi + \rho_V)}\\
    \Psi &= \max \left( \frac{ \psi^{-5}}\pi \nabla^2\psi(x) \right)
\end{align}
In particular, we choose $\psi$ of the form  
\be
\begin{split}
& \psi = \exp\left(\zeta/2 \right) 
\\
\zeta = \exp& \left( - \vec x^2/\sigma_\zeta^2 \right) ~, \quad \sigma_\zeta  \approx \frac{L}{10}
\end{split}
\ee
which generates a spherical $\Pi_{\rm M}=0$ contour centred in the simulations grid  
,  allowing positive and negative values of $\Pi_{\rm M}$.
\end{enumerate}
\clearpage

\tocless\section{Results  \label{p1_sec:results}}

In this section, we present the main results of our simulations.  The paradigm that we consider is a universe dominated by a single scalar field, initially conformally flat, and in absence of ECMs.  Nonetheless, ECMs will develop during the time evolution and affect the dynamics of the system. Such modes are then   \textit{self-genereted} by the inhomogeneities present during the evolution.
We also emphasize that in models with a plateau potential like Starobinsky or Higgs inflation, the role of the potential on the highly inhomogeneous dynamics of preinflation is marginal and only provides the reference energy scale $H_{\rm {Inf}} $ at which inflation begins.  Because we chose a mean scalar field value in the slow-roll plateau, one has $H_{\rm {Inf}} \approx 10^{-6} M_{\rm pl}$.

It is important to notice that $H_{\rm {Inf}}^{-1}$ \textit{is not} related to the size of the Hubble horizon prior to the inflation onset.  During preinflation, this depends on the total energy content in given volumes, and for our choice for the initial conditions, the initial Hubble scale is ${ H_{\rm ini} \approx \langle \rho \rangle^{1/2} }$.

By considering fluctuations of size similar to the lattice size $L$, one can distinguish between super-Hubble ($L > H_{\rm ini}^{-1}$) and sub-Hubble ($L \lesssim H_{\rm ini}^{-1}$) initial conditions.  Only the Planck scale limits the initial energy density, and so the amplitude of field fluctuations.  In our simulations, we only consider energy densities up to two orders of magnitudes from it.
\ \\

\tocless\subsection{Homogenisation phase}  

\begin{figure}[ht!]
\begin{center}
\hspace*{-0.75cm}
\includegraphics[width=0.990\textwidth]{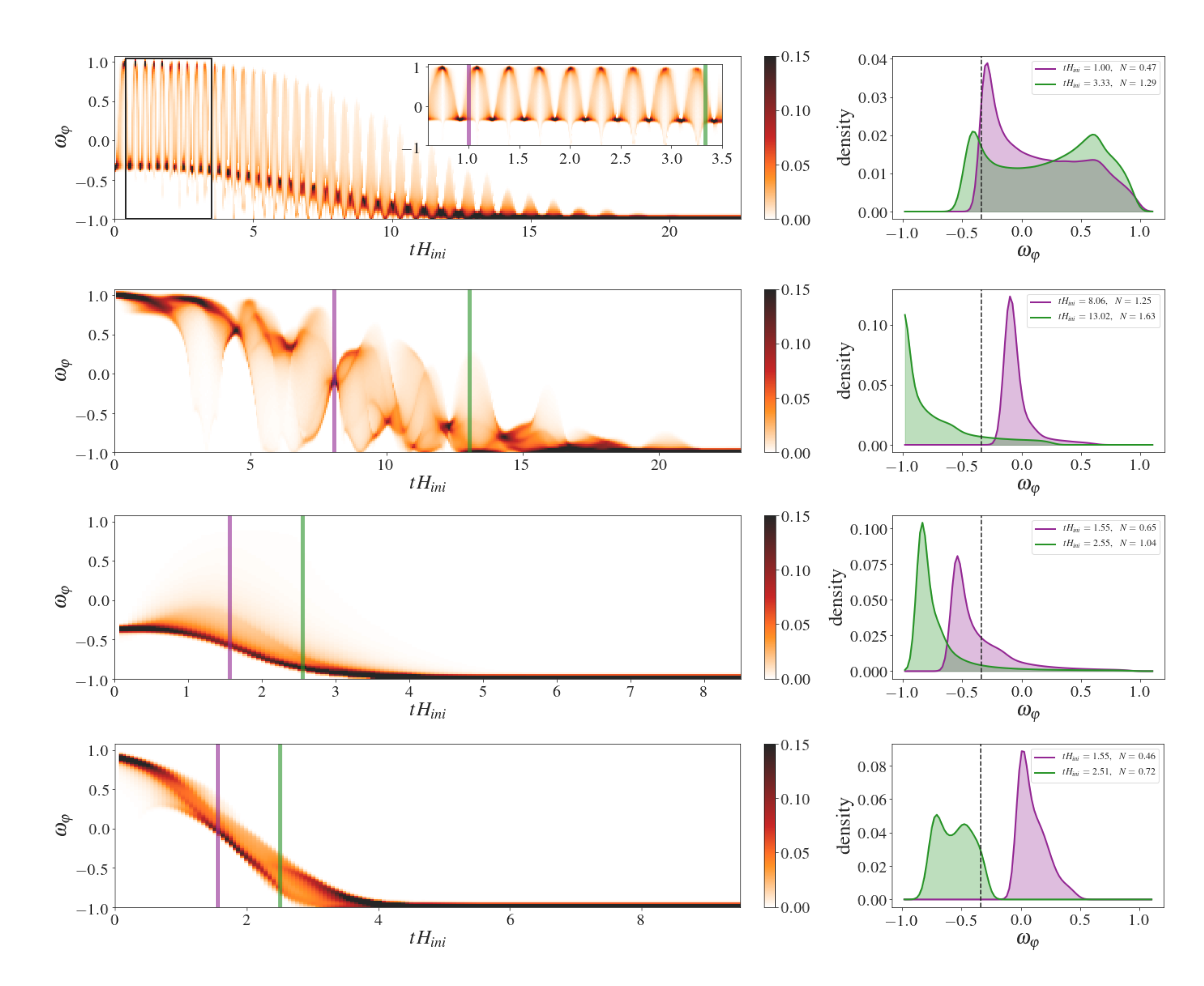}
\caption{ 
Illustrative examples of the dynamics of the equation of state for sub-Hubble and gradient and kinetic initial conditions (Top, centre-top), and super-Hubble with gradient and kinetic initial conditions (centre-bottom, bottom). In the left, the distributions in the physical grid are represented over Hubble times. Vertical purple and green lines within indicate selected times for which the distribution is represented in the right panels. Vertical dotted lines indicate the $\omega_\varphi = - 1/3$ threshold. 
\label{p1_fig:EOS_sample}}
\end{center}
\end{figure}

The scalar field gradients in principle decay like ${\rho_{\rm grad} \propto a^{-2}}$, while
its kinetic energy
scales like ${\rho_{\rm kin} \propto a^{-6}}$. However,
these two contributions are observed to alternatively dominate the total energy content, and shortly after the onset of the simulations, the effective scaling goes like
\be
\langle \rho_{\rm grad} +  \rho_{\rm kin} \rangle \propto a^{-4} ~.
\ee
This mixing behaviour, to some extent, is generic for both gradient and kinetic dominated initial conditions.  But this takes place at a slower rate in the case of a large and dominant background field velocity.  Illustrative examples of this evolution are provided in Figure~\ref{p1_fig:EOS_sample}. These  results are consistent with the analytical predictions in Refs.~\cite{Goldwirth:1990pm,Goldwirth:1991rj} and later confirmed in Ref.~\cite{East:2015ggf}.

In the gravity sector, the breaking of the initial ``staticity" (allowing $S_i \neq 0$), triggers perturbations in the extrinsic curvature. This is manifested by a growing variance of its components, i.e. with ${\langle \Delta K \rangle ^2 = \langle \left(K - \langle K\rangle \right)^2 \rangle > 0}$, and similarly  $\langle \Delta \mathbf{\aleph} \rangle >0 $. After a short period of time, a new inhomogeneous equilibrium is reached and a  ``re-homogenisation" phase begins leading approximately to the following scaling relations during preinflation,  

\be
\begin{split}
\langle  K \rangle \propto \sqrt{\langle\rho \rangle} ~, \quad
\langle \Delta K \rangle \propto a^{-2} ~, \quad
\\
\langle \mathbf{\aleph}  \rangle
\propto a^{-2}  ~, \quad
\langle \Delta \mathbf{\aleph} \rangle \propto a^{-2} ~.
\end{split}
\ee

The dynamics vary depending on particular examples. Such complexity is due to the combination between source terms, and the Hubble friction present in eq.~(\ref{p1_eqn:dtAij2}). At early times, the contribution from the $R\ij^{TF}$-term is dominant, which initially corresponds to the spatial variation in the scalar curvature. 

\clearpage

\vspace*{2cm}

\clearpage

\vspace*{3cm}

\begin{figure}[ht!]
\begin{center}
\hspace*{-0.5cm}
\includegraphics[width=0.990\textwidth]{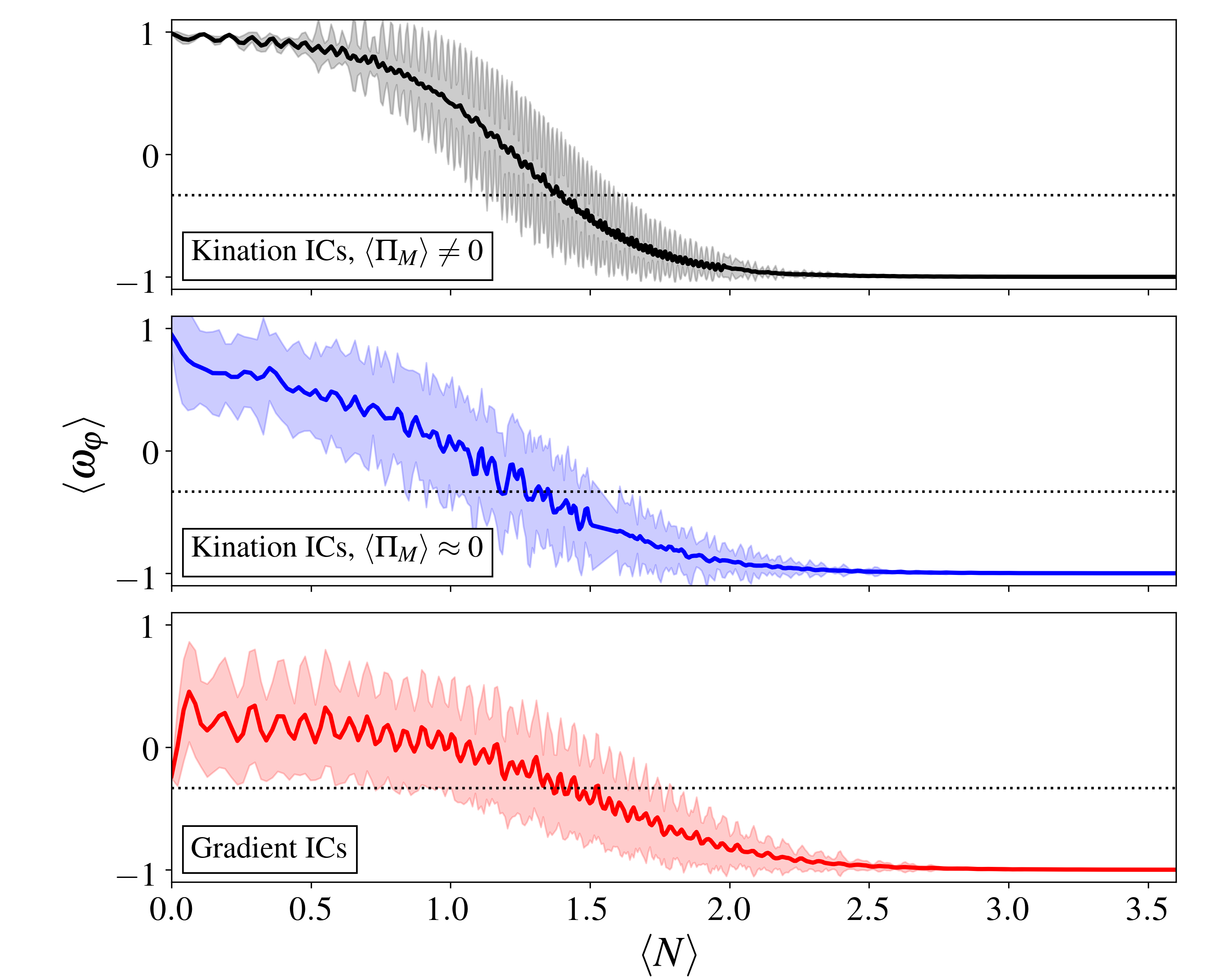}
\caption{ Mean equation of estate (solid line) and standard distribution (showed region) over mean efold of expansion. Black and Blue lines represent cases with sub-Hubble Kinetic initial conditions, Red line corresponds to sub-Hubble initial conditions in the form of gradients. Dotted horizontal line marks the threshold $\omega_\varphi = -1/3$.
\label{p1_fig:mean_EoS}}
\end{center}
\end{figure}

\clearpage

\vspace*{3cm}
\begin{figure}[ht!]
\begin{center}
\includegraphics[width=0.990\textwidth]{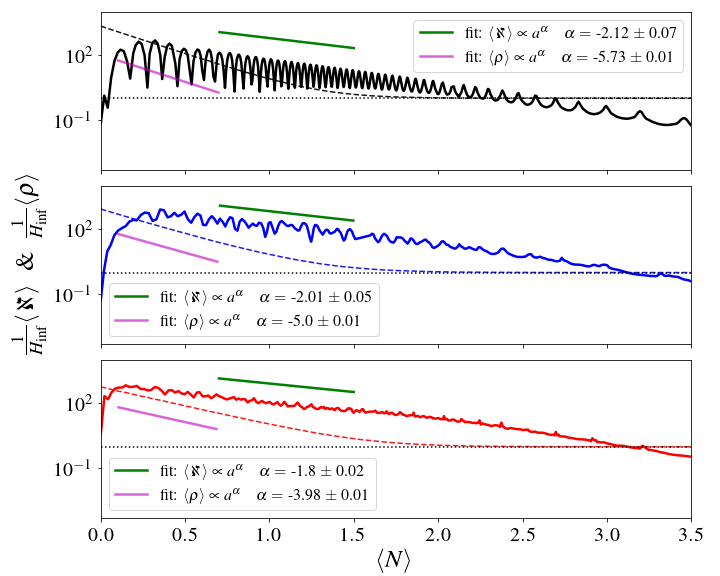}
\caption{ Same cases as in figure \ref{p1_fig:mean_EoS}, but for means of the metric (solid line), and scalar field (dash line) energy density. Green and pink lines correspond to linear fits of the scaling relations of the scalar-field and metric energy densities, respectively.  In all plots, y-axis have been normalized over the inflationary scale $H_{\rm INF}$, marked in the plot with a horizontal dotted line.
\label{p1_fig:mean_ECM}}
\end{center}
\end{figure}

\clearpage

\tocless\subsection{Oscillatory equation of state}

As we discussed in Section \ref{p1_sec:cors_universe},
the onset of inflation, in the sense of accelerated expansion, is conditioned on the equation of state being $\langle\omega_\varphi \rangle < - 1/3$.
In the following, we show that the evolution of the equation of state, for all the different sets of initial conditions, generically lead to inflating regions.

For sub-Hubble modes, Fig \ref{p1_fig:EOS_sample} (top panel) displays the time evolution of the distribution over the physical lattice of $\omega_\varphi$, for an illustrative case initially dominated by gradient energy density due to a single sinusoidal mode for the scalar field. It puts in evidence the oscillatory behaviour of  $\omega_\varphi$.  During this phase, the effective equation of state corresponds to a radiation dominated universe with  $\langle \omega_{\varphi} \rangle_{\rm eff}\approx 1/3$, which transits to potential domination and almost de-Sitter expansion, after roughly $\Delta N \approx 2$ efolds, where $\langle \omega_{\varphi} \rangle_{\rm eff} \approx -1$.  In this case, short-lasting contracting regions appear, whose integrated ADM-mass is above the Planck mass, but no black hole has been found by using an apparent-horizon finder code.

Conversely, for the super-Hubble gradient scenario shown in the third panel, the overall region tends to smoothly transit into the inflation, where most of the energy is found to be dissipated in a few Hubble times. However, part of the initial gradient energy is feeding a relativity small but strongly contracting region, undergoing kination, and forming a black hole (of mass $M_{\rm BH} \approx 10^5 M_{\rm pl}$).

Figure \ref{p1_fig:EOS_sample} also shows examples of kinetic-dominated initial conditions.  In the second panel, we consider (large) sub-Hubble fluctuations in the initial field velocity on top of a non-vanishing value.  It shows a strongly varying equation of state. Long-lasting contracting regions appear after four Hubble times but still with no black hole found. The latter panel corresponds to a similar case but now initial kination at super-Hubble scales. The kinetic energy density is dissipated entirely with almost no mixing into gradients. This particular case smoothly transition into inflation and without forming black holes.

Additional cases are displayed in Figure \ref{p1_fig:mean_EoS}, it shows the mean evolution of the equation of state with sub-Hubble inhomogeneities. The first two examples respectively correspond to kination with and without dominant background, and the third one corresponds to a configuration with gradients. Remarkably, the dynamics of mean averages of inhomogeneous kination are in very good agreement with the semi-analytical solution of the \textit{homogeneous} case (e.g. see section II-C in Ref. ~\cite{Chowdhury:2019otk}).
Figure~\ref{p1_fig:mean_EoS} can be complemented with Figure \ref{p1_fig:mean_ECM}, where the mean metric and scalar-field energy densities are displayed for the same examples (see colour-code).
\ \\

\tocless\subsection{Generation of extrinsic curvature modes}

The observed dynamics in the equation of state does not go unnoticed in the gravity sector.  
We find that ECMs are generated in the case of (large) sub-Hubble scalar field inhomogeneities.  This is not the case for super-Hubble fluctuations, for which there is no oscillation between the energy in the metric and the other energy components.  Instead the metric energy smoothly grow in contracting regions that end up in the formation of black holes. The rise of $\langle \mathbf{\aleph}\rangle$ can therefore be seen as a byproduct of preinflation black hole formation.  But since this happens on super-Hubble scales, they do not have a significant impact on the expansion dynamics elsewhere.  

For oscillatory sub-Hubble ECMs the situation is different. These modes are strongly sourced at early times as a product of non-vanishing anisotropic stress tensor and Ricci tensor. The $\langle \mathbf{\aleph} \rangle$ reaches a maximal mean value in less than $\Delta N\approx 0.5$ efolds, and then it decays approximately as $\langle \mathbf{\aleph} \rangle \propto a^{-2}$ during the rest of preinflation (see Figure \ref{p1_fig:mean_ECM}).  Because the decay rate of ECMs is lower than for other energy components,
they can potentially provide, in average, a dominant contribution to the preinflation dynamics. Nonetheless, this scenario is restricted to very large sub-Hubble perturbations in the matter sector. One typically needs dominant perturbation sizes of $ \lambda/ H^{-1} \gtrsim 0.1 $  for the effect to be significant. Yet, this condition is weaker at higher energy scales ({${\langle \delta\rho \rangle \sim 0.01 ~ M_{\rm pl}}$}), because of a longer duration of the homogenisation phase and given that $\langle \mathbf{\aleph} \rangle / \langle \rho \rangle \propto a^2$.

In Figure \ref{p1_fig:mean_postinf_ECM}, the late time scaling is shown, once accelerated expansion has begun. During the transition towards inflation, their decaying rate gradually strengthen, reaching up to $\langle \mathbf{\aleph} \rangle \propto a^{-4}$. For sub-Hubble ECMs, the oscillation frequency also gradually decreases, until they eventually ``freeze-out" at horizon exit. 
\ \\

\tocless\subsection{Contracting regions and black holes}

Even if the condition $\langle K \rangle < 0$ is always satisfied,
over-densities may generate local contracting regions, themselves embedded in expanding ones. 
Figure \ref{p1_fig:subHub_BH} shows examples of sub-Hubble and super-Hubble initial conditions, where time-histograms of expanding and contracting regions are plotted in the top and bottom panels, respectively. Contracting regions may develop because of over-densities in either the scalar-field or metric energy densities.  

When they occur because of the matter sector, they are formed inside kinetic dominated regions driven by strong scalar field Laplacians that overcome the Hubble friction in Eq.~({\ref{p1_eq:dtPi}). Once contraction is triggered, the Hubble friction turns into a kind of Hubble boost for the kinetic energy, enhancing the formation of black holes.

When contracting regions develop because of over-densities in the metric sector, they originate from the ECMs sourced by the sub-Hubble scalar field dynamics. %
The formation of such contracting regions allows us to speculate on a distinct mechanism for (low mass) black hole production,  consisting of the gravitational collapse of ECMs in the sub-Hubble regime. 
However, we have not been able to confirm this by finding the apparent horizon with our codes. This could happen when the radius of the forming black hole is smaller than the resolution of the grid. Another possibility is that black hole formation is aborted if the equation of state rapidly approaches the inflationary attractor {$\omega_\varphi \approx -1$}. Determining whether or not these contracting regions develop into black holes is left for a future study.

For scalar field over-densities on super-Hubble scales, the above described kination-trigger mechanism occurs, and we confirm the results of previous works reporting on the formation of PIBHs \cite{Clough:2016ymm}. 
An apparent horizon is found and so the black hole mass can be inferred. In our simulations, the mass range of such black holes varies from $10$ to $10^5 ~M_p$.  Like in \cite{Clough:2016ymm}, the PIBH mass saturates at a given perturbation threshold, and decreases for very high amplitudes of the perturbations.  
It is also interesting to simulate the black hole formation with slightly more rigid gauge drivers, in order to have an alternative picture of the collapsing region. In Figure \ref{p1_fig:HExt_Sample} we provide such an example where, starting from super-Hubble gradient-dominated initial conditions, we find a bifurcation in the energy density between the expanding and collapsing region occurs after $N \sim 0.6$ efold. Interestingly, we notice that homogenisation is enhanced in the neighbourhood of the collapsing region, due to the gravitational pull of the yet-to-form black hole.  This suggests that super-Hubble scalar field fluctuations, while forming PIBHs, may also facilitate the onset of inflation due to a higher level of homogenisation around the black hole, which also ``traps'' most of the generated metric energy.
\vspace*{1cm}
\begin{figure}[ht!]
\begin{center}
\includegraphics[width=0.990\textwidth]{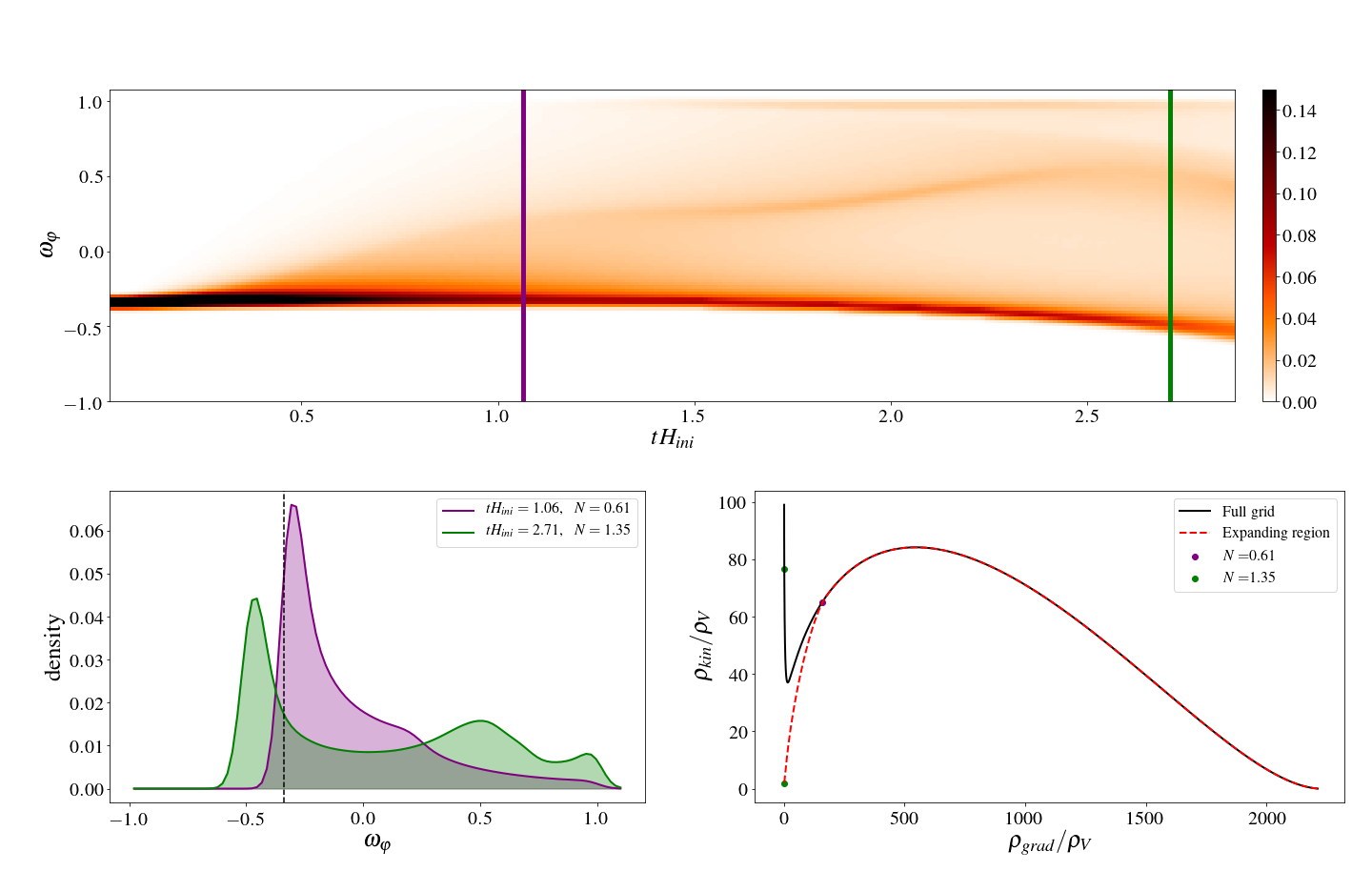}
\caption{ 
Top and bottom-left are similar than figure \ref{p1_fig:EOS_sample}, but for a critical simulation with super-Hubble gradient initial conditions evolved short after a black hole is formed in the $\omega_\varphi \approx 1$ region. Lower-right plots indicates trajectories in terms of the means of  kinetic and gradient energy, both normalized over the potential. Solid-black line corresponds to the whole grid averaging while dotted-red line only averages on expanding regions. 
\label{p1_fig:HExt_Sample}}
\end{center}
\end{figure}

\begin{figure}[t!]
\begin{center}
\includegraphics[width=0.48\textwidth]{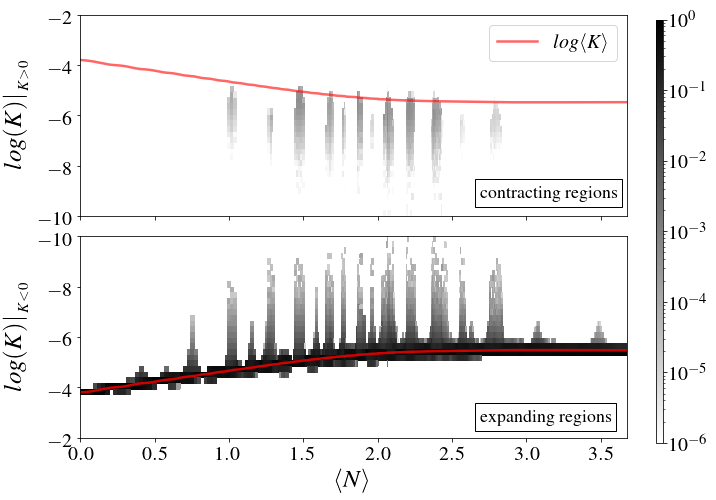} 
\includegraphics[width=0.48\textwidth]{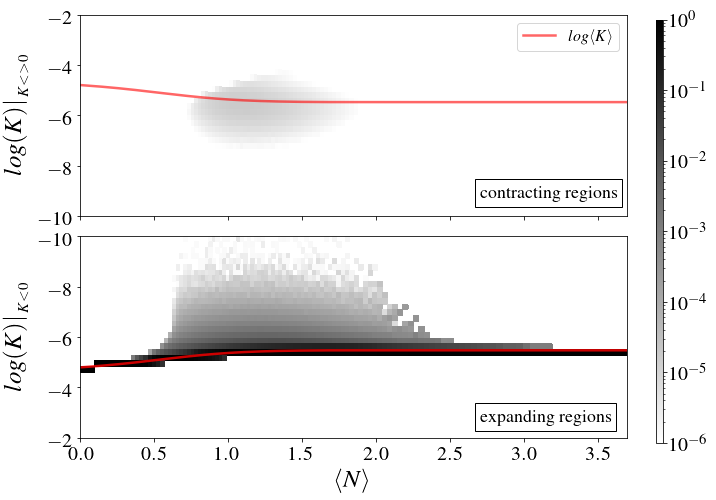}
\caption{ Distribution (logarithmic scales) of $|K|$ on the physical grid over mean number of efolds,  discerned into contracting (top) and expanding regions (bottom). The red-solid line indicates the mean of $K$. The plots corresponds to the simulations with sub-Hubble gradient initial conditions (left) and  with super-Hubble gradients (right)  both shown in fig. \ref{p1_fig:EOS_sample}. For the super-Hubble case, a black hole formation of mass $ M_{BH} \approx 2.6 \cdot 10^4 M_p$ is confirmed. After $N\approx 2$, the black hole falls down from the resolution grid of our simulation and we lose track of it. 
\label{p1_fig:subHub_BH}
}
\end{center}
\end{figure}

\begin{figure}[b!]
\begin{center}
\includegraphics[width=0.70\textwidth]{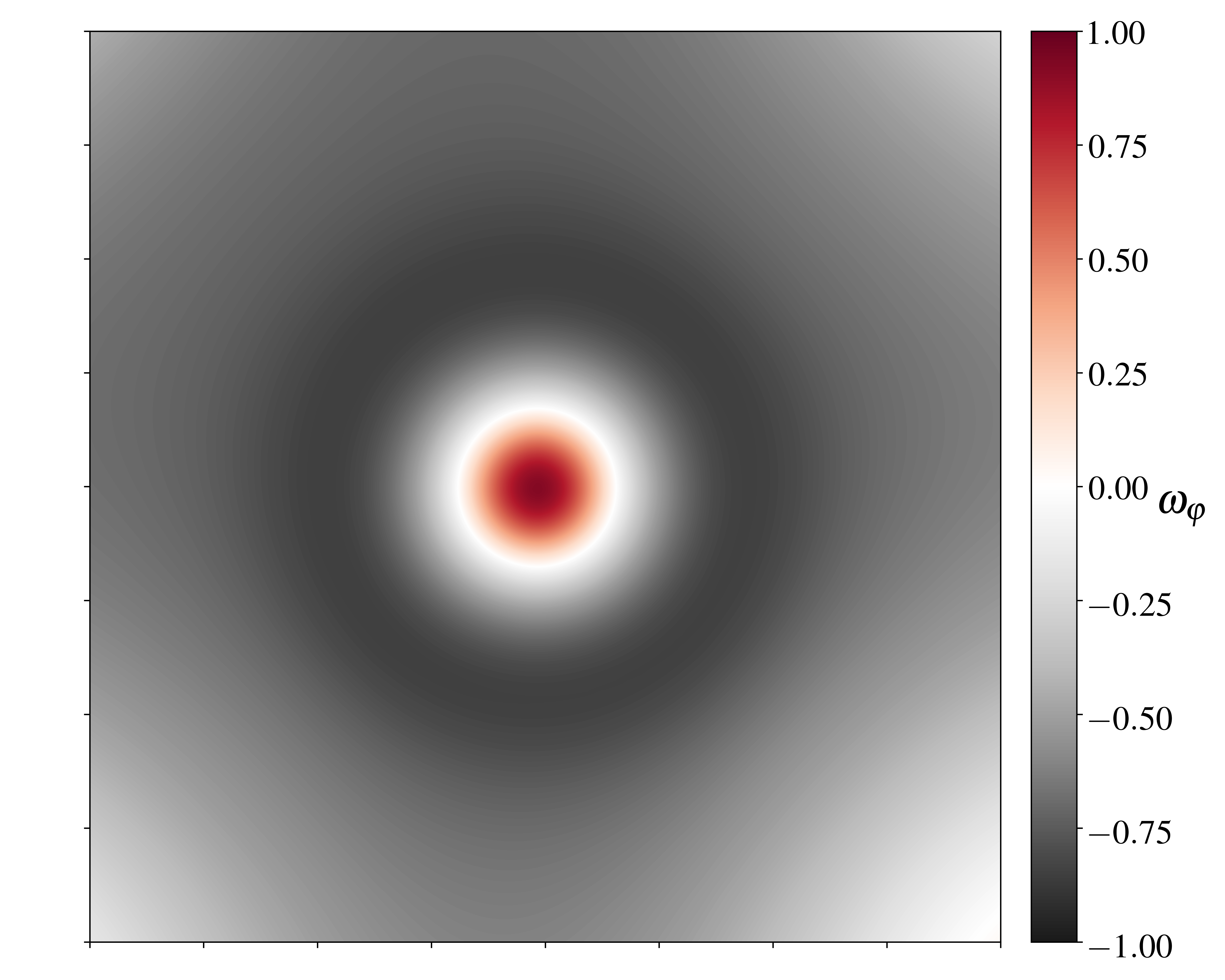}
\caption{
Equation of state in the black hole equatorial slice, corresponding to the example shown in Figure \ref{p1_fig:subHub_BH} (right)  at $N \simeq 1$.  The central red disk corresponds to the kination region and coincides with the region where the black hole forms.  The surrounding darker shell is a region where the equation of state is close to ${\omega_\varphi=-1}$, and so to the inflationary attractor. This suggests that the PIBH formation acts as a catalyst of inflation. }
\label{p1_fig:BH_slc}
\end{center}
\end{figure}

\clearpage

\tocless\section{Discussion on preinflation \label{p1_sec:discussion}}

The results of our lattice simulations have a series of implications, discussed below, about the naturalness of (chaotic) inflation, the emergence of classical conditions from a suspected quantum-gravity regime, the dynamics and duration of the preinflation era, the formation of preinflationary black holes.  
\ \\

\tocless\subsection{On the universal dynamics}

Our analysis extends previous works by considering not only large inhomogeneities in field gradients but also in the field velocity.  In both cases, the simulations for sub-Hubble fluctuations have produced similar dynamics for the preinflation era, with oscillations between kinetic and gradient terms in the density.  This result reinforces the robustness of inflation to its initial conditions, also in the case of non-linear sub-Hubble inhomogeneities.  Even if our initial conditions still assume a homogeneous expansion rate and conformal flatness, this configuration quickly evolves towards more general inhomogeneous configurations for all BSSN variables, including tensor and vector modes in the metric.  Therefore, our simulations also suggest that such behaviour is universal and insensitive to the exact configuration of the initial conditions.  
\\

Nevertheless, we have also identified a regime named \textit{kination}, in which the inhomogeneity in the field velocity are on top of a background velocity.  In such a case, the density is dominated by the kinetic term and the effective equation of state remains close to $\omega_\varphi=1$.  One can therefore wonder if drastic inhomogeneous configurations for the extrinsic curvature, together with gradient and kinetics in the scalar field,
could also lead to radically different regimes.  This problem is left for future work, because solving numerically the Hamiltonian and momentum constraints in such a case is still today a challenge \cite{Garfinkle_2020}.
\ \\

\tocless\subsection{ On the characteristic scale of the inflaton} 
The Hubble scale is one of the fundamental notions in order to understand gravitational interactions at large scale, because modes freeze at super-Hubble distances. However, in inhomogeneous scalar field cosmologies the Hubble scale depends, in part, on the contribution of gradient and kinetic energies into the total energy density. Therefore, assuming a universe dominated by gradients, the largest sub-Hubble modes (i.e. those within the realm of classical Gravitation) are bounded.  Indeed, by identifying the gravitational energy roughly with 
\be
\rho_{\rm grad} \approx  \frac 12 \left( \frac{ \delta\varphi}{ \lambda}\right)^2~,
\ee
where $\delta\varphi \equiv \varphi_{\rm max} -\varphi_{\rm min}$ is the allowed scalar field variation and $\lambda$ the mode wavelength of size 
$\lambda = H^{-1}_{\rm ini}$, and by means of the usual Friedman equation, one finds that 
\be
 \delta\varphi  \lesssim  \sqrt\frac 3 {4\pi}  \approx  \frac12 ~ m_{\rm p}~.
\ee

For plateau-like models where the field difference between the slow-roll region and the bottom of the potential is  
Super-Planckian,
the potential gradient is in general small.  The average value of the field follows the Klein-Gordon equation with a strong friction term $H_{\rm ini}$, and almost negligible driver $V'(\varphi)$ everywhere. Thus, during preinflation the trajectory of the mean scalar field down the potential is strongly suppressed, ensuring a large number of efolds after inflation begins. This is not the case for small-field potentials along the sub-Planckian slow-region~\cite{Clough_2015, Aurrekoetxea_2020}. This is the same reason why large-field models are in general more robust to the initial conditions than the small-field ones. 
\\

A similar reasoning can be made when the field 
has inhomogeneous kinetic initial conditions, when 
kinetic and gradient terms mix and mimic a radiation domination era.  In the case of a dominant background (super-Hubble) field velocity, the system is analogous to an homogeneous kination phase discussed in Ref. \cite{Chowdhury:2019otk}, and in agreement with our simulations, it shows that Starobinski inflation produces sufficient amounts of efolds during inflation when starting at super-Planckian field values, i.e. $\langle \varphi\rangle_{\rm ini} \gtrsim m_{\rm p}$.  
\ \\

\tocless\subsection{On the preinflationary black hole formation}

The transition from gradient to kinetic-dominated energies can give rise to contracting regions, typically where the scalar field Laplacian is the largest.  When this happens, the equation of state is $\omega_\varphi \approx 1$, one has locally
\be
-K < 0, \hspace{1cm} \partial_t (-K) < 0
\ee 
and there is simply no possibility for this region to expand again.  Therefore, this situation generically leads to the formation of PIBHs and the use of the puncture gauge guarantees that the lattice only probes the region surrounding those black holes.  For sub-Hubble fluctuations, there are many oscillations, and we observe the formation of a contracting region at similar locations.
If these regions form smaller PIBHs, those oscillations can thus either feed previously formed black holes, which would then grow in mass until inflation takes place, or  produce new PIBHs.  Their mass is of the order of the Planck mass for sub-Hubble fluctuations, so they would evaporate relatively quickly and produce a particle bath that is not taken into account in simulations, as well as eventual Planck relics.  In any case, the subsequent phase of inflation dilute them such that the density of these relics in our Hubble volume would be extremely small or vanish.
\ \\

\begin{figure}[ht!]
\vspace*{4cm}
\begin{center}
\includegraphics[width=0.80\textwidth]{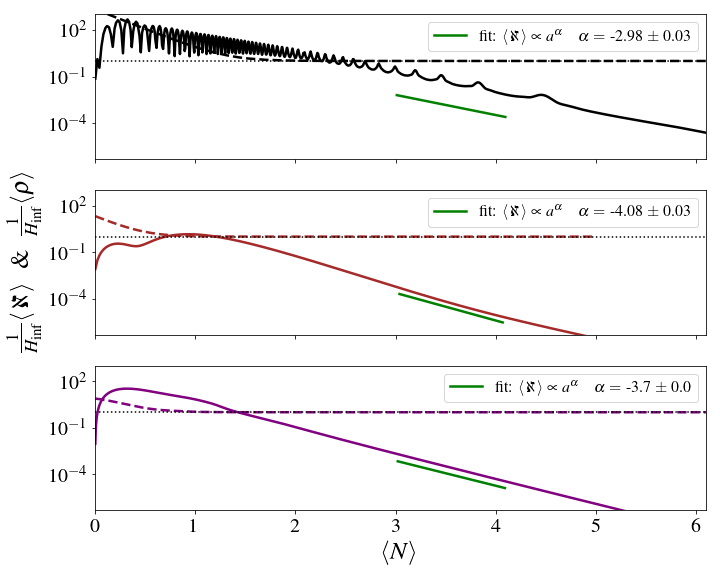}
\caption{ Similar as in figure \ref{p1_fig:mean_ECM}, but for late time dynamics. Top corresponds to a sub-Hubble kination ICs, while middle and bottom panel does to super-Hubble kination and gradient ICs, respectively.
\label{p1_fig:mean_postinf_ECM}}
\end{center}
\vspace*{4cm}
\end{figure}

\tocless\subsection{On the duration of the preinflation era}

A common picture of the very early Universe is the one of a classical phase, described by General Relativity, emerging from an unknown regime of Quantum Gravity at the Planck scale.  During this phase, there is no reason supporting homogeneity on scales larger than the initial Hubble radius.  For a while, this has been considered as a potential harm for inflation.  But simulations in full numerical relativity, like the ones presented in this work, show that non-linear Hubble-sized and sub-Hubble inhomogeneities do not prevent the onset of inflation after the density is damped by expansion at the level of the potential energy of the scalar field.  Several regimes have been identified for the preinflation era, characterized by different effective equation of states. We have emphasized the importance of  the $\mathbf{\aleph}$-term in the dynamics of the expansion.

This allows to set a limit on the maximal duration of the preinflation phase, expressed in terms of an averaged number of e-folds of expansion $\Delta N_{\rm pre-inf}$.  If the energy stored in field gradients, damped like $\rho_{\rm grad} \propto  \exp(-2N)$, were dominant all during preinflation, one would get  $\Delta N_{\rm pre-inf} \approx - \ln \left( \rho_{\rm V}\right) /2$.  However, because of the wobbling between kinetic and gradient contributions, the density is effectively damped like ${\rho \propto \exp(-4N)}$ on average, and thus $\Delta N_{\rm pre-inf} \approx \log \left( \rho_{\rm V}\right) / 4 \simeq 7 $, for Higgs or Starobinsky inflation.

If one now allows the energy to be stored in the form of ECMs, preinflation can last up to $\Delta N_{\rm pre-inf} \approx 14$. These numbers increase if inflation took place at lower energy, and inversely.  It can also be reduced if the energy scale of quantum gravity is lower than the Planck scale.  
\ \\

\tocless\subsection{On the possible observable signatures and the minimum size of the Universe}

In the case where the time of Hubble exit of cosmological scales occurs only a few e-folds (denoted $N_{\rr i*}$ ) after the onset of inflation, preinflation should leave observable signatures in the CMB temperature anisotropies and polarization.  Indeed, in such a case the gradient and kinetic energy are not damped well below the scalar field potential energy and still slightly impacts the expansion rate during inflation.  The $\aleph$-term can also have a similar effect.  This modifies the evolution of the Hubble-flow parameters
\be
\epsilon_1 \equiv - \frac{\rr d \ln H}{\rr d N}, \hspace{1cm} \epsilon_2 \equiv \frac{\rr d \ln \epsilon_1}{\rr d N}~,
\ee
and in turn modify the predicted scalar spectral index $n_{\rm s}$ and tensor-to-scalar ratio $r$,
\be
n_{\rm s} = 1 - 2 \epsilon_1 - \epsilon_2, \hspace{1cm} r = 16 \epsilon_1~,
\ee
at first order in slow-roll parameters. This certainly requires a significant amount of tuning for these effects to be observable, allowing for just enough inflation, without totally spoiling the (almost) scale invariance of the primordial scalar power spectrum.  

Indeed, the dynamics of the expansion rate is not only driven by the potential and the kinetic energy of the field, but is also somehow impacted by gradient and metric terms.  One has to consider an effective energy density $\rho_{\rm eff} = \rho_V + \delta \rho + \aleph / 16 \pi$, and take $\delta \rho + \aleph / 16 \pi \approx \rho_V \exp(-\eta N_{\rm i*})$ after the onset of inflation, with the effective value of $\eta$ depending on which preinflation regime the ICs belong to.  If the dynamics of $\epsilon_1$ is dominated by these terms coming from the pre-inflaiton era, one gets $ \epsilon_1 \approx \eta \exp(- \eta N_{\rr i*} )$ and $\epsilon_2 = - \eta^2 $.  An important fine-tuning of the initial conditions of the scalar field is therefore required to get a value of $N_{\rr i*}$ that can explain the observed scalar spectral index.  Furthermore, this would unavoidably generate a tensor-to-scalar ratio $r \simeq 16 \epsilon_1$ larger than unity, which is excluded. 

If one assumes that the dynamics of $\epsilon_1$ is governed by the potential, which gives $\epsilon_1 \simeq 2 \times 10^{-4}$ for Higgs/Starobinsky inflation, gradient or metric terms may still govern the dynamics of $\epsilon_2$.  One then gets $\epsilon_2 \approx - \eta^2 \exp(-\eta N_{\rr i*}) / \epsilon_1 $. 
As a consequence, the pre-inflaiton era could have an observable effect on the primordial power spectrum in a \textit{just enough} inflation scenario and could even give the correct value of the scalar spectral index, at the price of a significant fine-tuning of the initial scalar field.  If the effects on the second Hubble-flow parameter, that is determined at a  $20\%$ level, are not detectable, one gets from the current limits on the tensor-to-scalar ratio, a lower bound $N_{\rr i*} \gtrsim 3 $. An interesting corollary is that, in the above-mentioned scenario, the non-detection of the imprints of preinflation implies that the Universe is at least $\exp[3 (N_{\rm pre-inf}+ N_{\rm i*})] \gtrsim 10^{13}$ times larger than our current Hubble volume.
\ \\

\tocless\subsection{Is Linde's picture right?} 

In all the considered cases and regimes, inflation is a generic outcome, at least in most parts of the lattice. 
This conclusion applies to all expanding (in average) initial conditions with scalar field fluctuations of any size, as long as there exist a Hubble patch that its mean field value is in the slow-roll region of the potential.
On the contrary, inflation cannot be triggered from sub-Hubble fluctuations around the bottom of the potential, even if apparently the energy density initially stored in field gradients or velocity is much larger than the potential barrier to reach the slow-roll region.  The reason is that field gradients act as a damping term in the Klein-Gordon equation for the scalar field while large gradients in the field potential 'drags" the scalar field down towards the bottom, as seen in Ref. \cite{Aurrekoetxea_2020}.
One should not interpret this as an issue, since a large background field value may have emerged in an initially flat and compact Universe, with a 3-torus topology, as suggested by A. Linde \cite{Linde:2004nz,Linde:2017pwt,Linde:1983gd,Goncharov:1983mw,Linde:1983mx,Kofman:1985aw}, in support of chaotic inflation.  The second possibility is a much larger inhomogeneous Universe, in which it is sufficient to have a single super-Hubble patch with a large mean-field value superimposed by arbitrary sub-Hubble fluctuations, to naturally lead to inflation.  
\\

One could also wonder how the topology, the initial curvature and shape of the universe could influence preinflation and even prevent inflation.  Despite that we have been using periodic boundary conditions on a cubic lattice, reflecting the topology of a 3-torus, in the presence of large inhomogeneities the Universe is formed by locally open and closed regions as explained in Ref. \cite{East:2015ggf}. Therefore, even if an homogeneous closed universe would collapse when the energy content is dominated by curvature \cite{Linde:2004nz, Linde:1987aa}, in the largely inhomogeneous case, local open regions may still be expanding and eventually lead to inflation. 
However, scenarios with globally large positive curvature have not been tested yet.  This is left of a future work. 
\ \\

\tocless\section{Conclusions  \label{p1_sec:ccl}}

Fully relativistic lattice simulations in 3+1 dimensions provide the ideal method to study the possible non-linear inhomogeneous dynamics of the preinflation era.   In this work, we have extended previous analysis and considered new realizations of the initial conditions satisfying the Hamiltonian and momentum constraints, with an inhomogeneous scalar field velocity.  Even if in general the preinflation era does not leave distinguishable signatures in observations, it is related to fundamental questions such as how natural or fine-tuned are the initial conditions of inflation and how generic is the transition from a semi-classical regime emerging from some quantum gravity period, towards inflation.  In particular, the realisation of inflation starting from non-linear and sub-Hubble field fluctuations has been a long-standing and debated problem.

Besides confirming recent results on the robustness of inflation to sub-Hubble inhomogeneities, our simulations reveal a richer preinflation dynamic than expected, as well as some universal behaviours.  A new regime in which the expansion is driven by the traceless part of the extrinsic curvature tensor has been identified, which impacts the equation of state and the duration of the preinflation era.  We have also found that initial conditions with highly inhomogeneous scalar field velocity give rise to a regime in which the density remains dominated by the kinetic term, denominated \textit{kination}.  Otherwise, sub-Hubble and Hubble-sized inhomogeneities give rise to oscillations between gradient and kinetic dominated periods.  For super-Hubble sized fluctuations, our analysis also confirms the emergence of contracting regions, leading to the formation of preinflation primordial black holes, those being subsequently diluted by inflation.  Our findings are summarized in Fig.~\ref{p1_fig:scale_diagram} showing the possible outcomes starting from different regions of the parameter space.

The chosen set of initial conditions, even if extended to highly inhomogeneous field velocities, still corresponds to very specific cases.  Indeed, the initial extrinsic curvature tensor is taken with a vanishing traceless part $A_{ij}$ and a homogeneous negative trace $K$.  Nevertheless, we point out that for Hubble-sized and sub-Hubble inhomogeneities, the extrinsic curvature tensor rapidly becomes highly inhomogeneous, thereby exploring various other configurations, until it eventually drives the overall expansion dynamics.  Our initial data also correspond to a Ricci curvature that remains small on average (but it can be very large locally).   But the issue of solving the momentum and Hamiltonian constraints on the initial hypersurface for more general cases remains a major computational challenge that still limits the range of applicability of the simulations of the preinflation era in full 3+1 numerical relativity.} We also did not consider the eventual impact of additional scalar of matter fields on the preinflation dynamics, or other topological choices than periodic boundary conditions.  

Finally, we have focused on the Higgs - Starobinsky inflation model and the phenomenology of the preinflation era is marginally impacted by the field potential that only gives a sub-dominant contribution to the energy density during most of the preinflation phase. The potential only starts to dominate just before the onset of inflation and, therefore, our conclusions should remain valid for any plateau-type scalar field potential at super-Planckian characteristic values, these being currently favoured by CMB observations.  

In summary, our work contributes to paving the way to a better and more precise understanding of the rich phenomenology of the preinflation era.  It enlarges the diversity of initial conditions that have been considered so far.  But future work and the development of more advanced numerical methods will be needed to go beyond the assumption of initial conformal flatness, and to include inhomogeneous initial configurations for the extrinsic curvature in both $K$ and $\tA\ij$.  

As perspectives, we may also consider the effect of additional scalar or matter fields, analyze the dynamics for small-field potentials, and study more deeply the formation process of PIBHs.  In a semi-classical description of the preinflation era, it may also be possible to consider the non-linear effects of quantum scalar field fluctuations by adding a stochastic term in the Klein-Gordon equations.
It may be also interesting to study the case where preinflation leaves observable signatures in CMB observations, even if this requires a significant fine-tuning.

Finally, our work emphasizes that in absence of these signatures and assuming that classical preinflation emerged at the Planck scale, the Universe is at least ten thousand billions times larger than our current Hubble volume, usually referred as our observable Universe.

\chapter{Multi-field Higgs inflation: preinflation and preheating}
\label{p2_chap:prepaper2}
\pagestyle{fancy}

\section{Preamble}

In this section, we present another original research article where we expand the previous work, in the context of Higgs inflation, by considering the case of a multifield dynamical settings. In order to implement the reheating, additional degrees of freedom are necessary to allow for the energy transfer, via parametric resonance, from the inflaton condensate into the ultrarelativistic particles of the Hot Big Bang plasma. This phase occurs at the end of inflation, once the scalar field kinetic energy becomes comparable to the potential energy and breaks down the slow-roll conditions. Then, the field typically starts oscillating around the potential minimum, which enhances the fluctuations from the matter sector. This phase is known in the literature as the preheating epoch. 

The goal of the following article is two-fold:  First, \textit{(i)} we study the non-perturbative dynamics of preheating in both the scalar-field and the metric sectors, and we point out the emergence of structure formation already during the preheating. We note that full numerical relativity simulations for studying the preheating have been prohibited until recently by limited computational resources. The first work dates to 2019 in \Ref{PhysRevD.100.063543} which compared results from using the (linearized) CPT and (full-gravity) BSSN formalisms. Moreover, numerical relativity simulations on non-minimally coupled fields, and particularly Higgs preheating, are first considered in our paper.   Secondly, \textit{(ii)} we study the preinflationary era with inhomogeneous multifield configurations to cross-check that the presence of additional fields does not hinder the beginning of inflation. In there, we show that the non-minimal coupling between the Higgs field and gravity results in a stabilization mechanism that enhances the inflaton dynamics, facilitating the transition to inflation.

As a remark, in the context of preheating, the non-perturbative dynamics of the field have been studied extensively using lattice simulations with CPT, i.e. using linearised Einstein gravity, under the assumption that, at first order, metric back-reactions could be ignored. With this formalism, in the case of Higgs inflation, it was shown that self-interactions of the field alone, without the need for additional fields or other matter sources, were enough to quickly create abundant Higgs particles that dominate the dynamics (see Ref.~\cite{Sfakianakis2019}). However, our results indicate the opposite, and while non-linearities of the metric can enhance the structure formation in the final stages of the broad resonance period, the resonance's efficiency on particle production becomes diminished when metric perturbations are considered. In consequence, in our work, self-resonances of the Higgs alone cannot rapidly preheat the universe, and couplings to other field/matter sources are necessary. 

The following article has also been published in \textit{Physical~Review~D} in Ref.~\cite{Joana:2022uwc}. In order to facilitate its readability, we reproduce the article as published, although this includes some redundancies with the previous chapters of this thesis.

\newpage

\section*{Research article}
\addcontentsline{toc}{section}{Research article}

 \vspace*{10mm}
 
 \begin{center}
 \textbf{ \large 
Gravitational dynamics in Higgs inflation: \\ Preinflation and preheating with an auxiliary field
 }
\\[5mm]
 Cristian Joana
\end{center}

\vspace{5mm}

\hspace*{0.025\textwidth} 
 \parbox[t]{0.95\textwidth}
 {
The dynamics of both the preinflationary and the preheating epochs for a model consisting of a Higgs inflaton plus an additional auxiliary field are studied in full General Relativity. The minimally coupled auxiliary field allows for parametric-type resonances that successfully transfer energy from the inflaton condensate to particle excitations in both fields. 
Depending on the interaction strengths of the fields, the broad resonance periods lead to structure formation consisting of large under/over-densities, and possibly the formation of compact objects. Moreover, when confronting the same model to multifield inhomogeneous preinflation, the onset of inflation is shown to be a robust outcome. 
At relatively large Higgs values, the non-minimal coupling acts as a stabilizer, protecting the dynamics of the inflaton, and significantly reducing the impact of perturbations in other fields and matter sectors.  These investigations further confirm the robustness of Higgs inflation to multifield inhomogeneous initial conditions, while putting in evidence the formation of 
structure during the reheating.

}

\vspace{10mm}

\tocless\section{Introduction \label{p2_sec:intro}}

Cosmic inflation \cite{STAROBINSKY198099, PhysRevD.23.347, 10.1093/mnras/195.3.467, LINDE1982389} is the current paradigm of the early universe. It postulates an early phase where the universe underwent over a large period of accelerated expansion. Such period provides an explanation for today's large scale homogeneity and flatness of the Universe.
During inflation, quantum fluctuations became red-shifted, exiting the Hubble horizon at the time, and leading to a scale-invariant power spectrum of cosmological perturbations which can be matched to current observations \cite{Akrami:2018odb,Ade:2015lrj}. At later times, they provide the seeds needed for structure formation. 

In a universe governed by the Einstein's field equations, the accelerated expansion of the universe is obtained when the effective equation of state is strictly smaller than $\omega < -1/3$. In the slow-roll inflationary paradigm, this is typically achieved by postulating a universe dominated by a scalar field (slowly) rolling down its potential.  
%
Assuming homogeneity and isotropy, from the shape of the potential, the slow-roll conditions can be derived. When these conditions are satisfied, the energy budget is dominated by the potential energy (i.e it keeps an $\omega \approx -1$), and a sustained period of slow-roll inflation occurs. Surely, assuming homogeneity and isotropy for the initial conditions of the universe is one of the main problem inflation is supposed to solve, so the initial conditions required for inflation have often been a topic of controversy, e.g. in Refs.~\cite{Goldwirth:1989pr, Goldwirth:1990pm, Laguna:1991zs,KurkiSuonio:1993fg,Deruelle:1994pa} and more recently in Refs.~\cite{Martin:2013nzq, Ijjas_2013,Guth_2014,Easther:2014zga, Ijjas_2016,Chowdhury:2019otk}. Thus, the remaining question is: Can generic (inhomogeneous) preinflationary scenarios successfully lead to enough cosmic inflation ($\sim 60$ \text{efolds})?  

The issue of initial conditions for inflation has been studied extensively using analytical, semi-analytical and numerical approaches (for a review see Ref.~\cite{Brandenberger:2016uzh}). Full numerical relativity simulations have also been used to explore the dynamics of the preinflationary era beyond the perturbative regime. These have consisted of scenarios with a highly inhomogeneous scalar field \cite{East:2015ggf,Clough:2017ixw} and large tensor perturbations \cite{Clough_2018}.  The effects of concave and convex potential shapes were also studied in Ref.~\cite{Aurrekoetxea_2020}.

In our previous paper \cite{Joana2020}, the case of (single field) Higgs/Starobinsky preinflation was considered, containing large field gradients and inhomogeneous kinetic energies across Hubble scales. We have shown that for this model, gravitational shear and tensor modes can potentially delay the onset of inflation, but never prevent it. The question of the implications of adding extra fields is, however, still open. 

The (p)reheating epoch is a necessary phase occurring after the end of inflation. It starts once the slow-roll conditions are violated and the inflaton condensate begins to oscillate around the minimum of its potential. These oscillations transfer energy to the matter sector, through parametric resonances, originating in the hot big bang plasma \cite{PhysRevD.42.2491,1990Dolgov172D}. Usually, in the literature, the phase when particles are produced is known as ``preheating'', while the term ``reheating'' is left for when the inflaton has effectively decayed and the thermalization phase begins. 
The reheating process has direct implications on the Cosmic Microwave Background (CMB), and current measurements are sensitive to it \cite{Martin2010,Martin2015,Martin2016}. 
For a more elaborate review on the topic, see Refs.~\cite{Brandenberger2010, Tenkanen:2020cvw}.  

The dynamics of the initial stages of preheating have been extensively studied throughout the last decade. Perturbative approaches \cite{PhysRevD.42.2491,1990Dolgov172D,Kofman1994,Kofman1997, Tsujikawa1999A, Tsujikawa1999B} and numerical lattice simulations \cite{Tomislav1997,Gary2001,Bellido2003,Amin2010, Frolov2010,Lozanov2014,Repond2016,Lozanov2019} have been used extensively while assuming linearized Einstein gravity.  Reheating involving non-minimally coupled scalar fields has also been of large interest \cite{Tsujikawa1999A, Tsujikawa1999B, Bruck2017, DeCross2018A,DeCross2018B, DeCross2018C,Sfakianakis2019,Nguyen2019,Rubio2019,Vis2020,Ema2017,Ema2021}, and includes studies of Higgs inflation \cite{Repond2016,Sfakianakis2019,Rubio2019,Hamada2021}. While lattice simulations have been capable to preserve the non-perturbative dynamics associated with inhomogeneous scalar fields, they do not consider the fully non-linear gravitational counterparts \cite{Bassett1999}, whose effects on the structure formation might lead to the early formation of black holes \cite{Jedamzik2010A,Jedamzik2010B,Zihan2020}. Recently in 2019, Giblin \& Tishue, in Ref.~\cite{Giblin2019} presented the first preheating simulations in full general relativity for the canonical $m^2\varphi^2$ inflationary model. 
While their results disfavor the formation of compact structures, for that particular model, it shows the potential of numerical relativity to clarify the role of gravitational backreactions in the early universe, complementary to the standard cosmological perturbation theory. A year later, Kou \textit{et. al.} in Ref.~\cite{Kou:2019bbc} (see also \cite{Kou:2021bij}) presented numerical relativity simulations for an alternative inflationary model which allowed the formation of oscillons during the preheating potentially collapsing into black holes. 

In this paper, I present a set of full general relativity simulations concerning both the preinflationary and preheating epochs. The non-minimally coupled Higgs inflation model has been considered in the presence of an auxiliary scalar field. 
With the help of these simulations, we first ask ourselves how a full general relativistic treatment affects the resonant dynamics of preheating, and how the coupling strength of the fields affect the formation of structures during the broad resonance phase. Then, we check whether similar dynamics can be present during the preinflationary phase and, importantly, if these can undermine  the beginning of inflation in the first place.  I show that, in the presence of additional fields, the non-minimal coupling  to gravity of the Higgs field allows for an efficient preheating process; Large amount of particles are produced and the formation of complex structures occurs. However, during preinflation, at large enough Higgs field values,  the non-minimal coupling always acts as a stabilizer that protects the dynamics of the inflaton from inhomogeneities in other fields, ensuring the success of starting cosmic inflation. 

The organization of the manuscript is as follows: in Section \ref{p2_sec:Formalism} the generalized covariant formalism is introduced while in Section \ref{p2_sec:HiggsModel}  focus on the Higgs model. %
Section \ref{p2_sec:NumStrategy} explains the numerical strategy of the simulations. The results for preheating and preinflation are presented in Sections \ref{p2_sec:SimsReheat} and \ref{p2_sec:SimsPreinf}, respectively. Additional information on the notation, code performance, initial data sets and supplementary figures are available in the appendix.  
\ \\

\tocless\section{Covariant formalism \label{p2_sec:Formalism}}

In this section, we consider a universe containing an arbitrary number of scalar fields $\bar \phi^I$, labelled by Latin capital letters $I, J, K = 1, 2, ... , N$. We consider a metric tensor $\bar g^{\mu\nu}$ in $3+1$ dimensions where Greek letters are used to label spacetime indices $\mu, \nu = 0, 1, 2, 3$, using the ``mostly plus metric'' sign convention $(-+++)$. The variables with an upper-bar or ``hat'' are being described in the Jordan frame. In this kind of models, the action in the Jordan frame is given by 
\beq %
\begin{split}
S = \int d^4 x & \sqrt{-\bar{g} } \Big[ f (\bar\phi^I ) \bar{R} 
- \frac{\Mpl^2}{2} \delta_{IJ} \bar{g}^{\mu\nu} \partial_\mu \bar\phi^I \partial_\nu \bar\phi^J - U(\bar\phi^I ) \Big] ~ ,
\end{split}
\eeq
where $\Mpl$ is the reduced Planck mass, $\bar g$ is the determinant of the metric, $\bar R$ is the Ricci scalar, $U(\bar\phi^I)$ is the scalar field potential, and $f(\bar\phi^I)$ contains the fields non-minimal coupling gravity $\xi_I$, so that
\beq \label{p2_eq:f_func}
 f (\bar\phi^I ) = \frac {\Mpl^2} 2 \left[ 1 + \xi_K \left(\bar\phi^K\right)^2 \right] ~.
\eeq

The dynamical analysis of such systems is easier to deal with in the Einstein frame. This is done by rescaling the metric tensor, under the Weyl transformation 
\bea
\bar{g}_{\mu\nu} (x) \rightarrow g_{\mu\nu} (x) = \frac{2}{M_{\rm pl}^2} f \big(\bar\phi^I\big) \> \bar{g}_{\mu \nu} (x) ~.
\eea

Thus, now in the Einstein frame, the action reads 
\beq 
\begin{split}
S = \int d^4 x \sqrt{-g} \Big[ R & - \frac{\Mpl^2}{2} {\cal G}_{IJ} (\bar\phi^K ) g^{\mu\nu} \partial_\mu \bar\phi^I \partial_\nu \bar\phi^J
- V (\bar\phi^I ) \Big] \, ,
\end{split}
\label{p2_eq:action_E}
\eeq 

Where ${\cal G}_{IJ} (\phi^K)$ is a field-space metric containing the mixing with the non-minimal coupling, 
\beq
{\cal G}_{IJ} (\bar\phi^K ) = \frac{ M_{\rm pl}^2 }{2 f (\bar\phi^K) }\left[ \delta_{IJ} + \frac{3}{ f (\bar\phi^K) } \frac{\partial f} {\partial \bar\phi^I} \frac{\partial f} {\partial \bar\phi^J} \right] ~,
\label{p2_eq:G_IJ}
\eeq

and the field potential has been redefined as 
\beq
V (\bar\phi^I) = \frac{\Mpl^4}{ 4 f^2 (\bar\phi^I) } U(\bar\phi^I ) .
\label{p2_VE}
\eeq

Varying the action of Eq.~(\ref{p2_eq:action_E}) with respect to $\phi^I$, one can find the stress tensor and the field's equations of motion:
\beq
T_{\mu\nu} = {\cal G}_{IJ} \partial_\mu \bar\phi^I \partial_\nu \bar\phi^J - g_{\mu\nu} \left[ \frac{1}{2} {\cal G}_{IJ} \partial_\alpha \bar\phi^I \partial^\alpha \bar\phi^J + V (\bar\phi^I ) \right] ~,
\label{p2_eq:Tmn_J}
\eeq
\beq
\Box \bar\phi^I + g^{\mu\nu} \Gamma^I_{JK} \partial_\mu \bar\phi^J \partial_\nu \bar\phi^K - {\cal G}^{IJ} \frac{\partial }{\partial {\bar\phi}^J} V (\bar\phi^K) = 0 ,
\label{p2_eq:eomsf_J}
\eeq

where $\Box$ is the Alambertian operator, and $\Gamma^I_{JK} (\phi^L)$ are the Christoffel symbols constructed from the field-space metric ${\cal G}_{IJ}$.

The canonical Einstein fields denoted by $\Phi^I$ are defined by solving the following system of equations 
\beq \label{p2_eq:convert_frame}
\frac{\Mpl^2}{2} {\cal G}_{IJ} g^{\mu\nu} \partial_\mu \bar\phi^I \partial_\nu \bar\phi^J = {\delta}_{IJ} g^{\mu\nu} \partial_\mu \Phi^I \partial_\nu \Phi^J ~.
\eeq
This transformation further simplifies the action in Eq.~(\ref{p2_eq:action_E}). However, finding the solution to such a system of equations is not always straightforward.  In the Einstein frame, the field equations of motion are reduced to the classical Klein-Gordon equations of the form 
\beq
\Box \Phi^I -\frac{\partial }{\partial \Phi^I} V (\Phi^K) = 0 ~.
\eeq
\ \\

\tocless\section{Higgs Inflation \label{p2_sec:HiggsModel}}

In this work we consider the model of (non-minimally coupled) Higgs inflation which
is one of the most favored slow-roll inflation models by the latest CMB data from Planck~\cite{Martin:2013nzq}. We consider a dynamical system consisting of two scalar fields and gravity. Interaction between the Higgs field and other Standard Model particles, particularly in the electroweak sector, have been ignored. The evolution has been treated classically, therefore radiative loop corrections have also been neglected. Section \ref{p2_subsec:SingleHiggs}, 
briefly reviews the formalism for the single-field paradigm, assuming the unitary gauge, while  the implications of adding extra scalar fields are discussed in subsection  \ref{p2_subsec:AuxField}. 
\ \\

\tocless\subsection{The single-field case } \label{p2_subsec:SingleHiggs}

The Higgs inflation model~\cite{Bezrukov:2007ep} postulates that the inflaton is the Higgs field from the Standard Model of particle physics, with a non-minimal coupling to gravity. The Standard Model Lagrangian, therefore, includes an extra term $\xi H^\dagger H R$, where $R$ is the Ricci scalar, and $H$ is the Higgs field in the unitary gauge \cite{Bellido2009}, 
\beq
H = \frac{\Mpl}{\sqrt{2}}
\begin{pmatrix} 
0 \\ h
\end{pmatrix} ~,
\eeq
and, $\xi_h$ is the only free  parameter of the model. This term is somehow expected as it is  naturally generated by quantum corrections in curved spacetime \cite{1970Callan}. 

In the Einstein frame, the Higgs potential reads  
\beq \label{p2_VHI_pot}
V(h) = \Mpl^4 \frac{\lambda \left( h^2 - \frac{v^2}{\Mpl } \right)^2}{4 \left( 1 + \xi_h h^2\right)^2} ~,
\eeq
the shape of which is illustrated in Fig. \ref{p2_fig:Higgs_EJ}. For the single field case, using Eq.~(\ref{p2_eq:convert_frame}), one can  convert from the Jordan frame field $h$  to the canonical inflaton in the Einstein frame $\varphi$  by solving

\beq 
 \frac 1 {\Mpl} \frac{\text{d} \varphi} {\text{d} h} = \sqrt{{\cal G}_{hh}} = \frac{\sqrt{1+\xi_h(1+6\xi_h)h^2}}{1+\xi_h h^2} ~,
\eeq
which leads to the known expression \cite{MARTIN201475}

\beq \label{p2_eq:varphi_of_h}
\begin{split} 
\frac{\varphi}{ \Mpl} = &\sqrt{\frac{1+6\xi_h}{\xi_h}} \arcsinh\left[{h \sqrt{\xi_h (1+6\xi_h)}}\right] \\ &- \sqrt{6} \arctanh\left[{\frac{\xi_h\sqrt{6} h}{\sqrt{1+\xi_h(1+6\xi_h)h^2}}}\right] ~.
\end{split}
\eeq

\afterpage{
\vspace*{2cm}

\begin{figure}[!h]
\begin{center}
\hspace*{-5mm}
\includegraphics[width=0.990\textwidth]{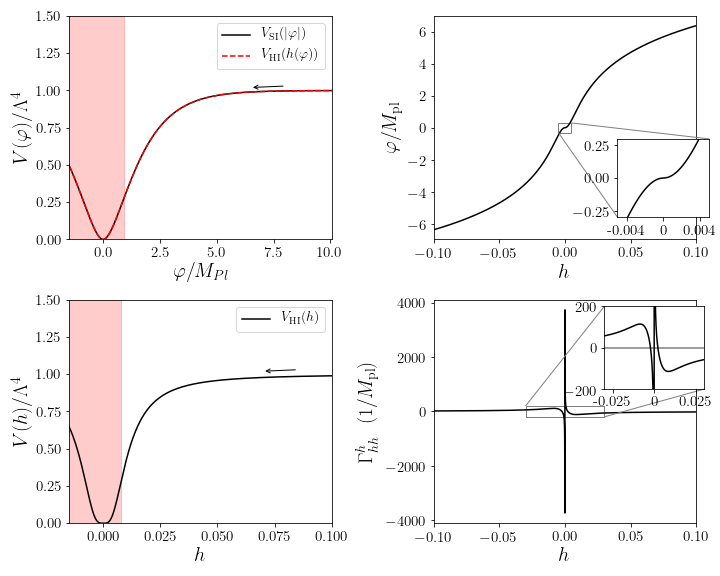}
\caption{Left panels illustrate the Higgs potential in the Einstein frame in terms of the inflaton $\varphi$ (top-left)  and the $h$-field (bottom-left). Slow-roll inflation runs from right to left as indicated by the arrow; the red-shaded area indicates the post-inflationary period, after the first slow-roll parameter becomes larger than unity.  The top-right panel shows the conversion between $h$ and $\varphi$, and in the bottom-right panel, the field-space Christoffel symbol $\Gamma^h_{hh}$ is plotted to illustrate the kinematic factor felt by $h$ due to the non-minimal coupling, as seen in the Einstein frame, (see equation  \ref{p2_eq:eomsf_J}). The large fluctuations around $h\approx 0$ result into the Riemann spikes seen during the evolution.}
\label{p2_fig:Higgs_EJ}
\end{center}
\end{figure}

\clearpage
}

Expanding the above expression and substituting it in the potential (\ref{p2_VHI_pot}), one gets, in terms of the $\varphi$-field,  

\beq \label{p2_eq:V_SI}
V_{\rm SI}(\varphi) \approx \Lambda^4 \left( 1 - {\rm e}^{-\sqrt{2/3} |\varphi| / \Mpl} \right)^2~,
\eeq
where
\beq
\Lambda^4 \equiv \Mpl^4 \lambda / (4 \xi_h^2) ~,
\eeq
is the overall amplitude of the potential. 

The energy scale of inflation is given by the amplitude of the potential, $H^2_{\rm inf} \approx \Lambda^4/(3\Mpl)$. Assuming that  the observable modes exited the Hubble radius at $N_\star = 55$ efolds  before the end of inflation, the scalar and tensor perturbations of the CMB power-spectrum lead to  $\Lambda \simeq 3.1 \times 10^{-3} \Mpl $ ~\cite{Martin:2013nzq}. Thus, the ratio between the Higgs self-coupling and the non-minimal coupling must obey
\beq \label{p2_eq:ratio_coupling}
\frac \lambda {\xi_h^2} \simeq 5 \cdot 10^{-10} ~.
\eeq
The value of the Higgs self-coupling is measured by collider physics to be 
$\lambda \simeq 0.13$ \cite{PhysRevD.98.030001}. 
Therefore, {{Eq.~(\ref{p2_eq:ratio_coupling})}} fixes the value of the Higgs non-minimal coupling to ${\xi_h \approx 1.8 \cdot 10^4}$.

At leading order, the firsts two slow-roll parameters read 
\begin{align}
\epsilon_1 &\simeq \frac{\Mpl^2}2 \left( \frac{\partial_h V}{V}\right)^2   ~,
\\ %
\epsilon_2 &\simeq 2\Mpl^2 \left[ \left(\frac{\partial_h V}{V}\right)^2 - \frac{\partial^2_h \, V}{V} \right] ~, 
\end{align}
and as long $\epsilon_1 <1$ (homogeneous) inflation is granted. In other words, the inflationary trajectory ends when $\epsilon = 1$, corresponding to an equation of state $\omega = -1/3 $. Assuming that inflation lasted, at least, the minimum amount to explain the CMB observations, ${\Delta N \simeq 55}$ efolds, this implies that  it should have started at a field value of  $\varphi_* \gtrsim 5.5\ \Mpl$ ($h_* \gtrsim 0.1$). Once cosmic inflation takes place, the field slowly rolls down the potential until the kinetic energy breaks the slow-conditions. The end of inflation occurs approximately at $\varphi_{\rm end} \approx 0.94\> \Mpl$ ($h_{\rm end} \approx 0.008$) ~\cite{MARTIN201475}, signifying the beginning of the reheating epoch.
\ \\

\tocless\subsection{Higgs with an auxiliary field }\label{p2_subsec:AuxField}

Let us consider now the addition of an auxiliary field $s$, into the Higgs inflation model. To keep within the spirit of the original model~\cite{Bezrukov:2007ep}, in this paper, we restrict ourselves to the case where this new field is minimally coupled to gravity (${\xi_s=0}$).  On the other hand, an interaction term is added in the action of Eq.~(\ref{p2_eq:action_E}), 
\beq \label{p2_eq:L_int}
{\cal L}_\text{int} = - {\mathsf{g}\,} h^2s^2 ~,
\eeq
where $\mathsf{g}$ is the field-field coupling constant. This term is necessary for a parametric-type preheating to occur at the end of inflation. After this  modification, the potential in {Eq.~(\ref{p2_eq:V_SI})} becomes
\beq \label{p2_eq:V_HIaux}
V(h, s) ={\Mpl^4} \frac { \left[ \frac{\lambda}4 \left( h^2 - {v^2}/{\Mpl^2} \right)^2 + {\mathsf{g}\,} h^2 s^2 \right]} 
{ \left(1 + \xi_h h^2  \right)^2}~.
\eeq

It is  relevant to note that  the effect of the non-minimal coupling $\xi_h$ on the potential is crucial. While in the Jordan frame the potential becomes larger $U (h, s) \rightarrow \infty$ at large Higgs-values $ h \rightarrow \infty$, in the Einstein frame the potential tends to the constant plateau $ V(h,s) \rightarrow \Lambda^4$, effectively suppressing the interaction term and stabilizing the dynamics. This effect applies as well to any other possible coupling between the inflaton and other matter sources, including high energy new physics \cite{Branchina2019}, which remarkably generalizes the dynamics at large field values, thus during preinflation \cite{Linde:1983mx,Linde:2017pwt}. 

In this extension of the model, the canonically normalised fields in the Einstein frame are denoted by $\varphi,~\chi$. Where $\varphi$ represents the inflaton, and $\chi$ the auxiliary field.  
\ \\

\tocless\subsection{Conversion between the fields in the Jordan and Einstein frame notation}  

The fact that the $s$-field is assumed to be minimally coupled, 
facilitates the analysis as it simplifies the mixing between the fields and gravity. Indeed, under this assumption, the field-space metric becomes diagonal, i.e. $ \mathcal{G}_{IJ} = \mathrm{diag} (\mathcal{G}_{hh}, \mathcal{G}_{ss})$.  This is convenient because allows us to easily infer the momentum of the Einstein framed fields ($\Pi_\varphi,\ \Pi_\chi$)  the Jordan ones ($\Pi_h,\ \Pi_s$), by  (no index-summation implied) 
\beq
\Pi_\varphi^2 = {\cal G}_{hh} \Pi_h ^2 
~, \qquad 
\Pi_\chi^2 = {\cal G}_{ss} \Pi_s ^2 ~.
\eeq

In principle, the conversion of the field values should be done by solving {Eq.~(\ref{p2_eq:convert_frame})}. 
However, at small-field values the conversion can be well approximated by solving 
\begin{align} \label{p2_eq:approx_fieldconvers}
  \frac 1 {\Mpl} \frac{\partial \varphi} {\partial h} \approx \sqrt{{\cal G}_{hh}}  ~, \qquad 
   \frac 1 {\Mpl} \frac{\partial \chi} {\partial s}  \approx  \sqrt{{\cal G}_{ss}}  ~,
\end{align}
recovering  {Eq.~(\ref{p2_eq:varphi_of_h})} for the inflaton, while the auxiliary  field in that frame 
is approximately given by
\beq \label{p2_eq:phi_of_s}
\frac \chi \Mpl \approx \left( 1 + \xi_h h^2\right)^{-1/2} s ~.
\eeq

Note that, as shown in appendix \ref{ApB_ApConversionJE}, these approximations are not valid in some parts of the field-space when $s \gtrsim 0.1$, therefore they cannot be used when large field excursions are present, such as when considering preinflationary scenarios (see section \ref{p2_sec:numdetails}). 
\ \\

\tocless\section{Numerical strategy \label{p2_sec:NumStrategy}}

The end goal of this paper is to test if Higgs inflation in the presence of an auxiliary field can begin from inhomogeneous initial conditions, and samewise if it is able to preheat the universe after the end of inflation via parametric preheating. To that end, I will be using the GRChombo numerical relativity code to simulate the pre- and post-inflationary dynamics in full general relativity. %

In the 3+1 decomposition of General Relativity the line element is written as 
\bea \label{p2_timeline}
\rr d s^2 = - \alpha^2 \rr d t^2 + \frac1\chi \tgm\ij (\rr d x^i + \beta^i \rr d t)(\rr d x^j + \beta^j \rr d t)~,
\eea
where it has been used the conformal decomposition of the metric, $\gamma\ij = \frac 1\chi \tgm\ij$. The lapse and shift gauge parameters are given by $\alpha$ and $\beta^i$, {respectively}. 
In this section, $\chi$ is the metric conformal factor which relates to the cosmological scale factor as $\chi = 1/a^{2}$.  
The extrinsic curvature $K\ij$ is also split into its conformal traceless part $\tA\ij$ and the trace $K$, 
\bea
K\ij = \frac1\chi \left( \tA\ij +\frac13\tgm\ij K\right)~.
\eea
It relates with the Hubble rate $H$, in the homogeneous case, as
\bea
 H = - \frac { K} 3 ~ .
\eea

The energy-momentum tensor can be decomposed into the scalar fields' energy density $\rho_{\rm sf}$, momentum density $S_i$ and anisotropic tensor $S_{ij}$, 
\bea \label{p2_3+1sources}
 \rho_{\rm sf}&=& n_\mu n_\nu T^{\mu\nu} ~,\\
 S_i &=& -\gamma_{i\mu} n_\nu T^{\mu\nu} ~,\\ 
 S_{ij} &=& \gamma_{i\mu} \gamma_{j\nu} T^{\mu\nu} ~,\\ 
 S &=& \gamma\IJ S\ij ~,
\eea
where $n^\mu=(1/\alpha, -\beta^i/\alpha)$ is the unit normal vector to the three-dimensional slices. In analogy to the perfect fluid case with pressure $p =S/3$, the effective equation of state can be defined by 
\bea 
 \omega \equiv \frac{p}{\rho_{\rm sf}}= \frac 13 \frac{S}{\rho_{\rm sf}}~.
\eea

In the gravity sector, the energy associated with gravitational vector and tensor modes is given by 
\bea
  \rho_{\rm shear} = \frac {\Mpl^2}{2} \tilde A\ij \tilde A\IJ \propto \partial_t\tgm\ij \partial_t\tgm\IJ ~,
\eea
and the curvature contribution to the energy budged is written in terms of the Ricci scalar (of the 3-dim metric) 
\bea  
\rho_R = \frac{\Mpl^2}{2} R ~.
\eea

Then one can write the Hamiltonian and Momentum constraint equations as 
\begin{align}
\mathcal{H} & =  \frac{\Mpl^2}{3} K^2 + \frac{\Mpl^2}{2} R  - \frac{\Mpl^2}{2} \tilde A_{ij}\tilde A^{ij} - \rho_{\rm sf}, \label{p2_eqn:Ham} \\
  & = 3\Mpl^2 H^2 + \rho_R - \rho_{\rm shear} - \rho_{\rm sf}= 0\, , \nonumber \\[1mm] 
\mathcal{M}_i & = D^j (K_{ij} - \gamma_{ij} K) - 8\pi S_i =0\, . \label{p2_eqn:Mom}
\end{align}

From the  Arnowitt-Deser-Misner formalism, it can also be shown that the conditions to have an accelerated expansion of the universe are given when 
\begin{gather}
\omega < -\frac13 %
~, \qquad 
\rho_{\rm shear} < \left| \frac{3\rho_{\rm sf}} 4 \left( \frac13 + \omega \right) \right| ~.
\label{p2_eq:inf_cond} 
\end{gather}

Averaging overall space, we can use these conditions to determine the beginning of inflation after the preinflationary era, as well as to set the time of which preheating starts.
\\

In the following analyses, the mean value of variable at a given time is denoted with $\langle ... \rangle$ brackets. {For instance, for a given variable $\theta$ }
\be \label{p2_eqn:Kini}
\langle \theta \rangle \equiv \frac{1}{{\cal V}} \int \theta \, \rr d {\cal V}~,
\ee
where ${\cal V}$ is the spatial volume. Similarly, the root-mean-square (rms) and the standard deviations (std)  are computed like 
\be
{\rm rms} (\theta) = \sqrt{\langle \theta^2 \rangle}
~, \qquad 
{\rm std}(\theta) = \sqrt{\langle \theta^2 \rangle - \langle \theta \rangle^2 } ~,
\ee
These identities are used to assess the level of inhomogeneity in variable $\theta$, as well as the scope of local overdensities. Some example includes the density contrast $\delta_{\rho_{\rm sf}}$ and curvature contrast $\delta_{R}$   which are  given by 
\be \label{p2_eq:deltarho}
\delta_{\rho_{\rm sf}}= \frac{ \rho_{\rm sf}- \langle \rho_{\rm sf}\rangle}{3\Mpl^2 H^2} 
~, \qquad 
\delta_{R}= \frac{ \rho_{R}- \langle \rho_{R}\rangle}{3\Mpl^2 H^2} ~.
\ee
The scalar curvature $\zeta$, as well as the mean number of efolds $\langle N \rangle $ are computed with 
\beq
 \langle N \rangle  = \langle \ln(a) \rangle  
 ~, \qquad 
 \zeta = {\rm std}\left[\ln(a)\right] ~,
\eeq
with $a = 1/\sqrt{\chi} $. Here, $\chi$ denotes the metric conformal factor in Eq.~(\ref{p2_timeline}).
\\

\tocless\subsection{Computational details \label{p2_sec:numdetails} }

All simulations are done  in a grid composed by $(128)^3$ to $(156)^3$ cells with an initial grid-size $L$ which is of the order of the Hubble size. The topology is of a 3-dimensional torus with periodic boundary conditions in all dimensions. The initial configurations assume conformal flatness, (e.g. $\tilde\gamma\ij = {\rm diag}(1,1,1)$ and $\tilde A\ij = 0$),
where inhomogeneities are contained in the form of scalar field gradients, which are then compensated by the conformal factor (i.e. gravitational scalar curvature). The valid sets of initial data have been computed by solving the Hamiltonian constraint iteratively, as in most of the previous works \cite{Clough_2015,Aurrekoetxea_2020,Joana2020,Garfinkle_2020}.  The evolution of the system is computed in the Einstein frame by numerical integration of the BSSN equations \cite{PhysRevD.52.5428,Baumgarte_1998,10.1143/PTPS.90.1} in 3+1 dimensions, implemented in the GRChombo code. A more detailed explanation of the structure and validation of the code can be found in the appendix and in Refs.\cite{Clough_2015,Andrade2021}.

Two different evolution schemes have been used for the numerical evolution of the fields. Simulations on the preheating epoch are evolved using the canonical Einstein fields $\varphi,\> \chi$. Therefore, at each timestep,  the approximate conversions of Eq.~(\ref{p2_eq:approx_fieldconvers}) are used to recover the values of $h,\> s$ needed to evaluate the potential, Eq.~(\ref{p2_eq:V_HIaux}), and its derivatives.  This is done to solve dynamical instabilities occurring at $h\approx 0$, where the $h$-field experience transients accelerations as a result of the presence of $\Gamma^h_{hh}$-term in the evolution equations Eq.~(\ref{p2_eq:eomsf_J}) (and see fig. \ref{p2_fig:Higgs_EJ}). Despite that this issue can be overcome by shortening the time integration during the coherent linear phase, the code becomes very unstable during the broad resonance period, when the $h$-field inhomogenizes.  This issue is solved when the system is evolved using the Einstein-frame notation, and it allows us to continue the simulations for a longer time. As shown in section \ref{p2_sec:SimsReheat}, both evolution schemes give numerically equivalent results. 
On the other hand, simulations on the preinflationary era are done using the (exact) formalism with the Jordan-framed  $h,s$-fields. This does not represent an issue, as $h$ does not continuously oscillate around zero and therefore the instability is not present. 

\vspace*{1cm}

\begin{figure}[!h]
\hspace*{1.8cm} \textbf{Evolution in $\varphi$ and $\chi$} \hspace*{2cm} \textbf{Evolution in $h$ and $s$}
\begin{center}
\includegraphics[width=0.990\textwidth]{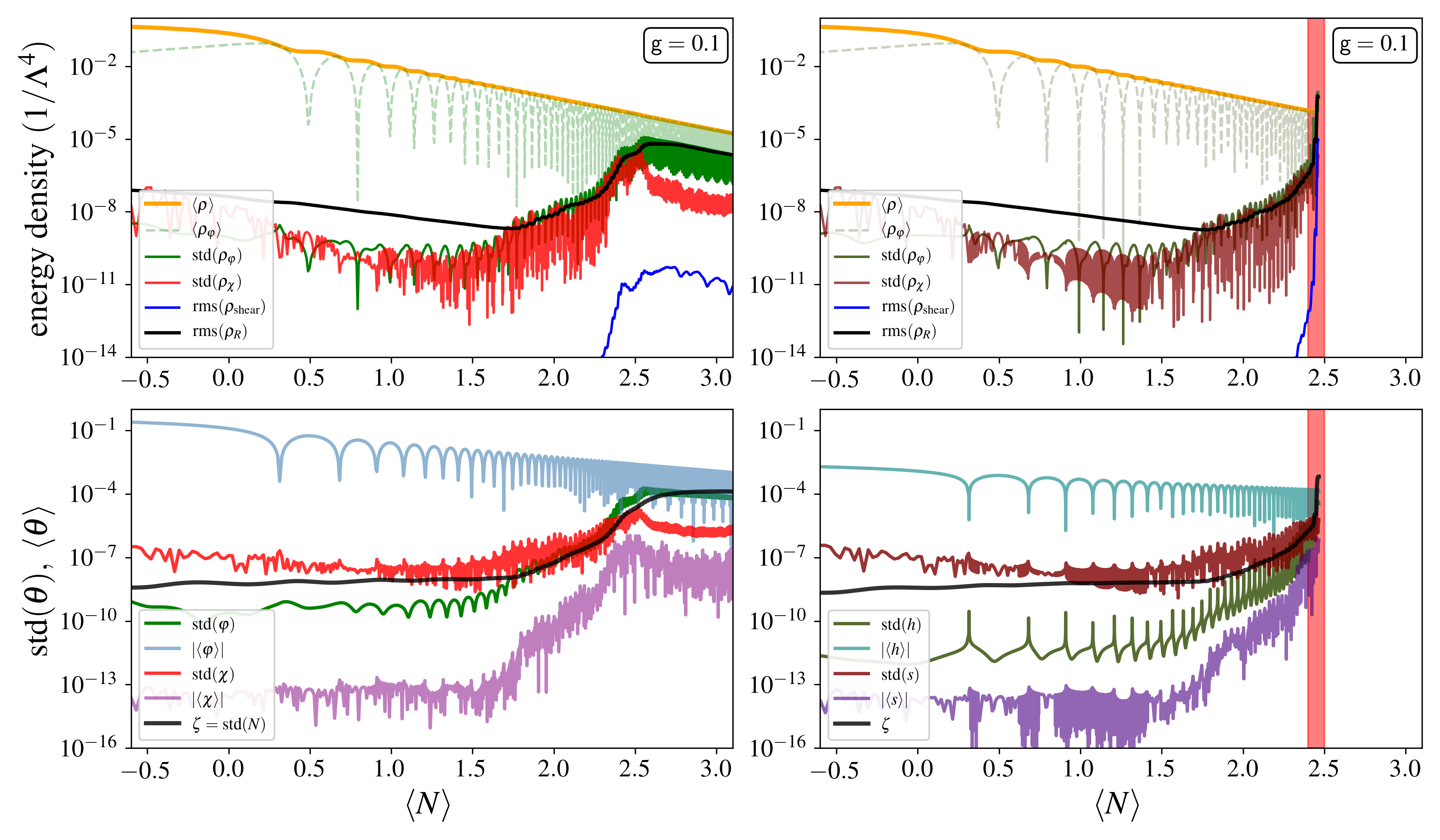}
\end{center}
 \caption{  
 Comparison of the two evolution schemes; Left panels use the canonical Einstein field $\varphi, \chi$ during the numerical evolution, right panels uses the Jordan-frame field $h, s$. The top panels show the time evolution of the standard deviations for the kinetic inflaton energy $\rho_\varphi$ (green line), auxiliary field kinetic energy $\rho_\chi$ (red line), and the root-mean-square values for the gravitational shear  $\rho_{\rm shear}$ (blue line), and curvature contributions $\rho_{\rm R}$ (black line). In both plots the mean energy densities, $\langle\rho_{\rm sf}\rangle $ (in orange) and $\langle\rho_{\varphi}\rangle $ (in dotted green line) have been added as a  reference.  The bottom panels show the evolution of scalar curvature $\zeta$ (in black) and the std  (in green and red lines) and mean values (blue and purple lines, respectively) of the scalar fields in the Jordan (left panel) and Einstein (right panel) frame notation. The fields are represented in units of Planck mass ($m_p$). Red shaded area indicates the region when dynamical instabilities in the $h$-field evolution raises large  violations in the constraint equations~(\ref{p2_eqn:Ham}).  
 \label{p2_fig:reheating}
}
\end{figure}

\clearpage

\tocless\section{Dynamics of preheating \label{p2_sec:SimsReheat} } 

At the end of inflation, the inflaton field starts a period of coherent oscillations around the potential minimum. The large amplitude of the oscillations justifies a classical treatment of the inflaton field. The simulations start about $N_{\rm ini} \approx -1$ $e$-folds before the end of inflation, thus the field is considered to be initially homogeneous\footnote{
Simulations containing initial perturbations in the Higgs field has also been considered without significant changes in the resonance dynamics. See the Appendix \ref{app:3}. 
} 
in field value at the edge of the plateau, and is rolling down the potential with a background kinetic term. The initial values are set to
\beq
h \approx  1.1 \cdot 10^{-2} 
~, \qquad 
\Pi_h  \approx - 8.1 \cdot 10^{-9} ~\Mpl 
\eeq
what is equivalent to 
\beq
\varphi \approx 2.0 ~\Mpl 
~, \qquad 
\Pi_\varphi \approx -1.2 \cdot 10^{-6} ~\Mpl^2  
\eeq
where $\Pi_h,\> \Pi_\varphi$ correspond to the fields momentum. \\

On the other hand, the auxiliary field is assumed to be in its vacuum state due to the redshift caused during inflation, where fluctuations of the field are of quantum origin. The initial state of the field is set by
\beq
\begin{split}
s(\vec x) = \ &\langle s_0 \rangle +
\sum_{n=1}^{N_m}\sum_{i=1}^3\frac{\Delta_{n}}3  \cos  \left( \frac{2\pi n x}\lambda   + \theta_n \right) 
\\
&\text{with  } ~ \langle s_0 \rangle = 0~, \quad \Delta_{n} = \frac{\pi n}\lambda
\end{split}
\eeq
where $\lambda \approx L/10$ is the largest perturbation size,  $\theta_n$ is a random phase, and the number of modes $N_m$ set between $10$ and $50$.  
The momentum of the $s$-field (and $\Pi_\chi$) is initially set to zero. 
\ \\

\tocless\subsection{Parametric resonances}

In the analysis, the evolution of the gravitational  and scalar field sector are considered, i.e. $\rho_{\rm shear},\> \rho_R,\>\text{and } \rho_{\rm sf}$. The later one is further decomposed into the inflaton and auxiliary field parts, $\rho_\varphi,\> \rho_\chi$ by assuming
\beq      
   \rho_\varphi \approx \frac12 \Pi^2_\varphi       
   ~, \qquad 
   \rho_\chi \approx \frac12 \Pi^2_\chi       
\eeq
where the kinetic energy of the fields is used as a proxy to estimate their energy contribution to the global dynamics.

Because we can use two evolution schemes for the fields, namely using the Jordan or the Einstein frame definitions,  let us first compare both schemes and ensure that we obtain equivalent results. In Fig. \ref{p2_fig:reheating}, this is done by direct comparison of the evolution of the energy densities shown in the top panels.
The mean energy density is shown with the orange lines,  while the standard deviation of the fields kinetic energies $\rho_\varphi$ and $\rho_\chi$, corresponding to the green and red lines, respectively. Additionally, the densities from the gravitational curvature (black lines) and shear (blue lines) are also shown.
In the bottom panels, the mean and standard deviation of the fields and scalar curvature $\zeta$ are shown. It's interesting to note several differences in the evolution of the fields: Because the shape of the potential is different in both representations of the field, this is reflected in the scaling of mean values of the fields \cite{Martin2010}. In particular, these simulations show that the mean of Higgs field scales like  $\langle h\rangle \propto a^{-3/4}$. On the other hand, the canonically normalized inflaton scales like $\langle\varphi\rangle \propto a^{-3/2}$. The latter is analogous to a quadratic potential around its minimum \cite{Martin2010,Giblin2019}. Differences are also noticeable when looking at the field excitations (standard deviations): in the evolution of perturbations in $h$, the Riemann spikes that occur when $h\approx 0$ are clearly visible, while for perturbations in $\varphi$ they are hidden because of the mixing with $R$, in the Einstein frame. In both cases, though, the scalar perturbation $\zeta$ closely follow the fluctuations of the  $h,\varphi$ fields. On the other hand, the auxiliary fields behave very similarly in both schemes, and we can clearly relate the $s$ and $\chi$ fields. This is not surprising as the $s$-field is chosen to be minimally coupled to gravity and therefore the mixing after the Weyl transformation is predominantly between $h$ and $R$. All in all, we see that, for this particular case, the broad resonance of the fields occurs within $N \approx 1.5 - 2.5$ efolds, where the excitations of the fields (in all frames) grow exponentially.

In a similar way, {Fig.$\>$\ref{p2_fig:reheating2}} shows simulations for different values in $\mathsf{g}$. During the broad resonance period, we find that curvature grows strictly following the excitations of the fields (the std values). The efficiency of resonances is conditioned by the $\mathsf{g}$-coupling, as the interaction term (Eq.~\ref{p2_eq:L_int}) can be interpreted as the effective mass terms of the fields. 
For large couplings,  $\mathsf{g} \gtrsim 1$, fluctuations of the field are largely suppressed during the last efolds of inflation, pushing the field down to zero. This overdamping, which is partly due to  the classical treatment of the initial gradients, impedes the resonance periods at  later times and preheating fails.  Coupling strengths in the range of  $ 0.1 \gtrsim \mathsf{g} \gtrsim 10^{-4}$ does allow preheating. However, while at larger coupling values the broad resonant phase starts earlier, the produced fluctuations saturate at lower energies resulting in lower energy transfer from the background field.   

For even lower coupling values, $\mathsf{g} \lesssim 10^{-4}$, the particle production  becomes inefficient  (at least during the first 3-4 efolds post-inflation) and the energies associated with them fail to co-dominate the dynamics. 
The preheating of the universe is therefore presumably  delayed to later times, but we cannot numerically explore this region.  

In summary, within the assumptions of the model, a successful and fast preheating of the Universe occurs for a range in the field-field strength coupling of $ 1 \gtrsim \mathsf{g} \gtrsim 10^{-4}$, with a peak efficiency of around $ \mathsf{g}  \approx 10^{-3}$. This is a surprising result because other studies on Higgs inflation \cite{Sfakianakis2019} found that self-resonances from the Higgs, alone, effectively preheat the universe when considering linearized gravity. These simulations show that this is no longer the case when considering full gravity. 

\afterpage{
\vspace*{2cm}

\begin{figure}[h!]
\begin{center}
\includegraphics[width=0.990\textwidth]{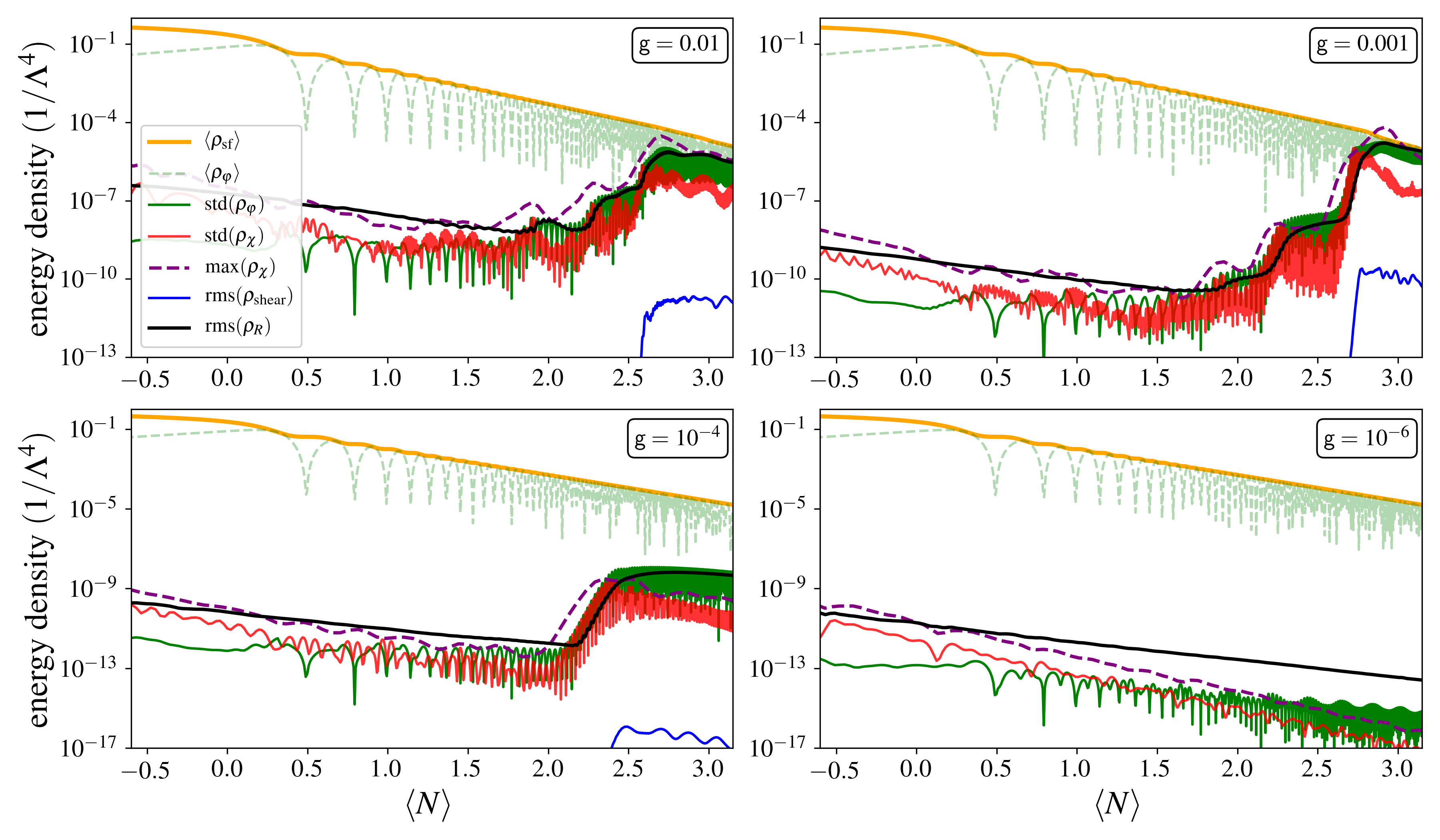}
\end{center}
\vspace*{-0.5cm}
\caption{  \label{p2_fig:reheating2}
{
Same as in the top left panel in figure \ref{p2_fig:reheating} for several simulations with different choices in $\mathsf{g}$. Dashed purple lines denote the maximum values of the auxiliary field's kinetic energy. %
}
}
\end{figure}

\clearpage
}

\clearpage

\tocless\subsection{Structure formation}

\begin{figure}[h!]
\begin{center}
\hspace*{-0.5cm}
\includegraphics[width=0.80\textwidth]{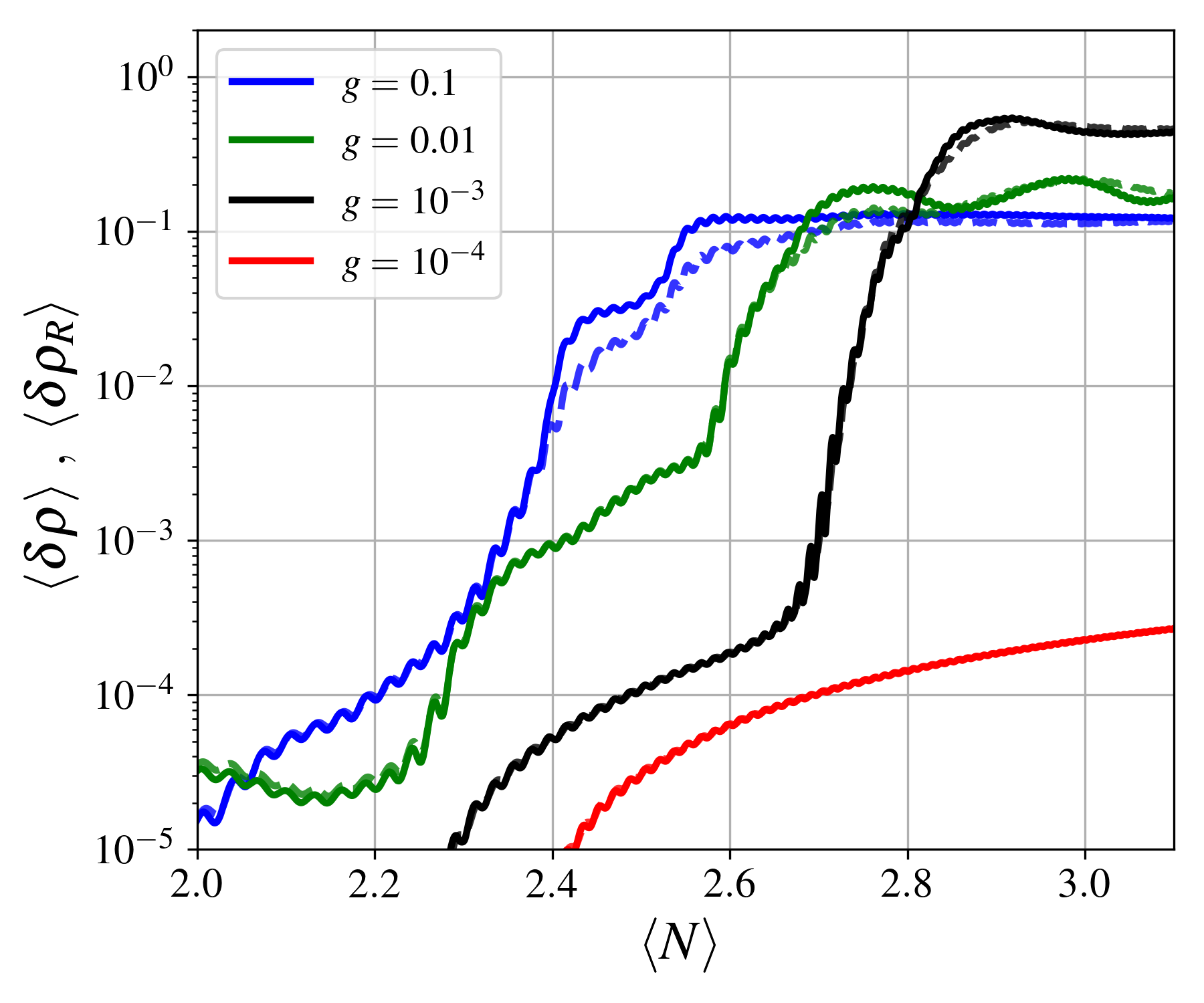} 
\includegraphics[width=0.90\textwidth]{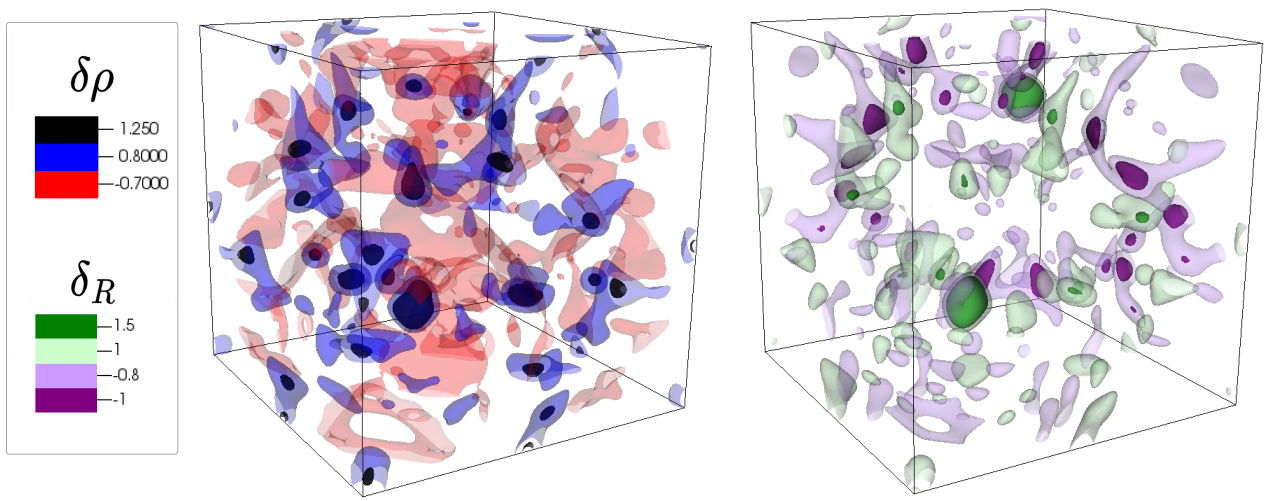}
\end{center}
\vspace*{-0.5cm}
\caption{  \label{p2_fig:reh_structure}
Structure formation during reheating. Top panel shows the evolution of the global density contrasts for $\rho_{\rm sf}$ (solid line) and $\rho_R$ (dashed lines). Bottom plots show contours of under/overdense regions: $\rho_{\rm sf} = 1.25$ (black), $\rho_{\rm sf} = 0.8$ (blue), and $\rho_{\rm sf} = -0.7$ (red), as well for negative/positive curvature: $\rho_{R} = 1.5$ (dark green), $\rho_{R} = 1.0$ (light green), $-0.8$ (light purple), $\rho_{R} = -1.0$ (dark purple).  These 3D representations correspond to the simulation with $\mathsf{g} = 0.001$ shown above,  at {${N \approx 2.9}$}. 
}
\end{figure}

Structure formation starts when the energy fluctuations of the fields grow comparable to the background energy density. In our simulations, this occurs around $N \gtrsim 2.5$ efolds after the end of inflation, flagging the highly non-linear phase in both matter and gravitational sectors.  As shown in Fig. \ref{p2_fig:reh_structure}, the structure consists of the region of space containing both under- and  overdense energies. Overdense (underdense) scalar-field regions coexist with large local positive (negative) Ricci scalar fluctuations of the order of $\delta\rho, \delta\rho_R \approx 1$ ($\delta\rho, \delta\rho_R \approx -1$), reaching even larger values for low-mass particles (i.e. $  \mathsf{g} \approx 10^{-3} $). The type of structure formed in these simulations resembles to what was reported in other works as oscillons (or \textit{transfers} \cite{Lozanov2019}). 
Because during the structure formation the dominant energies are shifted to smaller scales, our simulations can not accurately run long enough to confirm the formation of black holes. However, other works have shown that instabilities on such oscillon-like objects, can lead to the formation of primordial black holes through self-collapse \cite{Kou:2019bbc,Nazari2021,de_Jong_2022}.  This will be studied with dedicated simulations in future works.    

\vspace*{1cm}

\clearpage

\tocless\section{Dynamics of preinflation \label{p2_sec:SimsPreinf}}
The preinflationary scenario is dependent on the initial conditions of the universe, and therefore, its properties are unknown. Arguably, in the classical regime, the primordial universe can be thought of as an inhomogeneous inflaton field that successfully leads to inflation when the kinetic and gradient energies fall below the field's potential energy. The necessary conditions to trigger exponential expansion are a negative effective equation of state that $\langle \omega \rangle < -1/3$, and a subdominant contribution of gravitational modes, i.e. Eq.~(\ref{p2_eq:inf_cond}). If these conditions are satisfied quickly enough, so that the mean field values are still in the flat part of the potential, then inflation starts. 

The particular case of (single field) Higgs inflation model was considered in our previous paper, Ref.~\cite{Joana2020}. The model showed to be robust to large inhomogeneities at sub- and super-Hubble scales. Our simulations showed that highly dynamical field fluctuations source large gravitational (shear and tensor) modes that can eventually dominate the energy budget. The energy density associated with field fluctuations decays like radiation, $\rho_{\rm sf}\propto a^{-4}$, and these gravitational modes like 
$\rho_{\rm shear} \propto a^{-2}$. In any case, both scalar-field and gravitational excitations eventually become subdominant in just a few $e$-folds and inflation begins. In the following, these analyses are expanded by adding an auxiliary field. 

The simulations on preinflation initially contain field gradients in  both the inflaton and auxiliary fields, with perturbation in sub- and super-Hubble configurations. The initial mean value of the Higgs (inflaton) is always considered to be beyond $ \langle h \rangle  > 0.5$,  $(\langle \varphi \rangle /\Mpl > 6)$, so it is deeply located in the flat region of the potential. For the auxiliary field, cases with zero and non-zero mean values have been considered. The selection of these cases have been chosen so that the overall mean energy density is a few orders of magnitude larger than the energy scale of inflation, i.e. $\langle \rho_{\rm sf}(t_0) \rangle >  \Lambda$. Thus, all considered cases contain inhomogeneities well beyond the linear regime. 

\afterpage{
\vspace*{2cm}

\begin{figure}[h!]
\hspace*{1.1cm} \textbf{Super-Hubble perturbations } \hspace{0.5cm} \textbf{Sub-Hubble perturbations }
\begin{center}
\includegraphics[width=0.450\textwidth]{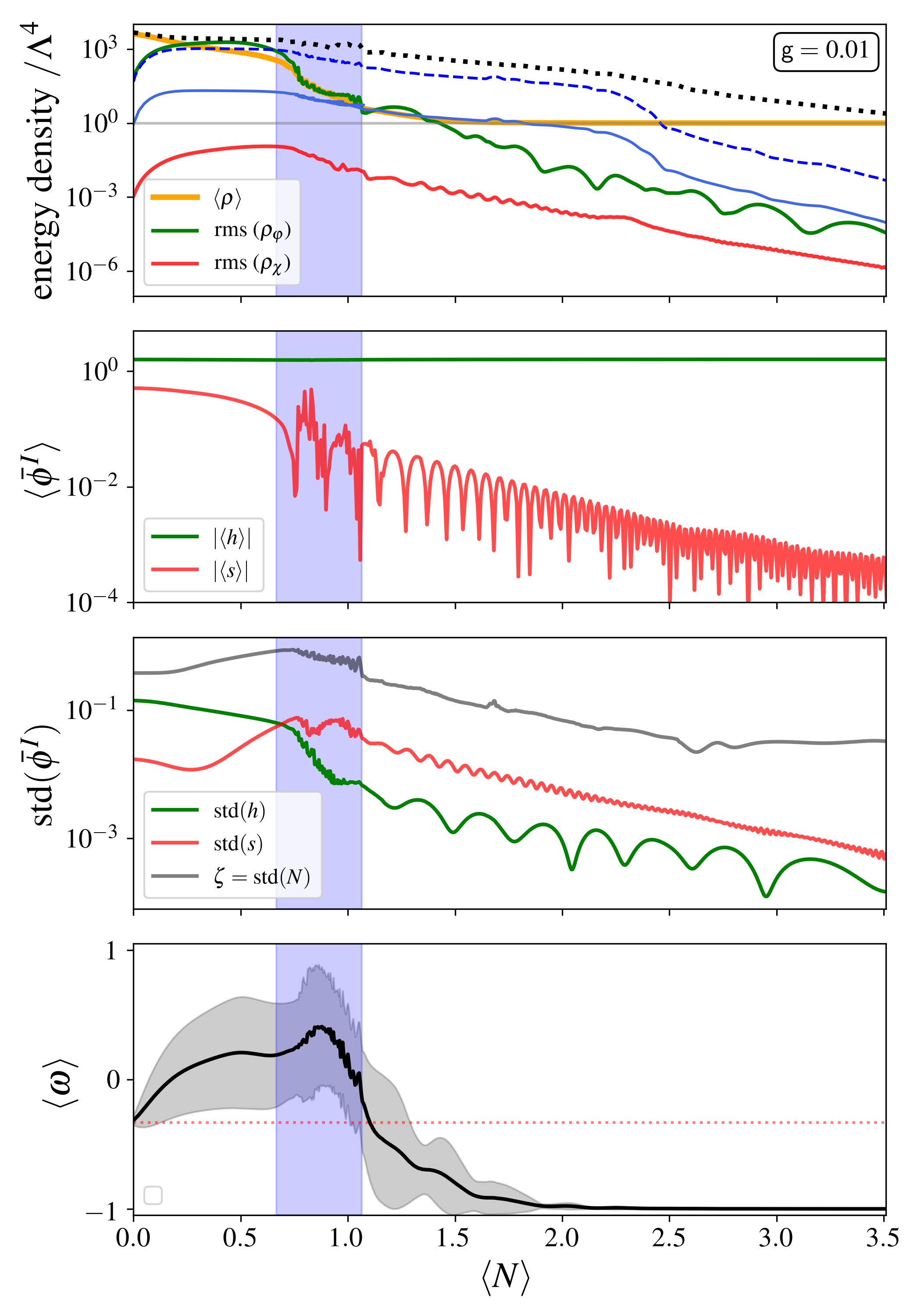}
\includegraphics[width=0.45\textwidth]{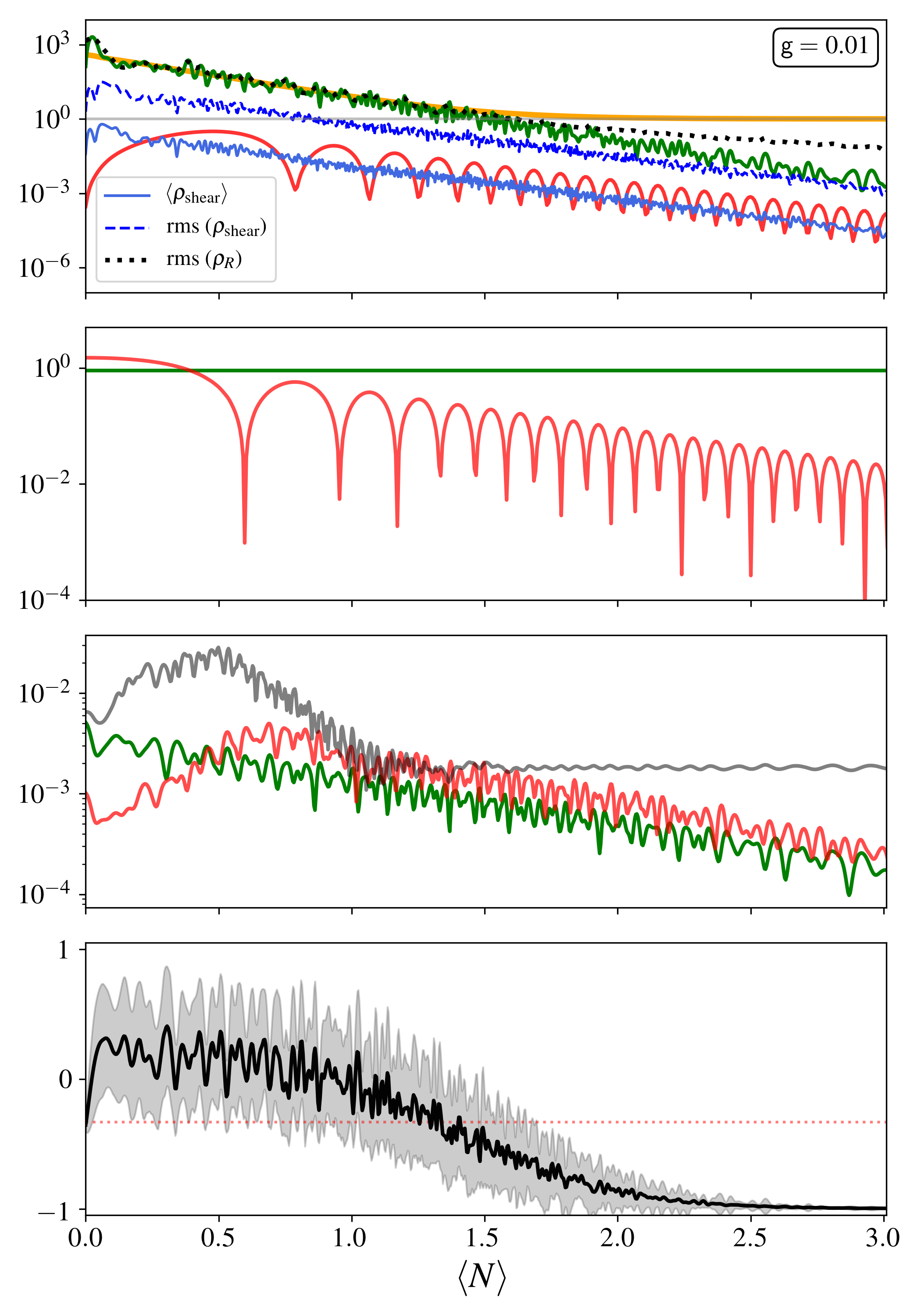}
\caption{ \label{p2_fig:preinflation} 
Dynamics of two example simulations initially at super-Hubble (left) and sub-Hubble (right) perturbations. Top panels show the evolution of scalar field's energy density (orange line), and shear (solid blue line). The rms values for the Higgs (green line), auxiliary field (red line) as well as for the gravitational shear (dotted blue line) and curvature densities (dotted black lines). The mean field evolution (upper-middle panels) and std values (lower-middle panels) for the Higgs (green line) and auxiliary (red line) fields. The scalar curvature perturbation $\zeta$ is shown in gray lines. Bottom panels show the evolution of the equation of state (solid black line), with plus/minus std values in the shaded gray area. The red-dotted line denotes the {${\omega < -1/3}$} threshold necessary for accelerated expansion of Universe.  
}
\end{center}
\end{figure}

\clearpage
}

Figure \ref{p2_fig:preinflation} show two example cases, at super-Hubble (left) and sub-Hubble (right) scales. In both cases, the preinflationary phase consists of a homogenization period driven by the (in average) positive expansion of the Universe. Similarly as shown in Ref.~\cite{Joana2020}, super-Hubble initial conditions tend to form trapped surfaces, or preinflationary black holes (PIBHs), after Horizon crossing. Because these black holes are  always (much) smaller than the Hubble radius,  instead of impeding inflation, they tend to facilitate it by trapping the overdense regions thereby, fastening the homogenization. On the other hand, at sub-Hubble scales, perturbation modes transit back-and-forth between gradients and kinetic energies, effectively making the energy density scale like radiation $\rho_{\rm sf}\propto a^{-4}$. In that scenario, the optimal conditions for triggering inflation look like a dynamical attractor, and cosmic inflation starts within a few efolds.  

Interestingly, in the presence of the auxiliary (spectator) field,  these oscillations also trigger energy transfer between the (Jordan framed-) Higgs and the auxiliary field. 
However, these dynamics do not originate in enhancement of structures like in preheating, because now at field values $ h > 0.02$, the non-minimal coupling of the Higgs has the effect of significantly reducing the impact of the (minimally coupled) auxiliary field when seen in the Einstein frame. This effect can be observed in Fig. \ref{p2_fig:preinflation}, where even when perturbations in the $s$-field are larger than in the $h$-field (see Fig.~\ref{p2_fig:preinflation} middle panels), the dynamical term is always orders of magnitude smaller in the auxiliary field $\chi$.  
The suppression effect comes from the mixing, which is introduced by the field-space metric lower than unity, i.e. ${\cal G}_{ss} = 1/({1+\zeta_h h^2}) \leq 1$ and becomes orders of magnitude smaller for large enough $h$-field values (e.g. ${\cal G}_{ss}\ll 1$ for $h > 0.02$). 
This suppression factor should apply to all other possible matter components that are minimally coupled to gravity.
\ \\

\tocless\subsection{On the initial conditions for inflation \label{p2_sec:OnInitialConditions}}

In this paper we have extended previous works on testing the initial conditions for inflation by including the interplay of an extra (minimally coupled) scalar field. We have tested cases when the initial configuration of the inflation is deeply inhomogeneous but with its mean-field value inside the slow-roll region of the potential.  In the context of Higgs inflation, this corresponds to a mean value close to the plateau. Initial states where the mean-field is in the non-inflationary region (i.e. in the bottom of the potential) have not been considered as these cases should not lead to inflationary regions, as tested in Ref.~\cite{Joana2020}. This is because gradients terms make the fields oscillate around the mean value, thus these inhomogeneities are not capable  of driving the field up to the plateau. One could still consider large field inhomogeneities which spans the scalar field into the potential's plateau, however these perturbations are necessarily super-Hubble (at sub-Planckian gradient energies) and, thus, these regions can be treated as separate Universes. This is  particular to Higgs inflation, as the plateau starts at $\varphi > \Mpl$. 

Our initial settings have also assumed a conformally flat expanding Universe. These scenarios corresponds to the case with only scalar perturbations in the gravitational part, (i.e. without vector and tensor gravitational modes). This is  related to the choice of considering a null kinetic term in the initial hypersurface, which can be seen as a rather ``special" slicing choice at the instantaneous initial time  where scalar-field kinetic terms have been gauged away, trivially satisfying the momentum constraint Eq.~(\ref{p2_eqn:Mom}). Because these initial setting are highly dynamical, this kind of slicing is not stable and once the system is time-evolved both gravitational modes and scalar field kinetic terms are quickly generated, leading to a less symmetric inhomogeneous system. In particular, one could have chosen an analogous situation with initially homogeneous field values but with largely inhomogeneous kinetic terms which would rise scalar-field inhomogeneities in the immediate time evolution \cite{Joana2020}. Because the minimally coupled fields are energetically subdominant at high enough Higgs values, the previous picture still holds beyond the single-field case. 

Nonetheless, there are still several limitations with such initial settings. The assumption on conformal flatness only allow for gravitational perturbations risen by the scalar field evolution, and therefore independent large tensor metric perturbation are ignored. Studying these cases requires solving (non-trivially) both the Hamiltonian and Momentum constraints and future works will deal with this challenge. In addition, this work has assumed that only the Higgs field has a non-minimal coupling to gravity, serving as a reference for other more specific models like quintessential Higgs inflation \cite{Es-haghi:2020oab}, two Higgs doublet models \cite{Lee:2021rzy}, etc. 
Still, systems with two or more non-minimally coupled fields can show a much richer dynamical evolution during (pre-)inflation; Exploring complex trajectories in field space and possibly including multiple inflationary phases at distinct energy stages. All these considerations are left to future works.

Under the previous considerations, in all the considered cases, we find common dynamical patterns of the Higgs preinflationary era. This phase can be described as a homogenization era with a varying inhomogeneous equation of state, which effectively  correspond to a radiation dominated universe $\langle \omega \rangle \approx 1/3$. 
Once the scalar-field falls below the energy scale of the inflationary potential, the equation of state tends to a de-Sitter Universe with $\omega \approx -1$, satisfying the first condition for inflation, i.e. Eq.~(\ref{p2_eq:inf_cond}). It has also been shown that strong field dynamics near the Hubble scales develop large gravitational modes that potentially influence the expansion of the Universe until they become subdominant. These modes effectively delay the beginning of inflation, but do not prevent it. Moreover, during the preinflationary era, lasting $N \approx 3\text{-}7$ efolds, the variation on the average value of the inflaton $\varphi$ is negligible, which prevents the ``overshooting'' problem seen in other models \cite{Aurrekoetxea_2020}. 
All these considerations make me conclude that, under the considered settings, the Higgs inflation model is very robust to the inhomogeneous multifield initial conditions of the preinflationary era.
\ \\

\tocless\section{Conclusions \label{p2_sec:Conclusion}}

In this paper, I have used fully general relativistic simulations to investigate the robustness of the Higgs inflation model to inhomogeneous multifield initial conditions. Specifically, in the presence of additional field couplings, these being necessary for a parametric-type reheating. It is shown that, at large enough Higgs values,  the non-minimal coupling of the Higgs protects the dynamics of the inflaton by diminishing the impact of couplings to other fields and matter sectors.  And, as shown in Ref.~\cite{Joana2020}, the dynamics from gravitational shear and tensor modes can only delay, but not prevent, cosmic inflation.

Additionally, I also presented simulations on the preheating dynamics of the two-field system where full gravitational backreactions in the metric have been considered. As expected, it is shown that the efficiency of the preheating is conditioned to coupling strength between the fields. In particular, for such a simple model, I found the preheating of the Universe within the first $3-4$ efolds post-inflation to occur for couplings in the range of $ 0.1\lesssim \mathsf{g} \lesssim 10^{-4}$. On the other hand, self-resonance from the Higgs alone, fails to reheat the Universe within the first $3-4$ efolds after inflation. 

These simulations have also shown the formation of complex structures during the preheating, consisting in large under/overdensities as well as strong positive/negative (local) curvature regions, suggesting the possibility of (seeding) later formation of compact structures like primordial black holes. Nonetheless, these results should be taken cautiously as further investigations, including dedicated numerical simulations, are necessary to accurately resolve these highly non-linear objects. 

Future works are also necessary to study more realistic preheating scenarios, including the Higgs couplings to the Standard Model particles. These are important, because they potentially could shorten the preheating within one $e$fold after inflation \cite{Ema2017,Ema2021}, if metric backreactions allow it. Other interesting aspects to be studied are the emission of gravitational waves  - and possible amplification effects\cite{Chunshan2016,Lozanov2019,Zihan2020,Cai2021}. Importantly, some of the described  phenomena are expected to be in the observable range of future gravitational wave experiments. 

\chapter{Additional work}
\label{ch:addwork}
\pagestyle{fancy}

In addition to the work published in the two articles presented before, I have also spent part of my PhD on developing, extending and adapting existing codes. During the doctoral program, I became a member of the GRChombo code community, which has allowed me to benefit from all the work already done by my colleagues, as well as to contribute with novel extensions to the code, and the development of Python scripts for post-simulation analysis and visualization. 

For the first paper, in Chap.~\ref{p1_chap:prepaper1},  I have extensively used the already available scalar-field implementation in the GRChombo code, as well as the initial condition solver to satisfy the Hamiltonian constraint for initial data with large gradients. In that case, an adaptation of the code to replace the pre-existing inflation model for the Starobinsky model was needed. Additionally, a novel methodology for solving the Hamiltonian constraint containing large inhomogeneous kinetic energies was developed, as explained in Chap.~\ref{p1_chap:prepaper1}. 

For the second publication of Chap.~\ref{p2_chap:prepaper2}, the covariant formalism of a two-field system with an arbitrary non-minimal coupling to gravity was implemented within the BSSN scheme. The initial conditions solver has also been adapted to consider the two-field system in the Jordan and Einstein frames. In the current format, these codes are written for the case of Higgs inflation, but they are easily adaptable so that, in future uses, it will be straightforward to consider other types of potentials.  

Another major contribution to the GRChombo code has been the development of a package to solve the gravitational hydrodynamics equations for simulations containing fluids within the BSSN formalism. This new code allows full gravitational treatment for a perfect fluid with a barotropic equation of state, and it is used in current ongoing studies of primordial black hole (PBH) formation through the collapse of curvature perturbations during the early Universe evolution (see below). 

Last, I also became an active member of the Laser Interferometer Space Antenna (LISA) collaboration, and I contributed to the elaboration of the LISA Cosmology white paper of  Ref.~\cite{LISACosmologyWorkingGroup:2022jok}. In addition, two other papers are currently in preparation consisting of an exhaustive review of the PBH literature available to date (midst of 2022), and the development of a Python toolkit to be used for PBH analytical studies and the forecasting of cosmological observables, with particular focus on the gravitational wave signatures. My main contribution to the code has been the homogenization of its structure and implementation of the algorithm described in Ref.~\cite{Musco:2020jjb}, used to compute the PBH formation threshold for a given power spectrum. The paper documenting the code is currently being written.

\section{Gravitational hydrodynamics}

We proceed to give a brief introduction to the field of general relativistic hydrodynamics, with particular focus on the simplest case, described by the  \textit{perfect fluid}\cite{Alcubierre:1138167}. We remind the reader that energy-momentum is given by 
\be
T_{\mu\nu} = (\rhofl + p) u_\mu u_\nu + p\> g\munu  ~,
\ee
where $\rhofl$ and $p$ are the fluid's energy density and pressure, respectively, and $u_\mu$ is the $4$-velocity vector. 

The fluid's energy density is commonly separated into the \textit{rest energy density} $\rhozero$ and the \textit{interal energy} $\varepsilon$, such that 
\be
\rhofl = \rhozero ( 1 + \varepsilon)~. 
\ee
In that way, for an ideal fluid corresponding to the ultra-relativistic matter, we have $\varepsilon \gg 1$, and so $\rhofl \simeq \rhozero \varepsilon$. And vice-versa, for the non-relativistic matter, we have $\rhofl \simeq \rhozero$. Also, we cautiously remark that the $\rhofl$ is in general \textit{not} the same quantity as the ADM term (see section \ref{ADM_form}) 
\be
\begin{split}
\rho & \equiv  n^\mu n^\nu T\munu
\\  & = (\rhofl + p) W^2 - p 
\\  & = \rhozero h W^2 - p 
~, 
\end{split}
\ee
where, above,  we have used the definitions for the Lorentz factor $W$ and the enthalpy term $h$~\cite{Alcubierre:1138167},  
\be 
W \equiv -n^\mu u_\mu =   \frac 1 {\sqrt{ 1-v^2 }}~,   \qquad   h \equiv \left(1 + \varepsilon + \frac p \rhozero \right) ~, 
\ee
with $v^i$ being the 3-velocity component of the fluid,  i.e.  $v^i = u^i / W$.  Indeed, $\rhofl$ and $\rho$ are only equal when $W=1$, and thus, when the local coordinates of an Eulerian observer follow the fluid element. 

The state of the fluid is described at any time by the six \textit{primitive variables}: $\rhozero$, $\varepsilon$, $p$, $v^i$. The evolution equations of such a system can be derived by using the conservation laws 
\be
 \nabla_\mu \left(  \rhozero u^\mu \right) = \partial_\mu \left(  \sqrt{-g}\> \rhozero u^\mu  \right)  = 0 ~,
\ee
and
\be
\nabla_\mu T^\mu_\nu  = \partial_\mu \left( \sqrt{-g}\> T^\mu_\nu \right)  - \sqrt{-g}\> \Gamma^\alpha\munu  T^\mu_\alpha  = 0 ~. 
\ee

In here, we skip the full derivation of the equations of motions as these can be found in full detail in Alcubierre's book, Ref.~\cite{Alcubierre:1138167}. 
This procedure leads to a system of equations where a new set of variables are defined to be used during the evolution. These are the so-called \textit{conserved variables}: $D$, $S_i$ and $\mathcal{E}$,  and they are defined in terms of the primitive variables, 
\begin{align}
 D &= \rhozero W  ~,  \label{eq:D}
\\
 S_i &= \rhozero h W^2 v_i  ~,   \label{eq:Si}  
\\
\mathcal{E} &= \rhozero h W^2 -p - D     \label{eq:calE}   ~.  
\end{align}

Thereafter, the evolution equations written in a 3+1 form read 
\begin{align}
\left( \partial_t -  \mathcal{L}_\beta \right) D  & =   - D_k (\alpha D v^k) + \alpha K D              ~,   \label{eq:evoD}
\\
 \left( \partial_t - \mathcal{L}_\beta \right) S^i  &=    - D_k \left[ \alpha  \left( S^i v^k  + \gamma^{ik} p \right) \right]    - (\mathcal{E} + D) D^i\alpha  + \alpha K S^i         ~,     \label{eq:evoSi}
\\
\left( \partial_t - \mathcal{L}_\beta \right) \mathcal{E} & =     - D_k \left[ \alpha v^k  \left( \mathcal{E} + D \right) \right]   + \alpha K (\mathcal{E} + p)   \nonumber \\
 & \> +  (\mathcal{E} + D + p) (\alpha v^mv^n K_{mn} - v^m \partial_m \alpha)~,    \label{eq:evoE}
\end{align}
where $\mathcal{L}_\beta$ denotes the Lie derivative with respect to the shift vector $\beta^i$.

The previous system of equations gives the evolution of the gravitating fluid in terms of the conserved variables. However, the recovery of the primitive variables is not an easy task in general. This is because, in their definition,  Eqs.~(\ref{eq:D}-\ref{eq:calE})  depend explicitly on the velocity squared (i.e. hidden inside $W$) and in the pressure, and therefore in the particular equation of state. For a generic equation of state, one often requires a root-finding technique that obtains the value of $p$ and $v$, prior to the recovery of the other variables. Furthermore, the recovery of the primitive variables is needed at every time step during the evolution, as their evolution equations (\ref{eq:evoD}-\ref{eq:evoE}) also depend explicitly on $v$ and $p$. This has the inconvenience of largely increasing the computational cost of the simulations, particularly for systems of fluids with a complex equation of state. However, for cases with a simple equation of state, the recovery of $v$ and $p$ can be done analytically by finding the physical root of a high-order polynomial. For a barotropic fluid with $p = \omega \rhofl $, this implies solving a second-order polynomial, which diminishes the simulations' computational burden.

In the next section, we present a current application to the code, where we investigate the formation of primordial black holes during the radiation-dominated epoch of the Universe.

\section{Formation of primordial black holes}

One of my ongoing projects is the study of the formation of primordial black holes (PBHs) in the early Universe using fully general relativistic simulations. 
For that goal, I implemented a gravitational hydrodynamics module in the GRChombo code using the formalism described above. %
In this work, we assume scenarios where the Universe can be represented by a perfect fluid, on an FLRW background with an overlying super-Hubble curvature perturbation. In our preliminary investigations, we have tested the code by running simulations with quasi spherically-symmetric initial conditions and a Gaussian-shaped curvature fluctuation, 
\be
  \zeta = \zeta_{\star} \exp \left( - \frac{r^2}{2\sigma^2}  \right) ~.
\ee
Here, $r$ is the radial coordinate, $\zeta_{\star}$ is the initial amplitude of the curvature perturbation and $\sigma$ is the typical fluctuation's size. The initial density perturbation $\rho$ can be found numerically by solving the Hamiltonian constraint, under the assumption of spatial conformal flatness, i.e. $\tilde\gamma\ij =  \delta\ij$ and $\tilde A\ij = 0$, at super-Hubble scales. 

\begin{figure}[h]
\vspace*{20mm}
\includegraphics[width=0.98\textwidth]{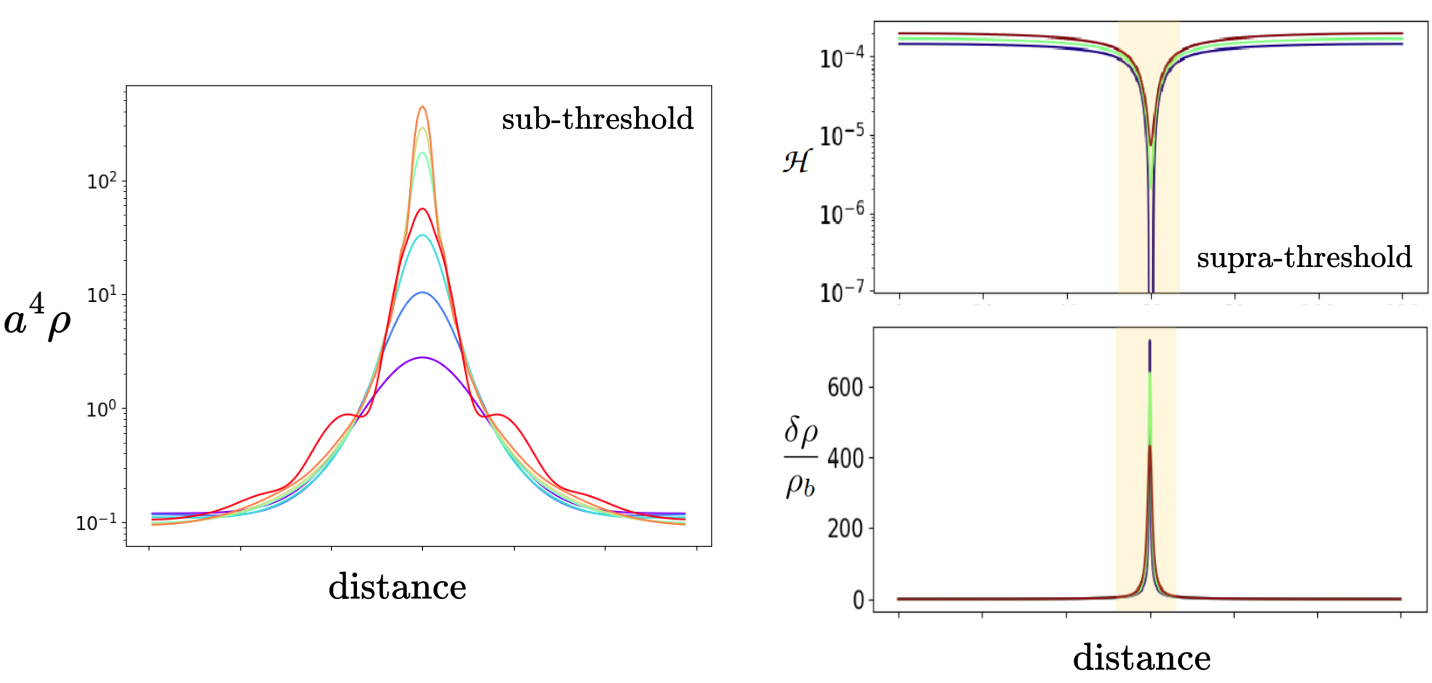}
\caption{ Simulations of ultra-relativistic fluid system with a spherically symmetric curvature perturbation. The left panel shows the evolution of the energy density for a sub-threshold amplitude example, i.e. $\zeta_* = 0.75$, which fails to collapse into a black hole. Different times slices are indicated in the colour code (from blue to red). The right panels show the local Hubble rate (top) and energy density contrast (bottom) for and with an initial supra-threshold amplitude, i.e. $\zeta_* = 0.85$, at the time when the black hole is already formed (approximate radius indicated with yellow band).   
}
\label{Fig:PBH_080}
\vspace*{20mm}
\end{figure}

\begin{figure}[!ht]
\vspace*{20mm}
\includegraphics[width=0.990\textwidth]{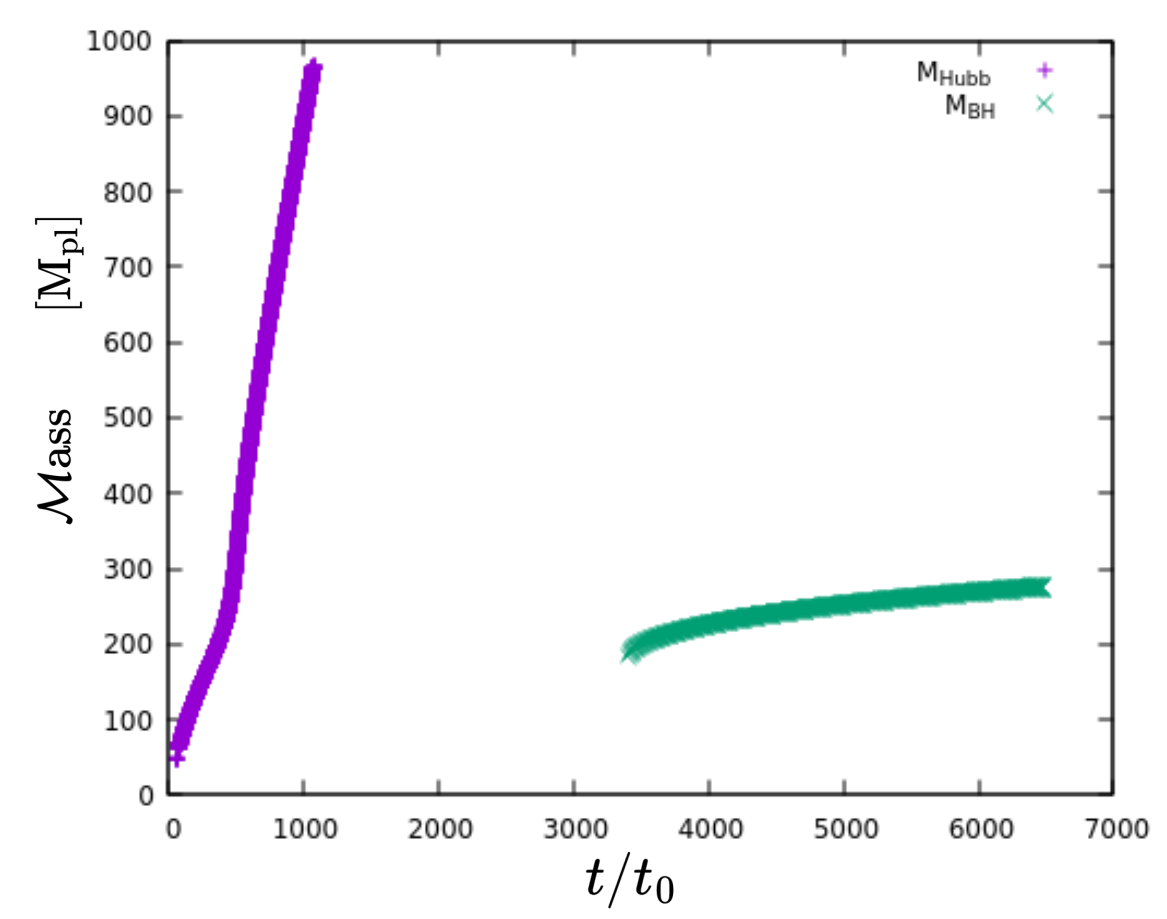}
\caption{Formation of a primordial black hole. The solid lines show the evolution of the mass associated with the apparent horizons corresponding to the Hubble radius (purple) and the formed PBH (green). The PBH mass is proportional to the mass of the Hubble radius when the curvature perturbation re-enters the Hubble volume, but the black hole is formed at a later time. The PBH mass still grows up to a factor 2 until it stabilizes.  
}
\label{Fig:PBH_massEvo}
\vspace*{20mm}
\end{figure}

In Figure \ref{Fig:PBH_080}, we replicate the results shown in Ref.~\cite{Yoo:2020lmg}. We confirm that the threshold for PBH formation is around $\zeta_{\star} \approx 0.80$  for initial perturbations at $\sigma^2 = 5/H_{\rm ini}$, where $H_{\rm ini}$ is the initial Hubble parameter. In addition, we continue our simulations after the black hole is formed to investigate the evolution of its mass. Figure \ref{Fig:PBH_massEvo} shows the evolution of the background Hubble volume's mass and PBH mass during the simulation. We find that PBH mass experiences a logarithmic growth, up to a factor 2, until the final mass stabilizes.

\clearpage

In extensions of this work, we aim to study scenarios beyond spherical symmetry and investigate possible changes in the formation threshold, and the related variations in the characteristic masses and spins of forming black holes. Another important consequence of the non-sphericity is the expected emission of gravitational waves that should occur during the collapse, with possible observational evidence in current and future gravitational wave experiments.

\begin{figure}[htbp!]
\vspace*{10mm}
\centering 
\includegraphics[width=0.51\textwidth]{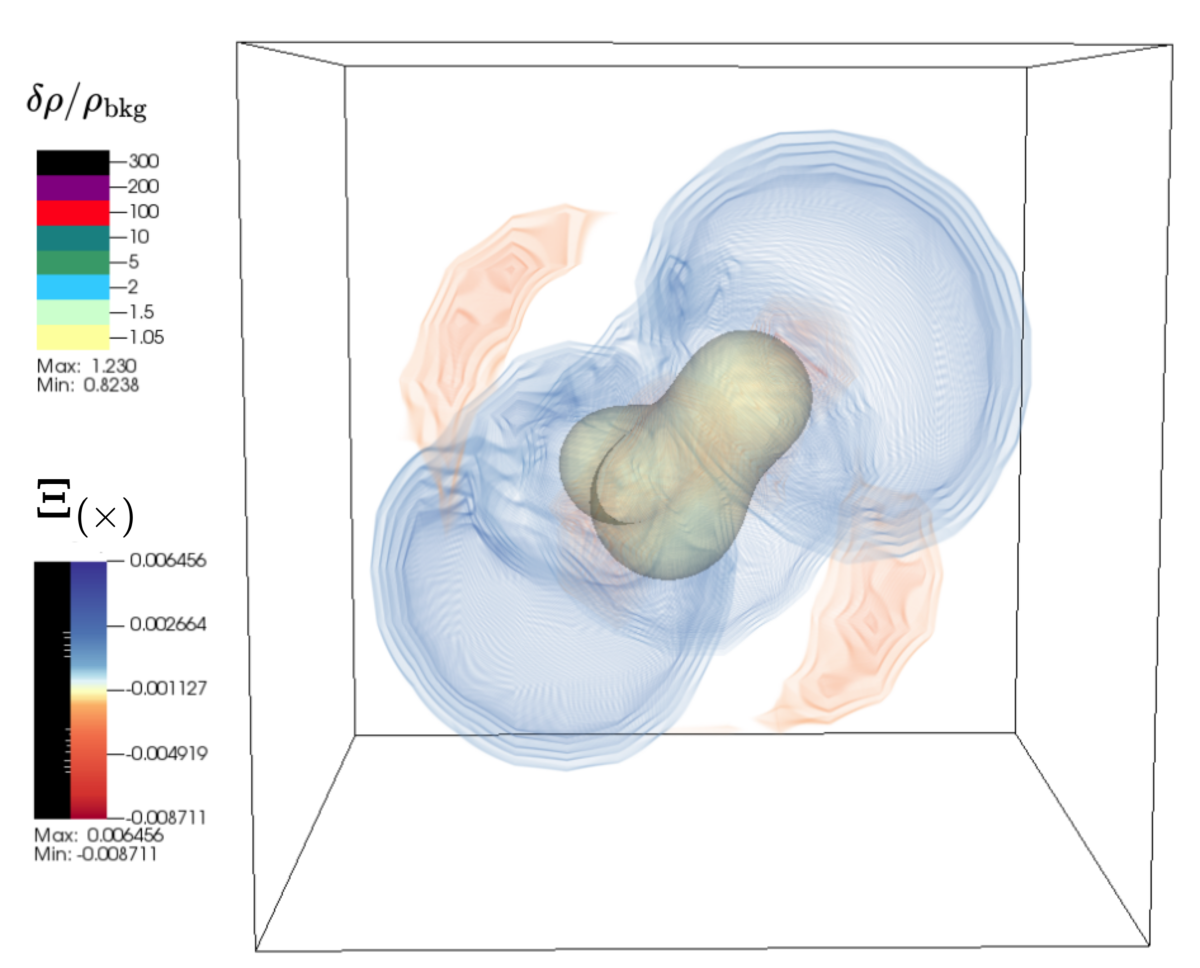} 
\vspace*{2mm}
\includegraphics[width=0.495\textwidth]{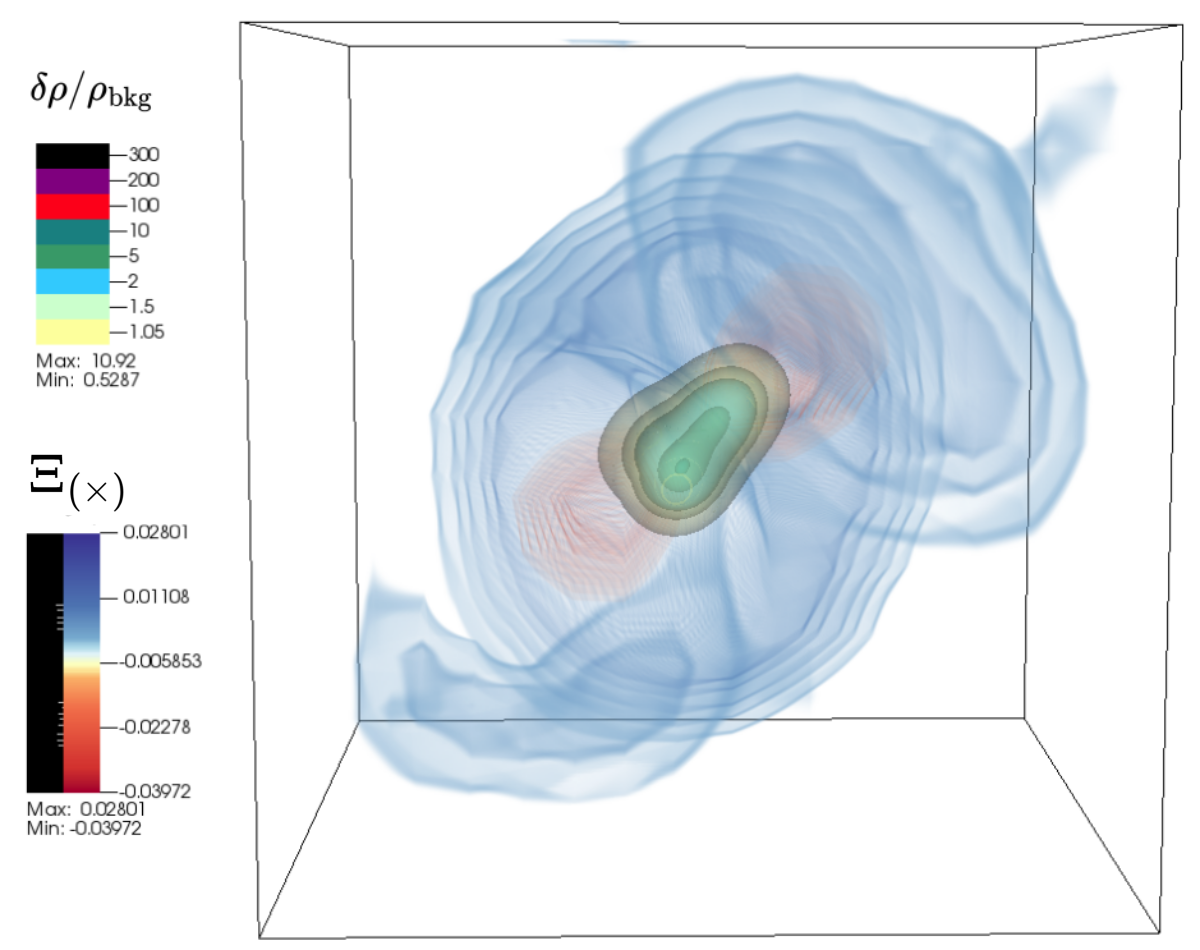}
\includegraphics[width=0.495\textwidth]{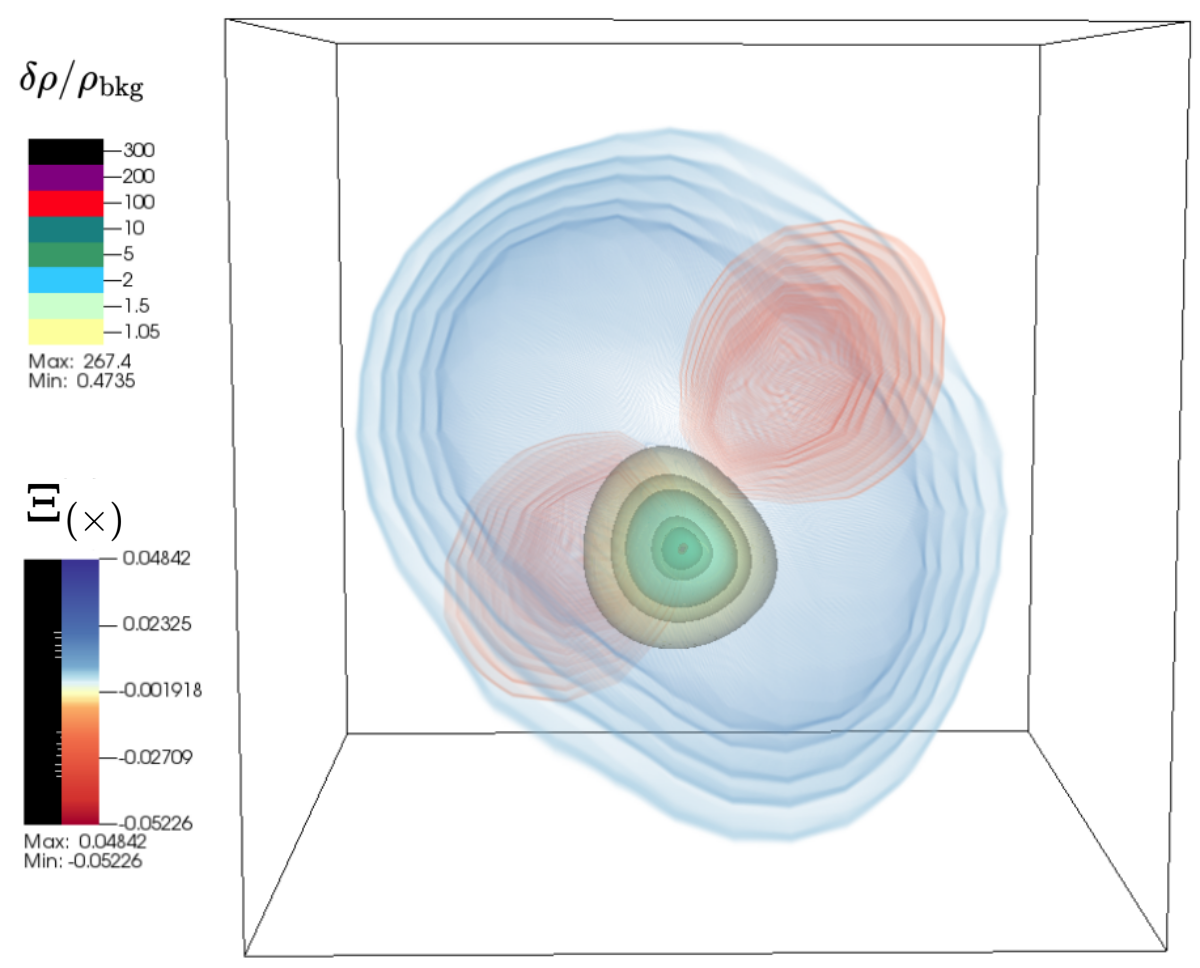}
\caption{
Snapshots of simulations on gravitational collapse of asymmetric curvature perturbation in radiation domination. The boxes correspond to three different times, chronologically ordered from top to bottom-right. Contour lines indicate the energy contrast in the fluid $\delta\rho/{\rho_{\rm bkg}}$, while the volume rendering surfaces in blue and red colours indicates the sum  of the non-diagonal metric perturbations, $\Xi_{(\times)}$. 
\label{Fig:PBH_asim}
}
\end{figure}

In our preliminary results, in Fig.~\ref{Fig:PBH_asim},  we show a simulated example of asymmetric PBH formation, which initial data consists on three overlaying Gaussian perturbations with origin displaced from the centre. The profile is the following,  
\be 
\zeta = \sum_{i=1}^{3} = \Delta_i \exp\left[ -\frac12 \frac{\left(r-c_i\right)^2}{\sigma^2}   \right] ~,
\ee
where $\sigma = 5 H_{\rm ini}^{-1}$,  the gaussian amplitudes $\Delta_i$ and displacement from the grid's center $\delta c_i = (r_{\rm center} - c_i) /H_{\rm ini}^{-1}$ are 
\be
\begin{split}
 \Delta_1 = 0.7, & \quad  \Delta_2 = 0.5, \quad \Delta_1 = 1.3 \\
 \delta c_1 = \left(3, 3, 0\right) \quad \delta c_2 &= \left(-5, -2, -5\right) \quad \delta c_3 = (-2, -3, 5)  ~,
\end{split}
\ee
with an initial grid size $L = 60 H_{\rm ini}^{-1}$, and $r_{\rm center} = \left( L/2, L/2, L/2\right)$. 
The figure illustrates three snapshots: the top box corresponds to an early time before the perturbation re-enters the Hubble volume, the bottom left and right boxes to later times during the collapse of the originating fluid overdensity, i.e. $\delta\rho/\rho_{\rm bkg} \gg 1$. We use a proxy variable to estimate the amiplitude of metric  perturbations defined as 
\be
\Xi_{(\times)} \equiv  \frac 13 \left(\tgm_{xy} +\tgm_{xz} +\tgm_{yz}\right) ~, 
\ee
which is shown in blue and red surfaces and, in the figure, we can appreciate a coplanar surface (in blue) and an axial region (in red). This might hint at the development of a non-vanishing associated spin to the forming black hole. We stress that this variable corresponds to a mixture of tensor and scalar modes and extracting the purely tensor modes, i.e. GWs, is currently a work in progress. At the time of the black hole formation, or more precisely, at the time when the black hole's apparent horizon is detected first, the black hole's  mass $M_{\rm PBH}$ and dimensionless spin $\bar S_{\rm PBH}$ are 
\be
M_{\rm PBH} \approx 0.27 H_*^{-1}, \qquad   \bar S_{\rm PBH} \approx -3 \cdot 10^{-3} ~, 
\ee
where $H_*^{-1}$ 
is the size of the Hubble radius at reentering. We are also investigating the evolution of these quantities at later times. 

%
%
%
%

\chapter{Conclusions and Outlook}
\label{chap:conclusions}
\pagestyle{fancy}

We are reaching the end of this manuscript and it is now time to summarize our work.  We started by reviewing the historical and current status of cosmology in the first introductory chapters. In chapter~2, we introduced the standard model of cosmology within the paradigm of Einstein's General Relativity and the FLRW Universe. We have treated the historical Big Bang problems, namely the horizon and flatness problems, and how an early phase of cosmic inflation was proposed to solve them. Chapter 3 was focus on slow-roll inflation, and within the framework of cosmological perturbation theory, we described how the fields' quantization provides for the generation mechanism of the primordial scalar and tensor fluctuations.

The inhomogeneous Universe was treated in chapter 4, and a fully general relativistic framework based on the 3+1 formalism of General Relativity was presented. We introduced the BSSN system of equations which are widely used for running numerical simulations. 
We have discussed the applications of this formalism to cosmological simulations, putting particular emphasis on the description of scalar fields, which was later used in chapters~5 and 6 in the context of inflation. 

Chapter 5 has presented our investigations on inhomogeneous single-field preinflationary scenarios. We have shown rich dynamical simulations considering the Higgs and Starobinsky inflation models, where we have identified the requirements for inflation to start from highly inhomogeneous conditions. Our work shows a variety of possible scenarios that could have played a role in the Universe prior to inflation. These consist of a preinflationary epoch of up to 7 efolds of scalar-field dominated universes with equations of state of kination and radiation. Other scenarios consist of an even longer period of preinflation dominated by gravitational vector/tensor modes lasting up to 14 efolds of decelerated expansion. This suggests that our Universe is at least one billion times larger than the volume of our observable Universe today. Finally, we conclude that any local Hubble region during preinflation having an average scalar field value in the slow-roll part of the potential will eventually lead to inflation, no matter how inhomogeneous the initial conditions are.

In Chapter 6, we expanded our previous investigations by including an auxiliary scalar field and further tested the robustness of the non-minimally coupled Higgs Inflation during the preinflation and the preheating eras. The inclusion of the auxiliary field allows for parametric-type resonances that, during preheating, successfully transfer energy from the inflaton condensate into localized excitations of the fields. After the broad resonance period, we showed that field excitations begin to form structures consisting of large under/over-densities regions. On the other hand, in the context of preinflation,  we showed that the non-minimal coupling acts as a stabilization mechanism that enhances the inflaton dynamics. This effect significantly reduces the impact of perturbations from other fields and matter sectors on the global dynamics, and therefore, it facilitates the beginning of inflation. 

In summary, in the previous investigations, we have shown that the Higgs (and Starobinsky) models are very robust to the considered initial conditions, and we have probed the preheating mechanism also in full General Relativity. However, there have been some limitations in these studies, particularly on the assumptions taken when generating the initial data for the simulations. In the case of preinflation, we assumed scenarios with a conformally flat metric which only allowed us to consider initial scalar fluctuations in the metric; and the tensor modes sourced by the posterior evolution. Future works could relax this assumption by including large tensor modes, \textit{ab initio}, from an independent origin, or even with non-flat background geometries.  This approaches typically requires of a full initial data solver that iteratively solves both the Hamiltonian \textit{and} momentum constraint. There has been recent theoretical and numerical advances in this regards 
\cite{Aurrekoetxea:2022mpw}.  
Similarly,  the initial data in our preheating studies has been constructed by the superposition of several sinusoidal modes in the scalar field,  while we have assumed no initial tensor perturbations. Instead, future works might want to investigate more realistic configurations by constructing initial data in the Bunch-Davies vacuum for the scalar fields, as well as for the gravitational waves. The inclusion of tensor modes could induce the amplification of gravitational waves during the resonance periods, which could be of observational interest to future GW experiments. We plan to investigate these phenomena in future works.

Numerical relativity is rapidly becoming an important tool for studying cosmological settings beyond the perturbative regime.  
Despite that, here, we have been focusing on the initial condition problem of inflation; many other situations exist where the nonlinearity of the Einstein equations should not be neglected. The effects of gravitational backreactions on the resonance structure during the preheating have been one example, and future works are necessary to satisfactorily comprehend these dynamics. 
The same occurs with other hypothetical processes that might have developed in the early Universe, and numerical investigations allow us to search for detectable signatures in future experiments. 
Indeed, in a similar manner that numerical relativity has been essential to precisely compute GW waveform from astrophysical binary mergers \cite{LIGOScientific:2016aoc,LIGOScientific:2018mvr}, these same tools can now be used to compute the signals from previous cosmological phase transitions and the search for evidence of exotic gravitating objects like cosmic strings, boson stars, primordial black holes, among others. 

In the context of late-time cosmology, there are intense lines of research regarding the nature of dark matter and dark energy. However, in recent years, substantial experimental evidence suggests that the current paradigm, which considers that the Universe is homogeneous and isotropic at ``suitably" large scales (i.e. the so-called Cosmological principle) should be put under scrutiny. The evidence includes the observation of large non-linear structures and voids at scales above 100 Mpc, the tensions on the $H_0$ measurement between early and late cosmological probes (i.e. between the CMB and supernova), and ultimately, the need for dark matter and dark energy in our cosmological standard model. In the years to come, we will have access to  new, unprecedented cosmological observations and data (e.g. large scale structure surveys, 21cm signal, gravitational-waves...) that will potentially challenge our current understanding of the Universe. To answer these questions and test the raising hypothesis, methods of numerical relativity will probably be a valuable resource, opening new lines of research.

\begin{appendix}
%
%
%
%

\addcontentsline{toc}{chapter}{Appendix}
\chapter{Considered inflation models}\label{app: Considered inflation models}

\section{Starobinsky inflation}

The Starobinsky model of inflation is built in the context of $f(R)$ modified theories of gravity. The action in the Jordan frame reads  
\beq \label{eq:S_SI_JF}
S = \frac{\Mpl^2}2 \int \ud^4 x \sqrt{-\bar g} \> f (\bar R) ~, \quad \text{with} \quad f(\bar R) = \bar R + \alpha \bar R^2 ~,
\eeq
where $\bar R$ is the Ricci scalar and $\alpha$ is a parameter with dimensions of inverse mass squared. We use the notation with the ``bar'' to  label  the variables in the Jordan frame. Next, we want to rewrite the action in the Einstein frame, i.e. a frame linear in $R$, by applying the following  metric transformation
\beq
g\munu = \Omega^2 \bar g\munu~, \quad  \text{and} \quad \sqrt{-g} = \Omega^4 \sqrt{-\bar g} ~,
\eeq
where, for the moment, $\Omega$ is an unknown conformal factor that needs to be determined. This transformation also redefines the scalar curvature according to 
\beq
\begin{split}  \label{transform_R}
\bar R &= \Omega^2 \left( R  + 6\, \square\phi -  6\, g^{\mu\nu} \partial_\mu \phi \partial_\nu \phi \right) ~,   \\
       \text{with}  \quad \phi & = \ln \Omega, \qquad   \square \phi = \frac 1 {\sqrt{-g}} \partial_\mu \left(\sqrt{-g} \> g^{\mu\nu}  \partial_\nu \phi \right)~. 
\end{split}
\eeq

We now proceed on calculating $\Omega$.  The action in Eq.~(\ref{eq:S_SI_JF}) can be rewritten in terms of a function of a scalar term $\psi$, such that
\beq
S = \frac{\Mpl^2}2 \int \ud^4 x \sqrt{-\bar g} \>  \left[  f (\psi)  + (R - \psi) F (\psi) \right] ~,
\eeq
where $F (\psi)$ is an arbitrary function that acts as a Langrange multiplier. After applying the variational principle with respect $\psi$, we obtain
\be
  \frac{\delta S}{\delta \psi} \quad \Rightarrow \quad 0  =  \frac{\partial f}{\partial \psi} + (R - \psi) \frac{\partial F}{\partial \psi} - F  ~,
\ee
which directly give us the solution%
\footnote{There is also an alternative solution, with $\frac{\partial F}{\partial \psi} = \text{const.}$ that corresponds to $f(\bar R) \propto \bar R $, thus to Einstein's  General Relativity. This solution is therefore excluded from the analysis.},  
\beq
F = \frac{\partial f}{\partial \psi} \quad \text{and} \quad \psi = R ~. 
\ee

Now, rewriting and manipulating the action we obtain  
\beq
\begin{split}
 S  &= \frac{\Mpl^2}2 \int \ud^4 x \sqrt{-\bar g} \>  \left[  F(R) R  + f(R)  - F(R) R \right]  ~,\\
    & = \frac{\Mpl^2}2 \int \ud^4 x \sqrt{ g} \>  \left[  F \Omega^{-2}  \left( R  + 6\, \square\phi -  6\, g^{\mu\nu} \partial_\mu \phi \, \partial_\nu \phi \right)  +  \Omega^{-4}\left( f  - FR\right) \right] ~, 
\end{split}
\eeq
where we have used  \Eq{transform_R} to replace the first term. Because we now wish to move to the Einstein frame, and therefore linear in $R$, we must choose 
${ \Omega^{2} = F}$. 

Moreover, by defining a new scalar-field $\varphi$ such that 
\be \label{def_varphi}
\frac \varphi\Mpl = \sqrt{\frac{3}2} \ln F = -\sqrt{6}\, \phi ~,
\ee
the action in the Einstein frame becomes 
\be
S  =  \int \ud^4 x \sqrt{- g} \>  \left[  \frac {\Mpl^2}2 R - \frac 12 g^{\mu\nu}\partial_\mu \varphi \, \partial_\nu \varphi - V \right]  ~,
\ee

\ \\

where we have discarded the total derivative term $6\, \square\phi$, and  defined a potential such that 
\be
\begin{split}
V(R) & = \frac{\Mpl^2}2 \frac{FR - f}{F^2}  ~, \\ 
     & =  \frac{\Mpl^2}2   \frac{ \alpha R^2}{\left(1 + 2\alpha R\right)^2} ~.
\end{split}
\ee

Using the definition in  \Eq{def_varphi}, we find that
\be
R =  \frac{1}{2\alpha} \left(  e^{\sqrt{\frac 23} \frac{\varphi}\Mpl} - 1  \right)~,
\ee

and then, the potential can be written in terms of the scalar field, which reads

\be
 V(\varphi)  = \Lambda^4 \left(  1 - e^{-\sqrt{\frac 23} \frac{\varphi}\Mpl} \right)^2~,   
\ee

with $\Lambda^4 = \frac{\Mpl^2} {8 \alpha}$. As we see later in Fig.~\ref{fig_SI}, it corresponds to potential of a plateau-shape. 

Calculating the slow-roll parameters, we obtain 
\begin{align}
\epsilon_1 = \frac 43 \left( 1 - e^{-1\sqrt{2/3} \varphi/\Mpl }  \right)^{-2} \\
\epsilon_2 = \frac 23 \left[ \sinh\left(\frac{\varphi}{\sqrt{6}\Mpl } \right) \right]^{-2} ~. 
\end{align}

And thus, we find that the end of inflation, which occurs when at $\epsilon_1 =1$, corresponds to 
\be
\varphi (\epsilon_1 = 1)  \simeq 0.94 \Mpl~,
\ee
while the slow-roll conditions stop being satisfied at  $\epsilon_2 = 1$, corresponding to 
\be
\qquad 
\varphi (\epsilon_2 = 1)  \simeq 1.83 \Mpl~. 
\ee

\clearpage
\section{Higgs inflation}

In this scenario inflation is driven by the Higgs field of the standard model of particles physics, which in the unitary gauge is given by
\beq
H = \frac{\Mpl}{\sqrt{2}}
\begin{pmatrix} 
0 \\ h
\end{pmatrix}  ~,
\eeq
where $h$ is a dimensionless scalar quantity. 

The action in the Jordan frame read 
\beq
\begin{split}
S =  \frac{\Mpl^2}{2} \int d^4 x & \sqrt{-\bar{g} } \Big[ \left( 1 + \xi_h h^2   \right) \bar{R}  -  \bar{g}^{\mu\nu} \partial_\mu h \partial_\nu h - 2 U( h ) \Big] ~ ,
\end{split}
\label{p2_eq:action_J}
\eeq
with the potential being 
\beq
U(h)  =  \Mpl^2 \frac \lambda 4 \left(h^2 - v^2\right)^2 \simeq  \Mpl^2 \frac\lambda 4 h^4  ~,
\eeq
where $v$ is the Higgs vacuum expectation value, which can be safely neglected during inflation.

The transformation to the Einstein frame is done after replacing 
\beq
 g\munu = \Omega^2 \bar g\munu~, \qquad \text{with}\quad \Omega^2 = \frac{\Mpl^2}{2\left( 1 + \xi_h h^2 \right)} ~, 
\eeq
which let to the following action in the Einstein frame, 
\begin{align}
 S &= \int d^4 x  \sqrt{-{g} } \Big[ \frac{\Mpl^2}{2}  {R}  - \frac{\Mpl^2}{2}  \mathcal{G}_{hh} {g}^{\mu\nu} \partial_\mu h \partial_\nu h - V(h) \Big] ~ ,
\\
 & \simeq \int d^4 x  \sqrt{-{g} } \Big[ \frac{\Mpl^2}{2} {R}  - \frac{1}{2}  {g}^{\mu\nu} \partial_\mu \varphi \partial_\nu \varphi - V(\varphi) \Big] ~ ,
\end{align}
where in the last line we introduced the (Einstein frame) canonical scalar-field $\varphi$. The approximation used, here denoted by the $\simeq$ sign, is not directly related to the conversion of the field, from $h \rightarrow \varphi$, as an exact solution is found after solving the identity 
\beq \label{p2_eq:diff_varphi_of_h}
 \frac 1 {\Mpl} \frac{\text{d} \varphi} {\text{d} h} = \sqrt{{\cal G}_{hh}} = \frac{\sqrt{1+\xi_h(1+6\xi_h)h^2}}{1+\xi_h h^2} ~,
\eeq
which leads to exact  expression \cite{MARTIN201475}
\beq
\begin{split} \label{p2_eq:varphi_of_h}
\frac{\varphi}{ \Mpl} = &\sqrt{\frac{1+6\xi_h}{\xi_h}} \arcsinh\left({h \sqrt{\xi_h (1+6\xi_h)}}\right) \\ &- \sqrt{6} \arctanh\left({\frac{\xi_h\sqrt{6} h}{\sqrt{1+\xi_h(1+6\xi_h)h^2}}}\right) ~.
\end{split}
\eeq

On the other hand, the conversion in the opposite direction, i.e. from $\varphi \rightarrow h$, can not be analytically inverted exactly. This conversion is needed in order to express the potential in terms of the canonical field, $V(\varphi)$, and thus one must rely on an approximate expression of $\varphi(h)$. In this case, in the limits where $\xi_h \gg 1$ and $\xi_h h \gg 1$, one can use the expansion in the trigonometric functions 
\be
\arcsinh \theta = \ln \left( \theta \right)~, \qquad \arctanh \theta = \frac 12 \ln \left[\frac{1+\theta}{1-\theta}\right]~. 
\ee
After expanding the arguments of the logarithms in $1 / {\xi_h}$ and $1 / ({\xi_h} h^2) $, one finds 
\be
\frac{\varphi}{\Mpl} \simeq \sqrt{\frac{3}{2}} \ln \left( 1 + \xi_h h^2 \right). 
\ee

\begin{figure*}[t!]
\begin{center}
\includegraphics[width=0.8\textwidth]{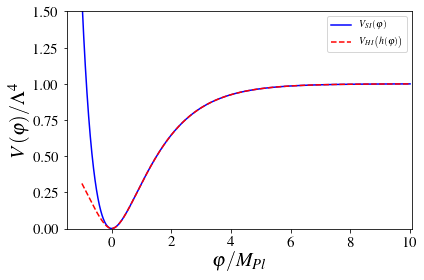}
\includegraphics[width=0.83\textwidth]{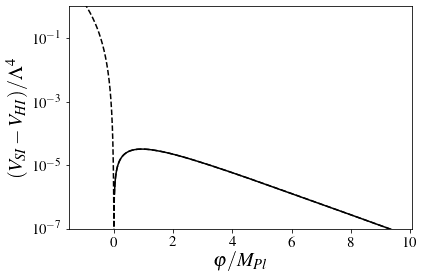}
\caption{ 
Starobinsky and Higgs potential (top panel), and difference between exact and approximate solution in Higgs inflation (bottom panel). 
\label{fig_SI}
}
\end{center}
\end{figure*}

Finally, the field potential can now be written as 
\begin{align}
 V(h) & \equiv  \Mpl^2 \frac{ U(h)}{ \left( 1 + \xi_h h^2 \right)^2} ~,  \label{eq:HI_exact}
 \\
 \Rightarrow \qquad V(\varphi) & \simeq \Lambda^4 \left( 1 - {\rm e}^{-\sqrt{2/3} |\varphi| / \Mpl} \right)^2~,  \label{eq:HI_app}
\end{align}
with $\Lambda^4 = \Mpl^4  \lambda / ({4 \xi_h^2})$. With that approximation, we recovered the same potential shape as in Starobinsky inflation.  Figure~\ref{fig_SI} illustrates the potential, and the (small) deviation between the exact and approximate expressions, i.e.  Eqs.~(\ref{eq:HI_exact}) and (\ref{eq:HI_app}),  respectively. On the other hand, the two expressions diverge from each other at low and negative field, what leads to distinct predictions for Starobinsky and Higgs inflation during the reheating \cite{MARTIN201475}.

\clearpage

\chapter{GRChombo code}  \label{ApB_app:2} 

In this thesis, the numerical relativity simulations are done using the open source \texttt{GRChombo} numerical relativity code ~\cite{Clough_2015,Andrade2021}. The code uses a block-structure adaptive mesh refinement (AMR) for solving the partial differential equations.   The key features of the code are the following: 

\begin{itemize}
 \item C++ class structure code, with AMR implementation using the \texttt{Chombo} libraries~\cite{Chombo}.\\
 \item OpenMP and MPI parallelization capable to scale efficiently up to several thousand CPU-cores per run. \\
 \item BSSN formalism of numerical relativity. Allows for a flexible gauge choice, including the popular ``moving puncture gauge''. \\
 \item Additional diagnostic variables such the Hamiltonian and Momentum constraint or other user-defined variables of interest.  \\
 \item Standardized Input/Output file system using HDF5 format, which is supported by popular visualization tools such as Visit, Paraview or YT (python library). Checkpoint and diagnostic output capabilities with a flexible output of diagnostic variables.  \\
 \item 4th order spatial stencils for spatial derivatives and 4th order Runge-Kutta for the time evolution. \\
 \item Kress-Oliger dissipation is used to control numerical errors rising from the truncation and interpolation algorithms during the regridding. \\
 \item Boundary conditions implementations includes periodic boundaries, reflective boundaries or Sommerfeld-type boundaries. \\
 \item Apparent horizon finder algorithm, which can be called within the simulation time of run.  
\end{itemize}

For more a detail information, we refer to Refs.~\cite{Clough_2015,Andrade2021}.  Below, we summarize the numerical formalism used in this thesis.

\section{BSSN evolution equations}

In the context of the 3+1 decomposition of General Relativity, the line element reads
\be\label{ApB_timeline_AP}
\rr d s^2 = - \alpha^2 \rr d t^2 + \gamma_{ij}(\rr d x^i + \beta^i \rr d t)(\rr d x^j + \beta^j \rr d t)
\ee
where $\gamma\ij$ is the metric of the 3-dimensional hypersurface, and the lapse and shift gauge parameters are given by $\alpha(t)$ and $\beta^i(t)$ {respectively}. A further conformal decomposition of the 3-metric follows,
\be
\gm\ij = \frac1\chi \tgm\ij = \psi^4\tgm\ij \quad \text{with } \text{ det}(\tgm\ij) = 1 ~, 
\ee
where $\chi$ and $\psi$ are two different parametrisations of the metric conformal factor. While the former is used during the temporal integration, the latter is preferred when constructing the initial conditions. The extrinsic curvature is split in $\tA\ij$ and $K$, respectively, the conformal traceless part and its trace, 
\be
K\ij = \frac1\chi \left( \tA\ij +\frac13\tgm\ij K\right)~.
\ee
In addition, the first spatial derivatives of the metric are promoted to dynamical variables
\be
\tilde\Gamma^i \equiv \tgm^{jk} \tilde\Gamma^i_{jk} = - \partial_j\tgm\ij ~,
\ee
where $ \tilde\Gamma^i_{jk} $ are the Christoffel symbols associated with the conformal spatial metric $ \tilde\gamma_{ij} $.

The evolution equations for the BSSN variables are then given by 
\begin{align} 
&\partial_t\chi=\frac{2}{3}\,\alpha\,\chi\, K - \frac{2}{3}\,\chi \,\partial_k \beta^k + \beta^k\,\partial_k \chi ~ , \label{ApB_eqn:dtchi2} \\
&\partial_t\tilde\gamma_{ij} =-2\,\alpha\, \tA_{ij}+\tilde \gamma_{ik}\,\partial_j\beta^k+\tilde \gamma_{jk}\,\partial_i\beta^k \nonumber \\
&\hspace{1.3cm} -\frac{2}{3}\,\tilde \gamma_{ij}\,\partial_k\beta^k +\beta^k\,\partial_k \tilde \gamma_{ij} ~ , \label{ApB_eqn:dttgamma2} \\
&\partial_t K = -\gamma^{ij}D_i D_j \alpha + \alpha\left(\tilde{A}_{ij} \tilde{A}^{ij} + \frac{1}{3} K^2 \right) \nonumber \\
&\hspace{1.3cm} + \beta^i\partial_iK + 4\pi\,\alpha(\rho_{\rm sf}+ S) \label{ApB_eqn:dtK2} ~ , 
 \end{align}
 \begin{align} 
&\partial_t\tA_{ij} = \left[- \chi D_iD_j \alpha + \chi \alpha\left( R_{ij} - 8\pi\, \,S_{ij}\right)\right]^\textrm{TF} \nonumber \\
&\hspace{1.3cm} + \alpha (K \tA_{ij} - 2 \tA_{il}\,\tA^l{}_j) \nonumber \\
&\hspace{1.3cm} + \tA_{ik}\, \partial_j\beta^k + \tA_{jk}\,\partial_i\beta^k \nonumber \\
&\hspace{1.3cm} -\frac{2}{3}\,\tA_{ij}\,\partial_k\beta^k+\beta^k\,\partial_k \tA_{ij}\, \label{ApB_eqn:dtAij2} ~, \\ 
&\partial_t \tilde \Gamma^i=2\,\alpha\left(\tilde\Gamma^i_{jk}\,\tA^{jk}-\frac{2}{3}\,\tilde\gamma^{ij}\partial_j K - \frac{3}{2}\,\tA^{ij}\frac{\partial_j \chi}{\chi}\right) \nonumber \\
&\hspace{1.3cm} -2\,\tA^{ij}\,\partial_j \alpha +\beta^k\partial_k \tilde\Gamma^{i} \nonumber \\
&\hspace{1.3cm} +\tilde\gamma^{jk}\partial_j\partial_k \beta^i +\frac{1}{3}\,\tilde\gamma^{ij}\partial_j \partial_k\beta^k \nonumber \\
&\hspace{1.3cm} + \frac{2}{3}\,\tilde\Gamma^i\,\partial_k \beta^k -\tilde\Gamma^k\partial_k \beta^i - 16\pi\,\alpha\,\tilde\gamma^{ij}\,S_j ~ , \label{ApB_eqn:dtgamma2}
\end{align} 
where the superscript $\rm{TF}$ denotes the trace-free parts of tensors, with $R\ij$ being the (3-dimensional) Ricci tensor.  The 3+1 decomposition of the energy-momentum tensor $T^{\mu\nu}$ gives
\bea \label{ApB_3+1sources_AP}
 \rho &=& n^\mu n^\nu T_{\mu\nu} ~,\\
 S_i &=& -\gamma^{\mu}_i n^\nu T_{\mu\nu} ~,\\ 
 S_{ij} &=& \gamma^{\mu}_i \gamma^{\nu}_j T_{\mu\nu} ~,\\ 
 S &=& \gamma\IJ S\ij ~,
\eea
where $n^\mu=(1/\alpha, -\beta^i/\alpha)$ is the unit normal vector to the three-dimensional slices.

The Hamiltonian and momentum constraints, 
\begin{align}
\mathcal{H} & = R + K^2-K_{ij}K^{ij}-16\pi \rho= 0\, , \label{ApB_eqn:HamSimp} \\
\mathcal{M}_i & = D^j (K_{ij} - \gamma_{ij} K) - 8\pi S_i =0\, , \label{ApB_eqn:MomSimp}
\end{align}
where $R$ is the Ricci scalar, are only solved explicitly when constructing initial data. However, they are monitored during the time evolution in order to ensure that the numerical integration scheme do not diverge significantly from General Relativity. \

\section{Gauge choice and singularity avoidance }%

The gauge parameters are initially set to $\alpha=1$ and $\beta^i=0$ and then evolved in accordance with the \textit{moving puncture gauge} \cite{Baker_2006, Campanelli_2006}, for which  the evolution equations are
\begin{eqnarray}
\partial_t \alpha &=& -\eta_\alpha \alpha \left( K - \langle K \rangle \right) + \,\beta^i\partial_i \alpha \ , \label{ApB_eqn:alphadriver}\\
\partial_t \beta^i &=& B^i\, \label{ApB_eqn:betadriver},\\
\partial_t B^i &=& \frac34\, \partial_t \tilde\Gamma^i - \eta_B\, B^i\ \,, \label{ApB_eqn:gammadriver}
\end{eqnarray}
where the constants $\eta_\alpha$ and $\eta_B$ are conveniently chosen to improve the numerical stability. This way, $\alpha$ and $\beta^i$ are boosted in the problematic regions with strongly growing extrinsic curvature and spatial derivatives of the metric $\tilde \gamma_{ij}$. 
The goal of this gauge is to prevent the code from resolving the central singularity of any black hole that may eventually form, as well as to prevent coordinate singularities on converging geodesics.

\section{Scalar field equations}
For the Einstein frame canonical scalar fields $\varphi^I$, the energy-momentum tensor is given by
\begin{equation}
T_{\mu\nu} = \delta_{IJ}\left( \partial_\mu \varphi^I\, \partial_\nu \varphi^J - \frac{1}{2} g_{\mu\nu}\, \partial_\lambda \varphi^I \, \partial^\lambda \varphi^J \right) - g_{\mu\nu} V(\varphi^K) \,
\end{equation}

The scalar field dynamics is governed by the the Klein-Gordon equation, split into two first order equations for the field and its momentum $\Pi_{\rm M}^I$
\begin{align}
\partial_t \varphi^I &= \alpha \Pi_{\rm M} +\beta^i\partial_i \varphi \label{ApB_eq:dtvarphi} ~ , \\
\partial_t \Pi_{\rm M}^I &= \beta^i\partial_i \Pi_{\rm M}^I + \alpha\partial^i\partial_i \varphi^I + \partial_i \varphi^I \, \partial^i \alpha \\
& \ +\alpha \left[ K\Pi_{\rm M}^I-\gamma^{ij}\Gamma^k_{ij}\partial_k \varphi^I - \frac{d}{d\varphi^I}V(\varphi^K) \right] ~ ,
\end{align} 

Still in the Einstein frame, but with the Jordan defined scalar fields $\bar\phi^I$, the energy momentum is written as

\beq
T_{\mu\nu} = {\cal G}_{IJ} \partial_\mu \bar\phi^I \partial_\nu \bar\phi^J - g_{\mu\nu} \left[ \frac{1}{2} {\cal G}_{IJ} \partial_\alpha \bar\phi^I \partial^\alpha \bar\phi^J + V (\bar\phi^I ) \right] ~,
\eeq

and the evolution equations read

\begin{align}
\partial_t \bar\phi^I &= \alpha {\bar\Pi}_{\rm M}^I +\beta^i\partial_i \bar\phi^I ~ , \\
\partial_t {\bar\Pi}_{\rm M}^I &= \beta^i\partial_i {\bar\Pi}_{\rm M}^I + \alpha\partial^i\partial_i \bar\phi^I + \partial_i \bar\phi^I \, \partial^i \alpha \\ \nonumber
& \ +\alpha \Big[ K {\bar\Pi}_{\rm M}^I- \gamma^{ij}\Gamma^k_{ij}\partial_k \bar\phi^I \\ \nonumber
& + \Gamma^I_{JK} \left( - \bar\Pi_{\rm M}^J\bar\Pi_{\rm M}^K  + \gamma\IJ\partial_i \bar\phi^J \partial_j \bar\phi^K \right)
- {\cal G}^{IJ} \frac{d }{d {\bar\phi}^J} V (\bar\phi^K) \Big] ~,
\end{align} 
where $\Gamma^I_{JK}$ is the affine connexion corresponding to the field-space metric ${\cal G}_{IJ}$.  

If, instead, the system is evolved using the Einstein frame notation for the scalar fields $\Phi^I$, the energy tensor simplifies to 

\beq
T_{\mu\nu} = {\delta}_{IJ} \partial_\mu \Phi^I \partial_\nu \Phi^J - g_{\mu\nu} \left[ \frac{1}{2} {\delta}_{IJ} \partial_\alpha \Phi^I \partial^\alpha \Phi^J + V (\Phi^I ) \right] ~,
\eeq

and, the evolution equations are given by

\begin{align}
\partial_t \Phi^I &= \alpha {\Pi}_{\rm M}^I +\beta^i\partial_i \Phi^I ~ , \\
\partial_t {\Pi}_{\rm M}^I &= \beta^i\partial_i {\Pi}_{\rm M}^I + \alpha\partial^i\partial_i \Phi^I + \partial_i \Phi^I \, \partial^i \alpha \\ \nonumber
& \ +\alpha \Big[ K {\Pi}_{\rm M}^I- \gamma^{ij}\Gamma^k_{ij}\partial_k \Phi^I  - \frac{d }{d {\Phi}^I} V (\Phi^K) \Big] ~ .
\end{align}

\section{Field-space metric and Christoffel symbols}
\label{ApB_sec:AppendixAFieldSpace}

Given $f ( {\phi_1} , {\phi_2} ) =  \frac 12 \left({M^2_{\rm pl}} + \xi_{1} {\phi_1}^2 + \xi_{2} {\phi_2}^2 \right)$ for a two-field model, with non-minimal couplings $\xi_{1},\> \xi_{2}$, the field-space metric in the Einstein frame, takes the form
\beq
{\cal G}_{IJ}=
\left( \frac{M_{\rm pl}^2}{4f^2} \right)
\begin{pmatrix} 
2f+6\xi_{1}^2 {\phi_1}^2 & 6 \xi_{1} \xi_{2} {\phi_1} {\phi_2} \\
6 \xi_{1} \xi_{2} {\phi_1} {\phi_2} & 2f+6\xi_{2}^2 {\phi_2}^2
\end{pmatrix} \, ,
\label{ApB_G_hh}
\eeq
\beq
{\cal G}^{IJ}=
\left( \frac{2f}{M_{\rm pl}^2C} \right)
\begin{pmatrix} 
2f+6\xi_{2}^2 {\phi_2}^2 & -6 \xi_{1} \xi_{2} {\phi_1} {\phi_2} \\
-6 \xi_{1} \xi_{2} {\phi_1} {\phi_2} & 2f+6\xi_{1}^2 {\phi_1}^2
\end{pmatrix} \, ,
\label{ApB_invG}
\eeq
where $C ({\phi_1} , {\phi_2} )$ is defined as
\beq
\begin{split}
C({\phi_1} , {\phi_2}) = 2f + 6 \xi_{1}^2 {\phi_1}^2 + 6 \xi_{2}^2 {\phi_2}^2 .
\end{split}
\label{ApB_C}
\eeq

The Christoffel symbols associated with this field space metric 
\beq
\begin{split}
\Gamma^{1}_{\>\> {1} {1} } &= \frac{\xi_{1} (1 + 6 \xi_{1} ) {\phi_1} }{C} - \frac{\xi_{1} {\phi_1} }{f}\, , \\
\Gamma^{1}_{\>\> {1} {2} } &= - \frac{\xi_{2} {\phi_2} }{2f} \quad = \Gamma^ {\phi_1} _{\>\> {2} {1} } \, , \\
\Gamma^{1} _{\>\> {2} {2} } &= \frac{\xi_{1} (1 + 6 \xi_{2} ) {\phi_1} }{C} , \\
\Gamma^{2} _{\>\> {2} {2} } &= \frac{\xi_{2} (1 + 6 \xi_{2} ) {\phi_2} }{C} - \frac{\xi_{2} {\phi_2} }{f} \, ,\\ 
\Gamma^{2} _{\>\> {2} {1} } &= - \frac{\xi_{1} {\phi_1} }{2f} \quad = \Gamma^ {\phi_2} _{\>\> {1} {2} } \, , \\ 
\Gamma^{2}_{\>\> {1} {1} } &= \frac{\xi_{2} (1 + 6 \xi_{1} ) {\phi_2} }{C} \, . 
\end{split}
\label{ApB_Gammas}
\eeq


\section{Low-field approximation between the Jordan and the Einstein frame \label{ApB_ApConversionJE} }

Transforming from the scalar fields in the Jordan frame $\bar\phi^I$ to the Einstein frame $\Phi^I$ is done by finding an approximate solution to the following system of equations

\beq \label{ApB_eq:AP_convert_frame}
{\cal G}_{IJ} g^{\mu\nu} \partial_\mu \bar\phi^I \partial_\nu \bar\phi^J = {\delta}_{IJ} g^{\mu\nu} \partial_\mu \Phi^I \partial_\nu \Phi^J ~.
\eeq

Assuming two Jordan scalar fields, the Higgs $h$ with non-minimal coupling $\xi_h$  and an auxiliary field $s$ with non-minimal coupling $\xi_s$, we search for a transformation into the Einstein frame such as $\varphi(h, s),\> \chi(h,s)$. Assuming $\xi_s = 0$ , the above mentioned system of equation simplifies to
\begin{align}
 \left( \frac{\partial \varphi}{\partial h} \right)^2 &+ \left( \frac{\partial \chi}{\partial h} \right)^2 = {\cal G}_{hh} \\  
  \left( \frac{\partial \varphi}{\partial s} \right)^2 &+ \left( \frac{\partial \chi}{\partial s} \right)^2 = {\cal G}_{ss} \\ 
  \left( \frac{\partial \varphi}{\partial h} \frac{\partial \varphi}{\partial s} \right) &+ \left( \frac{\partial \chi}{\partial h} \frac{\partial \chi}{\partial s} \right) = 0 ~. 
\end{align}

\begin{figure}[b]
\begin{center}
\hspace*{-2mm}
\includegraphics[width=0.90\textwidth]{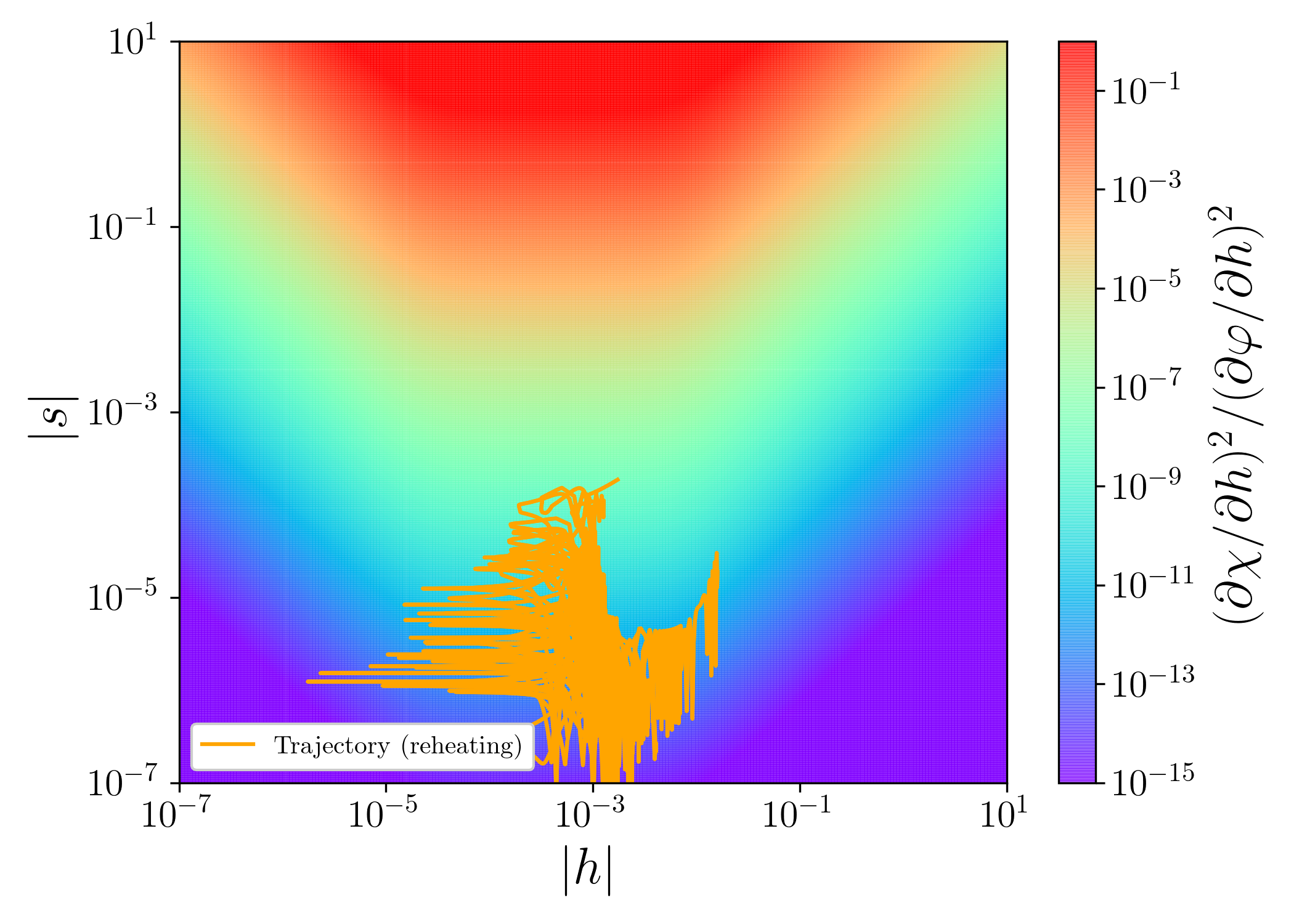}
\end{center}
\caption{ Error of the approximation (\ref{ApB_eq:AP_convert_frame}) to convert the fields between the Jordan and Einstein frame.  \label{ApB_fig:approx}}
\end{figure}

By assuming $ {\partial \varphi}/{\partial h} \approx  \sqrt{{\cal G}_{hh}}$ and $ {\partial \chi}/{\partial s} \approx  \sqrt{{\cal G}_{ss}}$ implies that, in the range of validity of this approximation, 
\begin{align}
\left(\frac{\partial \varphi}{\partial s}\right) ^2 &\ll \left(\frac{\partial \chi}{\partial s}\right)^2  \label{AP_cond1}
\\ 
\left(\frac{\partial \chi}{\partial h}\right)^2    &\ll
\left(\frac{\partial \varphi}{\partial h}\right)^2 
\label{AP_cond2} ~.
\end{align}
While the first identity is trivially satisfied, the second one,  Eq.~(\ref{AP_cond2}), it is not. The approximation proposes the solution to be {$s \approx \chi \sqrt{2 f(h)}$}, which range of validity depends on the region in consideration of the field space. This is illustrated in Fig. \ref{ApB_fig:approx}. In general, the parameter space when these assumptions are valid is when  $ |s| <10^{-2}$, and when  $ |s| < |h| / 100$.  These regions corresponds to $ 
\left(\frac{\partial \chi}{\partial h}\right)^2 / \left(\frac{\partial \varphi}{\partial h}\right)^2  < 10^{-5}$.

\chapter{Validation and testing}  \label{app:3} 

\section{Convergence tests}

The common validation test for the stability of the simulations is the monitoring of the Hamiltonian constraint equations. In order to infer whether there is a correct cancellation within its terms, a relative quantity between the Hamiltonian constraint violation and the norm of it is computed, so that  
\be
  \mathcal{H}_{\rm REL} = \frac {\mathcal{H} }{\left[\mathcal{H}\right] }  ~, 
\ee
where the denominator is defined as 
\be
\left[\mathcal{H}\right]  \equiv \Biggl[  \left( {R} \right) ^2 +  \left( \tilde A\IJ \tilde A\ij \right) ^2 
+ \left(  \frac23 K^2 \right) ^2  + \left( 16\pi \rho \right) ^2 \Biggr] ^{1/2} ~. 
\ee

This quantity has been computed for all simulations. Furthermore, it is also convenient to test the variation of these variables for different grid resolution sizes. Because the associated numerical errors should decrease at increasing box resolutions, we call these tests as ``convergence tests''. 

\begin{figure*}[!ht]
\begin{center}
\includegraphics[width=0.51\textwidth]{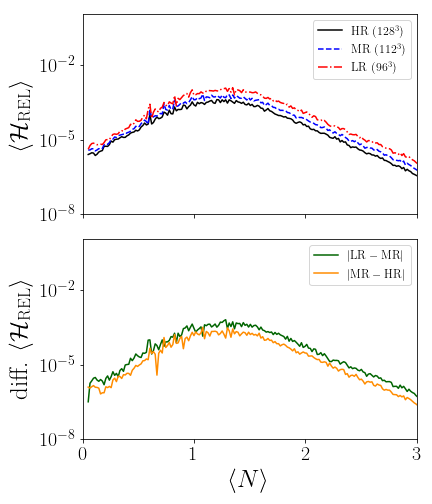}
\includegraphics[width=0.49\textwidth]{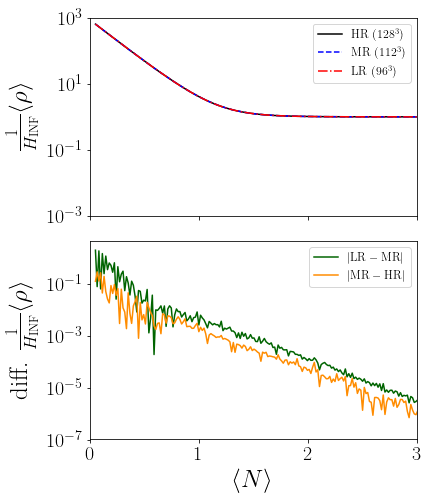}
\includegraphics[width=0.49\textwidth]{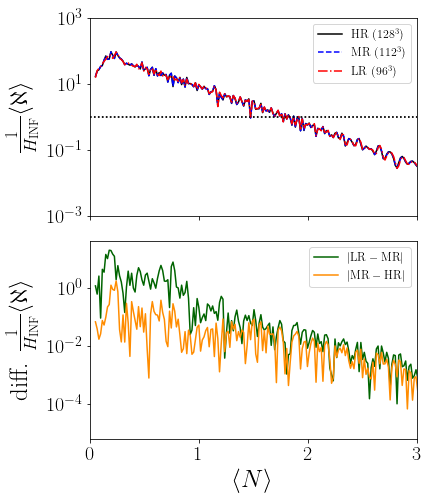}
\caption{ 
Convergence tests on the Hamiltonian constraint (left), mean scalar field energy density (center) and mean metric energy density (right).   Top panels: evolution of the mean values for low (LR, red), medium (MR, blue) and high (HR, black) resolutions. Bottom panels:  LR-MR (green) and MR-HR (orange) comparisons. These tests correspond to the work presented in Chapter~\ref{p1_chap:prepaper1}.
\label{fig:A2_convergence}
}
\end{center}
\end{figure*}

\begin{figure*}[!ht]
\begin{center}
\includegraphics[width=0.80\textwidth]{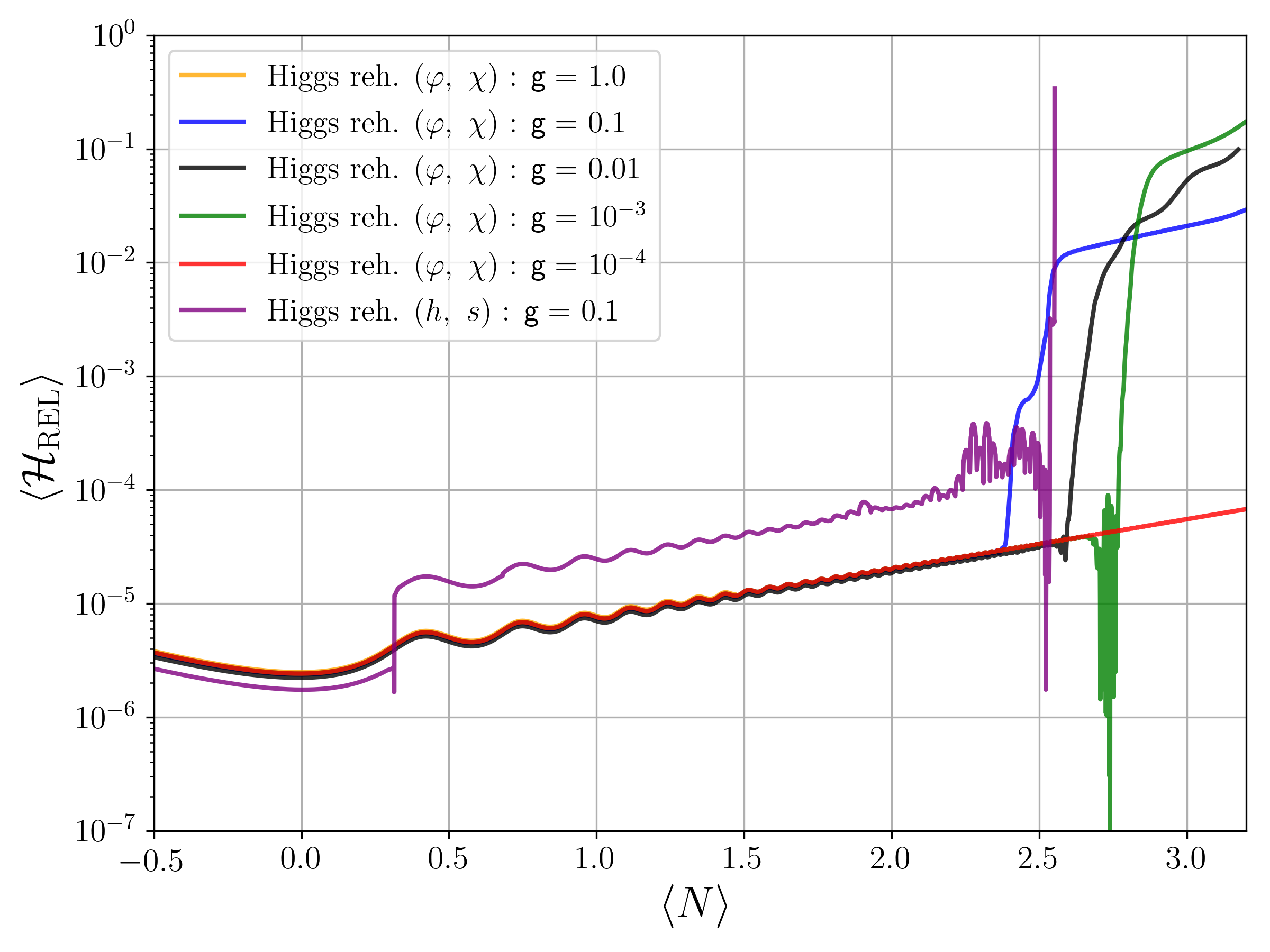}
\includegraphics[width=0.80\textwidth]{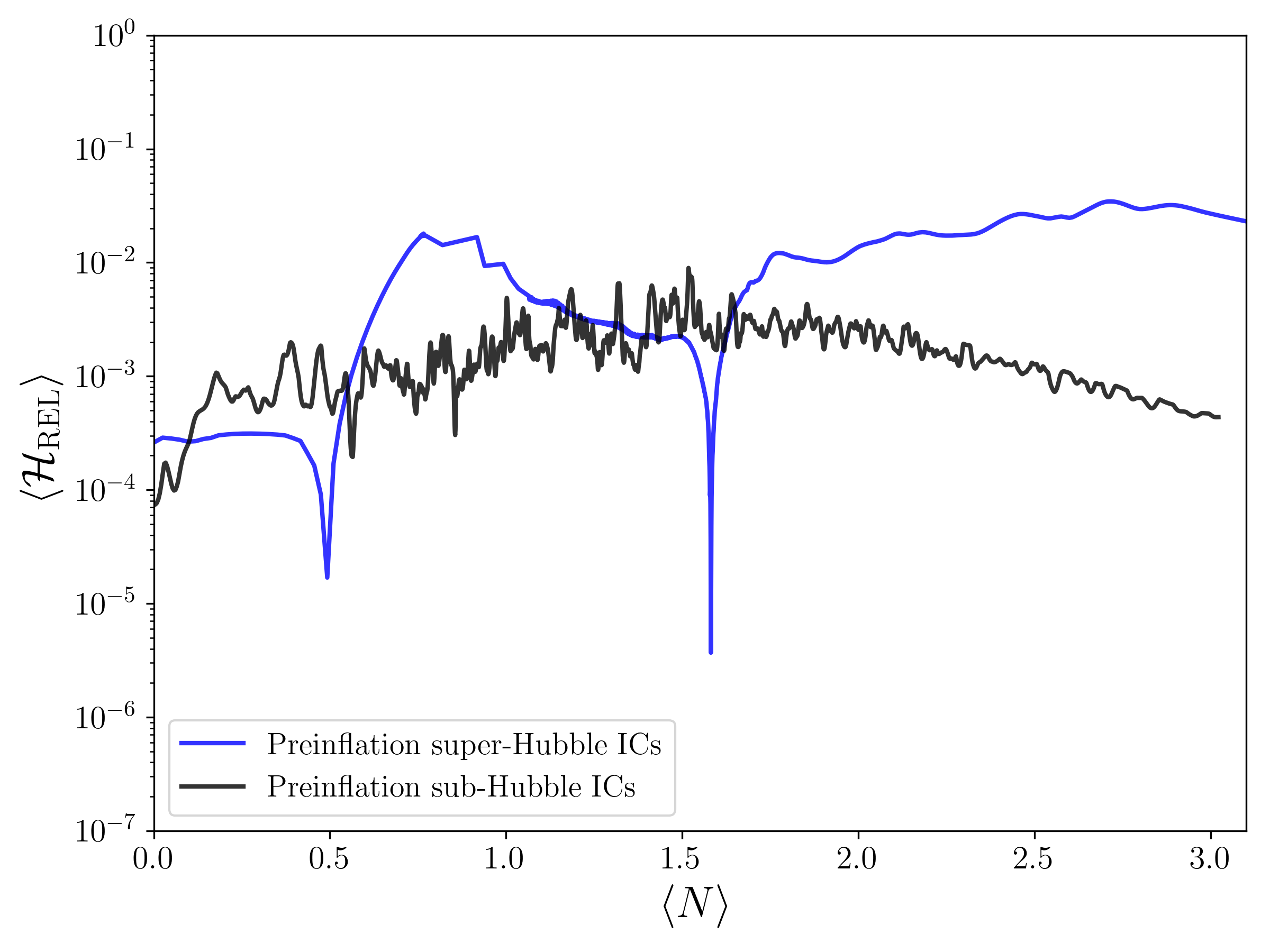}
\end{center}
\caption{ Relative Hamiltonian constraint for simulations on reheating (left panel) and on preinflation (right panel).  
These tests correspond to the simulations presented in Chapter~\ref{p2_chap:prepaper2}.
\label{fig:HamVal} }
\end{figure*}

\begin{figure*}[!ht]
\begin{center}
\includegraphics[width=0.48\textwidth]{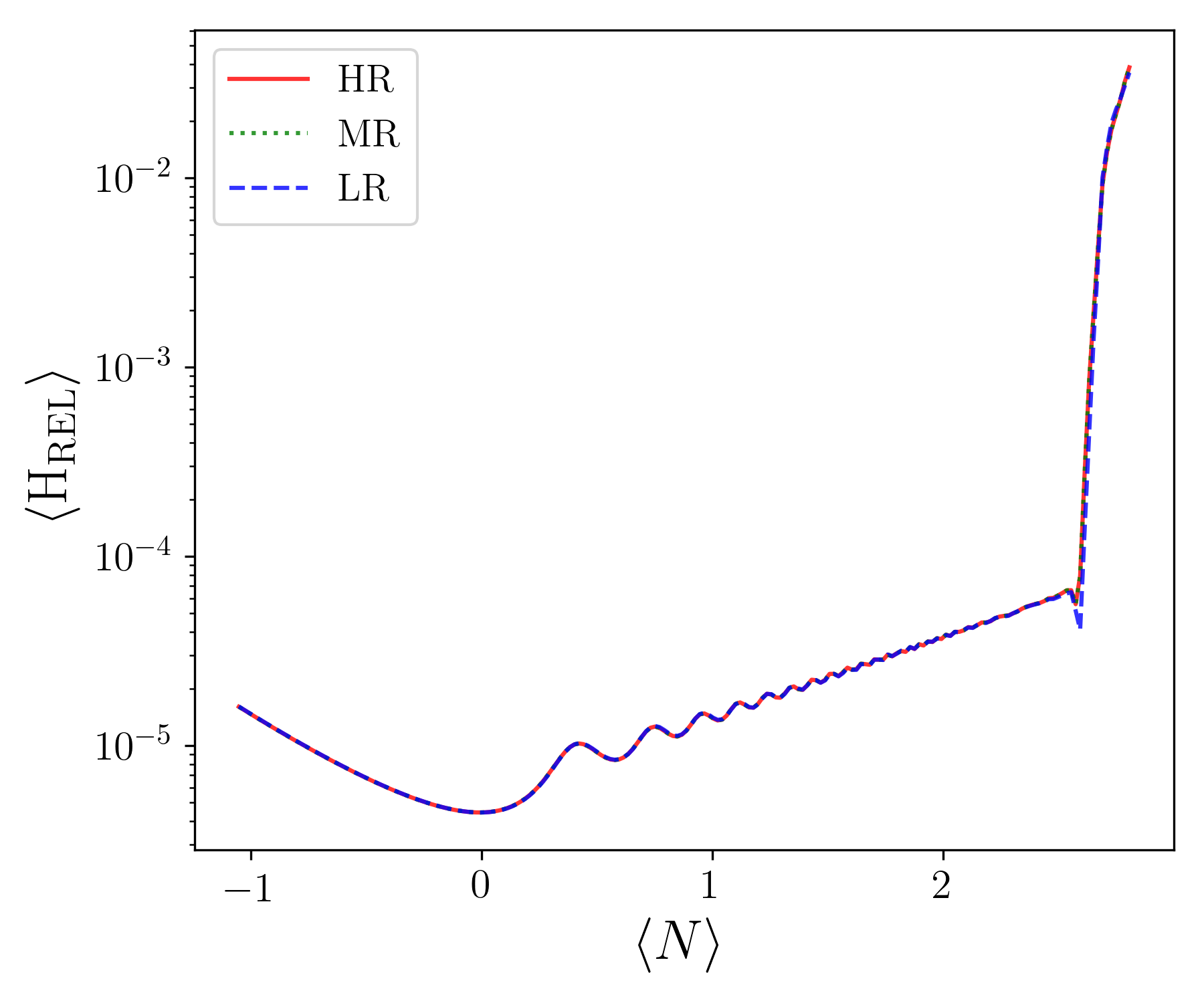}
\includegraphics[width=0.48\textwidth]{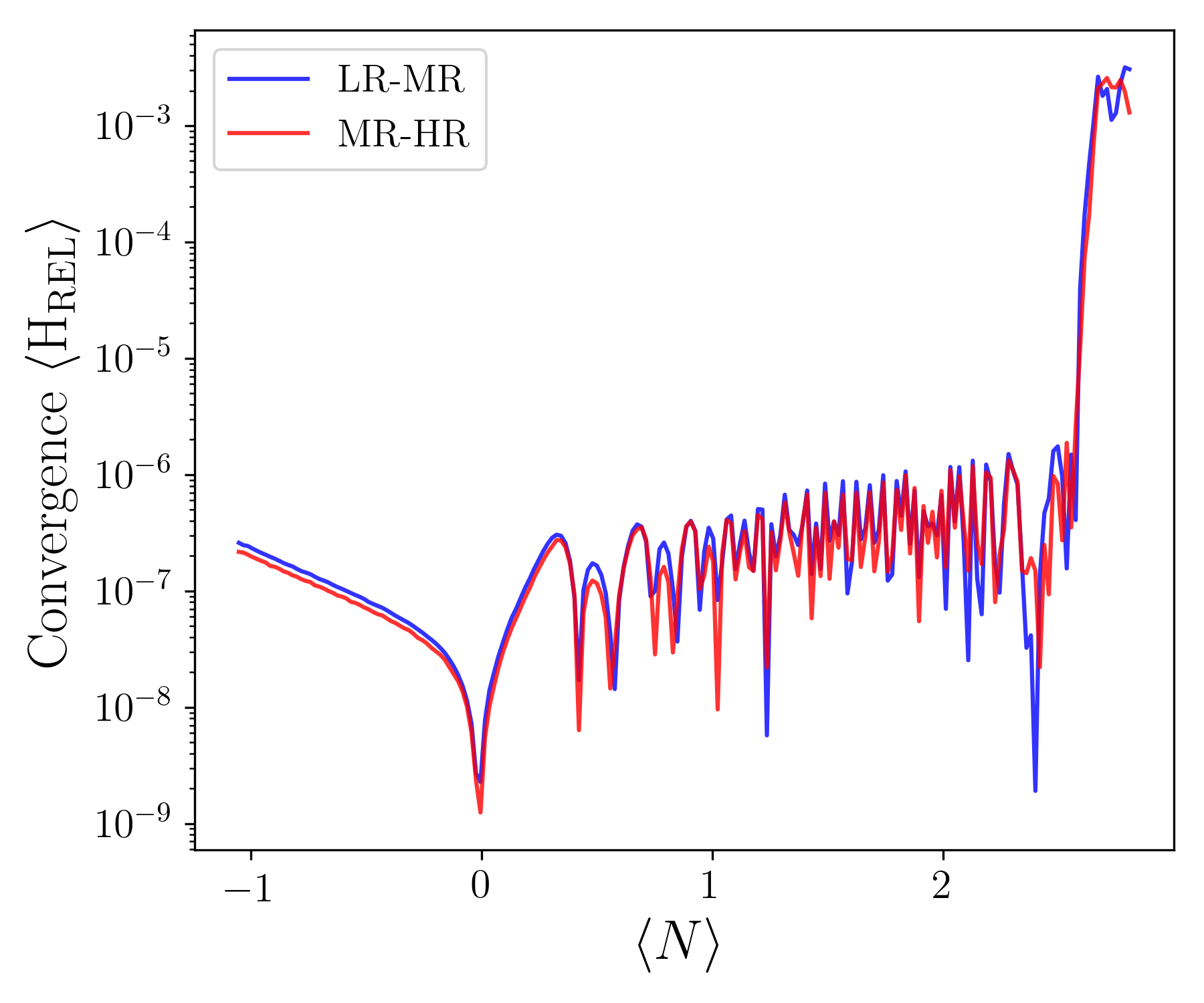}
\includegraphics[width=0.48\textwidth]{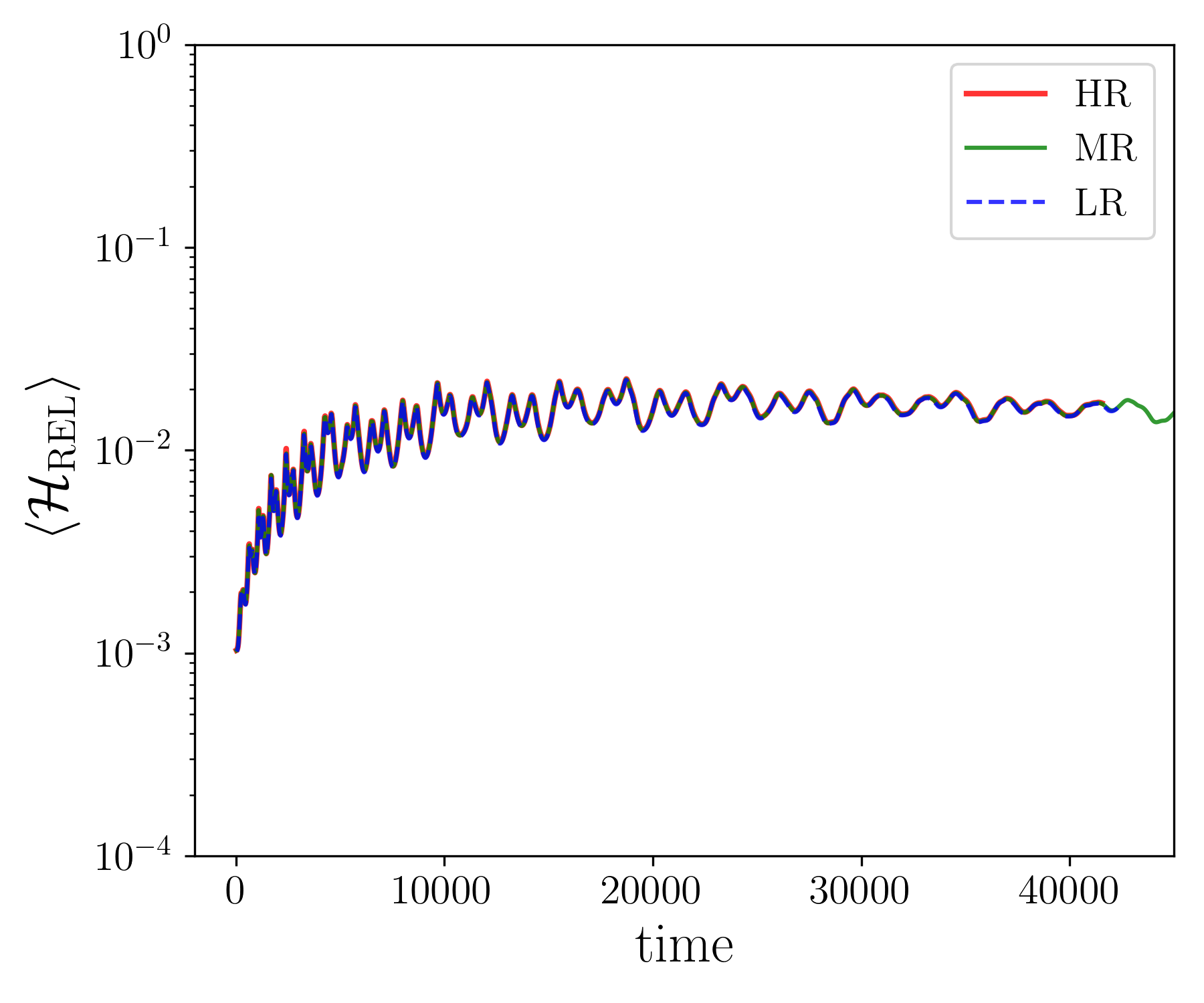}
\includegraphics[width=0.48\textwidth]{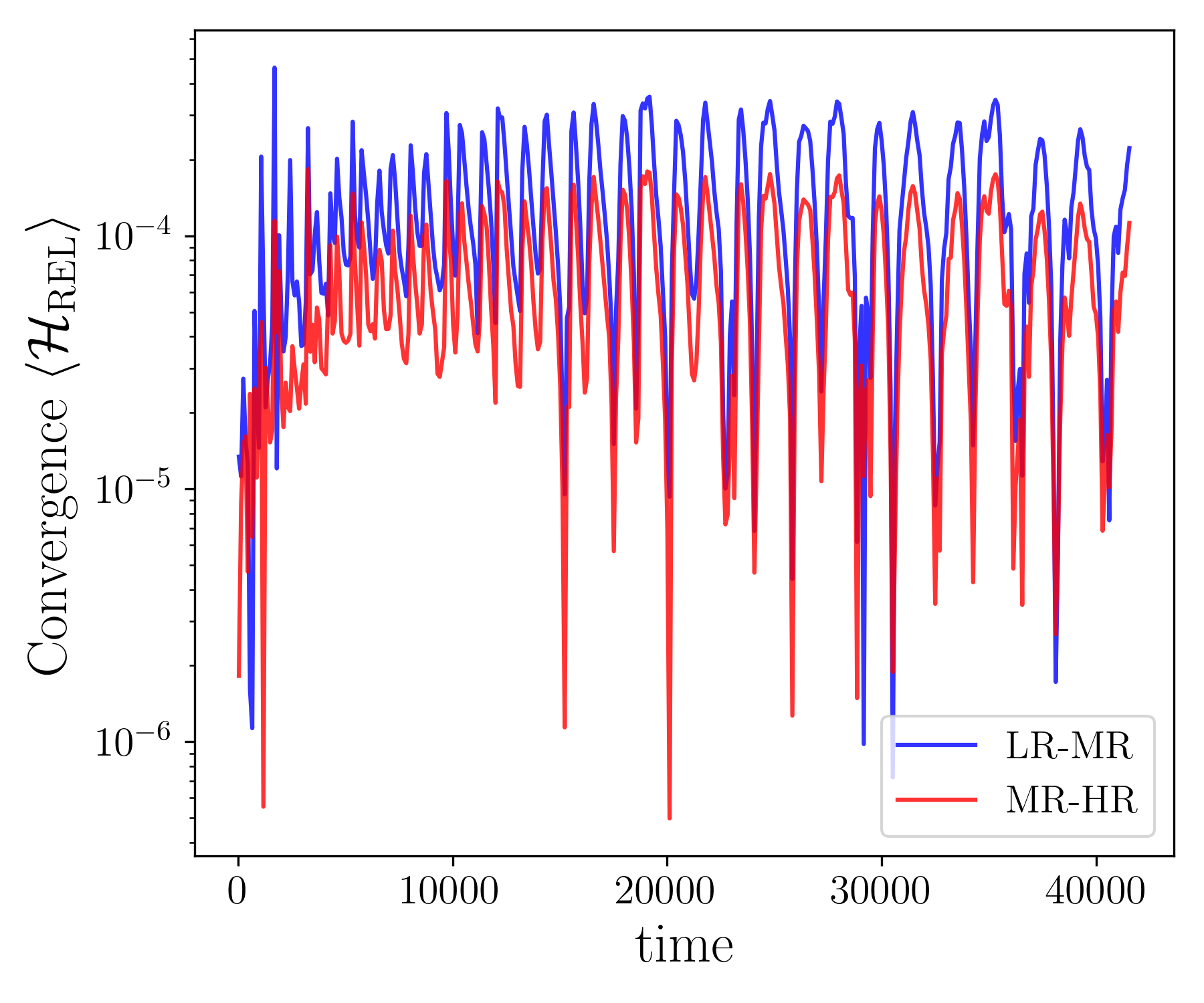}
\end{center}
\caption{ Convergence testing: Relative difference between low (LR), medium (MR) and high (HR) resolutions grids for simulations on preheating (top panels) and preinflation (bottom panels).  The size grid used is LR = $128^3$, MR = $144^3$, HR = $156^3$ for the case of preheating, and   LR = $128^3$, MR = $180^3$, HR = $220^3$ for the case of preinflation. These tests correspond to the simulations presented in Chapter~\ref{p2_chap:prepaper2}.
\label{fig:ctest} }
\end{figure*}

The convergence tests for single-field simulations from chapter~5 are shown in Fig.~\ref{fig:A2_convergence}, and for multifield simulations of chapter~6 are shown inin Figs.~\ref{fig:HamVal} and~\ref{fig:ctest}.


%

\clearpage

 \section{Consistency tests} 
 
 In addition, we performed additional simulations to test that non-physical parameters did not affect the outcome of the simulations. For this purpose, we have run analogous simulations as the ones shown in chapter~6 for both the preinflation and the preheating with some variation in the initial data. We call this kind of exercise ``consistency checks''. For simulations on preinflation, we tested for consistency after increasing the initial box size of the simulations, and thus for super-Hubble simulations searching for possible effects when increasing the number of modes capable of re-entering the causal domain. For simulations on preheating, we also checked the consistency after increasing the box size,  which can potentially modify the allowed resonant modes; and also, we tested if the inclusion of perturbations in the Higgs field in the initial conditions had a significant impact on the resonances.   
 
\begin{figure*}[!ht]
\vspace{30mm}
\begin{center}
\includegraphics[width=0.99\textwidth]{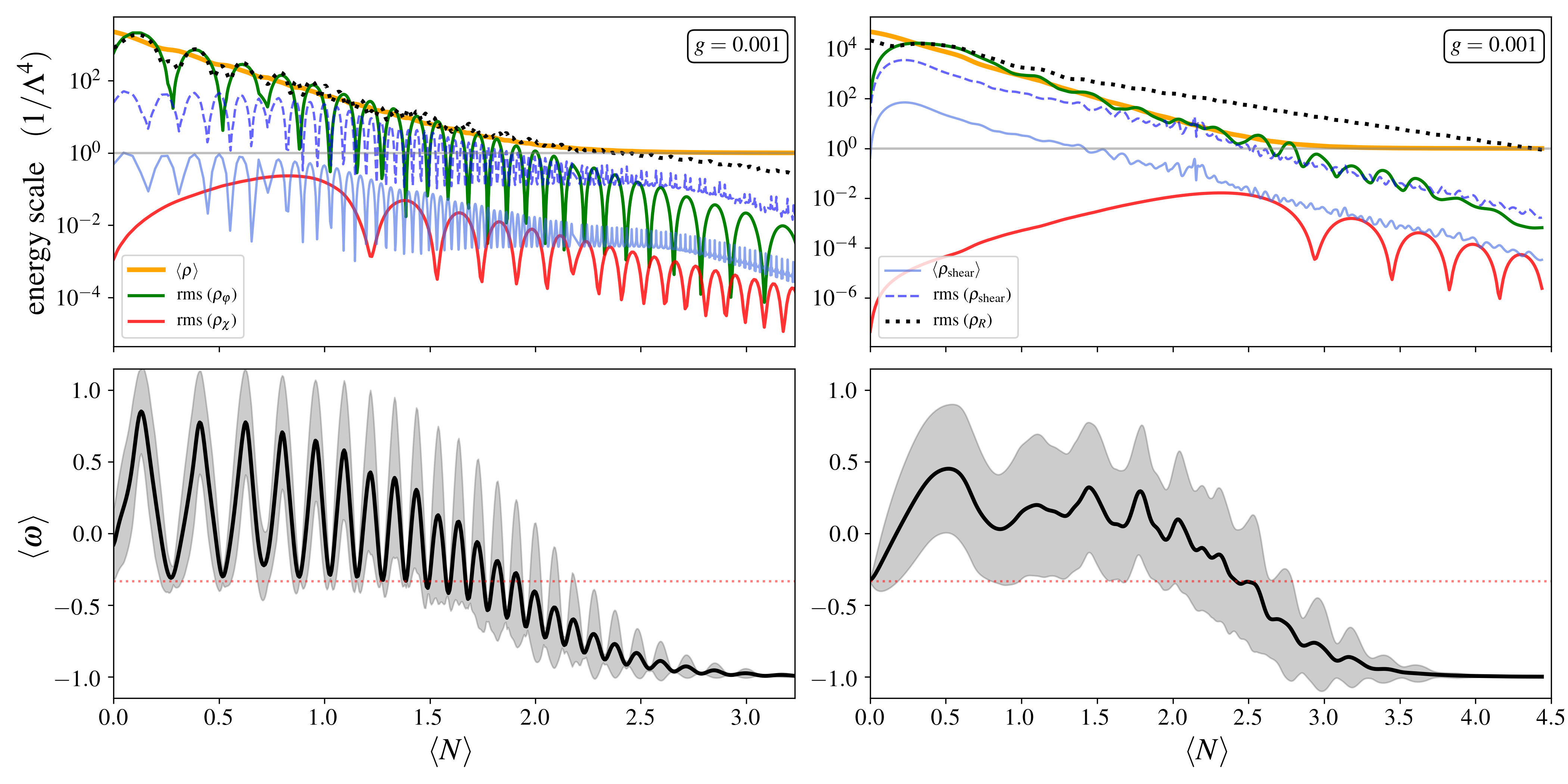}
\end{center}
\caption{ 
Same as in top and bottom panels of Fig.~\ref{p2_fig:preinflation}. It shows the dynamical evolution of sub-Hubble (left) and super-Hubble (right) energetically dominated initial conditions corresponding to the pre-inflationary era until the onset of inflation. The initial box size of the simulations corresponds to $L\approx 2 H^{-1}$ for the sub-Hubble case and $L\approx 10 H^{-1}$ for the super-Hubble one.
\label{p2_fig:aux_preinf}
}
\vspace{20mm}
\end{figure*}

\begin{figure*}[!ht]
\vspace{30mm}
\begin{center}
\includegraphics[width=0.99\textwidth]{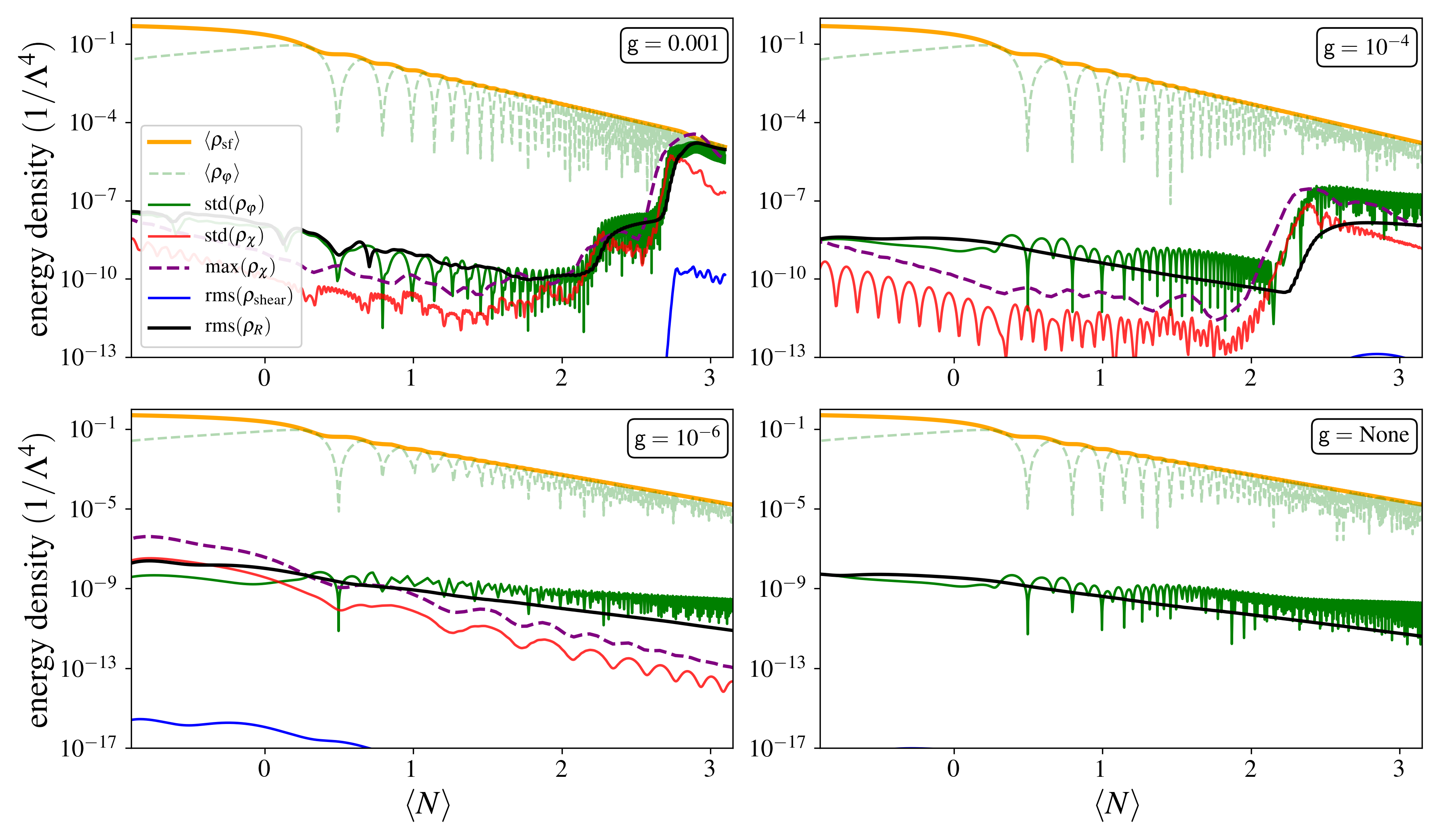}
\end{center}
\caption{evolution of the energy density with respect to the expansion history. These simulations correspond to scenarios of preheating with $\mathsf{g} \leq 10^{-3}$ from Fig.~\ref{p2_fig:reheating2}, but include perturbations in the initial state of the Higgs field. The bottom-right panel corresponds to the single-field case. The box size of the simulations at the end of inflation corresponds to $L\approx 5 H^{-1}$.
\label{p2_fig:aux_reheating}
}
\vspace{20mm}
\end{figure*}

 In Fig.~\ref{p2_fig:aux_preinf} for simulations of preinflation, and Fig.~\ref{p2_fig:aux_reheating} for the preheating, we show that there is no significant changes when these changes are included, and therefore these simulations lead to the same conclusions explained in chapter~\ref{p2_chap:prepaper2}.

 \clearpage
 
 \clearpage
 
%
\end{appendix}

\clearpage
\addcontentsline{toc}{chapter}{Bibliography}
\bibliographystyle{ieeepes}
\bibliography{bibliography}

\end{document}